\documentclass[preprint,12pt]{elsarticle}
\usepackage{amsfonts}
\usepackage{amsmath}
\usepackage{amsthm}
\usepackage{graphicx}
\usepackage{hyperref}
\usepackage{color}
\usepackage{amssymb}
\usepackage{dsfont}
\usepackage{verbatim}
\usepackage{mathrsfs}
\usepackage{sidecap}
\usepackage[T1]{fontenc}
\usepackage[utf8]{inputenc}
\usepackage[a4paper, total={6in, 9in}]{geometry}
\usepackage[nottoc,notlot,notlof]{tocbibind}
\usepackage{float}
\usepackage{ulem}
\usepackage{overpic}
\usepackage{url}
\usepackage{extarrows}

\hypersetup{
	colorlinks=true,
	linkcolor=black
}

\def\Z2{$\mathbb{Z}_{2}$}
\def\psb{\bar\psi}
\def\alb{\bar\alpha}

\def\mt#1{\mathcal{#1}}
\def\gt#1{\bar{\mathcal{#1}}}
\def\mmod#1{\textrm{mod}\left(#1\right)}
\def\dci#1{\hat{c}^{\dagger}_{#1}}
\def\ci#1{\hat{c}_{#1}}
\def\vac{|0\rangle}

\def\vacc{\big|0\big\rangle}
\def\bvacc{\big\langle 0\big|}
\def\dcip#1#2{\left(\dci{#1}\right)^{#2}}
\def\cip#1#2{\left(\ci{#1}\right)^{#2}}
\def\opn#1{\hat{n}_{#1}}
\def\ket#1{|#1\rangle}
\def\bra#1{\langle#1|}
\def\xib{\bar{\xi}}
\def\Gintbx#1{\int_{\xib_{#1}\xi_{#1}}\xi_{#1}^{p(j_{#1})}\xib_{#1}^{p(j_{#1})}}

\def\phif#1{\phi^{(#1)}}
\def\psif#1{\psi^{(#1)}}
\def\nn#1{\langle{#1}\rangle}
\def\lr#1#2{\langle{#1}|{#2}\rangle}
\def\lpr#1#2#3{\langle{#1}|{#2}|{#3}\rangle}
\def\psidu#1#2{\psi_{#1}^{(#2)}}
\def\psbdu#1#2{\psb_{#1}^{(#2)}}

\def\psbf#1{\bar{\psi}^{(#1)}}
\def\Tr{\textrm{Tr}~}
\def\mub{\bar{\mu}}
\def\nub{\bar{\nu}}
\def\thb{\bar{\theta}}
\def\psixp#1#2{\psi_{#1}^{p(#2)}}
\def\psbxp#1#2{\bar{\psi}_{#1}^{p(#2)}}

\begin{document}

\begin{frontmatter}

\title{\textbf{Grassmann tensor networks}}

\author[a]{Jian-Gang Kong}
\author[c,d]{Jia-Ji Zhu}
\author[a,b]{Z. Y. Xie}
\ead{qingtaoxie@ruc.edu.cn}
\address[a]{School of Physics, Renmin University of China, Beijing 100872, China}
\address[b]{Key Laboratory of Quantum State Construction and Manipulation (Ministry of Education), Renmin University of China, Beijing, 100872, China}
\address[c]{School of Electronic Science and Engineering, Chongqing University of Posts and Telecommunications, Chongqing 400065, China}
\address[d]{Chongqing Key Laboratory of Dedicated Quantum Computing and Quantum Artificial Inteligence, Chongging, 400065, China}

\begin{abstract}
Developing non-perturbative methods to reveal exotic properties of strongly correlated fermionic systems remains one of the most essential tasks of theoretical physics. Tensor network methods with Grassmann algebra offer powerful numerical tools for fermionic many-body systems in the coherent-state path-integral representation. Despite their vast potential for both condensed-matter and particle-physics communities, Grassmann tensor network methods are somewhat underexploited in practical simulations. In this work, we provide a detailed, self-contained introduction to Grassmann tensor network methods, from the basics of the Grassmann tensor operations to the Grassmannization of typical tensor network algorithms. Furthermore, the resulting Grassmann tensor network methods are validated in several interesting models in both particle physics and condensed matter physics.
\end{abstract}


\end{frontmatter}

%\newpage

\tableofcontents

\section{Introduction}

Fermions are exotic quantum particles that obey anticommutation relations. In condensed matter physics, interacting fermion models serve as key theoretical frameworks for unraveling a broad spectrum of emergent phenomena beyond few-body physics. For instance, the Hubbard model \cite{Hubbard1963Electron} and the $t$-$J$ model \cite{Zhang1988Effective} are foundational for understanding high-temperature superconductivity in cuprates; the Anderson impurity model \cite{Anderson1961Localized} describes magnetic impurities embedded in metals and underpins the physics of heavy fermion compounds; the Kondo model \cite{Kondo1964} captures the screening of local magnetic moments by conduction electrons, accounting for the anomalous low-temperature transport properties of dilute magnetic alloys; and the Heisenberg model, which can be derived from the Hubbard model in the Mott insulating limit \cite{AssaBook1994}, elucidates magnetic ordering and spin dynamics in strongly correlated insulators. Meanwhile, high-energy physics relies heavily on strongly interacting fermions coupled to non-Abelian gauge fields—most notably in quantum chromodynamics, the fundamental theory describing the strong nuclear force. These fermion-gauge field couplings govern the internal structure and dynamics of nucleons, as well as the non-perturbative phenomena critical to strong interaction physics, such as quark confinement \cite{Wilson1974} and chiral symmetry breaking \cite{Nambu1961}. 

Unfortunately, interacting fermion models remain some of the most complex systems to simulate in higher dimensions, mainly because of the so-called sign problem in quantum Monte Carlo \cite{NightBook1999}. Therefore, to fully appreciate the rich phenomena exhibited by fermions, it is crucial to develop other efficient methods. Among the various numerical methods, tensor network methods combine the tensor-network ansatz of the partition function \cite{SRGreview2010} and quantum wave function \cite{PEPS2004}, contraction strategy originates from the idea of renormalization group \cite{RanBook2020, XiangBook2023}, and optimization techniques developed in artificial intelligence more recently, and are drawing increasing attention in the past two decades. Since they are free of sign problem, and can directly deal with two-dimensional quantum lattice models in the thermodynamic limit, tensor network methods have already been successfully applied to study quantum magnetization \cite{Kagome2017, LWY2024, LY2025}, topological order \cite{Nobert2013, WR2024}, quantum field theory \cite{CMPS2010, CPEPS2019}, statistical models \cite{Levin2007, Xie2012HOTRG}, and so on. Especially, tensor networks emerge as a promising tool for directly addressing these challenges in fermionic problems, such as superconductivity \cite{YS2023, CorbozHubbard2016, CorbozTJ2014}. In this work, we focus only on the fermionic tensor network methods, which are briefly reviewed in the following.

A natural way to deal with fermionic systems is through the Jordan-Wigner transformation \cite{Jordan1928JW}, which maps a fermionic model to a bosonic system and works well in one dimension. However, in two or more dimensions, the Jordan-Wigner transformation yields non-local string operators that are not straightforward to handle. Nevertheless, there are some successful applications in this direction in some specific tensor network structures, such as the multiscale entanglement renormalization ansatz and tree tensor network \cite{Corboz2010Simulation}, where the non-local strings can be diminished efficiently through a dynamical reordering strategy \cite{Pineda2010Unitary}. 

To study fermionic systems on a general two-dimensional lattice, Kraus \textit{et al.} proposed the fermionic projected entangled pair states (PEPS) \cite{Kraus2010Fermionic}, a straightforward generalization of its bosonic version \cite{PEPS2004} that prevails in the study of quantum spin systems. In this proposal, auxiliary fermions are introduced at each lattice site, form maximally entangled states with other auxiliary particles residing at nearest-neighbor locations, and a projector mapping the direct-product space spanned by auxiliary fermions to the desired Hilbert space spanned by physical fermions is defined at each lattice site. By introducing additional indices for the local tensors, this wave function can be represented as a bosonic PEPS, enabling practical numerical simulations using established tensor network methods for spin systems.

Apart from the proposal by Kraus \textit{et al.}, several other groups \cite{Barthel2009Contraction,Pivzorn2010Fermionic,Corboz2009Fermionic,Corboz2010SimulationPEPS} have independently proposed fermionization methods applicable to any tensor network ansatz, without significantly increasing computational cost compared with their bosonic counterparts. In the context of fermionic operator circuits \cite{Barthel2009Contraction}, Barthel \textit{et al.} introduced the basic rules for contracting such structures, which can be extended straightforwardly to general fermionic tensor networks. In particular, Corboz \textit{et al.} introduced similar rules in fermionic tensor networks \cite{Corboz2009Fermionic, Corboz2010SimulationPEPS}: (i) All local tensors are parity invariant or $\mathbb{Z}_{2}$-symmetric, which is natural for a fermionic Hamiltonian that creates and annihilates particles in pairs, and this further requires that all the tensor operations should preserve $\mathbb{Z}_{2}$ symmetry. (ii) Replace all line crossings in the two-dimensional graphical projection of the tensor network states by the so-called fermionic swap gates, which give a minus sign when both the two lines carry an odd number of fermions, and play the central role for fermion exchange statistics. This swap gate can be conveniently absorbed into local tensors, and the resulting tensor network can be contracted as usual symmetric tensor networks \cite{Singh2011Tensor,Singh2012Tensor,Marek2025YASTN,Lukas2025TensorKit}. It can be shown \cite{RomanEPJB2014} that the wave function obtained by these rules is equivalent to the original fermionic PEPS proposal \cite{Kraus2010Fermionic}. Therefore, in practice, these rules constitute the central part of the fermionic PEPS algorithm in practical simulations \cite{CorbozTJ2014,Corboz2011Stripes,Corboz2012Comment,Wang2014Fermionic,Zheng2017Stripe,Ponsioen2019Period,Benedikt2021Beginner,Li2021Study,Ponsioen2023Superconducting,Chung2019SU3,Chen2024Orbital,Zhang2025Frustration}.

The Grassmann tensor network is a parallel but equivalent route to the fermionic tensor network state. It is well known that Grassmann algebra provides a natural mathematical framework for describing fermions in the coherent-state path-integral formulation of quantum field theory, thereby allowing a unified treatment of bosons and fermions in perturbative calculations and in supersymmetry theory. The first attempt to combine tensor networks with Grassmann algebra was proposed by Gu \textit{et al.} in Ref.~\cite{Gu2010Grassmann}. In that work, Grassmann numbers are introduced in the construction of the tensor network state ansatz and associated with the local tensors. Each physical index is assigned a definite parity, and it imposes constraints on the Grassmann numbers associated with the surrounding virtual indices via parity conservation. By enforcing the Grassmann numbers to reside on the same links with total even parity, and integrating out all the Grassmann variables, one obtains a wave function satisfying the fermionic anticommutation law. This construction is shown to be helpful in the study of fermionic string-net models and strongly correlated projective states \cite{BeriPRL2011}, e.g., fractional quantum Hall states. 
In the same work, the tensor entanglement-filtering renormalization group algorithm~\cite{Gu2008Tensor} is adapted for contracting Grassmann tensor networks to compute physical observables.
Later, the construction is simplified further to a more compact form via local decomposition and some redefinitions, and the corresponding imaginary-time evolution algorithm for determining the ground-state Grassmann representation of a fermionic system is also established \cite{Gu2013Efficient}. 
Since the Grassmann formulation does not require two-dimensional graphical projection of tensor network states, it manipulates sign factors in a local way which should be advantageous for complex lattice structures and higher-dimensional lattices. 

As a matter of fact, Grassmann tensor network states now serve as an alternative tool for studying superconducting properties in strongly correlated fermionic models~\cite{Gu2013Time,Gu2014Symmetry,Jie2015Combining,Gu2020Emergence,Miao2025Spin,Yue2024Pseudogap,Liu2025Accurate}. 
Shortly after the development of the refined Grassmann tensor product state approach~\cite{Gu2013Efficient}, it was applied to study the doped $t$-$J$ model on the honeycomb lattice~\cite{Gu2013Time}. At low doping, a novel $d+id$ superconducting (SC) phase that breaks time-reversal symmetry was discovered at $t/J = 3$, and found to coexist with antiferromagnetic (AFM) order. 
Later, the Grassmann approach was applied to the vanishing-$J$ limit of the same model, which corresponds to the infinite-$U$ Hubbard model, and revealed a $p+ip$ SC state coexisting with ferromagnetic order~\cite{Gu2020Emergence}. In contrast to the $p+ip$ superconductivity induced by the coexistence of AFM and $d+id$ SC order in Ref.~\cite{Gu2013Time}, the mechanism for triplet SC order here is argued to result from spin–charge separation~\cite{Anderson1987Resonating}.
Subsequently, Ref.~\cite{Miao2025Spin} mapped out a phase diagram of the honeycomb lattice $t$-$J$ model as a function of $t/J$ ratio and doping, revealing a $d+id$ SC phase, a coexistence region of SC and AFM order, and a non‑Fermi liquid phase. A slave‑fermion mean‑field theory was constructed and reproduces the phase diagram qualitatively. The attractive interaction among holons and spin–charge separation play vital roles in the emergence of $d+id$ SC.
Beyond the novel SC orders in Dirac systems, Ref.~\cite{Yue2024Pseudogap} employed Grassmann tensor network methods to study SC orders in the square‑lattice $t$-$J$ model, and discovered a pair‑density wave (PDW) superconducting state with propagation vector $Q = (\pi,\pi)$ at low doping. This PDW state is characterized by the vanishing of singlet pairing as the bond dimension $D$ increases and a weak breaking of $C_{4}$ rotational symmetry, and may hold significant relevance to the pseudogap phase in high‑temperature superconductors.

%Though the representation seems elegant, contracting a Grassmann tensor network to obtain the expectation value or the partition function requires extra care with the Grassmann algebra. In the original work of Grassmann tensor network \cite{Gu2010Grassmann,Gu2013Efficient}, the tensor entanglement-filtering renormalization group algorithm \cite{Gu2008Tensor} is adapted to Grassmann networks to perform the ground state calculation of free and interacting fermionic models on the honeycomb lattice. 
In the high-energy physics community, Kuramashi \textit{et al.} proposed the Grassmann version of a similar coarse-graining renormalization group method \cite{Levin2007,ShimizuGrassmann2014} to numerically investigate the lattice regularized Schwinger model with one flavor of the Wilson fermion, where the partition function of lattice field theory in path integral form is represented as a Grassmann tensor network. In view of the free of sign feature, to simulate the quantum chromodynamics on the lattice, the (3+1)-dimensional relativistic field theory describing quarks coupled with non-Abelian gauge interactions, another coarse-graining renormalziation group method based on higher-order singular value decomposition \cite{Xie2012HOTRG} was Grassmannized in (2+1)-dimensional \cite{Sakai2017Higher} and (3+1)-dimensional \cite{Akiyama2021Restoration} models involving relativistic fermions. More recently, Akiyama \textit{et al.} developed a general framework for constructing the Grassmann tensor network representation given the lattice action of fermionic field theory \cite{Akiyama2021More}, based on which a series of applications have been conducted. For a more complete introduction to the contraction of a Grassmann tensor network and its applications in the lattice field theory community, please refer to Ref.~\cite{Akiyama2024Tensor} and references therein. Notably, following a similar procedure, the path integral representation of the Hubbard model \cite{Akiyama2022Metal,Akiyama2021Tensor} can also be represented as Grassmann tensor networks, and the ground state properties can be obtained effectively \cite{Akiyama2021Restoration,Adachi2020Anisotropic}.

At last, providing an equivalent but more concise representation of the Grassmann tensor networks, tensor networks with $\mathbb{Z}_{2}$-graded vector space are drawing increasing attention. In a \Z2-graded tensor network state, each local tensor are associated with base kets defined in a \Z2-graded vector space (also referred to as super vector space), the direct sum of two ordinary vector space each of which is associated with different definite parity, and the direct product is anti-commutative for any two base kets carrying odd parity. Tensor network contractions defined under such constraints can easily capture fermion-exchange statistics. In the tensor network community, it is first adapted as a convenient mathematical language in the context of fermionic tensor networks \cite{Gu2014,Bultinck2017fMPS,Bultinck2017fPEPS} to classify symmetry-protected phases of interacting fermions, while large-scale numerical applications appear much later \cite{YS2023,Zheng2025Competing,Xu2024Global,Zheng2025Revealing}. More recently, a detailed description of the $\mathbb{Z}_{2}$-graded tensor network formulation, along with its numerical implementations, is reviewed in Ref.~\cite{Quinten2025Fermionic}. In particular, it is shown in that work that the fermionic $\mathbb{Z}_{2}$-graded structure is compatible with both Abelian and non-Abelian symmetry considerations \cite{Quinten2025Fermionic}.

In this work, we provide a pedagogical introduction to Grassmann tensor networks. We highlight the distinctive features of our presentation compared with existing reviews on fermionic tensor networks~\cite{Akiyama2024Tensor,Quinten2025Fermionic}: (i) Algorithmic development. Starting from basic Grassmann algebra, 
we illustrate in detail how tensor network algorithms can be Grassmannized and validate them in concrete examples. (ii) Unification of formalisms. We establish explicit connections between the Grassmann tensor networks and other fermionic tensor network formulations, including the $\mathbb{Z}_{2}$-graded space and swap‑gate. (iii) Unified perspective. Built on a simplified Grassmann notation, we present a unified view of Grassmann tensor networks for representing both fermionic states and partition functions. This perspective is relevant to both the lattice gauge theory and the condensed matter communities (including ground-state and thermodynamic studies).

The rest of the paper is organized as follows. We introduce the basic Grassmann tensor operations serving as the essential building blocks of Grassmann tensor network algorithms in Sec.~\ref {grassmann_op}, and the Grassmann tensor network representation of fermionic states, operators, as well as partition functions in Sec.~\ref {grassmann_tn}. The equivalence to other fermionic tensor network methods is also discussed in detail by examples. After these preparations, we Grassmannize several typical tensor network algorithms, such as the variational version of density matrix renormalization group (DMRG) \cite{WhiteDMRG1992, SchRev2011}, one-dimensional (1D) time-evolving block decimation (TEBD) \cite{VidalTEBD2003, VidalTEBD2007, RomanTEBD2008}, corner transfer-matrix renormalization group (CTMRG) \cite{CorbozTJ2014,Nishino1996Corner, Orus2009Simulation}, and adapted imaginary-time evolution of a two-dimensional quantum state in Grassmann representation \cite{Jiang2008Accurate, Xie2014Tensor}. The validity of the methods is demonstrated in several typical models spanning condensed matter to particle physics, including the 1D Hubbard model, thermodynamics in the 1D free fermion, a two-dimensional spinless fermion model, and the (1+1)-dimensional Gross-Neveu-Wilson model. Finally, in Sec.~\ref {conclusion}, we summarize and provide a prospect on Grassmann tensor network methods.

\section{Grassmann tensor operations} 
\label{grassmann_op}

In this section, we introduce the basic building blocks of any Grassmann tensor network algorithm. The most important object of the story is the Grassmann tensor. Operations on the Grassmann tensor will be described later, including contraction, fusion, and decomposition.

\subsection{Grassmann algebra}

The Grassmann algebra is generated by a set of Grassmann variables $\{\psi_{i}\}$ \cite{AltlandQFT2010}, which are anticommuting mathematical objects satisfying:
\begin{eqnarray} 
& & \label{grassmannalgebra1}
\psi_{i}\psi_{j} = - \psi_{j}\psi_{i}, \qquad \psi_i^2 = 0, \qquad \textrm{for any} ~i, j.
\end{eqnarray}
In many cases, by defining the zeroth power as one, i.e., $\psi^0=1$, one obtains the frequently used identity
\begin{eqnarray}
\psi_i^m\psi_j^n = (-1)^{mn}\psi_j^n\psi_i^m, \qquad \textrm{for}~ m, n = \{0, 1\}.
\label{GAdef}
\end{eqnarray}

Due to the nilpotency of Grassmann variables as described in Eq.~(\ref{grassmannalgebra1}), any functions of a single Grassmann variable can be truncated to (bi)linear terms after Taylor expansions. For example, the exponential function of Grassmann variables can be expanded as:
\begin{eqnarray} \label{grassmannalgebra7}
& &
{\rm e}^{\psi} = \sum_{n=0}^{\infty} \frac{\psi^{n}}{n!} = 1 + \psi.
\end{eqnarray}

The integrals over Grassmann variables, known as Berezin or Grassmann integrals \cite{BerezinInt1966}, are defined as (modified with a slightly different convention):
\begin{eqnarray} \label{grassmannalgebra3}
\int d\psi = 0, \qquad \int d\psi \psi = 1,
\qquad
\int d\psi_id\psi_j \left(\psi_j\psi_i\right) = - \int d\psi_id\psi_j \left(\psi_i\psi_j\right) = 1-\delta_{ij},
\end{eqnarray}
where, for consistency, the following relation holds for any $i$ and $j$: 
\begin{eqnarray}
d\psi_id\psi_j = - d\psi_jd\psi_i, \qquad \left(d\psi_i\right)\psi_j = - \psi_j\left(d\psi_i\right).
\label{GArule}
\end{eqnarray}

In practical applications, one usually introduces the dual Grassmann variable $\psb$, which should be treated as strictly independent of $\psi$. For simplicity, we introduce the following shorthand for any two different Grassmann variables $\psi_i$ and $\psi_j$ $(i\neq j)$, as also used in previous literature \cite{Liu2025Accurate,Akiyama2024Tensor,Atis2023GrassmannTN}:
\begin{eqnarray} 
\label{grassmanncontract0a}
\int_{\psi_i\psi_j}... \equiv \int d\psi_i d\psi_j ~{\rm e}^{-\psi_i\psi_j}...
\end{eqnarray}
Then using Eqs.~(\ref{grassmannalgebra1}-\ref{GArule}), one can easily verify the following identities:
\begin{eqnarray}
\int_{\psi_i\psi_j}
\psi_j^m
\psi_i^n 
= 
\delta_{m,n}, 
\qquad\textrm{for}~ m, n = \{0, 1\}, i\neq j.
\label{FundGA}
\end{eqnarray}
In particular, for $\psi$ and $\psb$, it reduces to 
\begin{eqnarray}
\int_{\psb\psi}\psi^{m}\psb^n = \int_{\psi\psb}\psb^{m}\psi^n = \delta_{m,n}, \qquad\textrm{for}~ m, n = \{0, 1\}. 
\label{grassmanncontract1a}
\end{eqnarray}
In this context, the Grassmann variables defined in a shared link are usually denoted by a dual pair, thus Eq.~(\ref{grassmanncontract1a}) can be regarded as a counterpart to a complete set but constructed by just two discrete variables, and provides the basics in Grassmann tensor contraction and decomposition introduced later. 

Another useful Grassmann integral identity is:
\begin{eqnarray}
{\rm e}^{\bar{\psi}\psi} = \int_{\alb\alpha} {\rm e}^{\psb\alb}{\rm e}^{\psi\alpha} = \int_{\alb\alpha}
{\rm e}^{\psb\alpha}
{\rm e}^{\alb\psi},
\label{grassmannalgebra6}
\end{eqnarray}
which is a reflection of the fact that only terms with exactly one $\alpha$ and $\alb$ in the integrant can survive due to Eqs.~(\ref{grassmannalgebra1}-\ref{grassmannalgebra7}). In fact, Eq.~(\ref{grassmannalgebra6}) can be regarded as a counterpart of Hubbard-Stratonovich transformation \cite{HubbardPRL1959}, in the sense that one can decouple Grassmann variables by performing Grassmann integrals of a pair of auxilliary Grassmann varibles $\alpha$ and $\alb$, which can be understood as bond/link variables connecting Grassmann tensors defined on neighboring lattice sites in the context of tensor networks. As we will see in Sec.~\ref{grassmann_tn}, it would be helpful when we derive the Grassmann tensor network representation of the fermionic partition function within the path integral formalism.

\subsection{Grassmann tensor}

Generally speaking, a rank-$m$ Grassmann tensor $\mt{T}$, denoted in calligraphic fonts in this work, has elements hosting the following structure
\begin{eqnarray}
\mt{T}_{i_1i_2...i_{m-1}i_m} = T_{i_1i_2...i_{m-1}i_m}~\psi_{1}^{p(i_1)}\psi_{2}^{p(i_2)}...\psi_{m-1}^{p(i_{m-1})}\psi_{m}^{p(i_m)},
\label{grassmanntensor1}
\end{eqnarray}
where each index $i_n$ of an ordinary tensor $T$ is associated with a Grassmann variable $\psi_n$ with a power $p(i_n)$, the parity of $i_n$, and is either 0 or 1. Here, the subscript $n$ denotes both the index and the species of Grassmann variables. The elements of $T$ are commuting numbers, and thus $T$ is just an $m$-dimensional array used in a bosonic tensor network. We stress that, in general, the ordering of Grassmann variables is not required to match the index ordering of $T$ or $\mt{T}$. In this work, unless explicitly defined, we assume the two orderings are identical. Note for simplification, a Grassmann tensor in Eq.~(\ref{grassmanntensor1}) is defined in a slightly different way from the original form in Refs.~\cite{Gu2010Grassmann, Gu2013Efficient}.

For each index of $T$, the dimension is divided into two parts, i.e., Grassmann-even ($p=0$) and Grassmann-odd ($p=1$). A Grassmann tensor satisfying the following constraint
\begin{eqnarray}
\mt{T}_{i_1i_2...i_m} = 0, \quad \textrm{as long as} \mod(p(i_1)+p(i_2)+...+p(i_m),2) = 1,
\label{grassmanntensor3}
\end{eqnarray}
is usually called Grassmann-even. The dual Grassmann tensor $\gt{T}$ of Eq.~(\ref{grassmanntensor1}) is defined as
\begin{eqnarray}
\gt{T}_{i_1i_2...i_{m-1}i_m} = T^*_{i_1i_2...i_{m-1}i_m}~\psb_{m}^{p(i_m)}\psb_{m-1}^{p(i_{m-1})}...\psb_{2}^{p(i_2)}\psb_{1}^{p(i_1)},
\label{DualDef}
\end{eqnarray}
where $(*)$ means complex conjugation. Note that the order of the dual Grassmann variables is reversed with respect to the indices of $T$ and $\mt{T}$. To make them consistent, employing Eq.~(\ref{GAdef}), one obtain the following identity
\begin{eqnarray}
\gt{T}_{i_1i_2...i_{m-1}i_m} = (-1)^{s}\times T^*_{i_1i_2..i_{m-1}i_m}~\psb_{1}^{p(i_1)}\psb_{2}^{p(i_2)}...\psb_{m-1}^{p(i_{m-1})}\psb_{m}^{p(i_m)},
\label{DualDefEq}
\end{eqnarray}
where $(-1)^s = \pm 1$ is responsible for the reordering of Grassmann variables
\begin{eqnarray}
s = \sum_{a=1}^{m}p(i_a)\sum_{b>a}p(i_b).
\label{SignDef}
\end{eqnarray}

Sometimes, in tensor networks, it is helpful to represent a Grassmann tensor graphically using arrows, where each arrow denotes a pair of Grassmann variables residing on the link. To distinguish a Grassmann tensor and its dual tensor, especially when the tensor has definite parity, in this work, we stick to this convention: for each Grassmann pair defined on a given link, the Grassmann variable is affiliated with the tensor where the arrow points to, and the corresponding dual variable belongs to the tensor where the arrow comes from. For example, a rank-4 tensor represented as 
\begin{eqnarray}  
\label{grassmanntensor4}
\mt{T}_{i_1i_2i_3i_4}  = T_{i_1i_2i_3i_4}~
\psi_{1}^{p(i_{1})}
\psi_{2}^{p(i_{2})}
\bar{\psi}_{3}^{p(i_{3})}
\psi_{4}^{p(i_{4})}
\equiv
\raisebox{-0.48\height}{\includegraphics[width=0.22\textwidth, page=1]{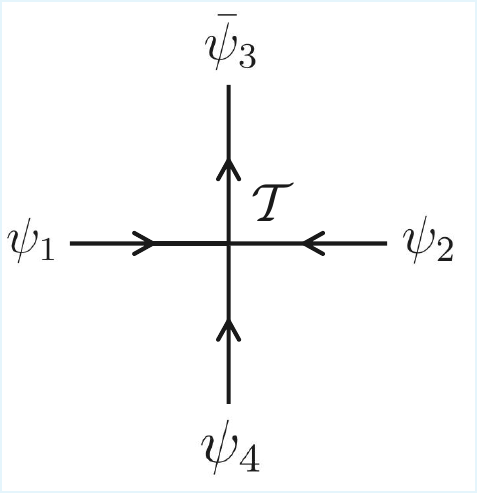}},
\end{eqnarray}
where the index $i_3$ is associated with a dual variable $\psb_3$, and other indices carry original variables, i.e., $\psi_1$, $\psi_2$, $\psi_4$. Essentially, there is no need to specify the category of the Grassmann numbers, since Grassmann numbers with different subscripts are completely different. The only necessity of defining the dual of a Grassmann number is to remind the fact that $\psi_i$ and $\psb_i$ are defined on the same link indexed by $i$. Then, following this, the dual tensor $\gt{T}$ can be represented as
\begin{eqnarray}  
\label{grassmanntensor5}
\gt{T}_{j_1j_2j_3j_4} = T^*_{j_1j_2j_3j_4}~
\psb_{4}^{p(j_{4})}
\psi_{3}^{p(j_{3})}
\psb_{2}^{p(j_{2})}
\psb_{1}^{p(j_{1})}
\equiv
\raisebox{-0.48\height}{\includegraphics[width=0.22\textwidth, page=1]{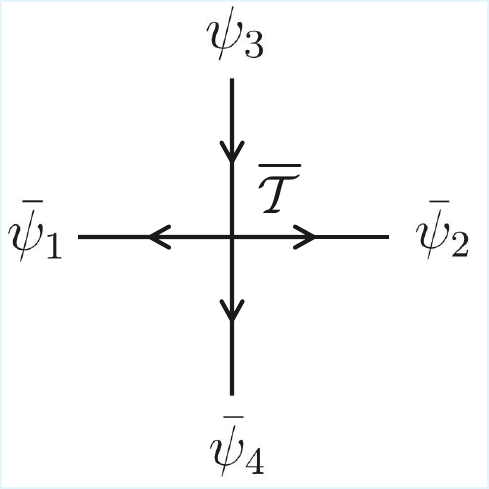}},
\end{eqnarray}
where all the arrows are reversed, compared to Eq.~(\ref{grassmanntensor4}). Following this convention, the Grassmann-even condition in Eq.~(\ref{grassmanntensor3}) can be understood as 
\begin{eqnarray}
\mmod{\sum_{n\in \textrm{in arrows}} p(i_n), 2} = \mmod{\sum_{m\in \textrm{out arrows}} p(i_m), 2},
\end{eqnarray}
since we can always divide the anti-commuting variables into two categories: the Grassmann variables and their duals. In this work, we always follow such conventions illustrated in Eq.~(\ref{grassmanntensor4}) and Eq.~(\ref{grassmanntensor5}).

We emphasize that we do not impose any specific order of Grassmann variables in reading the pictorial plot of the Grassmann tensor. Whatever order is given, the fermionic order is already encoded in the definition of Grassmann tensors $\mt{T}$ and $\gt{T}$. Different orders correspond to coefficient tensors that differ only by some signs, which provide constraints on the coefficient tensor and naturally reflect Fermi statistics. 

\subsection{Grassmann tensor contraction}

In this work, by contraction of two Grassmann tensors, we mean the summation over shared indices of the coefficient tensors, and more importantly, the integral over the corresponding Grassmann variables simultaneously. Any two Grassmann tensors hosting common indices can be contracted. 

Therefore, the contraction of two Grassmann tensors can be performed following a three-step procedure: (i) move the Grassmann variables specified by the Grassmann integral measure to adjacent positions, calculate the corresponding fermionic sign factor according to Eq.~(\ref{GAdef}); (ii) perform the Grassmann integral according to Eqs.~(\ref{FundGA}-\ref{grassmanncontract1a}), the resulting Kcronecker-$\delta$ function leads to a constrain on the corresponding parities; (iii) perform the contraction of the coefficient tensors, multiplied with the signs from (i), under the parity constrain obtained in (ii). 

To make it concrete, as an example, we consider the contraction of two rank-3 Grassmann tensors, $\mt{A}$ and $\mt{B}$, resulting in a rank-4 Grassmann tensor $\mt{T}$:
\begin{eqnarray}
\mt{T}_{i_2i_3i_4i_5} &=& \sum_{i_1}\mt{A}_{i_1i_2i_3}\mt{B}_{i_1i_4i_5}=\raisebox{-0.44\height}{\includegraphics[width=0.26\textwidth, page=1]{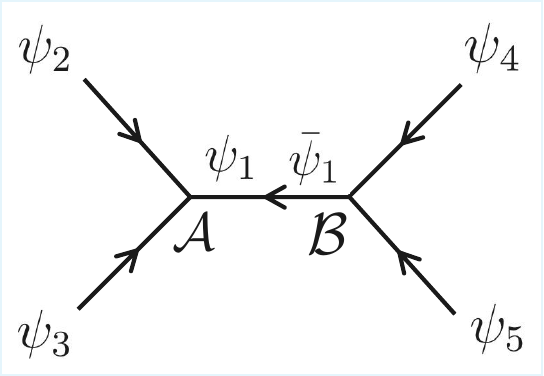}}, \nonumber \\
&\equiv&\sum_{i_1}A_{i_1i_2i_3}B_{i_1i_4i_5}\int_{\psb_1\psi_1}\left[\psi_{1}^{p(i_1)}\psi_{2}^{p(i_2)}\psi_{3}^{p(i_3)}\right]\left[\psb_{1}^{p(i_1)}\psi_{4}^{p(i_4)}\psi_{5}^{p(i_5)}\right],
\label{GContract}
\end{eqnarray}
where the Grassmann variable associated with $i_1$ in $\mt{B}$ is denoted by a dual variable $\psb_{1}$ to distinguish from the $\psi_{1}$ in $\mt{A}$. As a reminder, we emphasize that the Grassmann integral in Eq.~(\ref{GContract}) is defined in such a way that the relative order of the relevant Grassmann variables in the integral measure is reversed to that in the integrand, to carry out the integration according to Eq.~(\ref{FundGA}). By comparing the second line of Eq.~\eqref{GContract} with the first, we see that the connecting line, with an arrow pointing from $\mt{B}$ to $\mt{A}$, represents
\begin{eqnarray}
\raisebox{-0.28\height}{\includegraphics[width=0.16\textwidth, page=1]{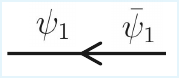}}
\equiv
\sum_{i}
\int_{\bar{\psi}_{1}\psi_{1}}.
\end{eqnarray}

To evaluate Eq.~(\ref{GContract}), one can follow the steps below:

(i) Make $\psb_{1}$ and $\psi_1$ neighbors. Here, we move $\psi_{1}$ to the exact front of $\psb_{1}$, which generates a sign factor $(-1)^s$ with
\begin{eqnarray}
s = p(i_1)\left[p(i_2) + p(i_3)\right]. \nonumber 
\end{eqnarray}

(ii) Perform the integral over $\psb_{1}$ and $\psi_{1}$, and obtain the following identity
\begin{eqnarray}
\int_{\psb_{1}\psi_{1}}\psi_1^{p(i_1)}\psb_{1}^{p'(i_1)} = \delta_{p(i_1),p'(i_1)}, 
\label{GrassID}
\end{eqnarray}
where the parity defined for $\mt{B}$ is denoted as $p'$ for clarity. This means the parity of the same index $i_1$ defined for $\mt{A}$ and $\mt{B}$ should equal, i.e., any two neighboring Grassmann tensors have the identical parity function $p$, and that is why we do not distinguish the parity functions defined for different Grassmann tensors in the whole work. 

In performing the Grassmann integral in Eq.~(\ref{GrassID}), we do not need to worry about the position of the integral measure $\int_{\psb\psi}$, since the measure is a bilinear form containing an even number of Grassmann variables, thus it commutes with any Grassmann numbers. 

(iii) Perform the ordinary numerical contraction of the coefficient tensor $A$ and $B$ accompanied by the sign factor obtained in step (i). That is, finally, we have
\begin{eqnarray}
\mt{T}_{i_2i_3i_4i_5} 
\equiv
T_{i_2i_3i_4i_5}
 ~\psi_{2}^{p(i_2)}\psi_{3}^{p(i_3)}\psi_{4}^{p(i_4)}\psi_{5}^{p(i_5)}, 
\quad
T_{i_2i_3i_4i_5}
=
\sum_{i_1}
(-1)^{s}A_{i_1i_2i_3}B_{i_1i_4i_5}.
\label{SGContract}
\end{eqnarray}
Comparing Eqs.~(\ref{GContract}) and (\ref{SGContract}), one find that in practical calculations, the contraction of $\mt{A}$ and $\mt{B}$ can be realized by numerical contraction of $A$ and $B$ but with a sign structure $(-1)^{s}$, and the corresponding Grassmann variables are obtained just by discarding the relevant Grassmann pairs and leaving only the remaining Grassmann variables with their relative order unchanged. This feature enables efficient contraction of two Grassmann tensors with definite parity, e.g., when $\mt{A}$ and $\mt{B}$ are Grassmann-even. In this case, Eq.~(\ref{SGContract}) can be evaluated by block matrix multiplication by combining blocks with the same parity together so that $(-1)^s$ is a common sign for each block, as done in ordinary symmetric tensor networks~\cite{Singh2011Tensor, Marek2025YASTN,Andreas2024QSpace,Singh2012Tensor,SCHMOLL2020168232,Lukas2025TensorKit}.

A specific example of Grassmann tensor contraction is the norm, whose square is defined in the following for a rank-$m$ Grassmann tensor $\mt{T}$,
\begin{eqnarray}
|\mt{T}|^2 = \sum_{i_1i_2...i_m}\gt{T}_{i_1i_2...i_m}\mt{T}_{i_1i_2...i_m}.
\label{NormDef}
\end{eqnarray}
Following the rule discussed above, we can show that
\begin{eqnarray}
|\mt{T}|^2 = \sum_{i_1i_2...i_m}T^*_{i_1i_2...i_m}T_{i_1i_2...i_m}\times g,
\label{NormDefG}
\end{eqnarray}
where $g$ is the Grassmann integral over all the Grassmann variables
\begin{eqnarray}
g=\int_{\psi_1\psb_1}\int_{\psi_2\psb_2}...\int_{\psi_m\psb_m}\psb_m^{p(i_m)}...\psb_{2}^{p(i_2)}\psb_1^{p(i_1)}\psi_1^{p(i_1)}\psi_2^{p(i_2)}...\psi_m^{p(i_m)}.
\end{eqnarray}
Notice the specific integral measure we have selected, such that we can employ the Grassmann integral relation, Eq.~(\ref{grassmanncontract1a}). One immediately finds $g = 1$ and this reduces Eqs.~(\ref{NormDef}-\ref{NormDefG}) to
\begin{eqnarray}
|\mt{T}| = |T| = \sqrt{\sum_{i_1i_2...i_{n-1}i_n}T^*_{i_1i_2...i_{n-1}i_n}T_{i_1i_2...i_{n-1}i_n}},
\end{eqnarray}
which is used for practical calculation.

\subsection{Grassmann tensor fusion} \label{Sec:fusion}
To make contraction more efficient, frequently one needs to fuse some indices into a single one and reshape a tensor to a matrix or vector so that high-level algebraic operations can be sped up by well-developed packages, such as the famous Basic Linear Algebra Subprograms (BLAS) \cite{BLAS} and Intel Math Kernel Library \cite{IntelMKL}.

In order to simplify the description, here we take the contraction of a rank-4 tensor $\mt{A}$ and a rank-5 tensor $\mt{B}$ expressed in the following as an example:
\begin{eqnarray}
\mt{T}_{i_1i_3i_5i_6i_7}=\sum_{i_2i_4}\mt{A}_{i_1i_2i_3i_4}\mt{B}_{i_5i_4i_6i_2i_7} = \sum_{i_2i_4}A_{i_1i_2i_3i_4}B_{i_5i_4i_6i_2i_7}\times g,
\label{ContractEq}
\end{eqnarray}
where
\begin{eqnarray}
g=\int_{\psb_2\psi_2}\int_{\psb_4\psi_4}
\left(
\psi_1^{p(i_1)}\psi_2^{p(i_2)}\psi_3^{p(i_3)}\psi_4^{p(i_4)}
\right)
\left(
\psi_5^{p(i_5)}\psb_4^{p(i_4)}\psi_6^{p(i_6)}\psb_2^{p(i_2)}\psi_7^{p(i_7)}  
\right).
\label{FusionEq}
\end{eqnarray}
Here, the Grassmann variables associated with $i_2$ and $i_4$ for $\mt{B}$ are denoted as dual variables.
To employ the level-3 optimization technique in BLAS, one needs to reshape the coefficient tensors $A$ and $B$ as matrices, which requires fusing $\{i_2,i_4\}$ as a single index $c$, $\{i_1,i_3\}$ as $a$, and $\{i_5,i_6,i_7\}$ as $b$. What matters in this context is that one must also fuse the Grassmann variables. A straightforward deduction of Eq.~(\ref{FusionEq}) gives
\begin{eqnarray}
g = \int_{\psb_2\psi_2}\int_{\psb_4\psi_4}(-1)^{s}\psi_1^{p(i_1)}\psi_3^{p(i_3)}\psi_2^{p(i_2)}\psi_4^{p(i_4)}\cdot(-1)^t\psb_4^{p(i_4)}\psb_2^{p(i_2)}\psi_5^{p(i_5)}\psi_6^{p(i_6)}\psi_7^{p(i_7)},
\label{FusionPoint}
\end{eqnarray}
where
\begin{eqnarray}
s = p(i_2)p(i_3), \qquad t = p(i_2)\left[p(i_5)+p(i_6)\right] + p(i_4)p(i_5).
\end{eqnarray}
If one introduces
\begin{eqnarray}
A'_{ac}&\equiv& A_{i_1i_2i_3i_4}, \quad B'_{cb}\equiv B_{i_5i_4i_6i_2i_7}, \nonumber \\
\psi_{a}^{p(i_a)}&\equiv& \psi_1^{p(i_1)}\psi_3^{p(i_3)}, \quad \psi_{c}^{p(i_c)}\equiv \psi_2^{p(i_2)}\psi_4^{p(i_4)}, 
\quad \psi_{b}^{p(i_b)}\equiv \psi_5^{p(i_5)}\psi_6^{p(i_6)}\psi_7^{p(i_7)},
\label{FusionDef2}
\end{eqnarray}
then the original problem in Eqs.~(\ref{ContractEq}-\ref{FusionEq}) can be formally reduced to
\begin{eqnarray}
\mt{T}_{ab} &=& \sum_{c}\mt{A}_{ac}\mt{B}_{cb}, \nonumber \\
\mt{A}_{ac} &=& (-1)^sA'_{ac}\psi_{a}^{p(i_a)}\psi_{c}^{p(i_c)}, \quad \mt{B}_{cb} = (-1)^tB'_{cb}\psb_{c}^{p(i_c)}\psi_{b}^{p(i_b)}.
\label{MatrixEq}
\end{eqnarray}
Note that the order of $\psb_4$ and $\psb_2$ in Eq.~(\ref{FusionPoint}) is arranged so that it is consistent with the definition of $\psi_c$ expressed in Eq.~(\ref{FusionDef2}) and the following integral identity:
\begin{eqnarray}
\int_{\psb_2\psi_2}
\int_{\psb_4\psi_4}
\psi_2^{p(i_2)}\psi_4^{p(i_4)}\psb_4^{p(i_4)}\psb_2^{p(i_2)} = 1 = \int_{\psb_c\psi_c}\psi_c^{p(i_c)}\psb_c^{p(i_c)},
\end{eqnarray}
where the first equal holds as a result of Eq.~(\ref{grassmanncontract1a}) and the second equal appears just as an introduction of the newly defined Grassmann variable $\psi_{c}$.

Furthermore, corresponding to Eq.~(\ref{MatrixEq}), the parity of the new Grassmann variables is defined formally in the following:
\begin{eqnarray}
p(i_a) &\equiv& \mmod{p(i_1)+p(i_3),2}, \quad p(i_c) \equiv \mmod{p(i_2)+p(i_4),2}, \nonumber \\
p(i_b) &\equiv& \mmod{p(i_5)+p(i_6)+p(i_7),2}.
\label{ParityDef}
\end{eqnarray}
Using Eqs.~(\ref{FusionDef2}-\ref{ParityDef}), the tensor contraction expressed in Eqs.~(\ref{ContractEq}-\ref{FusionEq}) can be formally reduced to matrix multiplications and performed efficiently by separating the Grassmann-even and Grassmann-odd parts as two individual blocks, after which well-established packages, such as BLAS, can speed up the ordinary block matrix multiplications. Once the contraction is done, one can safely recover the indices of $\mt{T}$ to original forms, i.e., finally
\begin{eqnarray}
\mt{T}_{i_1i_3i_5i_6i_7} = T_{i_1i_3i_5i_6i_7}\psi_1^{p(i_1)}\psi_3^{p(i_3)} \psi_5^{p(i_5)}\psi_6^{p(i_6)}\psi_7^{p(i_7)},
\end{eqnarray}
where the coefficient tensor is obtained previously as
\begin{eqnarray}\label{TAB_contract}
T_{i_1i_3i_5i_6i_7} = \sum_{c}(-1)^{s+t}A'_{a,c}B'_{c,b}.
\end{eqnarray}

In practical implementations, there is some freedom in defining the Grassmann fusion. However, any alternative definition must be accompanied by appropriate sign factors to ensure consistency. For instance, Eq.~(\ref{FusionPoint}) admits the following equivalent form:
\begin{eqnarray}
g = \int_{\psb_2\psi_2}\int_{\psb_4\psi_4}(-1)^{s}\psi_1^{p(i_1)}\psi_3^{p(i_3)}\psi_2^{p(i_2)}\psi_4^{p(i_4)}\cdot(-1)^{t+t_c}\psb_2^{p(i_2)}\psb_4^{p(i_4)}\psi_5^{p(i_5)}\psi_6^{p(i_6)}\psi_7^{p(i_7)},
\label{FusionPoint2}
\end{eqnarray}
where $t_c = p(i_2)p(i_4)$ is responsible to the extra exchange of $\psb_2$ and $\psb_4$. The composite Grassmann variables $\psi_c$ and $\varphi_c$ introduced via
\begin{eqnarray}
\psi_{c}^{p(i_c)}
\equiv
\psi_2^{p(i_2)}\psi_4^{p(i_4)},
\quad
\varphi_{c}^{p(i_c)} \equiv \bar{\psi}_2^{p(i_2)}\bar{\psi}_4^{p(i_4)},
\end{eqnarray}
are now strictly independent and are no longer related by the dual of the Grassmann product. With this choice, there is a sign during the contraction of $i_c$, i.e., Eq.~(\ref{MatrixEq}) must therefore be modified to 
\begin{eqnarray}
\mt{T}_{ab} 
&=& \label{MatrixEq2}
\sum_{c}\mt{A}_{ac}\mt{B}_{cb}
\cdot (-1)^{t_c}, 
\\
\mt{A}_{ac} &=& (-1)^sA'_{ac}\psi_{a}^{p(i_a)}\psi_{c}^{p(i_c)}, \quad \mt{B}_{cb} = (-1)^{t+t_{c}}B'_{cb}\psb_{c}^{p(i_c)}\psi_{b}^{p(i_b)}.
\end{eqnarray}
% The extra sign factor in Eq.~(\ref{MatrixEq2}) originates from enforcing the resolution of identity to hold for the composite Grassmann variables
The extra sign factor $(-1)^{t_c}$ arises when one enforces the resolution of identity to hold for the composite Grassmann variables
\begin{eqnarray}
1 \equiv \int_{\varphi_c\psi_c}\psi_c^{p(i_c)}\varphi_c^{p(i_c)}.
\end{eqnarray}
But note, the sign $(-1)^{t_c}$ in Eq.~(\ref{MatrixEq2}) shows up only when the contraction is performed, and cannot be absorbed into the Grassmann tensor. In other words, it is an auxiliary sign factor associated with the fused index itself and does not belong to any Grassmann tensor containing that index.

\subsection{Grassmann tensor decomposition}
\label{Sec:GTD}

\begin{figure}[htbp]
	\centering
	\includegraphics[height=3.4cm,width=9.0cm]{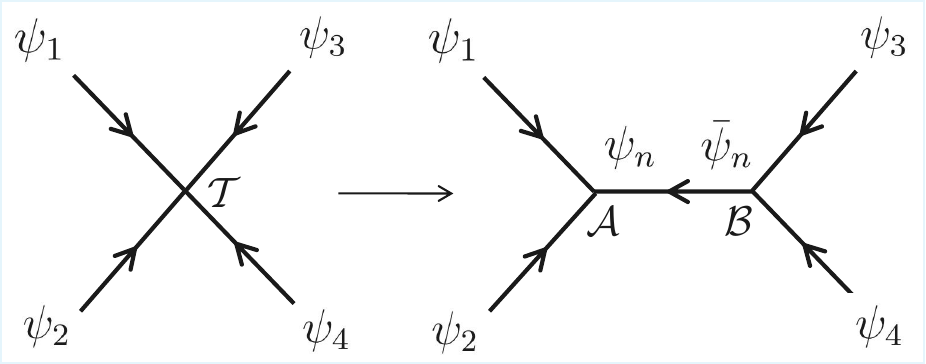}
	\caption{Decompose a rank-4 Grassmann tensor $\mt{T}$ to two rank-3 Grassmann tensors $\mt{A}$ and $\mt{B}$ by singular value decomposition, as expressed in Eqs.~(\ref{BipartT}-\ref{ABdef}).}
	\label{decomp}
\end{figure}

In evaluating tensor networks, one frequently encounters tensor decompositions, such as singular value decomposition (SVD). Suppose we want to decompose a Grassmann tensor $\mt{T}$ to two parts, $\mt{A}$ and $\mt{B}$, as illustrated in Fig.~\ref{decomp}.
By performing its SVD, we actually mean
\begin{eqnarray}
\mt{T}_{i_1i_2i_3i_4} = \sum_{n}\mt{U}_{i_1i_2n}\Lambda_{n}\mt{V}_{ni_3i_4},
\label{GSVD}
\end{eqnarray}
where the coefficient tensors in their matrix forms satisfy the ordinary matrix SVD 
\begin{eqnarray}
\tilde{T} = \tilde{U}\Lambda\tilde{V}, \quad\textrm{with}\quad \tilde{T}_{i_1i_2,i_3i_4}\equiv T_{i_1i_2i_3i_4}, \quad\tilde{U}_{i_1i_2,n}\equiv U_{i_1i_2n}, \quad\tilde{V}_{n,i_3i_4}\equiv V_{ni_3i_4}.
\label{SVD}
\end{eqnarray}
Note, in Eqs.~(\ref{GSVD}-\ref{SVD}), the $\Lambda$ is a positive definite diagonal matrix with no Grassmann counterpart defined, $\tilde{U}$ is a column-wise orthogonal matrix, and $\tilde{V}$ is row-wise orthogonal. Once Eq.~(\ref{GSVD}) is obtained, then we can absorb the ordinary matrix $\Lambda$ into $\mt{U}$ and $\mt{V}$ to obtain the desired bipartition as demonstrated in Fig.~\ref{decomp}
\begin{eqnarray}
\mt{T}_{i_1i_2i_3i_4} = \sum_{n}\mt{A}_{i_1i_2n}\mt{B}_{ni_3i_4},
\label{BipartT}
\end{eqnarray}
where $\mt{A}$ and $\mt{B}$ have identical Grassmann structures to $\mt{U}$ and $\mt{V}$, respectively, and their coefficient tensors are defined as
\begin{eqnarray}
A_{i_1i_2n} = U_{i_1i_2n}\sqrt{\Lambda_n}, \qquad B_{ni_3i_4} = \sqrt{\Lambda_n}V_{ni_3i_4},
\label{ABdef}
\end{eqnarray}

With the procedure discussed in Sec.~\ref{Sec:fusion}, we can introduce
\begin{eqnarray}
\psi_{r}^{p(i_r)}&\equiv& \psi_1^{p(i_1)}\psi_2^{p(i_2)}, \quad p(i_r) \equiv \mmod{p(i_1)+p(i_2),2}, \nonumber \\
\psi_{c}^{p(i_c)}&\equiv& \psi_3^{p(i_3)}\psi_4^{p(i_4)}, \quad p(i_c)\equiv \mmod{p(i_3)+p(i_4),2},  
\end{eqnarray}
and fuse the original Grassmann tensor $\mt{T}$ as
\begin{eqnarray}
\mt{T}_{rc} = \tilde{T}_{rc}\psi_{r}^{p(i_r)}\psi_{c}^{p(i_c)} = \sum_{n}\tilde{U}_{i_1i_2,n}\Lambda_n\tilde{V}_{n,i_3i_4}\psi_{r}^{p(i_r)}\psi_{c}^{p(i_c)},
\label{GSVD2}
\end{eqnarray}
where we used Eq.~(\ref{SVD}) for the second equal sign. For a Grassmann-even tensor $\mt{T}$, the coefficient matrix $\tilde{T}$ can be written as a block matrix by separating the Grassmann-even and Grassmann-odd parts as two individual blocks, and then the SVD in Eq.~(\ref{SVD}) can be sped up efficiently as discussed in Sec.~\ref{Sec:fusion}.

Comparing Eq.~(\ref{GSVD2}) and Eq.~(\ref{GSVD}), we can introduce an identity constructed by the Grassmann integral of $\psb_{n}\psi_{n}$ in Eq.~(\ref{GSVD}), or equivalently in Eq.~(\ref{GSVD2}), i.e.,
\begin{eqnarray}
\mt{T}_{rc} = \sum_{n}\tilde{U}_{i_1i_2,n}\Lambda_n\tilde{V}_{n,i_3i_4}\psi_{r}^{p(i_r)}\left[\int_{\psb_{n}\psi_{n}}\psi_{n}^{p(i_n)}\psb_{n}^{p(i_n)}\right]\psi_{c}^{p(i_c)}.
\label{Tinsert}
\end{eqnarray}
Recover in tensor form, we obtain
\begin{eqnarray}
\mt{T}_{i_1i_2i_3i_4} = \sum_{n}U_{i_1i_2n}\Lambda_nV_{ni_3i_4}\int_{\psb_{n}\psi_{n}}\left[\psi_1^{p(i_1)}\psi_2^{p(i_2)}\psi_{n}^{p(i_n)}\right]\left[\psb_{n}^{p(i_n)}\psi_3^{p(i_3)}\psi_4^{p(i_4)}\right],
\end{eqnarray}
from which we identify the following results
\begin{eqnarray}
\mt{U}_{i_1i_2i_n} = U_{i_1i_2n}\psi_1^{p(i_1)}\psi_2^{p(i_2)}\psi_{n}^{p(i_n)}, \quad \mt{V}_{ni_3i_4} = V_{ni_3i_4}\psb_{n}^{p(i_n)}\psi_3^{p(i_3)}\psi_4^{p(i_4)}.
\end{eqnarray}
Here the Grassmann variables $\psi_{n}$ and $\psb_{n}$ are associated with the new index $n$, therefore, its total dimension is determined by $\Lambda$ in the matrix SVD in Eq.~(\ref{SVD}), and its parity $p(i_n)$ is determined by the parity of $\mt{U}$ and $\mt{V}$, which are eventually governed by the parity of $\mt{T}$. For example, if $\mt{T}$ is Grassmann-even, then the following relation must hold
\begin{eqnarray}
p(i_n) = \mmod{p(i_1)+p(i_2),2} = \mmod{p(i_3)+p(i_4),2}.
\end{eqnarray}

It is important to note that for this specific example, we have employed the resolution of identity $\int_{\psb\psi}\psi^{p(i)}\psb^{p(i)} = 1$ in Eq.~(\ref{Tinsert}). This convention assigns an ordinary Grassmann variable to the $\mt{A}$ tensor and its dual variable to the $\mt{B}$ tensor, which corresponds graphically to an arrow pointing from $\mt{B}$ to $\mt{A}$, see also Fig.~\ref{decomp}. 
The arrow direction can, in principle, be reversed. To do so while leaving the integral measure formally unchanged, one introduces an additional sign factor via the relation:
\begin{eqnarray}
\int_{\psb_{n}\psi_{n}}
\psi_{n}^{p(i_n)}
\psb_{n}^{p(i_n)}
=
\int_{\psb_{n}\psi_{n}}
\psb_{n}^{p(i_n)}
\psi_{n}^{p(i_n)}
\times
(-1)^{p(i_{n})}.
\end{eqnarray}
For consistency, this sign factor must then be absorbed into the coefficient tensor of either $\mt{A}$ or $\mt{B}$. 

\section{Grassmann tensor network representations} \label{grassmann_tn}

In this section, we demonstrate how to represent fermionic states, operators, and partition functions as Grassmann tensor networks in the fermionic coherent-state representation. The Grassmann matrix product operator (MPO) form of fermionic Hamiltonians is also presented. In the last subsection, we discuss the relationships between the Grassmann formulation and other fermionic tensor network proposals, such as the $\mathbb{Z}_{2}$-graded space \cite{Bultinck2017fMPS, Bultinck2017fPEPS,Quinten2025Fermionic} and the fermionic swap-gate approaches \cite{Corboz2009Fermionic, Corboz2010SimulationPEPS}.

\subsection{Fermionic coherent states}
\label{Sec:FermiCoh}
To study fermionic states with Grassmann tensor networks, firstly, we need to introduce the many-body fermionic coherent states \cite{ColemanBook2015}, which are defined as 
\begin{eqnarray}
|\phi\rangle \equiv \exp\left[-\sum_{n}\psi_n\dci{n}\right]\vacc = \prod_{n}\left(1-\psi_n\dci{n}\right)\vac,
\label{FcoDef}
\end{eqnarray}
where $\psi_n$ is the Grassman variable associated with fermionic creation operator $\dci{n}$ for the $n$-th single-particle eigenstate. $\vac$ is the vacuum state, and by convention, we need the following relations for any integers $i$ and $j$
\begin{eqnarray}
\psi_i\dci{j} + \dci{j}\psi_i = \psi_i\ci{j} + \ci{j}\psi_i = 0, \quad \left(\psi_{i}\ci{j}\right)^{\dagger} \equiv \dci{j}\psb_{i}, \quad \left(\psi_{i}\dci{j}\right)^{\dagger} \equiv \ci{j}\psb_{i}, 
\label{AntiComDef}
\end{eqnarray}
which guarantees that all the terms in the exponent in Eq.~(\ref{FcoDef}) commute and thus $=$ holds. Moreover, Eq.~(\ref{AntiComDef}) means the order of the operators in $\prod$ in Eq.~(\ref{FcoDef}) does not matter and leads to the following identity:
\begin{eqnarray}
\ci{i}|\phi\rangle = \ci{i}\prod_{n}\left(1-\psi_n\dci{n}\right)\vac = \left[\prod_{n\neq i}\left(1-\psi_n\dci{n}\right)\right]\ci{i}\left(1-\psi_i\dci{i}\right)\vacc 
\label{EigenDef0}.
\end{eqnarray}
Using the following properties of the vacuum state
\begin{eqnarray}
\ci{i}\vac = 0, \quad \ci{i}\dci{i}\vac = \left(1-\dci{i}\ci{i}\right)\vac = \vac,
\end{eqnarray}
we identify Eq.~(\ref{EigenDef0}) as an eigenvalue equation
\begin{eqnarray}
\ci{i}|\phi\rangle = \left[\prod_{n\neq i}\left(1-\psi_n\dci{n}\right)\right]\psi_i\vacc = \left[\prod_{n\neq i}\left(1-\psi_n\dci{n}\right)\right]\psi_i\left(1-\psi_i\dci{i}\right)\vacc = \psi_i|\phi\rangle.
\label{EigenDef}
\end{eqnarray}
This means the Grassmann number $\psi_i$ is actually the eigenvalue of the fermionic annihilation operator $\ci{i}$. Generally using Eqs.~(\ref{grassmannalgebra1}), (\ref{AntiComDef}) and (\ref{EigenDef}), we can obtain
\begin{eqnarray}
\ci{1}\ci{2}...\ci{n}|\phi\rangle = \psi_{1}\psi_{2}...\psi_{n}|\phi\rangle,
\label{cprod}
\end{eqnarray}
To make it explicit, following the definition of a fermionic Fock state and its dual correspondence
\begin{eqnarray}
|\alpha_1\alpha_2\alpha_3...\alpha_n\rangle &\equiv& \dcip{1}{\alpha_1}\dcip{2}{\alpha_2}...\dcip{n}{\alpha_n}\vacc,  \nonumber\\
\langle \alpha_1\alpha_2\alpha_3...\alpha_n| &\equiv& \langle 0|\cip{n}{\alpha_n}...\cip{2}{\alpha_2}\cip{1}{\alpha_1}.
\label{FockDef}
\end{eqnarray}
where all the integers $\{\alpha_n\}$ can be only 0 or 1, sometimes one can introduce the following notation by writing the eigenvalues explicitly
\begin{eqnarray}
|\psi_1\psi_2...\psi_n\rangle \equiv |\phi\rangle,
\end{eqnarray}
which reduces Eq.~(\ref{cprod}) and its dual correspondence to
\begin{eqnarray}
\left(\ci{1}\ci{2}...\ci{n}\right)|\psi_1\psi_2...\psi_n\rangle 
&=& \left(\psi_{1}\psi_{2}...\psi_{n}\right)|\psi_1\psi_2...\psi_n\rangle, 
\nonumber \\
\langle\psb_1\psb_2...\psb_n|\left(\dci{n}...\dci{2}\dci{1}\right) &=& \langle\psb_1\psb_2...\psb_n|\left(\psb_n...\psb_2\psb_1\right).
\label{EigenGrass}
\end{eqnarray}

Furthermore, employing the dual correspondence 
\begin{eqnarray}
\langle \phi| = \bvacc\prod_{n}\left(1+\psb_{n}\ci{n}\right),
\end{eqnarray}
one immediately obtain $\langle \phi|0\rangle = 1 = \langle 0|\phi\rangle$, thus for any Fock states we have
\begin{eqnarray}
\langle \alpha_1\alpha_2...\alpha_n|\phi\rangle &=& \langle 0|\cip{n}{\alpha_n}...\cip{2}{\alpha_2}\cip{1}{\alpha_1}|\phi\rangle = \psi_{n}^{\alpha_n}...\psi_2^{\alpha_2}\psi_1^{\alpha_1}, 
\nonumber \\
\langle\phi |\alpha_1\alpha_2...\alpha_n\rangle &=& \langle\phi |\dcip{1}{\alpha_1}\dcip{2}{\alpha_2}...\dcip{n}{\alpha_n}|0\rangle = \psb_1^{\alpha_1}\psb_2^{\alpha_2}...\psb_n^{\alpha_n},
\label{OverLap}
\end{eqnarray}
from which we can justify the following overcompleteness as well as the inner product of the coherent states, i.e.,
\begin{eqnarray}
\int_{\psb\psi}|\psi_1\psi_2...\psi_n\rangle\langle\psb_1\psb_2...\psb_n| &=& 1, \qquad\textrm{with}\quad \int_{\psb\psi} \equiv \int_{\psb_1\psi_1}\int_{\psb_2\psi_2}...\int_{\psb_n\psi_n}, 
\nonumber \\
\langle \psb_1\psb_2...\psb_n|\psi'_1\psi'_2...\psi'_n\rangle &=& \exp\left[\sum_{i}\psb_i\psi'_i\right],
\label{OverCompEq}
\end{eqnarray}
where $\psi'_i$ is a Grassmann eigenvalue of $\ci{i}$ not necessarily equals $\psi_i$.

At last, an essential thing we would like to mention is that, using the fermionic coherent state to calculate the partition function or the statistical average of observables, the trace should be performed under special boundary conditions. This stems from the following identity: for $m, n = \{0, 1\}$ and $i\neq j$
\begin{eqnarray}
\delta_{mn} = \int_{\psi_i\psi_j}\psi_j^m\psi_i^n = \int_{\psi_i\psi_j}\left(-\psi_i\right)^n\psi_j^m = \int_{\psi_i\psi_j} \lr{-\psi_i}{n}\lr{m}{\psi_j}, 
\end{eqnarray}
where $\ket{m}$ and $\ket{n}$ are Fock basis. The second equal holds since $(-1)^{mn}\delta_{mn} = (-1)^n$. This gives the representation-independent trace for any operator $\hat{O}$, 
\begin{eqnarray}
\Tr\hat{O} = \sum_{mn}\lpr{m}{\hat{O}}{n}\delta_{mn} 
= \int_{\psi_i\psi_j}\sum_{mn}\lr{-\psi_i}{m}\lpr{m}{\hat{O}}{n}\lr{n}{\psi_j}.
\end{eqnarray}
Using the complete relation of the Fock basis, one arrives at the boundary conditions, sometimes referred to as anti-periodic, of trace in fermionic coherent state representations
\begin{eqnarray}
\Tr\hat{O} = \int_{\psi_i\psi_j}\lpr{-\psi_i}{\hat{O}}{\psi_j}\equiv -\int d\psi_id\psi_j
{\rm e}^{\psi_i\psi_j}\lpr{\psi_i}{\hat{O}}{\psi_j},
\label{ABC}
\end{eqnarray}
where the equivalent sign ($\equiv$) comes from a redefinition: $-\psi_i\rightarrow\psi_i$. This means that when performing trace in coherent state representations, the integral measure should use ${\rm e}^{\psi_i\psi_j}$ instead of the usual ${\rm e}^{-\psi_i\psi_j}$, and the matrix elements should be multiplied by a sign (-1).

%and the matrix elements should be multiplied by a sign (-1) while tracing over the odd-parity sector of the index.

\subsection{Fermionic states} \label{Sec:GMPS}

With the above preparation, we can relate a fermionic wave function to a Grassmann tensor network state via the coherent-state representation. Suppose the base kets are constructed from $m$ single-particle eigenstates, then a general fermionic state can be expanded in the Fock basis
\begin{eqnarray}
|\Psi\rangle = 
\sum_{\alpha_1,\alpha_2,...\alpha_m}
C_{\alpha_1\alpha_2...\alpha_m}|\alpha_{1}\alpha_{2}...\alpha_{m}\rangle
=
\sum_{\alpha_1,\alpha_2,...\alpha_m}
C_{\alpha_1\alpha_2...\alpha_m}
(c_{1}^{\dagger})^{\alpha_{1}}
(c_{2}^{\dagger})^{\alpha_{2}}
\cdots
(c_{m}^{\dagger})^{\alpha_{m}}
|0\rangle,
\label{FstateFock}
\end{eqnarray}
where $\alpha_{i} = 0, 1$, denoting the occupation number on the $i$-th (spin-resolved) single-particle eigenstate. In Eq.~(\ref{FstateFock}), the second equal denotes that each basis is related to $\vac$ through Eq.~(\ref{FockDef}) and essentially a Slater determinant, thus any two subscripts can not be equal. In such a situation, using the following anticommuting relations for fermionic operators
\begin{eqnarray}
\cip{i}{\alpha_i}\cip{j}{\alpha_j} = \left(-1\right)^{\alpha_i\alpha_j}\cip{j}{\alpha_j}\cip{i}{\alpha_i}, \qquad 
\dcip{i}{\alpha_i}\dcip{j}{\alpha_j} = \left(-1\right)^{\alpha_i\alpha_j}\dcip{j}{\alpha_j}\dcip{i}{\alpha_i}.
\end{eqnarray}
%we can recover the fully-antisymmetric property of the coefficient $C$ with respect to any two  subscripts $\alpha_i$ and $\alpha_j$:
we can recover the fully-antisymmetric property of the Fock basis $|\alpha_{1}\alpha_{2}...\alpha_{m}\rangle$ with respect to any two single-particle states $\alpha_i$ and $\alpha_j$:
\begin{eqnarray}
\nonumber
(\hat{c}_{1}^{\dagger})^{\alpha_{1}}
\cdots
(\hat{c}_{i}^{\dagger})^{\alpha_{i}}
\cdots
(\hat{c}_{j}^{\dagger})^{\alpha_{j}}
\cdots
(\hat{c}_{m}^{\dagger})^{\alpha_{m}}
|0\rangle
&=& 
(-1)^{s 
}
(\hat{c}_{1}^{\dagger})^{\alpha_{1}}
\cdots
(\hat{c}_{j}^{\dagger})^{\alpha_{j}}
\cdots
(\hat{c}_{i}^{\dagger})^{\alpha_{i}}
\cdots
(\hat{c}_{m}^{\dagger})^{\alpha_{m}}
|0\rangle,
\\ 
s 
&=&  \nonumber
\alpha_i\alpha_j +  
\left(
\alpha_i + \alpha_j
\right)
\times
\sum_{n}
\alpha_{n}, 
\quad
\textrm{for} ~ 
i < n < j.
\end{eqnarray}
For $\alpha_{i} = \alpha_{j} = 1$, we can easily see $\mmod{s, 2} = 1$. That is, the switch of any two occupied single-particle states will lead to a minus sign in the wave function, i.e., the superposition coefficient $C$.

Since the coherent states form an overcomplete set, we can expand the fermionic state $\ket{\Psi}$ in this set by inserting the identity in Eq.~(\ref{OverCompEq}) into Eq.~(\ref{FstateFock}), i.e.,
\begin{eqnarray}
\ket{\Psi} = \sum_{\alpha_1,\alpha_2,...\alpha_m}C_{\alpha_1\alpha_2...\alpha_m}\left[\int_{\psb\psi}\ket{\psi_1\psi_2...\psi_m}\bra{\psb_1\psb_2...\psb_m}\right]|\alpha_{1}\alpha_{2}...\alpha_{m}\rangle.
\end{eqnarray}
Using Eq.~(\ref{OverLap}), we obtain
\begin{eqnarray}
\ket{\Psi} 
&=& 
\int_{\psb\psi}
\left(
\sum_{\alpha_1,\alpha_2,...\alpha_m}
C_{\alpha_1\alpha_2...\alpha_m}
\psb_1^{\alpha_1}\psb_2^{\alpha_2}...\psb_m^{\alpha_m}
\right)
\ket{\psi_1\psi_2...\psi_m},
\label{GWF0}
\\
&=& \int_{\psb\psi}
\left(
\sum_{\alpha_1,\alpha_2,...,\alpha_m}
\mt{C}_{\alpha_1\alpha_2...\alpha_m}
\right)
\ket{\psi_1\psi_2...\psi_m},
\label{GWF}
\end{eqnarray}
where $\mt{C}$ is nothing but a rank-$m$ Grassmann tensor
\begin{eqnarray}
\mt{C}_{\alpha_1,\alpha_2,...\alpha_m} = C_{\alpha_1,\alpha_2,...\alpha_m}\psb_1^{p(\alpha_1)}\psb_2^{p(\alpha_2)}...\psb_m^{p(\alpha_m)}.
\label{CGrass}
\end{eqnarray}
In Eq.~\eqref{CGrass}, we introduce the parity function $p$ merely for notational consistency, even though in this case $p(\alpha) = \alpha$ holds trivially. The meaning of Eq.~(\ref{GWF}) is clear: the fermionic state can be obtained by an integral of the coherent-state basis, and for a given basis, the superposition coefficient is obtained by the self-trace (i.e., the summation of all indices) of a Grassmann tensor $\mt{C}$ constructed from the corresponding dual variables. In contrast to Eq.~\eqref{FstateFock}, where the Fermi statistics are naturally encoded in the Fock basis, or equivalently in the anticommutation relation of field operators, in Eq.~\eqref{GWF0} they are totally characterized by the Grassmann variables satisfying Eq.~(\ref{GAdef}), thereby enabling further numerical simulations that manipulate these objects locally.

Then, following the procedure discussed in Sec.~\ref{Sec:GTD}, we can represent the Grassmann tensor $\mt{C}$ appearing in Eqs.~(\ref{GWF}-\ref{CGrass}) as a matrix product state (MPS). To achieve this, similarly, we can firstly factorize the coefficient tensor $C$ into a usual MPS, e.g., with open boundary conditions
\begin{eqnarray} 
C_{i_1i_2...i_m} 
= \sum_{j_1j_2...j_{m-1}}T^{(1)}_{j_1}[i_1]~T^{(2)}_{j_1j_2}[i_2]~T^{(3)}_{j_2j_3}[i_3]~...~T^{(m-1)}_{j_{m-2}j_{m-1}}[i_{m-1}]~T^{(m)}_{j_{m-1}}[i_m],
\label{MPS}
\end{eqnarray}
whereas usually done for bosonic MPS, the physical indices are placed in the square brackets, and the introduced auxiliary indices $\{j\}$ are denoted as ordinary subscripts. Here, the subscript $j$ is a positive integer with a parity structure satisfying $p(j) = 0, 1$, and can be larger than one. To enforce the $\mathbb{Z}_{2}$ parity symmetry of the many-body wave function $C$, we require each local tensor $T$ to be Grassmann-even, which means
\begin{eqnarray}
T^{(1)}_{j_{1}}[i_{1}] &=& 0,\quad
\mmod{p(j_{1}) + p(i_{1}), 2} = 1, 
\\
T^{(a)}_{j_{a-1}j_{a}}[i_{a}] &=& 0,\quad
\mmod{p(j_{a-1}) + p(i_{a}) + p(j_{a}), 2} = 1,
\quad
\textrm{for}\quad 1<a<m,
\\
T^{(m)}_{j_{m-1}}[i_{m}] &=& 0
\quad
\mmod{p(j_{m-1}) + p(i_{m}), 2} = 1.
\end{eqnarray} 

Inserting an integral resolution for each $j$ between two adjacent $T$ tensors in Eq.~(\ref{MPS}), 
\begin{eqnarray} 
C_{i_1i_2...i_m} 
=\sum_{\{ j\}}&&T^{(1)}_{j_1}[i_1]
\left(\Gintbx{1}\right)
T^{(2)}_{j_1j_2}[i_2]
\left(\Gintbx{2}\right)
T^{(3)}_{j_2j_3}[i_3], \nonumber \\
&...&T^{(m-1)}_{j_{m-2}j_{m-1}}[i_{m-1}]
\left(\Gintbx{m-1}\right)
T^{(m)}_{j_{m-1}}[i_m],
\end{eqnarray}
and this reduces Eq.~(\ref{CGrass}) to the following Grassmann MPS form
\begin{eqnarray}
\mt{C}_{i_1i_2...i_m} &=& \sum_{\{j\}}~\mt{T}^{(1)}_{j_1}[i_1]~\mt{T}^{(2)}_{j_1j_2}[i_2]~...~\mt{T}^{(m-1)}_{j_{m-2}j_{m-1}}[i_{m-1}]~\mt{T}^{(m)}_{j_{m-1}}[i_m],
\nonumber \\
&=&
\raisebox{-0.46\height}{\includegraphics[width=0.4\textwidth, page=1]{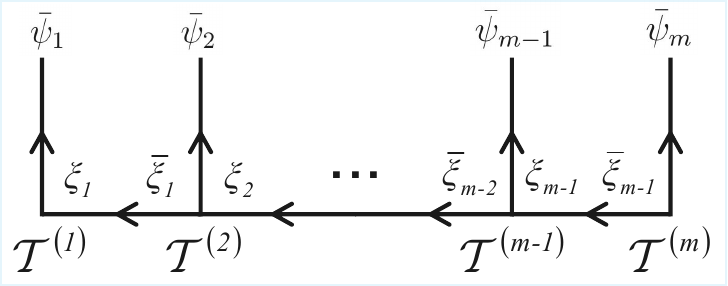}},
\label{GMPS}
\end{eqnarray}
where we have incoporated the Grassmann variables $\{\xib, \xi\, \psb, \psi\}$ into Grassmann tensors
\begin{eqnarray} \label{GMPS_def}
\mt{T}^{(1)}_{j_1}[i_1]&\equiv& T^{(1)}_{j_1}[i_1]~\psb_1^{p(i_1)}\xi_1^{p(j_1)}, \qquad\mt{T}^{(m)}_{j_{m-1}}[i_m]\equiv T^{(m)}_{j_{m-1}}[i_m]~\xib_{m-1}^{p(j_{m-1})}\psb_m^{p(i_m)}, \nonumber\\
\mt{T}^{(n)}_{j_{n-1}j_n}[i_n]&\equiv& T^{(n)}_{j_{n-1}j_n}[i_n]~\xib_{n-1}^{p(j_{n-1})}\psb_n^{p(i_n)}\xi_n^{p(j_n)}, \qquad \textrm{for}\quad 1<n<m.
\end{eqnarray}
As a reminder, the Grassmann integral measure is hidden in the definition of Grassmann tensor contraction expressed, e.g., in Eq.~(\ref{GContract}).

The MPS representation, Eq.~(\ref{GMPS}), of the Grassmann tensor in the fermionic wave function is suitable only for study in one dimension, but the construction can be straightforwardly generalized to two dimensions. 
A natural way to introduce translationally invariant infinite-size fermionic tensor network ansatz is via the projective construction, which is closer in spirit to the construction of their bosonic counterparts \cite{PEPS2004}, as done in Refs.~\cite{Kraus2010Fermionic,Quinten2025Fermionic}. Formally, the infinite projected entangled pair state (PEPS) on the square lattice in the Grassmann representation takes the following form:
\begin{eqnarray}
{\rm Tr}
\left(
\prod_{n}
\mathcal{T}^{(n)}_{\alpha\beta\gamma\delta}[i_{n}]
\right)
 =
\raisebox{-0.46\height}{\includegraphics[width=0.45\textwidth, page=1]{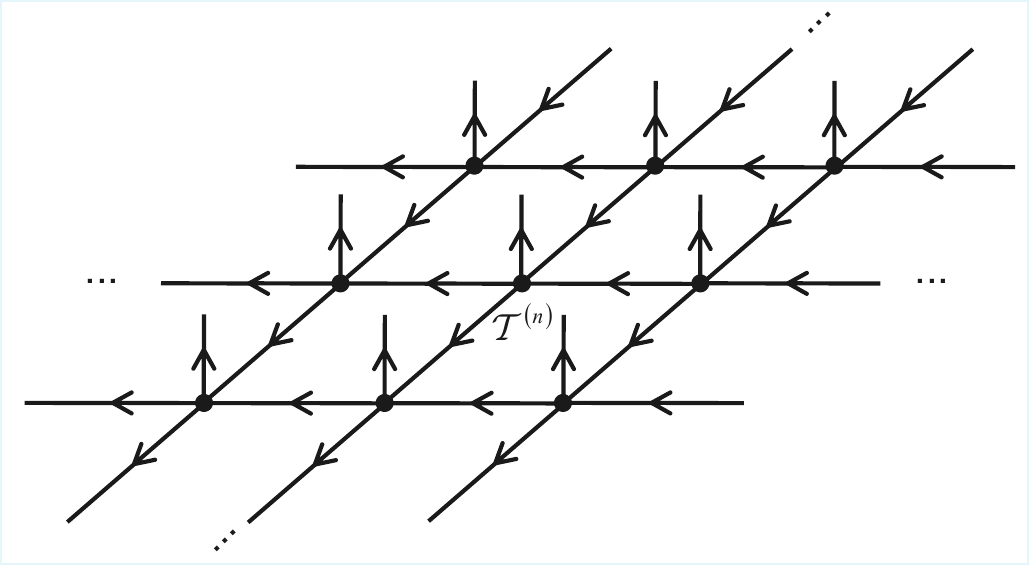}},
\end{eqnarray}
where the rank-5 local tensor $\mathcal{T}$ is defined as 
\begin{eqnarray}
\mathcal{T}_{\alpha\beta\gamma\delta}[i]
= 
\raisebox{-0.46\height}{\includegraphics[width=0.24\textwidth, page=1]{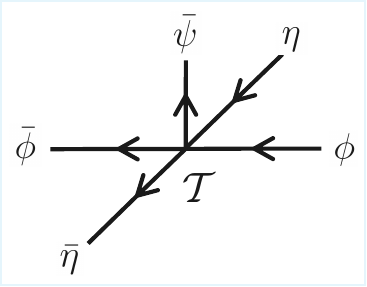}}
=
T_{\alpha\beta\gamma\delta}[i]
\bar{\psi}^{p(i)}
\bar{\phi}^{p(\alpha)}
\phi^{p(\beta)}
\eta^{p(\gamma)}
\bar{\eta}^{p(\delta)},
\end{eqnarray}
and the corresponding fermionic state can be expressed as
\begin{eqnarray}
| \Psi \rangle
=
\int_{\psb\psi}
\left[
\sum_{\{i\}}
{\rm Tr}
\left(
\prod_{n}
\mathcal{T}^{(n)}_{\alpha\beta\gamma\delta}[i_{n}]
\right)
\right]
\ket{\psi_1\psi_2...\psi_n...}.
\end{eqnarray}

\subsection{Fermionic operators} \label{Sec:ferop}

In the previous section, using the coherent-state representation, we reformulated any Grassmann wave function as the self-trace of a Grassmann tensor, which can eventually be represented as a tensor network state. Following the same route, we can also reformulate fermionic operators as Grassmann tensors. For fermionic Hamiltonians, the Grassmann tensor representation can further be cast into the Grassmann matrix product operator (MPO) form.

Let us take a two-body fermionic operator expanded in the Fock basis as the first example:
\begin{eqnarray} 
\hat{O} &=& \sum_{\beta_{1}\beta_{2}\alpha_{1}\alpha_{2}} 
O_{
\beta_{1}\beta_{2}
\alpha_{1}\alpha_{2}}
\ket{\beta_{1}\beta_{2}}
\bra{\alpha_1\alpha_2},  \\
&=& \nonumber
\int_{\bar{\varphi}\varphi}
\int_{\psb\psi}
\left(
\ket{\varphi_1\varphi_2}\bra{\bar{\varphi}_1\bar{\varphi}_2}
\right)
\left[
\sum_{\beta_{1}\beta_{2}\alpha_{1}\alpha_{2}} O_{\beta_{1}\beta_{2}\alpha_{1}\alpha_{2}}
\ket{\beta_{1}\beta_{2}}\bra{\alpha_1\alpha_2}
\right]
\left(\ket{\psi_1\psi_2}\bra{\psb_1\psb_2}\right),
\label{fermionic_op}
\end{eqnarray}
where the matrix element is $O_{\beta_{1}\beta_{2}\alpha_1\alpha_2}\equiv\bra{\beta_{1}\beta_{2}}\hat{O}\ket{\alpha_1\alpha_2}$. Employing the overlap between coherent state and Fock state, i.e., Eq.~(\ref{OverLap}), we obtain
\begin{eqnarray}
\hat{O} = 
\int_{\bar{\varphi}\varphi}
\int_{\psb\psi}
\left(\sum_{\{\alpha\}}
\sum_{\{\beta\}}
\mt{O}_{\beta_{1}\beta_{2}\alpha_1\alpha_2}
\right)
\ket{\varphi_1\varphi_2}\bra{\psb_1\psb_2}, 
\label{OpSum}
\end{eqnarray}
where the rank-4 Grassmann tensor $\mt{O}$ is defined as
\begin{eqnarray}
\mt{O}_{\beta_{1}\beta_{2}\alpha_1\alpha_2} 
\equiv O_{\beta_{1}\beta_{2}\alpha_1\alpha_2}
\bar{\varphi}_1^{\beta_1}
\bar{\varphi}_2^{\beta_2}
\psi_2^{\alpha_2} 
\psi_1^{\alpha_1}
=
\raisebox{-0.46\height}{\includegraphics[width=0.18\textwidth, page=1]{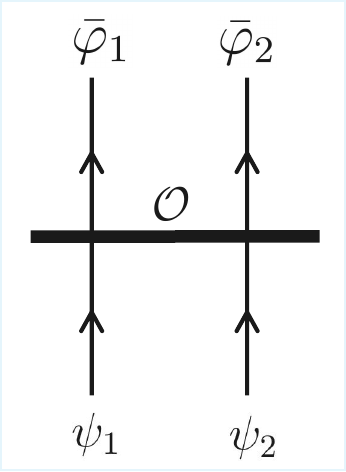}}.
\label{OpGtensor}
\end{eqnarray}
Note here $\alpha,\beta = 0, 1$, thus we can also write the exponents as $p(\alpha)$ and $p(\beta)$. The above procedure can be straightforwardly extended to many-body operators. For example, as done for Eq.~(\ref{fermionic_op}), we can expand a three-body Hamiltonian $\hat{H}$ in the coherent-state representation
\begin{eqnarray}
\hat{H} =
\int_{\bar{\varphi}\varphi} 
\int_{\psb\psi}
\left(
\sum_{\{\alpha\}}
\sum_{\{\beta\}}
\mt{H}_{\beta_1\beta_2\beta_3\alpha_1\alpha_2\alpha_3}
\right)
\ket{\varphi_{1}\varphi_{2}\varphi_{3}}
\bra{\psb_1\psb_2\psb_3},
\end{eqnarray}
where we have
\begin{eqnarray}\label{3sop}
\mt{H}_{\beta_1\beta_2\beta_3\alpha_1\alpha_2\alpha_3} = H_{\beta_1\beta_2\beta_3\alpha_1\alpha_2\alpha_3}~
\bar{\varphi}_1^{\beta_1}
\bar{\varphi}_2^{\beta_2}
\bar{\varphi}_3^{\beta_3}
\psi_3^{\alpha_3}
\psi_2^{\alpha_2}
\psi_1^{\alpha_1}
=
\raisebox{-0.47\height}{\includegraphics[width=0.18\textwidth, page=1]{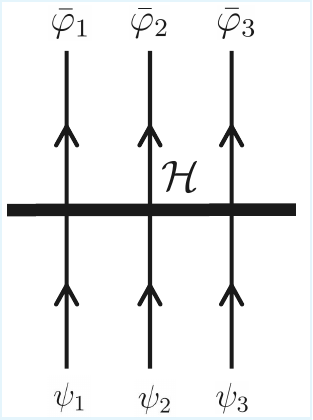}}.
\end{eqnarray}

We now present the Grassmann analogue of the MPO representation, which is essential for the modern variational formulation of the DMRG \cite{SchRev2011,Ostlund1995}, see e.g. Ref.~\cite{SchRev2011} for a review of bosonic MPOs. As a concrete illustration, we consider a three-site Hubbard model with open boundary conditions:
\begin{eqnarray}\label{3shubbard}
\hat{H} = 
-t\sum_{i=1,2}
\sum_{\sigma=\uparrow\downarrow}
\left(
\hat{c}_{i\sigma}^{\dagger}
\hat{c}_{i+1\sigma}
+
h.c.
\right)
+
U
\sum_{i=1}^{3}
\hat{n}_{i\uparrow}
\hat{n}_{i\downarrow}.
\end{eqnarray}

In Eq.~(\ref{3shubbard}), $\hat{n}_{i\uparrow}\equiv \hat{c}_{i\uparrow}^{\dagger}\hat{c}_{i\uparrow}$ and $\hat{n}_{i\downarrow}\equiv\hat{c}_{i\downarrow}^{\dagger}\hat{c}_{i\downarrow}$ denote the occupation number operators for spin-up and spin-down electrons, respectively. For the three-site Hubbard model, the corresponding MPO representation of $H$ in Eq.~(\ref{3sop}) is:
\begin{eqnarray} \label{MPOHubbard}
H_{\beta_1\beta_2\beta_3\alpha_1\alpha_2\alpha_3}
=
\sum_{l_{1}l_{2}}
(-1)^{r}
W^{(1)}_{\beta_1\alpha_{1}l_{1}}
W^{(2)}_{\beta_2\alpha_{2}l_{1}l_{2}}
W^{(3)}_{l_{2}\beta_3\alpha_{3}}.
\end{eqnarray}
The sign factor $(-1)^{r}$ to enforce the fermionic anti-commutation relations, see e.g. Ref.~\cite{Shimizu2014Application}, will be elaborated in detail later. We present the form of the local MPO tensors,
\begin{eqnarray} \label{W1}
W^{(1)} = 
\left(
\begin{array}{ccccccc}
U\mathbb{D}_{1} & t\mathbb{C}_{1\uparrow} & t\mathbb{C}_{1\downarrow} & -t\mathbb{C}_{1\uparrow}^{\dagger} &
-t\mathbb{C}_{1\downarrow}^{\dagger} &
\mathbb{I} 
\end{array}
\right),
\end{eqnarray}
\begin{eqnarray} \label{W23}
W^{(2)} = 
\left(
\begin{array}{ccccccc}
\mathbb{I} & 0 & 0 & 0 & 0 & 0 \\
\mathbb{C}_{2\uparrow}^{\dagger} & 0 & 0 & 0 & 0 & 0 \\
\mathbb{C}_{2\downarrow}^{\dagger} & 0 & 0 & 0 & 0 & 0 \\
\mathbb{C}_{2\uparrow} & 0 & 0 & 0 & 0 & 0 \\
\mathbb{C}_{2\downarrow} & 0 & 0 & 0 & 0 & 0 \\
U\mathbb{D}_{2} & t\mathbb{C}_{2\uparrow} & t\mathbb{C}_{2\downarrow} & -t\mathbb{C}_{2\uparrow}^{\dagger} & -t\mathbb{C}_{2\downarrow}^{\dagger} & \mathbb{I} \\
\end{array}
\right)
,\quad
W^{(3)} = 
\left(
\begin{array}{c}
\mathbb{I}  \\
\mathbb{C}_{3\uparrow}^{\dagger}  \\
\mathbb{C}_{3\downarrow}^{\dagger}  \\
\mathbb{C}_{3\uparrow}  \\
\mathbb{C}_{3\downarrow}  \\
U\mathbb{D}_{3}  \\
\end{array}
\right).
\end{eqnarray}
In Eqs.~(\ref{W1}-\ref{W23}), we have denoted
\begin{eqnarray}
\hat{1}_{i}
~\dot{=}~
\mathbb{I}_{i},\quad
\hat{c}^{\dagger}_{i\uparrow}
~\dot{=}~
\mathbb{C}_{i\uparrow}^{\dagger},\quad
\hat{c}^{\dagger}_{i\downarrow}
~\dot{=}~
\mathbb{C}_{i\downarrow}^{\dagger},\quad
\hat{c}_{i\uparrow}
~\dot{=}~
\mathbb{C}_{i\uparrow},\quad
\hat{c}_{i\downarrow}
~\dot{=}~
\mathbb{C}_{i\downarrow},\quad
\hat{n}_{i\uparrow}
\hat{n}_{i\downarrow}
~\dot{=}~
\mathbb{D}_{i},
\end{eqnarray}
where the symbol $(\dot{=})$ is used to denote that the fermionic operator $\hat{o}_{i}$ at the site $i$ is represented as a $4\times4$ matrix in the single-particle Fock basis,
\begin{eqnarray}
\hat{o}_{i} 
&=&  \nonumber
\sum_{\alpha_{i\uparrow}\alpha_{i\downarrow}=0}^{1}
\sum_{\beta_{i\uparrow}\beta_{i\downarrow}=0}^{1}
(\mathbb{O}_{i})_{
	(\beta_{i\uparrow}\beta_{i\downarrow})
	(\alpha_{i\uparrow}\alpha_{i\downarrow})
}
\vert \beta_{i\uparrow}\beta_{i\downarrow} \rangle
\langle \alpha_{i\uparrow}\alpha_{i\downarrow} \vert,
\\ \nonumber
&\equiv&
\sum_{\alpha_{i}=1}^{4}
\sum_{\beta_{i}=1}^{4}
(\mathbb{O}_{i})_{
	\beta_{i}
	\alpha_{i}
}
\vert \beta_{i} \rangle
\langle \alpha_{i} \vert,
\\ 
\vert \beta_{i} \rangle
&\equiv&
\vert \beta_{i\uparrow}\beta_{i\downarrow} \rangle
=
(\hat{c}_{i\uparrow}^{\dagger})^{\beta_{i\uparrow}}
(\hat{c}_{i\downarrow}^{\dagger})^{\beta_{i\downarrow}}
\vert 0 \rangle,
\quad
\langle\alpha_{i}\vert 
\equiv
\langle \alpha_{i\uparrow}\alpha_{i\downarrow} \vert
=
\langle 0 \vert
\hat{c}_{i\downarrow}^{\alpha_{i\downarrow}}
\hat{c}_{i\uparrow}^{\alpha_{i\uparrow}}.
\end{eqnarray}
We have introduced the composite index $\alpha_{i}$ (similarly for $\beta_{i}$) through index fusing:
\begin{eqnarray} 
& &\label{alpha_def1} \nonumber
\alpha_{i} = 1 \longleftarrow 
(\alpha_{i\uparrow} = 0,\alpha_{i\downarrow} = 0),\quad
\alpha_{i} = 2 \longleftarrow 
(\alpha_{i\uparrow} = 1,\alpha_{i\downarrow} = 1),\quad
\\
& &\label{alpha_def2}
\alpha_{i} = 3 \longleftarrow 
(\alpha_{i\uparrow} = 1,\alpha_{i\downarrow} = 0),\quad
\alpha_{i} = 4 \longleftarrow 
(\alpha_{i\uparrow} = 0,\alpha_{i\downarrow} = 1).
\end{eqnarray}
To proceed, we formulate $\mt{H}$ as a Grassmann MPO. Following the methodology of Sec.~\ref{Sec:GMPS}, we decompose the $\mt{H}$ from Eq.~(\ref{3sop}) into three distinct parts
\begin{eqnarray} \label{GMPO}
\mt{H}_{\beta_1\beta_2\beta_3\alpha_1\alpha_2\alpha_3} 
= 
\sum_{l_{1}l_{2}}
(-1)^{r}
(-1)^{\tilde{r}}
\mt{W}^{(1)}_{\beta_{1}\alpha_{1}l_{1}}
\mt{W}^{(2)}_{\beta_{2}\alpha_{2}l_{1}l_{2}}
\mt{W}^{(3)}_{l_{2}\beta_{3}\alpha_{3}},
\end{eqnarray}
where 
\begin{eqnarray}\label{GMPO_def}
\mt{W}^{(1)}_{\beta_{1}\alpha_{1}l_{1}}
&\equiv& 
W^{(1)}_{\beta_{1}\alpha_{1}l_{1}}
\bar{\varphi}_1^{p(\beta_1)}
\psi_1^{p(\alpha_1)}
\xi_1^{q(l_{1})}, \qquad
\mt{W}^{(3)}_{l_{2}\beta_{3}\alpha_{3}}
\equiv 
W^{(3)}_{l_{2}\beta_{3}\alpha_{3}}
\bar{\xi}_2^{q(l_{2})}
\bar{\varphi}_3^{p(\beta_3)}
\psi_3^{p(\alpha_3)},
\nonumber \\
\mt{W}^{(2)}_{\beta_{2}\alpha_{2}l_{1}l_{2}}
&\equiv& 
W^{(2)}_{\beta_{2}\alpha_{2}l_{1}l_{2}}
\bar{\xi}_1^{q(l_{1})}
\bar{\varphi}_2^{p(\beta_{2})}
\psi_2^{p(\alpha_2)}
\xi_2^{q(l_{2})}. 
\end{eqnarray}
Eqs.~(\ref{GMPO}-\ref{GMPO_def}) are obtained by substituting Eq.~(\ref{MPOHubbard}) into $\mt{H}$ and introducing resolutions of the identity to constrain the parity of the virtual bonds. The additional sign factors $(-1)^{\tilde{r}}$ in Eq.~(\ref{GMPO}), which stem from reordering the Grassmann variables, are explicitly given by:
\begin{eqnarray} \label{sign_r}
\tilde{r} = 
\left[p(\beta_{3}) + p(\alpha_{3})\right]
\times
\left[p(\alpha_{2}) + p(\alpha_{1})\right]
+
p(\alpha_{1})
\times
\left[
p(\alpha_{2}) + p(\beta_{2})
\right].
\end{eqnarray}
In Eq.~(\ref{GMPO_def}), we have introduced two distinct parity functions, $p$ and $q$, which act on the physical and virtual indices of the Grassmann MPO, respectively.
\begin{eqnarray}
p(\alpha) 
&=& \nonumber
0,
\quad 
\textrm{for}~\alpha = 1, 2, 
\\ \nonumber 
p(\alpha) 
&=& \nonumber
1, 
\quad
 \textrm{for}~\alpha = 3, 4,
\\ \nonumber
\label{def_q}
q(l) 
&=& \nonumber
0,
\quad 
\textrm{for}~l = 1, 6, 
\\  
q(l) &=& 1,
\quad 
\textrm{for}~l = 2, 3, 4, 5.  \label{q_def}
\end{eqnarray}
The definition of $p$ is evident from Eq.~(\ref{alpha_def1}). For $q$, we note that the matrices $\mathbb{C}_{i\sigma}$ and  $\mathbb{C}_{i\sigma}^{\dagger}$ with $\sigma=\uparrow,\downarrow$ are odd-parity matrices. Therefore, the specific form of $q$ given in Eq.~\eqref{q_def} is therefore designed to ensure that the tensors $W^{(1)}$, $W^{(2)}$ and $W^{(3)}$ are Grassmann-even as a whole. Given this Grassmann-even condition for  $W$ tensors, we have
\begin{eqnarray}
& & \nonumber
\mmod{p(\alpha_{2}) + p(\beta_{2}), 2} = \mmod{q(l_{1}) + q(l_{2}), 2},
\\ 
& &
\mmod{p(\alpha_{3}) + p(\beta_{3}), 2} =\mmod{q(l_{2}), 2}.
\end{eqnarray}
We can simplify the sign factor $(-1)^{\tilde{r}}$ in Eq.~(\ref{sign_r}) and obtain
\begin{eqnarray}
\tilde{r} = 
q(l_{1}) \times p(\alpha_{1})
+ 
q(l_{2}) \times p(\alpha_{2}).
\end{eqnarray}
After a moment of thought, we find this simplified sign factor is exactly the same as the $(-1)^{r}$ given in Ref.~\cite{Shimizu2014Application},
\begin{eqnarray}
& & \nonumber
(-1)^{r} = -1,
\quad\textrm{for}~
W^{(1)}_{\beta_{1}\alpha_{1}l_{1}}
~
\textrm{if}~l_{1} = 2, 3, 4, 5 
~\textrm{and}~
\alpha_{1} = 3, 4,
\\
& &
(-1)^{r} = -1, 
\quad\textrm{for}~
W^{(2)}_{\beta_{2}\alpha_{2}6l_{2}}
~
\textrm{if}~l_{2} = 2, 3, 4, 5 
~\textrm{and}~
\alpha_{2} = 3, 4,
\end{eqnarray}
and $(-1)^{r} = 1$ for other cases. Then we see $(-1)^{r}$ and $(-1)^{\tilde{r}}$ cancel each other, and the Grassmann MPO is simply given by:
\begin{eqnarray} \label{GMPO1}
\mt{H}_{\beta_1\beta_2\beta_3\alpha_1\alpha_2\alpha_3} =
\sum_{l_{1}l_{2}}
\mt{W}^{(1)}_{\beta_{1}\alpha_{1}l_{1}}
\mt{W}^{(2)}_{\beta_{2}\alpha_{2}l_{1}l_{2}}
\mt{W}^{(3)}_{l_{2}\beta_{3}\alpha_{3}}
=
\raisebox{-0.46\height}{\includegraphics[width=0.3\textwidth, page=1]{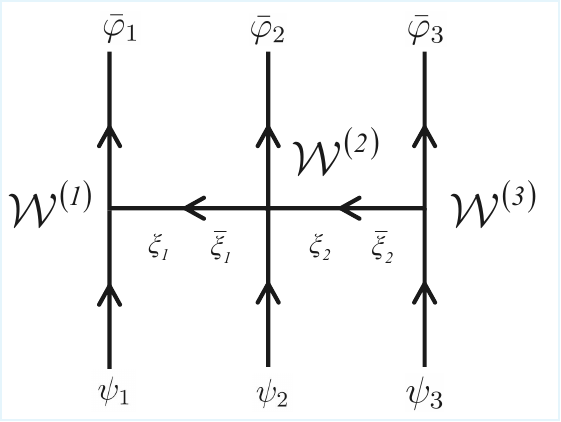}}.
\end{eqnarray}

Thus, Eq.~(\ref{GMPO1}) presents an intuitive and natural form of the Grassmann MPO. In this coherent-state representation, the local MPO tensors are constructed directly from fermionic operators in the single-particle Fock basis, requiring no auxiliary sign factors to enforce fermionic anti-commutation relations. Instead, the Fermi statistics are inherently encoded within the Grassmann variables associated with the MPO indices. It is crucial to note that, however, the explicit cancellation of the sign factors $(-1)^{r}$ and $(-1)^{\tilde{r}}$ occurs only when the Grassmann order is explicitly specified as the one in Eq.~\eqref{GMPO_def}.

\subsection{Fermionic partition functions} \label{fpf}

In the previous sections, we have shown that a fermionic wave function can be related to a Grassmann tensor network, and this is also true for fermionic operators. Moreover, it is known that the partition function of a fermionic Hamiltonian can be approximately converted to the summation of a Grassmann tensor network \cite{Akiyama2022Metal,Akiyama2021Tensor}, in a similar way as representing the partition function of a classical spin model as a tensor network summation {\cite{SRGreview2010}. In this section, we make this evident for 1D fermionic models. Note that although we focus on the non-relativistic fermionic models here, it is also known that the Grassmann tensor network representation of relativistic models that break particle number conservation can also be derived similarly \cite{ShimizuGrassmann2014,Akiyama2021More,Akiyama2024Tensor}. 

We start from the partition function of a 1D lattice fermion with Hamiltonian $\hat{H}$ represented as 
\begin{eqnarray}
Z = \textrm{Tr} \left(
{\rm e}^{-\beta H}
\right) = \textrm{Tr}\left({\rm e}^{-\tau H}\right)^m, \qquad\textrm{with}\quad m\tau = \beta,
\label{Zparts}
\end{eqnarray}
where the temperature is $1/\beta$. Here, we have divided $\beta$ into $m$ slices with a very small Trotter step $\tau$, e.g., $10^{-4}$ in most calculations in this work. For future purposes, we follow the convention adopted in Ref.~\cite{ColemanBook2015}, and use a different but equivalent superorthogonal relation for any two integers $a$ and $b$
\begin{eqnarray}
\int_{\phi}\ket{\phif{a}}\bra{\phif{b}} &\equiv& \int_{\psbf{b}\psif{a}}\ket{\psidu{1}{a}\psidu{2}{a}...\psidu{L}{a}}\bra{\psbdu{1}{b}\psbdu{2}{b}...\psbdu{L}{b}} = 1, 
\nonumber \\
\lr{\phif{a}}{\phif{b}} &=& \exp{\left[\sum_{i}\psb_i^{(a)}\psi_i^{(b)}\right]},
\label{GenOverComp}
\end{eqnarray}
where $L$ is the number of spin-resolved single-particle eigenstates used to construct the Fock basis. For cases where each lattice site contributes only a single eigenstate, e.g., the spinless fermion case, $L$ is simply the length of the chain. In this case, we express the trace in Eq.~(\ref{Zparts}) as
\begin{eqnarray}
Z = \int_{\phi}\lpr{-\phif{0}}{\left({\rm e}^{-\tau H}\right)^m}{\phif{1}},
\end{eqnarray}
where the anti-periodic condition is shown explicitly. We can insert more coherent-state basis $\{\phi^{(i)}\}$ between each two ${\rm e}^{-\tau H}$ in Trotter (imaginary-time) direction, i.e.,
\begin{eqnarray}
Z = \int_{\{\phi\}}\bra{-\phif{0}}{\rm e}^{-\tau H}\ket{\phif{m}}\bra{\phif{m-1}}{\rm e}^{-\tau H}\ket{\phif{m-1}}...\bra{\phif{2}}{\rm e}^{-\tau H}\ket{\phif{2}}\bra{\phif{1}}{\rm e}^{-\tau H}\ket{\phif{1}},
\label{Zorig}
\end{eqnarray}
where the superscript order of $\{\phi\}$ are arranged intentionally just for convention. 

The next task is to derive the matrix element of ${\rm e}^{-\tau H}$ in the coherent-state representation. For any Hamiltonian with only bilinear terms of creation and annihilation operators, i.e., $\hat{H} = \hat{H}(\dci{i}, \ci{j})$, expand the exponent by ${\rm e}^x\approx 1+x$, we have
\begin{eqnarray}
\lpr{\phif{n}}{{\rm e}^{-\tau H}}{\phif{n}} \approx \lpr{\phif{n}}{\left(1-\tau H\right)}{\phif{n}} = {\left(1-\tau H'\right)}\lr{\phif{n}}{\phif{n}},
\end{eqnarray} 
where we have used the eigen-equation expressed in Eq.~(\ref{EigenGrass}) and introduced the shorthand
\begin{eqnarray}
H' = H'(\psb_i, \psi_j) \equiv \hat{H}(\dci{i}\rightarrow\psb_i, \ci{j}\rightarrow\psi_j).
\end{eqnarray}
That is, $H'$ is obtained by replacing $\dci{}$ and $\ci{}$ in Hamiltonian by $\psb$ and $\psi$, respectively. Then, using Eq.~(\ref{grassmannalgebra7}) and Eq.~(\ref{GenOverComp}), we obtain
\begin{eqnarray}
\lpr{\phif{n}}{{\rm e}^{-\tau H}}{\phif{n}} \approx \exp\left[-\tau H' + \sum_{i}\psbdu{i}{n}\psidu{i}{n}\right].
\label{tauHformula}
\end{eqnarray}

To make it more concrete, in the rest of this section, let us consider the 1D spinless free fermion model with Hamiltonian
\begin{eqnarray}
\hat{H} = -t\sum_{i}\left(\dci{i}\ci{i+1}+h.c.\right) -\mu\sum_{i}\opn{i}.
\label{free_fermion_H}
\end{eqnarray}
Now, Eq.~(\ref{tauHformula}) reduces to ${\rm e}^{-S^{(n)}}$ with 
\begin{eqnarray}
S^{(n)} \equiv - \tau t\sum_{i}\left(\psbdu{i}{n}\psidu{i+1}{n} + h.c.\right) - (1+\tau\mu)\sum_{i}\psbdu{i}{n}\psidu{i}{n},
\end{eqnarray}
where for Grassmann variables, the $h.c.$ should be understood as their dual. 

\begin{figure}[htbp]
	\centering
	\includegraphics[height=5.4cm,width=6.0cm]{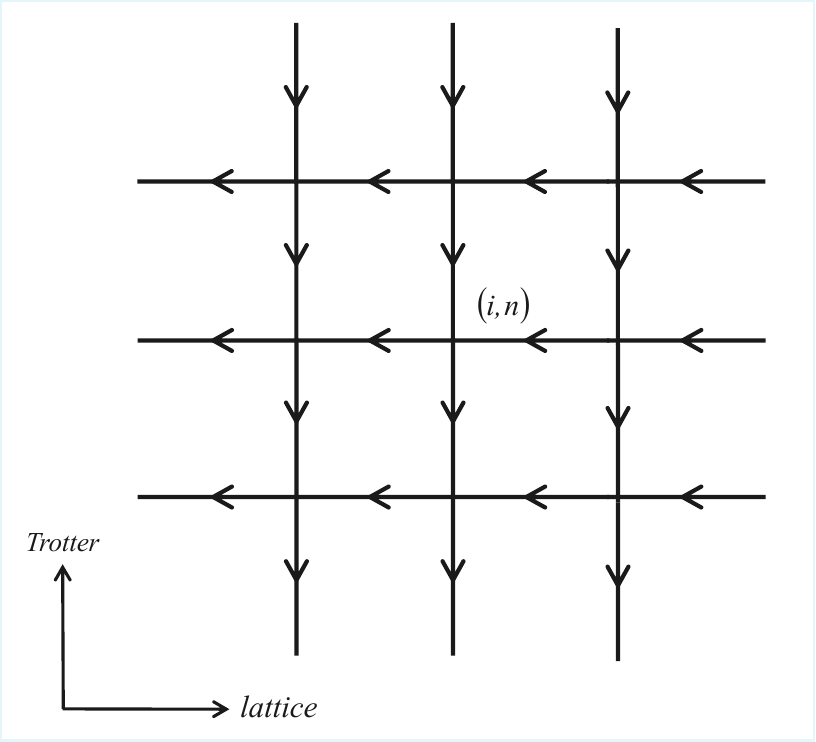}
	\caption{An effective (1+1)-dimensional lattice with coordinates $(i, n)$, the Trotter direction and lattice direction are specified.}
	\label{PFLattice}
\end{figure}

In the two-dimensional parameter space with coordinates $(i,n)$, where $i=1,2,...L$ is the lattice index and $n = 1, 2, ..., m$ is the index in Trotter direction, as shown in Fig.~\ref{PFLattice}, Eq.~(\ref{Zorig}) is similar to the partition function of a two-dimensional classical spin model. To see this, we write the integral measure of 
Eq.~(\ref{GenOverComp}) explicitly, 
\begin{eqnarray}
Z \approx \int (-1)^{L}\prod_{i,n}\left(d\psbdu{i}{n-1}d\psidu{i}{n}\right)\prod_{j,q<m}\left({\rm e}^{-\psbdu{j}{q}\psidu{j}{q+1}}\right)\prod_{j}\left({\rm e}^{\psbdu{j}{0}\psidu{j}{1}}\right)\cdot {\rm e}^{-\sum_{n}S^{(n)}},
\label{AddMeas}
\end{eqnarray}
where $(-1)^{L}$ denotes the sign from the anti-periodic boundary condition in Eq.~(\ref{ABC}) for all the $L$ sites in the 1D lattice, and we have $d\psbdu{i}{0}=d\psbdu{i}{m}$, $\psidu{i}{m+1} = \psidu{i}{1}$. Note the product of an even number of Grassmann variables commutes with any functions of Grassmann variables, thus the order of product over $i$ and $n$ in Eqs.~(\ref{AddMeas}) does not matter, thus we have
\begin{eqnarray}
\prod_{i,n}d\psbdu{i}{n-1}d\psidu{i}{n} &=& \prod_id\psbdu{i}{0}\left(d\psidu{i}{1}d\psbdu{i}{1}\right)\left(d\psidu{i}{2}d\psbdu{i}{2}\right)...\left(d\psidu{i}{m-1}d\psbdu{i}{m-1}\right)d\psidu{i}{m}, 
\nonumber \\
&=& 
(-1)^{(m-1)L} 
\int 
\mt{D}\psb\mt{D}\psi.
\label{ChanMeas}
\end{eqnarray}
where we have introduced the shorthand commontly adopted in standard textbooks \cite{ColemanBook2015,Negele2018Quantum}
\begin{eqnarray}
\int \mt{D}\psb\mt{D}\psi 
\equiv 
\prod_{i,n}\int d\psbdu{i}{n}d\psidu{i}{n}.
\label{PathMeas}
\end{eqnarray}
Note that the net factor $(-1)^{mL}$ contributed by Eq.~(\ref{AddMeas}) and Eq.~(\ref{ChanMeas}) is irrelevant in either the ground-state limit $(m\rightarrow\infty)$ or the thermodynamic limit $(L\rightarrow\infty)$, and can be elimilated by a redefinition of the Grassmann integral measure~\cite{Pierre2001}, hence it can be safely ignored hereafter. Pluging Eqs.~(\ref{ChanMeas}-\ref{PathMeas}) into Eq.~(\ref{AddMeas}), we finally obtain
\begin{eqnarray}
Z \approx 
\int\mt{D}\psb\mt{D}\psi 
\prod_{j,q<m}\left({\rm e}^{-\psbdu{j}{q}\psidu{j}{q+1}}\right)\prod_{j}\left({\rm e}^{\psbdu{j}{0}\psidu{j}{1}}\right)\cdot {\rm e}^{-\sum_{n}S^{(n)}} = 
\int\mt{D}\psb\mt{D}\psi~
{\rm e}^{-S}, 
\label{Zfinal}
\end{eqnarray}
with the total action defined as (setting $-\psi_i^{(0)} = \psi_i^{(m)}$)
\begin{eqnarray}
S 
&\equiv& \sum_{i,n}\left(\psbdu{i}{n}\psidu{i}{n+1}\right) + \sum_{n} S_n
\nonumber \\
&=& 
- \sum_{i,n}
\left[
\tau t\left(\psbdu{i}{n}\psidu{i+1}{n} + h.c.\right) - \left(\psbdu{i}{n}\psidu{i}{n+1}\right)
\right] - (1+\tau\mu)\sum_{i,n}\psbdu{i}{n}\psidu{i}{n}.
\label{Sfinal}
\end{eqnarray}

If we regard $\psidu{i}{n}$ and $\psbdu{i}{n}$ as two independent Grassmann (spin) variables locating at $(i,n)$-th site at the (1+1)-dimensional lattice, Eq.~(\ref{Sfinal}) can be roughly understood as an anisotropic \textit{classical} spin models with two species of interactions in lattice direction with strength $(-\tau t)$, one species of interaction in Trotter direction with unit strength, and some kinds of bilinear magnetic field with strength $-(1+\tau\mu)$. Here, by classical, we mean each two bilinear terms commute in Eq.~(\ref{Sfinal}), and thus ${\rm e}^{-S}$ in Eq.~(\ref{Zfinal}) can be evaluated in a very similar way to the case of classical partition functions, as discussed, e.g., in Ref.~\cite{SRGreview2010}. 

To simplify Eqs.~(\ref{Zfinal}-\ref{Sfinal}), we introduce the following notations similar to those frequently used in the spin models in condensed matter physics
\begin{eqnarray}
J \equiv \tau t, \qquad h \equiv 1+\tau\mu, \nonumber
\end{eqnarray}
\label{Jcoup}
and 
\begin{eqnarray}
\sum_{\nn{i,j}_r}\psb_i\psi_j &\equiv& \sum_{i,n}\psbdu{i}{n}\psidu{i+1}{n}, \qquad \sum_{\nn{i,j}_l}\psb_i\psi_j \equiv \sum_{i,n}\psbdu{i+1}{n}\psidu{i}{n}, 
\nonumber \\
\sum_{\nn{i,j}_u}\psb_i\psi_j &\equiv& \sum_{i,n}\psbdu{i}{n}\psidu{i}{n+1}, \qquad \sum_{i}\psb_i\psi_i \equiv \sum_{i,n}\psbdu{i}{n}\psidu{i}{n},
\label{SpinDef}
\end{eqnarray}
where the coordinates $\{n\}$ are incorporated into the new two-dimensional indices $\{i\}$, and $\sum_{\nn{i,j}}$ is used to denote the summation over all products of Grassmann pairs defined on nearest neighbors. The subscripts $l, r, u$ in $\nn{i,j}$ are used to distinguish the three kinds of products in Eq.~(\ref{Sfinal}). These notations immediately simplify the action as
\begin{eqnarray}
S = -J\sum_{\nn{i,j}_l}\psb_{i}\psi_{j} - J\sum_{\nn{i,j}_r}\psb_{i}\psi_{j} + \sum_{\nn{i,j}_u}\psb_{i}\psi_{j} - h\sum_{i}\psb_{i}\psi_{i},
\label{Scomp}
\end{eqnarray}
where $\psb_i$ and $\psi_i$ are two independent Grassmann variables defined on the (1+1)-dimensional lattice, and play a similar role to the Ising variables in classical spin models. Since all terms in Eq.~(\ref{Scomp}) are bilinear with respect to Grassmann numbers, they all commute, and this fact leads to the following representation of the partition function expressed in Eq.~(\ref{Zfinal}):
\begin{eqnarray}
Z \approx 
\int\mt{D}\psb\mt{D}\psi 
\prod_{\nn{i,j}_l}
{\rm e}^{J\psb_{i}\psi_{j}}\prod_{\nn{i,j}_r}
{\rm e}^{J\psb_{i}\psi_{j}}\prod_{\nn{i,j}_u}
{\rm e}^{-\psb_{i}\psi_{j}}\prod_{i}
{\rm e}^{h\psb_{i}\psi_{i}}.
\label{Zcomp}
\end{eqnarray} 
This means the partition function of a 1D spinless free model can be obtained by two steps: associate a pair of Grassmann variables to each site in the $(i,n)$ plane, relate each two neighboring Grassmann variables by defining exponents ${\rm e}^{c\psb_{i}\psi_{j}}$ on the links with $c$ defined properly and direction $(l,r,u)$ considered carefully, and place an extra bilinear exponent ${\rm e}^{\psb_{i}\psi_{i}}$ at each site; then perform Grassmann integral of the product of all these exponents over all all the Grassmann pairs. The only thing to be cautious about is that the \textit{classical} antiperiodic condition we have used in Eq.~(\ref{Sfinal}): 
\begin{eqnarray}
\psi_i^{(0)} = -\psi_i^{(m)}, \qquad \psb_i^{(0)} = -\psb_i^{(m)},
\end{eqnarray}
i.e., adding an exponent $\exp{\left(\psb^{(0)}\psi^{(1)}\right)} \equiv \exp{\left(-\psb^{(m)}\psi^{(1)}\right)}$ at each link crossing the boundary in Trotter direction.

At last, in order to construct a Grassmann tensor network with local tensors explicitly defined at sites on the $(i,n)$ plane as in the tensor network for spin models, we follow the procedure described in Ref.~\cite{Akiyama2021Tensor}, and finally represent Eqs.~(\ref{Zfinal}) and (\ref{Zcomp}) as a two-dimensional Grassmann tensor network, i.e., 
\begin{eqnarray}
Z \approx \Tr\prod_i\mt{T}^{(i)}_{l_ir_iu_id_i} = \raisebox{-0.48\height}{\includegraphics[width=0.34\textwidth, page=1]{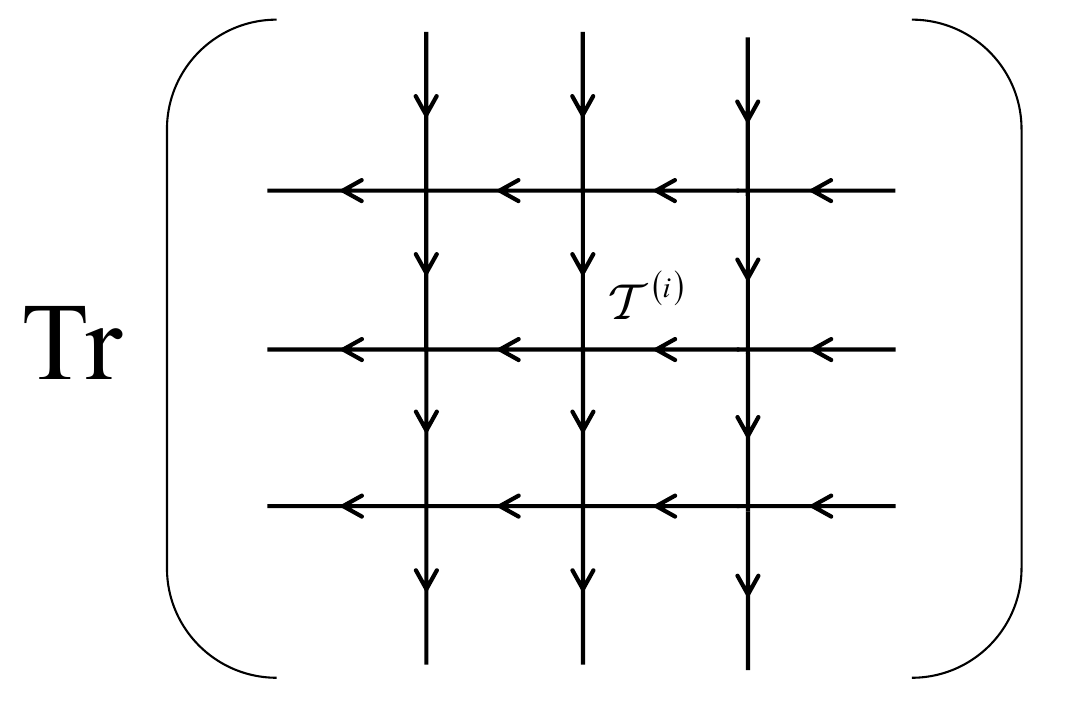}},
\end{eqnarray}
where $\textrm{Tr}$ means contraction by summing over all the shared indices in the Grassmann tensor network, which is discussed in Sec.~\ref{grassmann_op}, and $\mt{T}^{(i)}$ is the local Grassmann tensor defined on the $i$-th site with four link indices $l_i, r_i, u_i, d_i$.

To achieve this, we rely on the following relation ($c$ is a commuting number) for any two integers $i$ and $j$
\begin{eqnarray}
{\rm e}^{c\psi_i\psi_j} = 1 + c\psi_i\psi_j = \sum_{\alpha}\left(c\psi_i\psi_j\right)^{\alpha} = \sum_{\alpha}c^{\alpha}\psi_i^{\alpha}\psi_j^{\alpha}, \quad\textrm{for}\quad \alpha = 0, 1,
\end{eqnarray}
and then using the fundamental Grassman algebra, i.e., Eq.~(\ref{FundGA}), 
\begin{eqnarray}
{\rm e}^{c\psi_i\psi_j} = \sum_{\alpha}c_1^{\alpha}\psi_i^{\alpha}\left(\int_{\psb_k\psi_k}\psi_k^{\alpha}\psb_k^{\alpha}\right)c_2^{\alpha}\psi_j^{\alpha}
= \int_{\psb_k\psi_k}\sum_{\alpha}\left(c_1\psi_i\psi_{k}\right)^{\alpha}\left(c_2\psb_{k}\psi_j\right)^{\alpha},
\label{GrassDecom}
\end{eqnarray}
where we have decomposed the coefficient as $c=c_1c_2$. In fact, we just prove the second equal of the Grassmann integral identity in Eq.~(\ref{grassmannalgebra6}). That is, by introducing an auxiliary Grassmann pair $\{\psi_k,\psb_k\}$ and a discrete parity index $\alpha$ defined on the link, we have successfully separated the original Grassmann variables $\psi_i$ and $\psi_j$. Eq.~(\ref{GrassDecom}) is very important in representing Eq.~(\ref{Zcomp}) as a Grassmann tensor network, and is similar to the singular value decomposition used in the classical Ising model \cite{SRGreview2010} and the Fourier-like expansion used in the $q$-state clock model \cite{ClockLZQ2020}.

Using Eq.~(\ref{GrassDecom}), we decompose the exponents in Eq.~(\ref{Zcomp}) in the following:
\begin{eqnarray}
{\rm e}^{J\psb_i\psi_j} &\xlongequal{\nn{i,j}_l}& \int_{\mub\mu}\sum_{\alpha}\left(\sqrt{J}\psb_i\mu\right)^\alpha\left(\sqrt{J}\mub\psi_j\right)^\alpha = \int_{\mub\mu}\sum_{\alpha}\left(-\sqrt{J}\psb_i\mub\right)^\alpha\left(\sqrt{J}\mu\psi_j\right)^\alpha, 
\nonumber \\
{\rm e}^{J\psb_i\psi_j} &\xlongequal{\nn{i,j}_r}& \int_{\nub\nu}\sum_{\alpha}\left(\sqrt{J}\psb_i\nu\right)^\alpha\left(\sqrt{J}\nub\psi_j\right)^\alpha,
\nonumber \\
{\rm e}^{-\psb_i\psi_j} &\xlongequal{\nn{i,j}_u}& \int_{\thb\theta}\sum_{\alpha}\left(-\psb_i\theta\right)^\alpha\left(\thb\psi_j\right)^\alpha,
\label{ExpDecom}
\end{eqnarray}
where we have adopted the convention that $(\mu,\nu,\theta)$ (without bar) couple to the original Grassmann variables with smaller indices. An illustration is shown in Fig.~\ref{Fig:Tdecom}(a). Then reorder the exponent by collecting terms associated with the same site, and the partition function in Eq.~(\ref{Zcomp}) can be rewritten as
\begin{eqnarray}
Z 
&\approx& 
\int D_{\mu\nu\theta}\left[\psi\right] \sum_{\{\alpha\}} \prod_i \square_i, 
\label{Zcomplex}
\end{eqnarray}
where the symbol $\square_i$ is defined as ($\gamma = 0, 1$)
\begin{eqnarray}
\square_i
&=&
\sum_{\gamma}c\left(\psb_{i}\mub_{a_1}\right)^{\alpha_1}\left(\nub_{b_1}\psi_{i}\right)^{\alpha_2}\left(\mu_{a_2}\psi_{i}\right)^{\alpha_3}\left(\psb_i\nu_{b_2}\right)^{\alpha_4}\left(\psb_i\theta_{c_1}\right)^{\alpha_5}\left(\thb_{c_2}\psi_{i}\right)^{\alpha_6}\left(\psb_i\psi_i\right)^{\gamma}, \nonumber \\
&=& \left(\sum_{\gamma}c(-1)^{s_1}\psb_{i}^{\alpha_1+\gamma+\alpha_5+\alpha_4}\cdot(-1)^{s_2}\psi_i^{\alpha_2+\alpha_3+\alpha_6+\gamma}\right)\mub_{a_1}^{\alpha_1}\nub_{b_1}^{\alpha_2}\mu_{a_2}^{\alpha_3}\nu_{b_2}^{\alpha_4}\theta_{c_1}^{\alpha_5}\thb_{c_2}^{\alpha_6},
\label{square}
\end{eqnarray}
where 
\begin{eqnarray}
c &=& (-1)^{\alpha_1+\alpha_5}\left(\sqrt{J}\right)^{\alpha_1+\alpha_2+\alpha_3+\alpha_4}h^{\gamma}, \qquad
s_1 = \alpha_1(\gamma+\alpha_5+\alpha_4), 
\nonumber \\
s_2 &=& \gamma^2 + (\alpha_6+\gamma)(\alpha_4+\alpha_5+\alpha_6)+(\alpha_3+\alpha_6+\gamma)\alpha_3 + (\alpha_2+\alpha_3+\alpha_6+\gamma)(\alpha_1+\alpha_2), 
\nonumber \\
\end{eqnarray}
as illustrated in Fig.~\ref{Fig:Tdecom}(b). Before moving on, we mention that $\mu_{a_1,a_2}(\mub_{a_1,a_2})$ are the Grassmann variables used to decompose the left-moving terms $\nn{ij}_{l}$,  $\nu_{b_1,b_2}(\nub_{b_1,b_2})$ correspond to the right-moving terms $\nn{ij}_{r}$, and $\theta_{c_1,c_2}(\thb_{c_1,c_2})$ correspond to the up-moving terms $\nn{ij}_{u}$. This is clearly illustrated in Fig.~\ref{Fig:Tdecom}(a). $\alpha_i$ is the expansion power used in Eq.~(\ref{GrassDecom}), and can be seen as the parity of the auxiliary Grassmann variables. In Eq.~(\ref{Zcomplex}), the integral measure is defined as 
\begin{eqnarray}
\int D\mub D\mu &=& \prod_i\int_{\mub_i\mu_i}, \qquad
\int D\left[\mub, \nub, \thb~\right] \equiv \int D\mub D\mu\int D\nub D\nu\int D\thb D\theta, \nonumber \\
\qquad \int \mt{D}_{\mu\nu\theta} \left[\psi\right]&\equiv& 
\int\mt{D}\psb\mt{D}\psi~ D\left[\mub, \nub, \thb\right].
\end{eqnarray}
It is clear that Eq.~(\ref{Zcomplex}) involves integrals over both bond Grassmann variables $\{\mu, \nu, \theta\}$ and site Grassmann variables $\{\psi\}$, as well as the summation over discrete parity variables $\{\alpha\}$ defined also on the bonds. 

\begin{figure}[htbp]
	\centering
	\includegraphics[height=4.2cm,width=15.2cm]{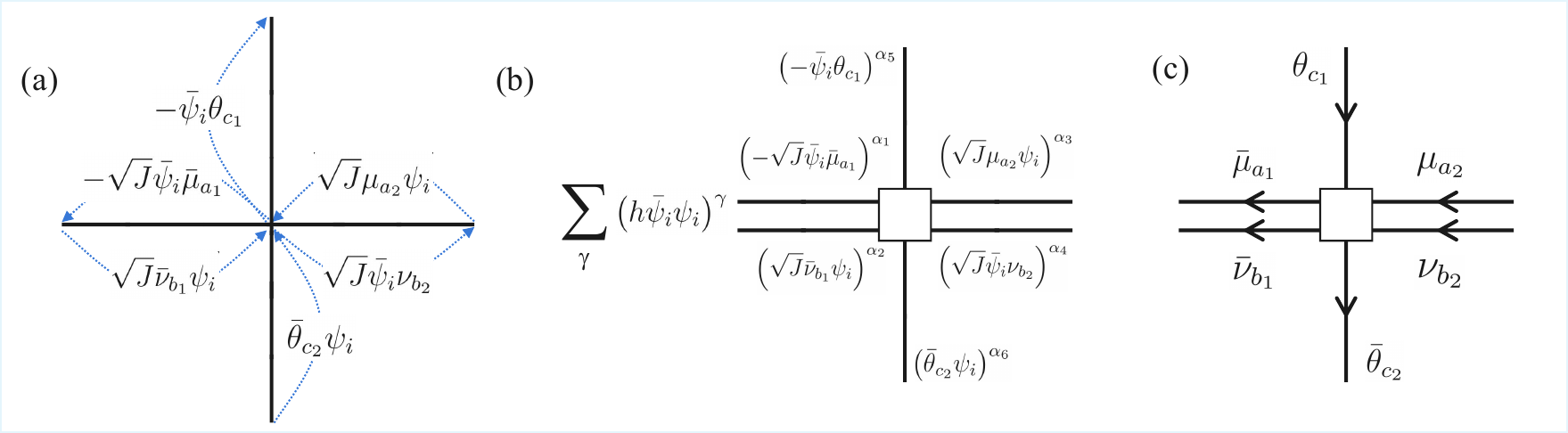}
	\caption{(a) The decomposition of the Grassmann exponents, as expressed in Eq.~(\ref{ExpDecom}), at a given lattice site. (b) Definition of $\square_i$ expressed in Eq.~(\ref{square}), integral of which leads to the (c) local Grassmann tensor $\mt{T}$, as defined by Eq.~(\ref{Tlocal}).}
	\label{Fig:Tdecom}
\end{figure}

Then, if we perform the integral of $\square_i$, we have
\begin{eqnarray}
\int d\psb_id\psi_i~
\square_i = \left(\sum_{\gamma}c(-1)^{s_1+s_2}\right)\delta_{\alpha_1+\alpha_4+\alpha_5,\alpha_2+\alpha_3+\alpha_6}\cdot\mub_{a_1}^{\alpha_1}\nub_{b_1}^{\alpha_2}\mu_{a_2}^{\alpha_3}\nu_{b_2}^{\alpha_4}\theta_{c_1}^{\alpha_5}\thb_{c_2}^{\alpha_6}.
\end{eqnarray}
Define a coefficient tensor $T$ in the following
\begin{eqnarray}
T_{\alpha_1\alpha_2\alpha_3\alpha_4\alpha_5\alpha_6} = \left(\sum_{\gamma}c(-1)^{s_1+s_2}\right)\delta_{\alpha_1+\alpha_4+\alpha_5,\alpha_2+\alpha_3+\alpha_6},
\label{Tcoef}
\end{eqnarray} 
which should be parity-even, clearly demonstrated by the Kronecker-$\delta$ function, then if we change the order of the integrals in Eq.~(\ref{Zcomplex}) and perform the integrals over site variables $\{\psb, \psi\}$ there first, we arrive at
\begin{eqnarray}
Z \approx \int \mt{D}\left[\mub, \nub, \thb~\right]\sum_{\{\alpha\}}\prod_i 
\mt{T}^{(i)}_{\alpha_1\alpha_2\alpha_3\alpha_4\alpha_5\alpha_6},
\label{Z1}
\end{eqnarray}
where the local Grassmann tensor $\mt{T}$ is defined as (see also Fig.~\ref{Fig:Tdecom}(c))
\begin{eqnarray}
\mt{T}_{\alpha_1\alpha_2\alpha_3\alpha_4\alpha_5\alpha_6} = T_{\alpha_1\alpha_2\alpha_3\alpha_4\alpha_5\alpha_6}~\mub_{a_1}^{\alpha_1}\nub_{b_1}^{\alpha_2}\mu_{a_2}^{\alpha_3}\nu_{b_2}^{\alpha_4}\theta_{c_1}^{\alpha_5}\thb_{c_2}^{\alpha_6},
\label{Tlocal}
\end{eqnarray}
with coefficient $T$ given by Eq.~(\ref{Tcoef}). Furthermore, if we use the contraction convention for Grassmann tensors, which includes both summation over all the discrete indices and integral over the associated Grassmann variables, then we finally obtain the desired form of the partition function expressed in Eq.~(\ref{Z1}) to 
\begin{eqnarray}
Z \approx \Tr\prod_{i}\mt{T}^{(i)}_{\alpha_{1}\alpha_{2}\alpha_{3}\alpha_{4}\alpha_{5}\alpha_{6}},
\label{Z0}
\end{eqnarray}
where $\mt{T}$ is a rank-6 Grassmann tensor defined on the $(i,n)$ plane. Note in Eq.~(\ref{Z0}), the index $\alpha$ is the parity used in the expansion as expressed in Eq.~(\ref{GrassDecom}) and can be only 0 or 1, thus the tensor $\mt{T}$ and its coefficient tensor $T$ is of $2^6$ elements satisfying the parity-even condition. However, as discussed in previous sections, the indices of the local tensor can have a much larger dimension than 2, provided the parity structure is maintained during the calculations. Therefore, we generally write Eq.~(\ref{Z0}) as
\begin{eqnarray}
Z \approx \Tr\prod_{i}\mt{T}^{(i)}_{i_1i_2i_3i_4i_5i_6},
\label{PFGrass}
\end{eqnarray}
which is exactly the same formula as the tensor network representation of the partition function of classical spin systems \cite{SRGreview2010}, but now the local tensor is a Grassmann tensor carrying Grassmann numbers.

For the 1D lattice fermion system, there are two indices in Eq.~(\ref{PFGrass}) for each link in the lattice direction, as shown in Fig.~(\ref{Fig:Tdecom}). In practical calculations, one may prefer to deal with a rank-4 Grassmann tensor instead, and this can be achieved by fusing the two Grassmann indices in lattice direction into a single index, as discussed in Sec.~\ref{Sec:fusion}. In this case, we can fuse $i_1$ and $i_2$ to $i_l$, $i_3$ and $i_4$ to $i_r$, i.e.,
\begin{eqnarray} 
\mt{T}_{i_li_ri_5i_6} = \raisebox{-0.46\height}{\includegraphics[width=0.5\textwidth, page=1]{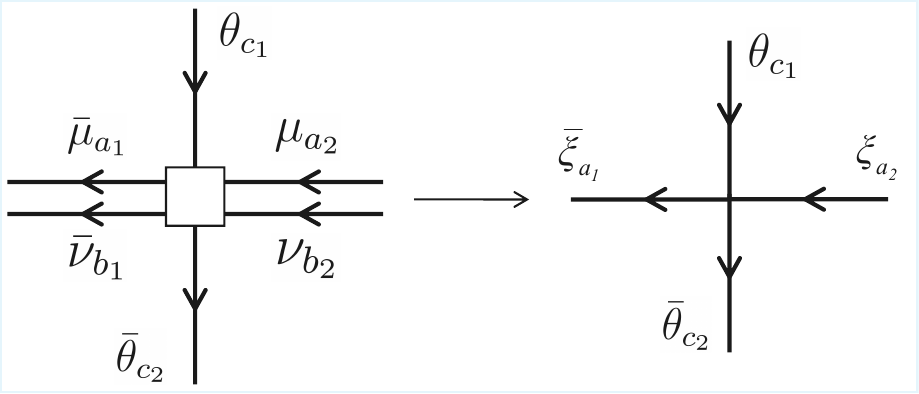}}.
\label{spf5}
\end{eqnarray}

The last point in this subsection we want to mention is the trace of a Grassmann matrix. As we explained in Sec.~\ref{Sec:FermiCoh}, we explicitly represent the correct anti-periodic boundary condition in Eq.~(\ref{ABC}). 
Considering the Grassmann tensor network construction outlined above, the anti‑periodic boundary condition influences the resulting Grassmann tensor through the Grassmann decomposition in Eq.~\eqref{ExpDecom}. More specifically, the factor ${\rm e}^{-\psb_i\psi_j}$ is replaced by ${\rm e}^{\psb_i\psi_j}$ as shown in Eq.~\eqref{ABC}, which introduces an extra minus sign into either $\theta$ or $\bar{\theta}$ in Eq.~(\ref{ExpDecom}). Consequently, when we trace over a Grassmann matrix $\mathcal{M}$ whose indices lie solely along the Trotter direction, we obtain
\begin{eqnarray}
\sum_{i}\mt{M}_{ii} = 
\sum_{i}
\left[
M_{ii}
\int_{\bar{\theta}\theta}
(-\theta)^{p(i)}
\bar{\theta}^{p(i)}
\right]
= \sum_{i}M_{ii}(-1)^{p(i)},
\label{APBC_again}
\end{eqnarray}
which is the trace of the even sector subtracted by that of the odd sector, quite different from the ordinary periodic boundary conditions. The boundary condition probably has negligible effects on the ground-state bulk properties, but it could affect finite-temperature calculations. 

\subsection{Correspondence to the swap-gate and $\mathbb{Z}_{2}$-graded space approaches}

The existing fermionic tensor network formulations are essentially equivalent \cite{RomanEPJB2014}, though the correspondence has seldom been discussed in detail. Here we discuss the relations between the Grassmann formulation \cite{Gu2010Grassmann,Gu2013Efficient} and the other two distinct approaches, i.e., the $\mathbb{Z}_{2}$-graded space approach \cite{Quinten2025Fermionic}, and the fermionic swap-gate method \cite{Corboz2009Fermionic,Corboz2010SimulationPEPS} in the Hamiltonian formalism. 

Firstly, it is straightforward to see that the rule of $\mathbb{Z}_{2}$-graded space has an almost one-to-one correspondence to the Grassmann algebra. As mentioned previously, the $\mathbb{Z}_{2}$-graded vector space $V$ has a direct sum structure $V = V^{0} \oplus V^{1}$ by definition, which naturally divides the basis states $\vert i \rangle$ to the even-parity states with $\vert i \vert = 0$ and odd-parity states with $\vert i \vert = 1$. The symbol $\vert \cdot \vert$ corresponds to the Grassmann parity function $p(\cdot)$. The corresponding basis vector in the dual space $V^{*}$ is denoted as bra state $\langle i \vert$, and plays a similar role as the dual of Grassmann numbers. For the most crucial fermion anticommuting statistics, the tensor product of any two vectors in $\mathbb{Z}_{2}$-graded spaces $V_{1}$ and $V_{2}$ can be swapped following the prescription:    
    \begin{eqnarray} \label{swap}
    V_{1} \otimes V_{2} \rightarrow V_{2} \otimes V_{1} :
    a \otimes b \rightarrow
    b \otimes a \times (-1)^{\vert a \vert\times\vert b \vert},
    \end{eqnarray} 
where $a \in V_1$ and $b \in V_2$, and both of them can be in either bra or ket notation. This rule simply corresponds to the definning property of Grassmann anticommuting relation $\psi_{i}^{p(i)}\psi_{j}^{p(j)} = (-1)^{p(i)p(j)}\psi_{j}^{p(j)}\psi_{i}^{p(i)}$, as expressed in Eq.~(\ref{GAdef}) similary. The canonical contraction map $\mt{C}$ between the basis vectors in $V$ and its dual space $V^{*}$ is introduced as:
    \begin{eqnarray} \label{contr_map}
    \mathcal{C}: V^{*}\otimes V \rightarrow \mathbb{C}: \langle i \vert j \rangle = \delta_{ij}, \qquad \mathcal{C}: V\otimes V^{*} \rightarrow \mathbb{C}: \ket{i}\bra{j} = (-1)^{|i|}\delta_{ij},
    \end{eqnarray}
which is equivalent to the Grassmann integral identity (\ref{FundGA}). The contraction map itself can be regarded as the Grassmann integral measure $\int_{\psb\psi}$, in this sense, the bra $\bra{i}$ corresponds to the dual Grassmann variable $\psi^{p(i)}$, while the ket $\ket{j}$ corresponds to the Grassmann variable $\psb^{p(j)}$. The Grassmann tensor (\ref{grassmanntensor4}) can now be directly translated to fermionic tensors $\mathscr{T}$ living in the tensor product of $\mathbb{Z}_{2}$-graded vector space $V_{1}^{*} \otimes V_{2}^{*} \otimes V_{3} \otimes V_{4}^{*}$: 
    \begin{eqnarray} \label{Z2tensor1}
    \mathscr{T}_{i_{1}j_{1}i_{3}i_{4}} = \sum_{i_1i_2i_3i_4}T_{i_{1}j_{1}i_{3}i_{4}} 
    \bra{i_1}\otimes\bra{i_2}\otimes\ket{i_3}\otimes\bra{i_4}.
    \end{eqnarray}
The even-parity fermionic tensor implies: 
\begin{eqnarray}
\mathscr{T}_{i_{1}j_{1}i_{3}i_{4}} = 0, \qquad \textrm{if}\quad \mmod{|i_1|+|i_2|+|i_3|+|i_4|,2} = 1,
\end{eqnarray}
which is identical to Eq.~(\ref{grassmanntensor3}) for Grassmann tensor. Correspondingly, the conjugated fermionic tensor $\bar{\mathscr{T}}$ living in the dual space of $\left(V_{1}^{*} \otimes V_{2}^{*} \otimes V_{3} \otimes V_{4}^{*}\right)$ is:
    \begin{eqnarray} \label{Z2tensor}
    \bar{\mathscr{T}}_{i_{1}j_{1}i_{3}i_{4}} = \sum_{i_1i_2i_3i_4}T^*_{i_{1}j_{1}i_{3}i_{4}} 
    \ket{i_4}\otimes\bra{i_3}\otimes\ket{i_2}\otimes\ket{i_1},
    \end{eqnarray}
where the dual correspondences of the bra/ket states are arranged in reversed order, and this is the same for the Grassmann tensor, as expressed in Eq.~(\ref{DualDef}).
    
The tensor contraction in this situation can be performed by the map $\mt{C}$. For example, we can contract the second index of a fermionic tensor $\mathscr{A}$ living in $(V_1^*\otimes V_2^*\otimes V_3^*)$ and the first index of $\mathscr{B}$ living in $(V_2\otimes V_4^*)$,     
    \begin{eqnarray}
    \mathcal{C}_{21}(\mathscr{A}\otimes \mathscr{B}) &=&  
    \mathcal{C}_{i_2i'_2} \left(\sum_{i_1i_2i_3}\left(A_{i_1i_2i_3}\bra{i_1}\otimes\bra{i_2}\otimes\bra{i_3}\right) \otimes \sum_{i'_2i_4}\left(B_{i'_2i_4}\ket{i'_2}\otimes\bra{i_4}\right)\right),  \nonumber \\
    &=& \sum_{\{i\}}A_{i_1i_2i_3}B_{i'_2i_4}(-1)^{|i'_2||i_3|}\bra{i_1}\otimes\left[\mathcal{C}_{i_2i'_2}\left(\bra{i_2}\otimes\ket{i'_2}\right)\right]\otimes\bra{i_3}\otimes\bra{i_4}, \nonumber \\
    &=& \sum_{i_1i_3i_4}\left(\sum_{i_2}A_{i_1i_2i_3}B_{i_2i_4}(-1)^{|i_2||i_3|}\right)\bra{i_1}\otimes\bra{i_3}\otimes\bra{i_4}.
    \label{Z2Contract}
    \end{eqnarray}    
%&=& \mathcal{C}_{i_2i'_2} \left(\sum_{\{i\}}A_{i_1i_2i_3}B_{i'_2i_4}\bra{i_1}\otimes\bra{i_2}\otimes\bra{i_3}\otimes\ket{i'_2}\otimes\bra{i_4}\right) \nonumber \\
In Eq.~(\ref{Z2Contract}), the $\mathcal{C}_{21}$ and $\mathcal{C}_{i_2i'_2}$ writes the contraction map between $i_2$ and $i'_2$ explicitly. To obtain the second equal $(=)$, the swap prescription expressed in Eq.~(\ref{swap}) is applied, which contributes $(-1)^{\vert i'_{2} \vert\times \vert i_{3} \vert}$. A $\delta$-function $\delta_{i_2i'_2}$ is derived from the contraction map in the second line, where Eq.~(\ref{contr_map}) is applied. We can see that the Grassmann variables and the graded vectors serve only as placeholders for Fermi statistics, and the numerical calculations rely mainly on the commuting coefficients. The procedure in Eq.~(\ref{Z2Contract}) has a one-to-one correspondence to the rules for Grassmann tensor contractions, as discussed in Sec.~\ref{grassmann_op}. 
    
Secondly, using two typical examples, we show that the fermionic swap-gate approach in the fermionic PEPS method can be derived exactly from Grassmann tensor networks by identifying the two-dimensional graphical projection in the former as a Grassmann tensor network and explicitly performing the Grassmann contractions.

    \begin{eqnarray} \label{swap_gate1}
    \langle
    \hat{O}_{13}
    \rangle
    \equiv
    \langle i_{1}i_{2}i_{3} \vert \hat{O}_{13} \vert j_{1}j_{2}j_{3}
    \rangle
    =
    \raisebox{-0.46\height}{\includegraphics[width=0.19\textwidth, page=1]{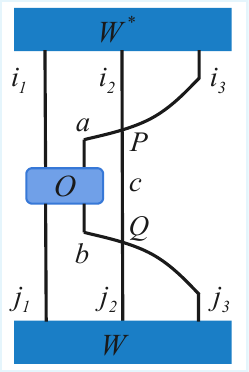}}
    =
    \quad
    \raisebox{-0.46\height}{\includegraphics[width=0.19\textwidth, page=1]{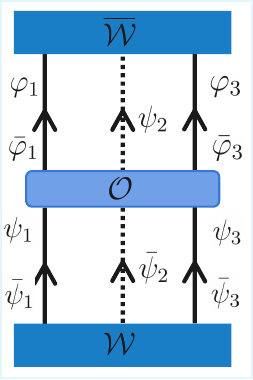}}
    \end{eqnarray}

One example is to compute the matrix element of a general two-site fermionic operator $\hat{O}_{13}$ in the three-site Fock basis, namely $\langle i_{1}i_{2}i_{3} \vert \hat{O}_{13} \vert j_{1} j_{2} j_{3} \rangle$ as shown in Eq.~(\ref{swap_gate1}). The expectation value calculation is a typical example frequently used in tensor network calculations. In Eq.~(\ref{swap_gate1}), the left diagram corresponds to the swap-gate representation, while the right one corresponds to the Grassmann representation of the matrix elements. The dashed line passes through the Grassmann operator $\mathcal{O}$. As already mentioned in the fermionization of MERA \cite{Corboz2009Fermionic}, the fermionic swap gates, $P$ and $Q$ introduced at line crossings in the left figure in Eq.~(\ref{swap_gate1}) and defined as follows, can indeed capture the fermion statistics in calculating $\langle\hat{O}_{13}\rangle$:
    \begin{eqnarray}
    & & \label{swap_gate0a}
    P_{i_2i_3ac} = \delta_{ai_3} \delta_{ci_2} \cdot (-1)^{p(i_2)p(i_3)}, \qquad
    Q_{j_2j_3bc} = \delta_{bj_3} \delta_{cj_2} \cdot (-1)^{p(j_2)p(j_3)}.
    \end{eqnarray} 
       
    Starting from the Grassmann representation of the three-site wave function and two-site operator as shown in the right figure in Eq.~(\ref{swap_gate1})
    \begin{eqnarray}
    \mt{W}_{j_1j_2j_3} &=& W_{j_1j_2j_3}~
    \bar{\psi}_{1}^{p(j_1)}
    \bar{\psi}_{2}^{p(j_2)}
    \bar{\psi}_3^{p(j_3)}, \qquad\bar{\mt{W}}_{i_1j_2i_3} = W^*_{i_1j_2i_3}~
    \varphi_{3}^{p(i_3)}
    \psi_{2}^{p(j_2)}
    \varphi_1^{p(i_1)}, 
    \nonumber \\
    \mt{O}_{i_1i_3j_1j_3} &=& O_{i_1i_3j_1j_3}~
    \bar{\varphi}_{1}^{p(i_1)}
    \bar{\varphi}_{3}^{p(i_3)}
    \psi_{3}^{p(j_3)}
    \psi_{1}^{p(j_1)},
    \end{eqnarray} 
    we have
    \begin{eqnarray}
    \nn{\hat{O}_{13}} = 
    \sum_{\{i,j\}}
    \bar{\mt{W}}_{i_1j_2i_3}
    \mt{O}_{i_1i_3j_1j_3}\mt{W}_{j_1j_2j_3} = \sum_{\{i,j\}}W^*_{i_1j_2i_3}O_{i_1i_3j_1j_3}W_{j_1j_2j_3}\cdot s, 
    \label{ExpEq}
    \end{eqnarray}
    where the Grassmann integral part gives
    \begin{eqnarray}
    s 
    &\equiv& 
    \int_{\psb\psi}
    \int_{\bar{\varphi}\varphi}
    \left(
    \varphi_{3}^{p(i_3)}
    \psi_{2}^{p(j_2)}
    \varphi_1^{p(i_1)}
    \right)
    \left(
    \bar{\varphi}_{1}^{p(i_1)}
    \bar{\varphi}_{3}^{p(i_3)}
    \psi_{3}^{p(j_3)}
    \psi_{1}^{p(j_1)}
    \right)
    \left(
    \bar{\psi}_{1}^{p(j_1)}
    \bar{\psi}_{2}^{p(j_2)}
    \bar{\psi}_3^{p(j_3)}
    \right),
     \nonumber \\
    &=& 
    (-1)^{p(j_2)p(i_3)}
    \cdot
    (-1)^{p(j_2)p(j_3)}.
    \end{eqnarray}
    If we express 
    \begin{eqnarray}
    O_{i_1i_3j_1j_3} 
    &=& \sum_{ab}O_{i_1aj_1b}\delta_{ai_3}\delta_{bj_3} ,\nonumber \\
    \sum_{i_2} &=& \sum_{i_2j_2}\delta_{i_2j_2} = \sum_{i_2j_2}\delta_{i_2j_2}\left(\sum_c\delta_{j_2c}\right) = \sum_{i_2j_2c}\delta_{i_2j_2}\delta_{j_2c} = \sum_{i_2j_2c}\delta_{i_2c}\delta_{j_2c},
    \end{eqnarray}
    then Eq.~(\ref{ExpEq}) becomes
    \begin{eqnarray}
    \nn{\hat{O}_{13}} 
    &=& 
    \sum_{\{i,j\}}\sum_{abc}
    W^*_{i_1j_2i_3}
    W_{j_1j_2j_3}
    \left(O_{i_1aj_1b}\delta_{ai_3}\delta_{bj_3}\right)\left(\delta_{i_2c}\delta_{j_2c}\right)
    (-1)^{p(j_2)p(i_3)}
    \cdot
    (-1)^{p(j_2)p(j_3)},
     \nonumber \\
    &=& 
    \Tr 
    W^*_{i_1j_2i_3}
    W_{j_1j_2j_3}
    O_{i_1aj_1b}\left[\delta_{ai_3}\delta_{i_2c}(-1)^{p(i_2)p(i_3)}\right]
    \left[
    \delta_{bj_3}
    \delta_{j_2c}(-1)^{p(j_2)p(j_3)}
    \right], 
    \nonumber \\
    &=& \Tr W^*_{i_1j_2i_3}
    W_{j_1j_2j_3}
    O_{i_1aj_1b}
    P_{i_2i_3ac}
    Q_{j_2j_3bc},
    \end{eqnarray} 
which is nothing but the left figure in Eq.~(\ref{swap_gate1}) regarded as a bosonic tensor network.

\begin{eqnarray} \label{swap_gate3}
\raisebox{-0.44\height}{\includegraphics[width=0.32\textwidth, page=1]{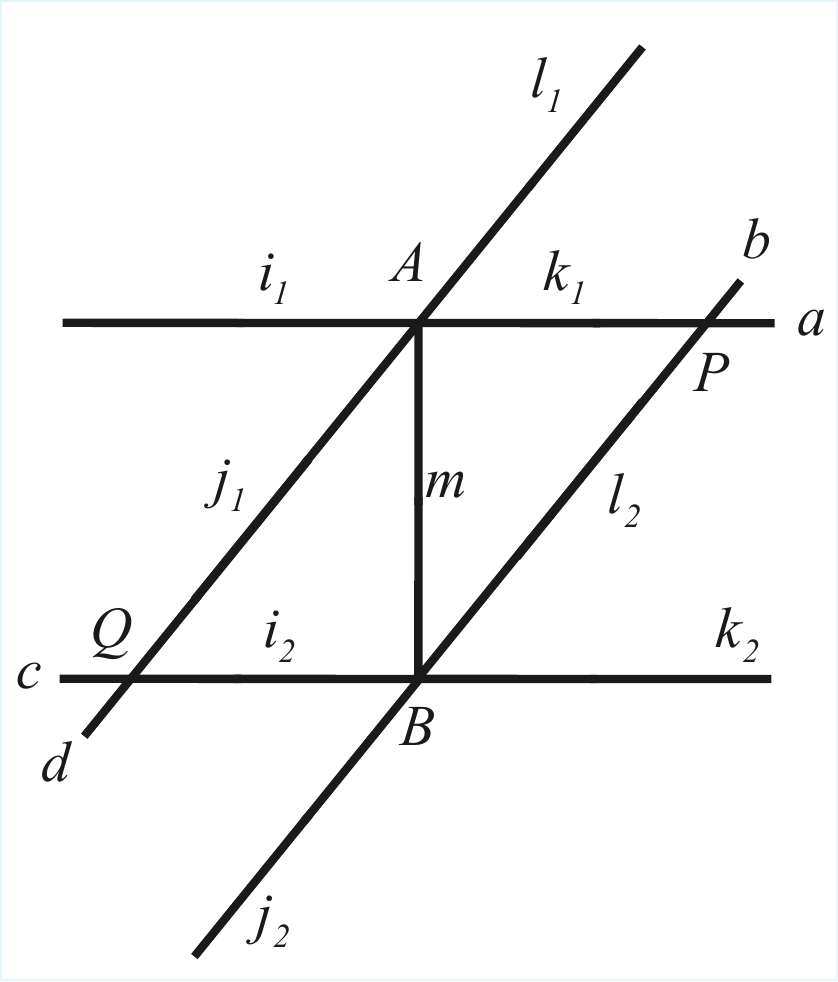}}
=
\quad
\raisebox{-0.44\height}{\includegraphics[width=0.3\textwidth, page=1]{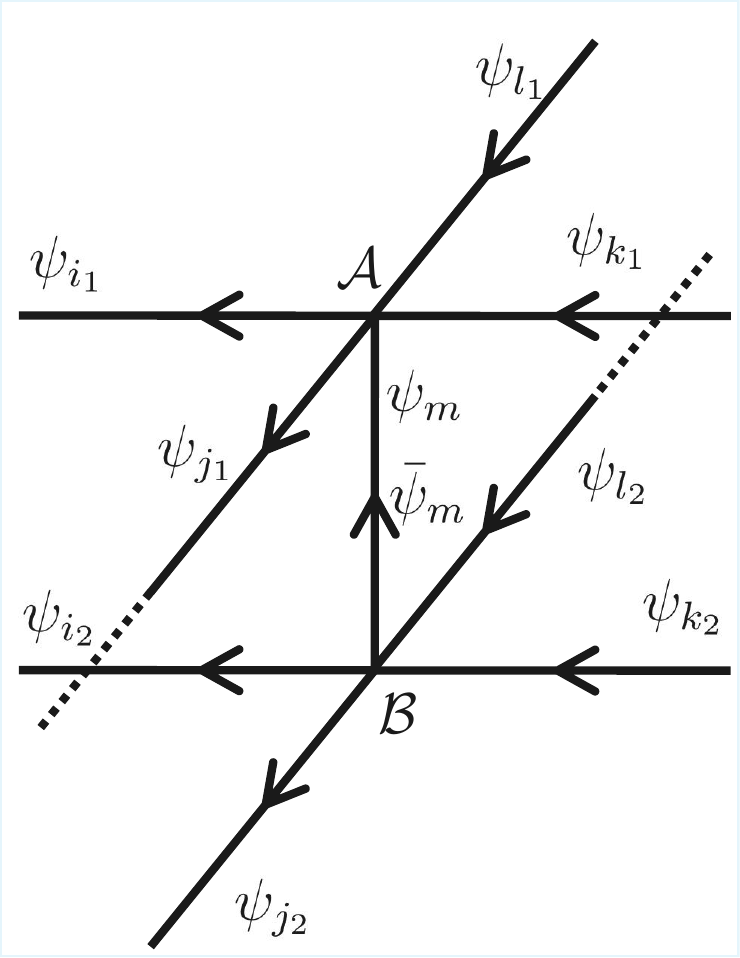}}
\end{eqnarray}

The other example is to construct the reduced tensor from rank-5 PEPS representations of two wave functions on a square lattice, as shown in Eq.~(\ref{swap_gate3}), which is frequently used in the context of tensor network study of a two-dimensional quantum lattice model, such as state norm and overlaps. Similar swap-gate operations are also discussed in previous literature, such as fermionic PEPS in Ref.~\cite{Corboz2010SimulationPEPS}. The two swap gates $P, Q$ are defined similarly as    
\begin{eqnarray} \label{swap_gate4}
P_{abk_1l_2} = \delta_{ak_1}\delta_{bl_2}\cdot(-1)^{p(k_1)p(l_2)}, \qquad 
Q_{cdi_2j_1} = \delta_{ci_2}\delta_{dj_1}\cdot(-1)^{p(i_2)p(j_1)}.
\label{PQdef}
\end{eqnarray}
For simplicity, suppose we have made the wave function a single-sublattice structure, e.g., by combining inequivalent local tensors, though a similar procedure can also be applied to multi-sublattice structures. To see the correspondence between swap-gate and Grassmann formulations more easily, we arrange all the indices of the local tensors in a counter-clockwise manner. Following a similar procedure, we start from the original Grassmann tensor representation,
\begin{eqnarray}
\mt{A}_{i_1j_1mk_1l_1} 
&=& A_{i_1j_1mk_1l_1}~\psixp{i_1}{i_1}\psixp{j_1}{j_1}\psixp{m}{m}\psixp{k_1}{k_1}\psixp{l_1}{l_1}, \nonumber \\
\mt{B}_{i_2j_2k_2l_2m} &=& B_{i_2j_2k_2l_2m}~\psixp{i_2}{i_2}\psixp{j_2}{j_2}\psixp{k_2}{k_2}\psixp{l_2}{l_2}\psbxp{m}{m}.
\end{eqnarray}
Note in the two-dimensional graphical projection, the indices order of $\mt{A}$ and $\mt{B}$ are different, as shown in Eq.~(\ref{swap_gate3}). Then the reduced tensor $\mt{T}$ can be obtained by Grassmann contraction over the index $m$, i.e.,  (indices are still counter-clockwisely arranged)
\begin{eqnarray}
\mt{T}_{i_1i_2j_1j_2j_2k_2k_1l_2l_1} = \sum_{m}\int_{\psb_m\psi_m}A_{i_1j_1mk_1l_1}B_{i_2j_2k_2l_2m} \square,
\label{TEq}
\end{eqnarray}
where the Grassmann product $\square$ is defined as
\begin{eqnarray}
\square &=& \psixp{i_1}{i_1}\psixp{j_1}{j_1}\psixp{m}{m}\psixp{k_1}{k_1}\psixp{l_1}{l_1}\psixp{i_2}{i_2}\psixp{j_2}{j_2}\psixp{k_2}{k_2}\psixp{l_2}{l_2}\psbxp{m}{m}, \nonumber \\
&=& (-1)^s \psixp{i_1}{i_1}\psixp{i_2}{i_2}\psixp{j_1}{j_1}\psixp{j_2}{j_2}\left(\psixp{m}{m}\psbxp{m}{m}\right)\psixp{k_2}{k_2}\psixp{k_1}{k_1}\psixp{l_2}{l_2}\psixp{l_1}{l_1},
\end{eqnarray}
where in the second line the order of the rest of the Grassmann variables is arranged according to the indices of $\mt{T}$, and
\begin{eqnarray}
s =&& p(i_2)\left[p(j_1)+p(m)+p(k_1)+p(l_1)\right] + p(j_2)\left[p(m)+p(k_1)+p(l_1)\right] 
\nonumber \\
&+& 
p(m)\left[p(k_1)+p(l_1)+p(k_2)+p(l_2)\right] \nonumber \\
&+& p(k_2)\left[p(k_1)+p(l_1)\right] + p(l_2)p(l_1).
\label{SignEq}
\end{eqnarray}
Note in fermionic PEPS algorithm, $A$ and $B$ respect $\mathbb{Z}_2$ symmetry, which means
\begin{eqnarray}
\mmod{i_1+j_1+k_1+l_1, 2} = m = \mmod{i_2+j_2+k_2+l_2, 2}.
\end{eqnarray}
Using this fact, and expanding $m$ in the third term in Eq.~(\ref{SignEq}), one can find 
\begin{eqnarray}
\mmod{s, 2} &=& \mmod{s_1+s_2, 2}, \nonumber \\
s_1 &=& p(i_1)+p(j_1)+p(k_1)+p(l_1), \nonumber \\
s_2 &=& p(j_1)p(i_2)+p(k_1)p(l_2),
\end{eqnarray}
which reduces Eq.~(\ref{TEq}) to
\begin{eqnarray}
\mt{T}_{i_1i_2j_1j_2k_2k_1l_2l_1} &=& T_{i_1i_2j_1j_2j_2k_2k_1l_2l_1}~\psixp{i_1}{i_1}\psixp{i_2}{i_2}\psixp{j_1}{j_1}\psixp{j_2}{j_2}\psixp{k_2}{k_2}\psixp{k_1}{k_1}\psixp{l_2}{l_2}\psixp{l_1}{l_1}, 
\nonumber \\
T_{i_1i_2j_1j_2k_2k_1l_2l_1} 
&=& \sum_{m}(-1)^{s_1}(-1)^{s_2}\cdot A_{i_1j_1mk_1l_1}B_{i_2j_2k_2l_2m}.
\label{Tfinal}
\end{eqnarray}
Note that $(-1)^{s_1}$ associate each link of the local tensor $A$ with a sign $(-1)^{p}$, since each bond is shared by two neighboring tensors $A$, thus $(-1)^p$ cancels and in this sense $(-1)^{s_1}$ can be regarded as a $Z_2$ gauge symmetry.  Therefore, we can ignore this sign in Eq.~(\ref{Tfinal}), thus renaming the indices of the Grassmann tensor $T$ such as
\begin{eqnarray}
j_1\rightarrow d, \qquad i_2\rightarrow c, \qquad k_1\rightarrow a, \qquad l_2\rightarrow b,
\end{eqnarray}
can be realized by
\begin{eqnarray}
\mt{T}_{i_1cdj_2k_2abl_1} &=& \sum_{j_1i_2k_1l_2}\mt{T}_{i_1i_2j_1j_2k_2k_1l_2l_1}\delta_{dj_1}\delta_{ci_2}\delta_{ak_1}\delta_{bl_2}, 
\nonumber \\
&=& T_{i_1cdj_2k_2abl_1}~\psixp{i_1}{i_1}\psixp{c}{c}\psixp{d}{d}\psixp{j_2}{j_2}\psixp{k_2}{k_2}\psixp{a}{a}\psixp{b}{b}\psixp{l_1}{l_1},
\label{MTfigure}
\end{eqnarray}
where the coefficient tensor
\begin{eqnarray}
&&T_{i_1cdj_2k_2abl_1} 
\nonumber \\
&=& 
\sum_{j_1i_2k_1l_2,m} \left[A_{i_1j_1mk_1l_1}B_{i_2j_2k_2l_2m}(-1)^{p(j_1)p(i_2)}(-1)^{p(k_1)p(l_2)}\right]\delta_{dj_1}\delta_{ci_2}\delta_{ak_1}\delta_{bl_2}, 
\nonumber \\
&=& \sum_{j_1i_2k_1l_2,m} A_{i_1j_1mk_1l_1}B_{i_2j_2k_2l_2m}\left[(-1)^{p(j_1)p(i_2)}\delta_{dj_1}\delta_{ci_2}\right]\left[(-1)^{p(k_1)p(l_2)}\delta_{ak_1}\delta_{bl_2}\right], \nonumber \\
&=& 
\sum_{j_1i_2k_1l_2,m} A_{i_1j_1mk_1l_1}B_{i_2j_2k_2l_2m}P_{abk_1l_2}Q_{cdi_2j_1}.
\label{Tfigure}
\end{eqnarray}
Eqs.~(\ref{MTfigure}-\ref{Tfigure}) means that we can define two ordinary tensors $P$ and $Q$ as expressed in Eq.~(\ref{PQdef}), and then the coefficient of the local Grassmann tensor arranged in Eq.~(\ref{MTfigure}) can be obtained by contracting the four tensors as done for a bosonic tensor network without extra fermion sign consideration. This is exactly the meaning of the left figure in Eq.~(\ref{swap_gate3}), which is used in the swap-gate approach of fermionic PEPS.

\section{Grassmann tensor network algorithms} \label{grassmann_tn_alg}

The previous sections outlined the essentials of Grassmann tensor operations and their representations of fermionic states and operators. In this section, we focus on the Grassmannization of several typical tensor network methods. While most parts of a given bosonic tensor network algorithm can be readily recycled by replacing ordinary tensor operations with their Grassmann counterparts, certain subtleties must be addressed. These will be elucidated below.

\subsection{Grassmann DMRG}\label{grassmann_dmrg}

\subsubsection{Method} \label{grassmann_dmrg_method}

The DMRG algorithm \cite{WhiteDMRG1992, SchRev2011} is the most widely used tensor network method for one- and two-dimensional quantum lattice systems. Within the MPO–MPS framework, the DMRG can be formulated as a variational optimization problem \cite{SchRev2011,Ostlund1995,Dukelsky1998Equivalence}, on which we focus in this work. 

For systems with fermionic degrees of freedom, a conventional approach is to apply Jordan-Wigner transformations to the fermionic Hamiltonian, thereby mapping it onto a hard-core bosonic system. In this subsection, we present a single-site Grassmann DMRG algorithm. This approach is based on the Grassmann MPS ansatz and the Grassmann MPO representation presented in Sec.~\ref{Sec:GMPS} and Sec.~\ref{Sec:ferop}, respectively, where Fermi statistics is encoded directly through Grassmann variables, thereby eliminating the need for Jordan-Wigner strings.

The objective is to find the ground state $\vert\Psi\rangle$ of a one-dimensional fermionic Hamiltonian $\hat{H}$ by minimizing the energy expectation value subject to the normalization constraint $\langle \Psi 
\vert \Psi \rangle = 1$. The corresponding loss function $f$ is:
\begin{eqnarray} \label{gdmrg2}
f = \langle \Psi \vert \hat{H} \vert \Psi \rangle
- 
\lambda \langle \Psi 
\vert \Psi \rangle,
\end{eqnarray}
where $\lambda$ is a Lagrange multiplier. To proceed, we express Eq.~(\ref{gdmrg2}) in the Grassmann coherent-state representation by inserting resolutions of the identity in terms of overcomplete sets. This yields
\begin{eqnarray} \label{gdmrg3}
f 
&=&
\sum_{i_{1}\cdots i_{m}j_{1}\cdots j_{m}}
\mt{\gt{C}}_{i_{1}\cdots i_{m}}
\mt{H}_{i_{1}\cdots i_{m}j_{1}\cdots j_{m}}
\mt{C}_{j_{1}\cdots j_{m}}
-
\lambda
\sum_{i_{1}\cdots i_{m}}
\mt{\gt{C}}_{i_{1}\cdots i_{m}}
\mt{C}_{i_{1}\cdots i_{m}},
\end{eqnarray}
where $m$ denotes the number of lattice sites. In Eq.~(\ref{gdmrg3}), the rank-$m$ Grassmann tensor $\mt{C}$ is parametrized as a Grassmann MPS via Eqs.~(\ref{GMPS}-\ref{GMPS_def}), with $\gt{\mt{C}}$ being its dual tensor. Similarly, the rank-$2m$ tensor $\mt{H}$ is represented as a Grassmann MPO, given by the $m$-site generalization of Eq.~(\ref{GMPO1}). As is evident from Eq.~(\ref{gdmrg3}), within the coherent-state representation, the Grassmann integrals effectively  couple the physical indices of the Grassmann MPO to those of the Grassmann MPS. 

The derivative of the loss function $f$ with respect to a single dual Grassmann tensor $\gt{\mt{T}}^{(n)}$ corresponds, diagrammatically, to removing (or \textit{punching a hole} at) the $n$-th lattice site in the corresponding Grassmann tensor network, as illustrated schematically below
\begin{eqnarray} \label{gdmrg4}
\dfrac{\partial f}{\partial \bar{\mt{T}}^{(n)}} = 
\raisebox{-0.45\height}{\includegraphics[width=0.78\textwidth, page=1]{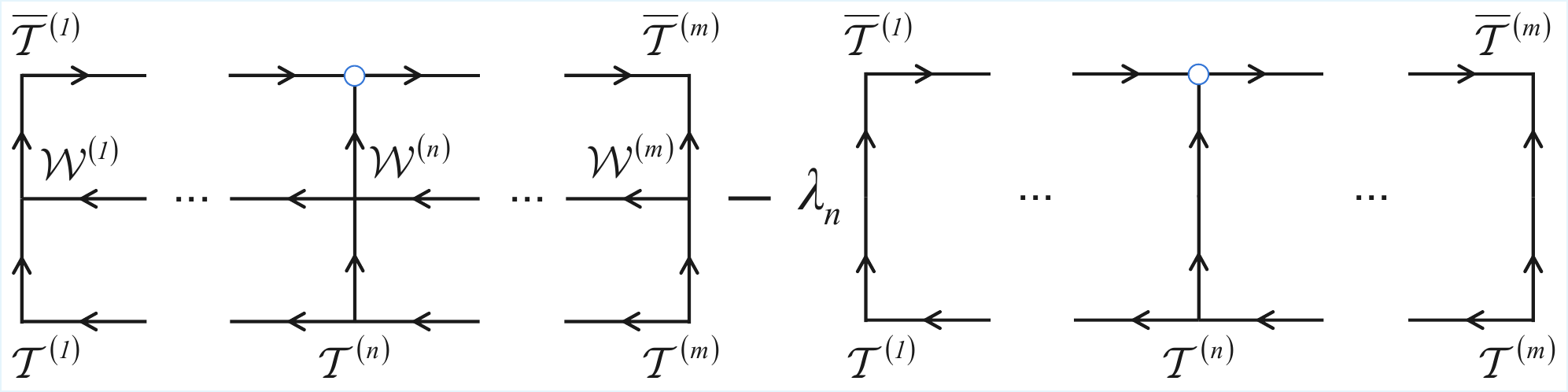}}.
\end{eqnarray}
To convert the tensor network in Eq.~(\ref{gdmrg4}) into a more local form, we exploit the site-canonical representation of the $m$-site Grassmann MPS.  This canonical form is readily obtained by applying successive Grassmann QR or LQ decompositions to a given Grassmann MPS, see e.g., section 4.4 in Ref.~\cite{SchRev2011} for constructing the canonical form of a bosonic MPS. For a Grassmann MPS in its canonical form centered at the $n$-th site, we have
\begin{eqnarray} \label{dmrg7}
\mt{C}_{i_{1}i_{2}\cdots i_{m}}
&=& \nonumber
\sum_{\{k\}}~
\mt{L}^{(1)}_{k_{1}}[i_{1}]
\mt{L}^{(2)}_{k_{1}k_{2}}[i_{2}]
\cdots
\mt{M}^{(n)}_{k_{n-1}k_{n}}[i_{n}]
\cdots
\mt{R}^{(m-1)}_{k_{m-2}k_{m-1}}[i_{m-1}]
\mt{R}^{(m)}_{k_{m-1}}[i_{m}],
\\ 
&=&
\raisebox{-0.4\height}{\includegraphics[width=0.6\textwidth, page=1]{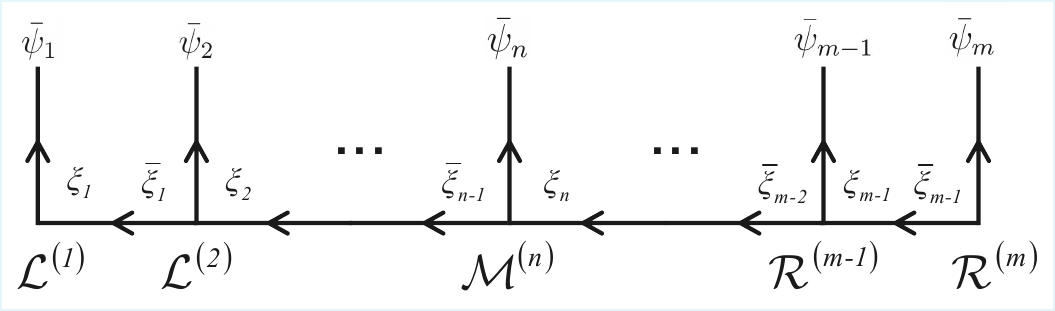}}.
\end{eqnarray}
The $\mt{L}$ tensors in Eq.~(\ref{dmrg7}) satisfy the Grassmann generalization of the left-canonical condition (see also Fig.~\ref{cano_cond}(a))
\begin{eqnarray}
\label{gdmrg6a}
\sum_{i_{1}}
\mt{\gt{L}}^{(1)}_{l_{1}}[i_{1}]
\mt{L}^{(1)}_{k_{1}}[i_{1}]
&=&
\mt{I}_{l_{1}k_{1}},
\\ \label{gdmrg6b}
\sum_{i_{a}}
\sum_{k_{a-1}}
\sum_{l_{a-1}}
\mt{\gt{L}}^{(a)}_{l_{a-1}l_{a}}[i_{a}]
\mt{I}_{l_{a-1}k_{a-1}}
\mt{L}^{(a)}_{k_{a-1}k_{a}}[i_{a}]
&=&
\mt{I}_{l_{a}k_{a}},~\textrm{for}\quad 1<a<n.
\end{eqnarray}
Similarly, the $\mt{R}$ tensors in Eq.~(\ref{dmrg7}) satisfy the Grassmann generalization of the right-canonical condition (see also Fig.~\ref{cano_cond}(b))
\begin{eqnarray}\label{gdmrg6c}
\sum_{i_{m}}
\mt{R}^{(m)}_{k_{m-1}}[i_{m}]
\mt{\gt{R}}^{(m)}_{l_{m-1}}[i_{m}]
&=&
\mt{P}_{k_{m-1}l_{m-1}},
\\ 
\label{gdmrg6d}
\sum_{i_{b}}
\sum_{k_{b}}
\sum_{l_{b}}
\mt{R}^{(b)}_{k_{b-1}k_{b}}[i_{b}]
\mt{P}_{k_{b}l_{b}}
\mt{\gt{R}}^{(b)}_{l_{b-1}l_{b}}[i_{b}]
&=&
\mt{P}_{k_{b-1}l_{b-1}},~\textrm{for}\quad n<a<m.
\end{eqnarray}
The Grassmann identity matrix $\mt{I}$ and the Grassmann parity matrix $\mt{P}$ are given by 
\begin{eqnarray}
& &
\mt{I}_{l_{a}k_{a}}
=
\delta_{l_{a}k_{a}}
\bar{\zeta}_{a}^{p(l_{a})}
\xi_{a}^{p(k_{a})},
\quad
\mt{P}_{k_{b}l_{b}}
=
(-1)^{p(l_{b})}
\delta_{k_{b}l_{b}}
\bar{\xi}_{b}^{p(k_{b})}
\zeta_{b}^{p(l_{b})}.
\end{eqnarray}
Compared to the Grassmann identity matrix $\mt{I}$, the coefficient tensor of $\mt{P}$ incorporates an additional sign factor that modifies the conventional Kronecker-$\delta$ function. The relations in Eqs.~(\ref{gdmrg6a}-\ref{gdmrg6d}) can be verified directly by evaluating the corresponding Grassmann integrals. This verification relies on the fact that the coefficient tensors of $\mt{L}$ and $\mt{R}$ satisfy the ordinary (non-Grassmann) left- and right-canonical conditions—a property that follows directly from their derivation via Grassmann QR and LQ decompositions.

\begin{figure}[htbp]
	\centering
	\includegraphics[height=6.2cm,width=15.0cm]{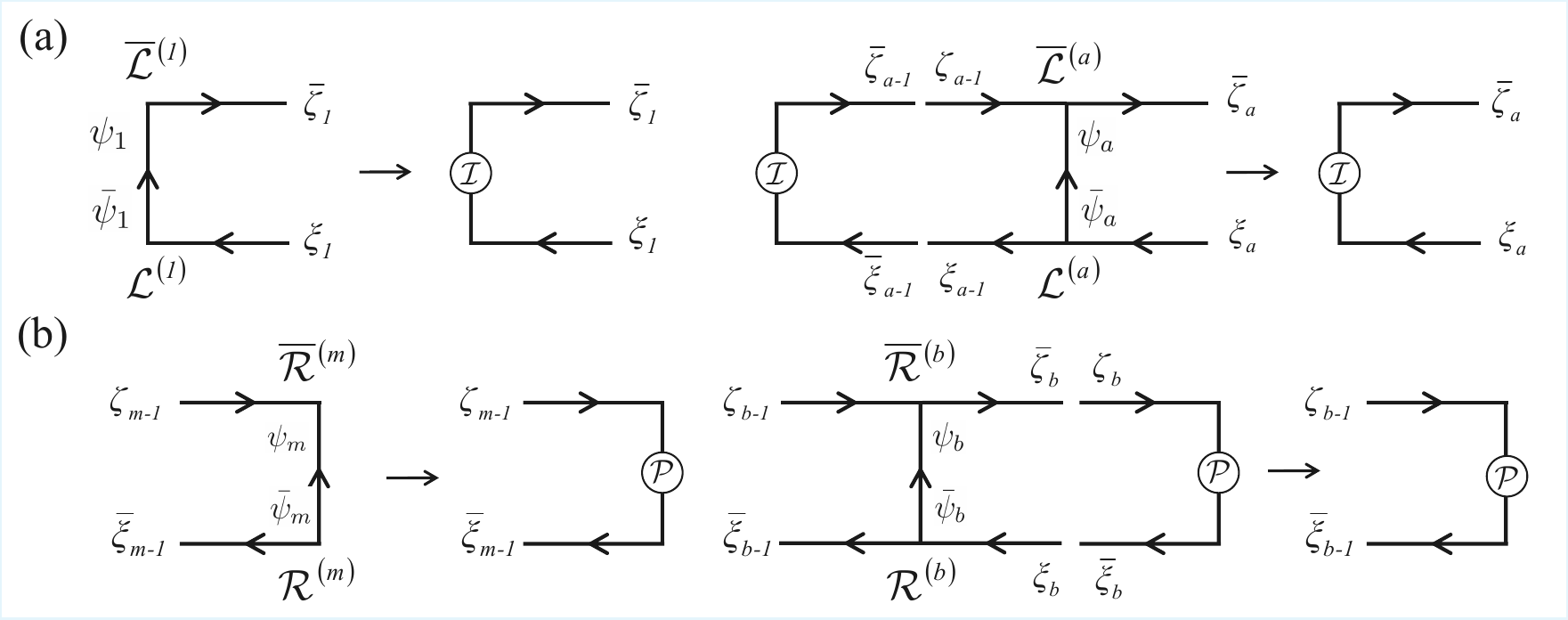}
	\caption{The Grassmann generalization of the (a) left-canonical condition and (b) right-canonical condiction.}
	\label{cano_cond}
\end{figure}

The site-canonical form introduced in Eq.~(\ref{dmrg7}) effectively channels the leftward and rightward environment information directly to that center site. Consequently, the wavefunction norm and the expectation value of local operators can be evaluated as contractions of a significantly reduced tensor network. By maintaining the Grassmann MPS in the site-canonical form throughout the optimization—via sequential Grassmann QR or LQ decompositions—the minimization of the loss function in Eq.~(\ref{gdmrg3}) yields a set of generalized eigenvalue equations. These equations are defined via Grassmann tensor contractions, given explicitly by
\begin{eqnarray}
\raisebox{-0.54\height}{\includegraphics[width=0.72\textwidth, page=1]{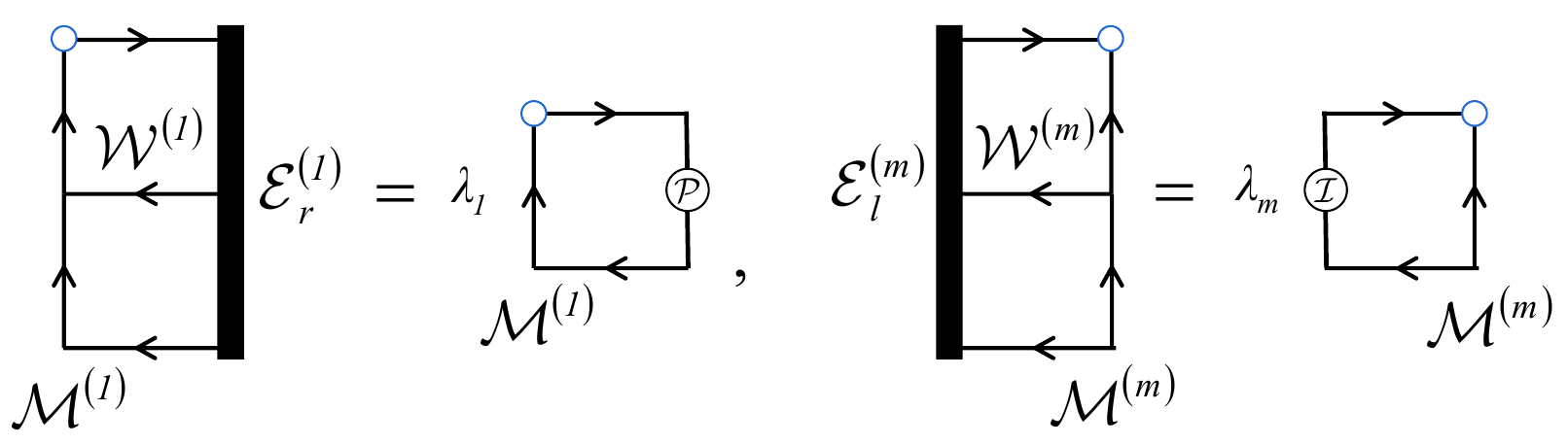}}
\label{gdmrg8a}
\\ 
\label{gdmrg8b}
\raisebox{-0.54\height}{\includegraphics[width=0.48\textwidth, page=1]{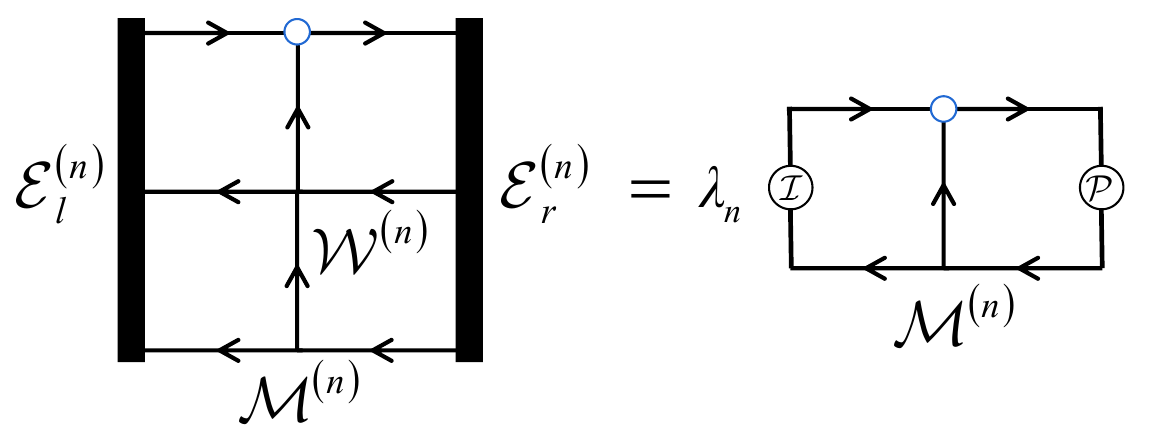}},
\quad
n = 2, \cdots, m-1
\end{eqnarray}
where the optimized tensors $\mt{M}^{(i)}$ correspond to the eigenstate with the lowest eigenvalue, both of which are available through the Lanczos method \cite{Lanczos1950Iteration}. The rank-3 Grassmann tensors $\mt{E}_{l}^{(n)}$ and $\mt{E}_{r}^{(n)}$ represent the effective left and right environments for the $n$-th site, respectively, which are constructed dynamically during the optimization sweep. To be precise,  consider a left-to-right sweep in which we optimize the $a$-th tensor for $a=1, \cdots, m-1$. Once the local tensor $\mt{M}^{(a)}$ is optimized according to Eq.~(\ref{gdmrg8a}) and (\ref{gdmrg8b}), we decompose it via a Grassmann QR factorization, $\mt{M}^{(a)} = \mt{L}^{(a)}\mt{X}^{(a)}$, where the physical index and left virtual index (if exist) of $\mt{M}^{a}$ are contained in $\mt{L}^{(a)}$. The left environment for the subsequent site $a+1$ is then updated according to:
\begin{eqnarray}
\raisebox{-0.42\height}{\includegraphics[width=0.72\textwidth, page=1]{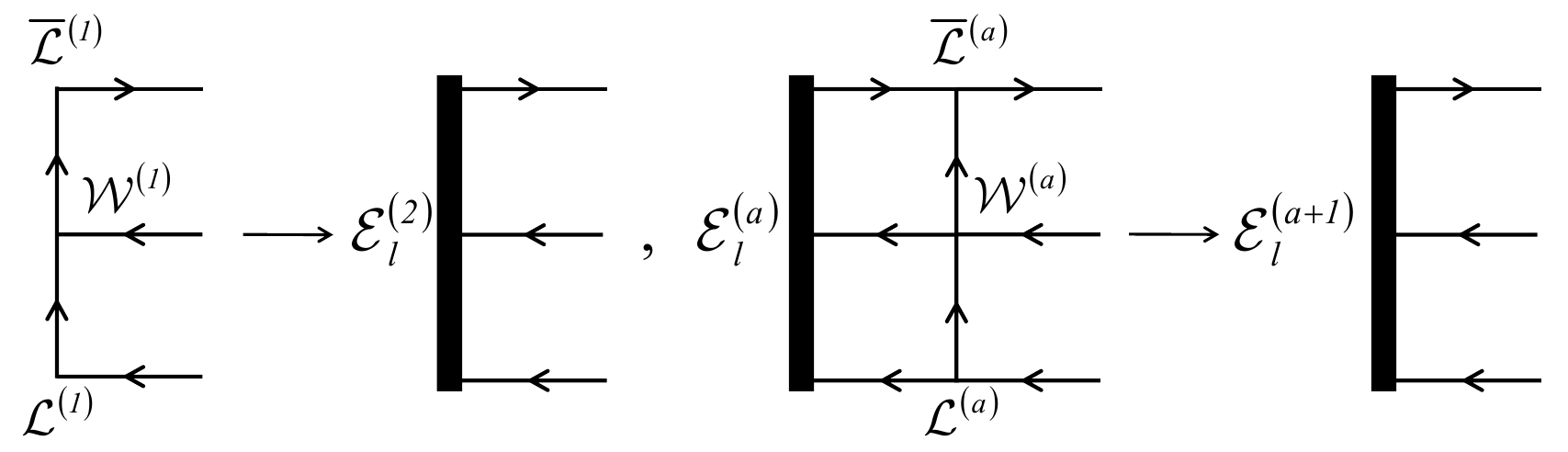}}
\end{eqnarray}
The tensor $\mt{X}^{(a)}$ is then contracted with $\mt{R}^{(a+1)}$ to produce $\mt{M}^{(a+1)} = \mt{X}^{(a)}\mt{R}^{(a+1)}$, shifting the canonical center from $a$ to $a+1$. Analogously, during a right-to-left sweep, the optimized tensor at site $b$ ($b = 2, \cdots, m$) is decomposed as $\mt{M}^{(b)} = \mt{Y}^{(b)}\mt{R}^{(b)}$, where the physical index and the virtual index on the right-side (if exist) are contained in $\mt{R}^{(b)}$, after which the right environment for site $b-1$ is updated as:
\begin{eqnarray}
\raisebox{-0.42\height}{\includegraphics[width=0.72\textwidth, page=1]{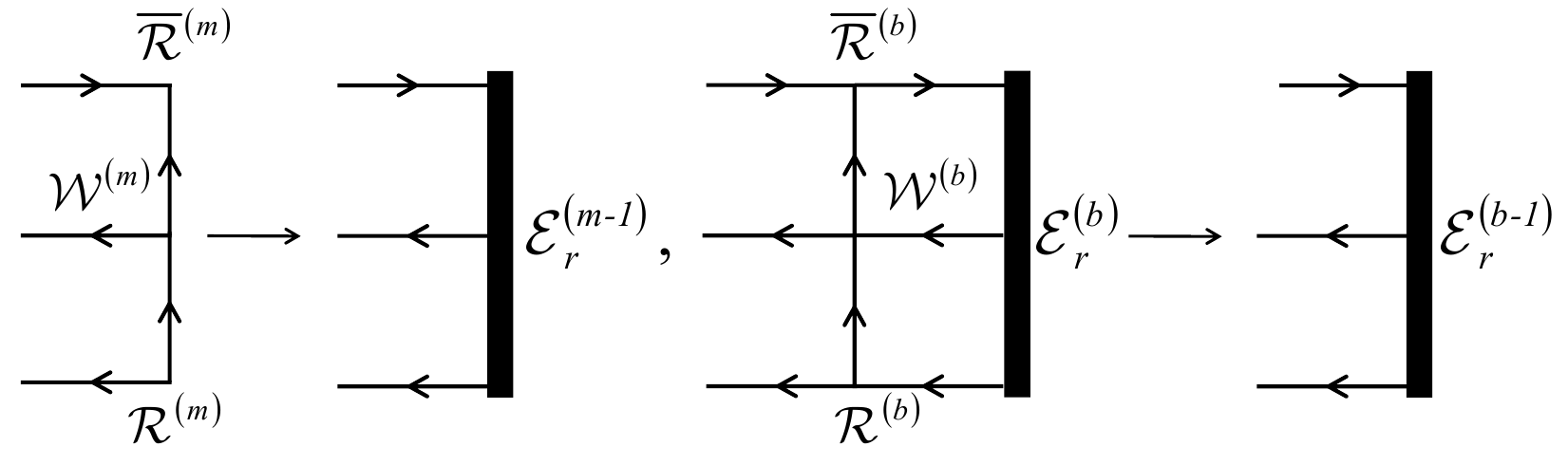}}
\end{eqnarray}
where $\mt{Y}^{(b)}$ combines with $\mt{L}^{(b-1)}$, leading to $\mt{M}^{(b-1)}$.

The single-site Grassmann DMRG algorithm can be extended directly to a two-site variant. We note that a fermionic two-site DMRG within a $\mathbb{Z}_{2}$-graded space framework has recently been proposed \cite{Quinten2025Fermionic}. In that formalism, a parity matrix—identical to the one in Eqs.~(\ref{gdmrg6c}) and (\ref{gdmrg6d}) — also appears as a key ingredient distinguishing it from a bosonic DMRG. Moreover, the Grassmann MPO/MPS formalism opens the door to exploring other tensor network algorithms. Promising candidates include the time-dependent variational principle (TDVP) \cite{Haegeman2011TDVP,Laurens2019Tangent}, whose Grassmann version can be readily adapted within the present framework.

\subsubsection{Example}

The Hubbard model \cite{Hubbard1963Electron,Arovas2022Hubbard,Qin2022Hubbard} is a paradigmatic model in condensed matter physics, displaying rich phenomena such as metal-insulator transitions, unconventional superconductivity, and various charge- and spin-ordered phases. In one dimension, it is exactly solvable via the Bethe Ansatz \cite{Lieb1968Absence} and therefore serves as an excellent platform for benchmarking numerical methods.

We apply the single-site Grassmann DMRG algorithm described in Sec.~\ref{grassmann_dmrg_method} to compute the ground-state wave function of the one-dimensional Hubbard chain, focusing on the particle-hole-symmetric case. Starting from the $m$-site generalization of the Hubbard Hamiltonian in Eq.~(\ref{3shubbard}) augmented by an additional chemical potential term $-\mu\hat{n}$, we enforce particle-hole symmetry by adding a constant $U/4$ and setting the chemical potential to $\mu = U/2$. The resulting Hamiltonian reads:
\begin{eqnarray} \label{half_fill_H}
\hat{H} = 
-t
\sum_{i=1}^{m-1}
\sum_{\sigma=\uparrow,\downarrow}
\left( 
\hat{c}_{i,\sigma}^{\dagger}\hat{c}_{i+1,\sigma} + 
h.c. \right) 
+ 
\dfrac{U}{4}
\sum_{i=1}^{m}
\left(
1-2\hat{n}_{i\uparrow}
\right)
\left(
1-2\hat{n}_{i\downarrow}
\right).
\end{eqnarray}

\begin{figure}[htbp]
	\centering 
	\begin{minipage}{0.48\textwidth}
		\centering 
		\begin{overpic}[height=4.4cm,width=6.7cm]{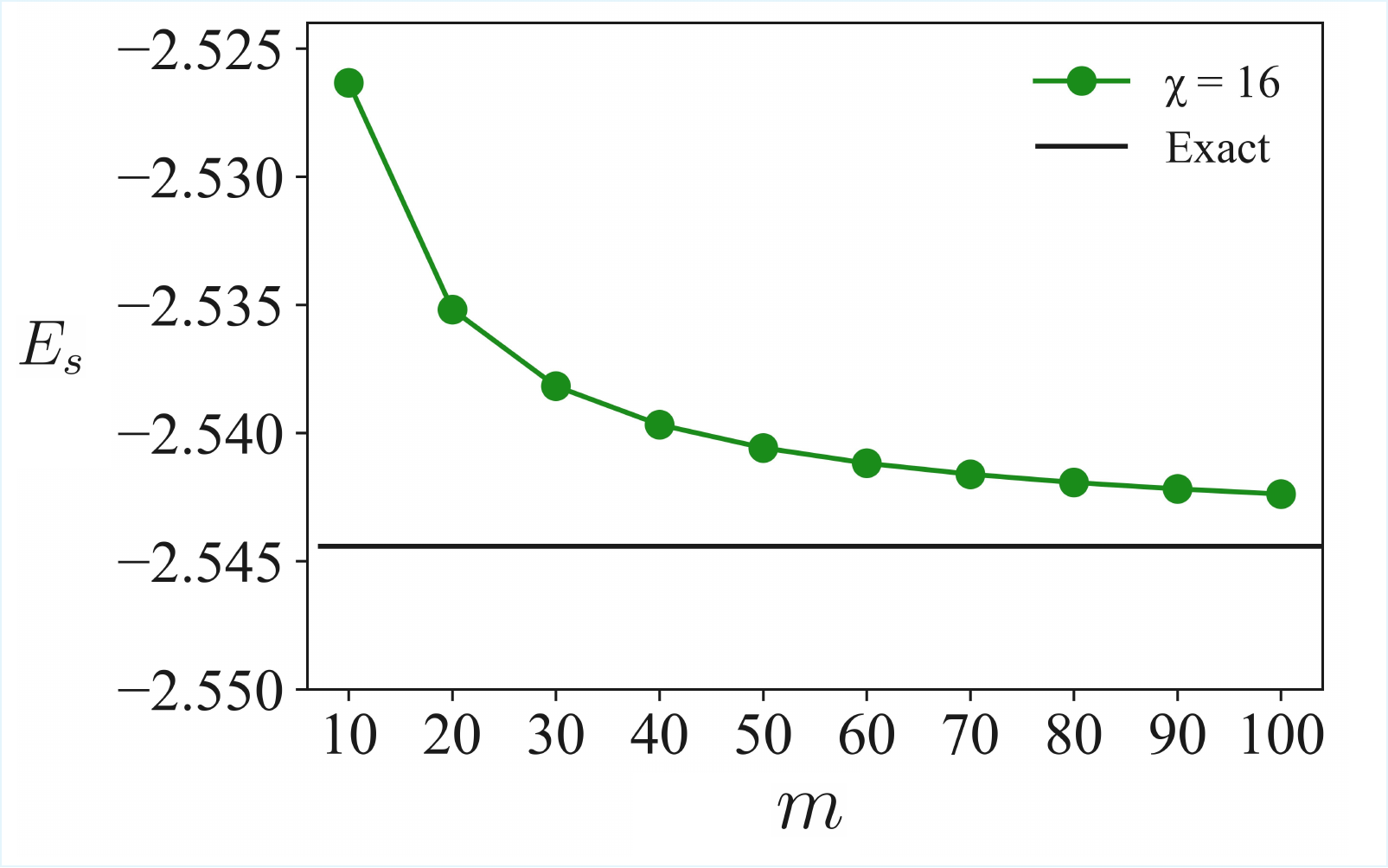}
			\put(1,68){(a)}
		\end{overpic}
		\label{gdmrg_hubbard1}
	\end{minipage}
	\begin{minipage}{0.48\textwidth}
		\centering 
		\begin{overpic}[height=4.4cm,width=6.7cm]{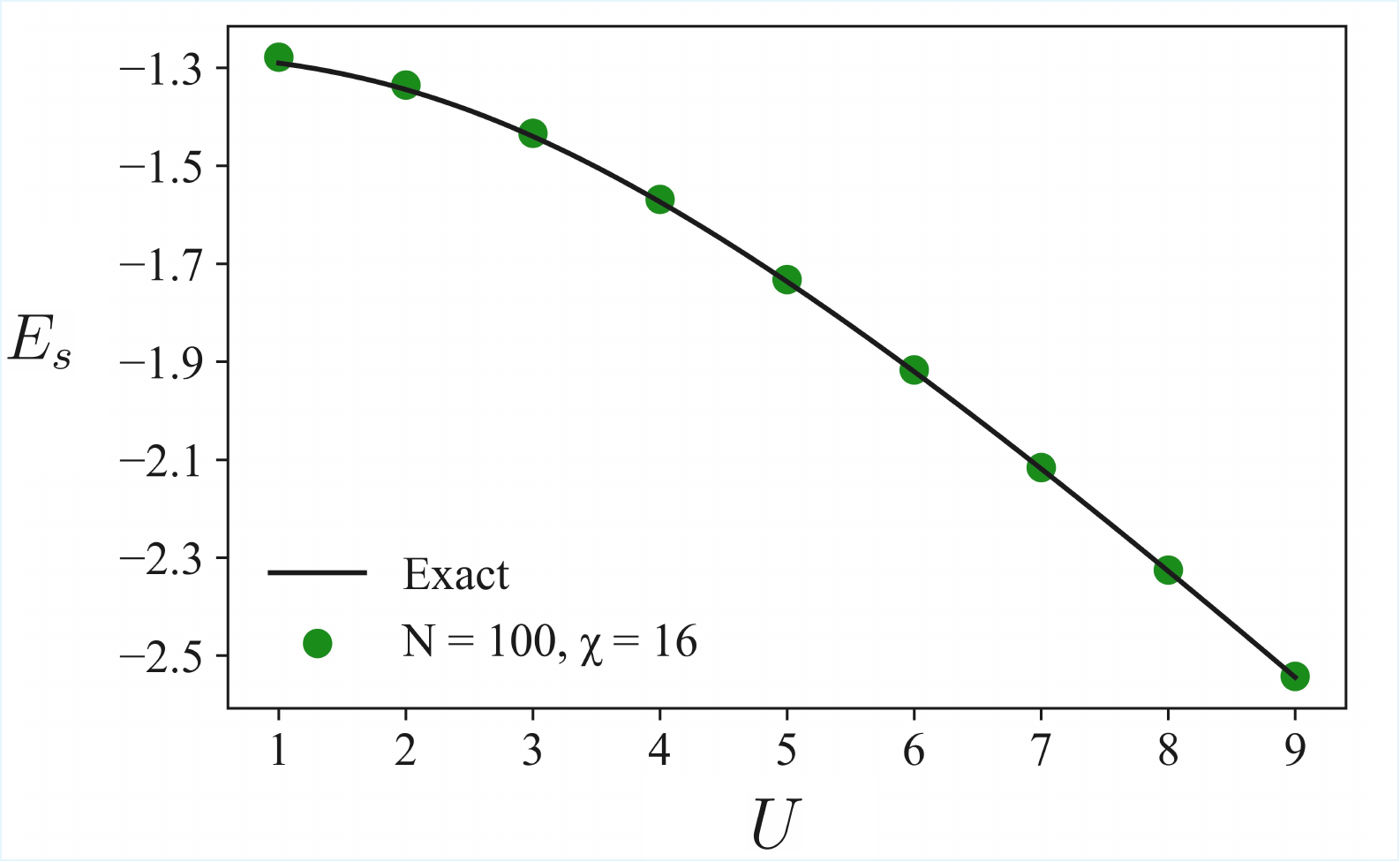}
			\put(-6,66){(b)}
		\end{overpic}
		\label{gdmrg_hubbard}
	\end{minipage}
	\begin{minipage}{0.48\textwidth}
		\centering 
		\begin{overpic}[height=4.4cm,width=6.7cm]{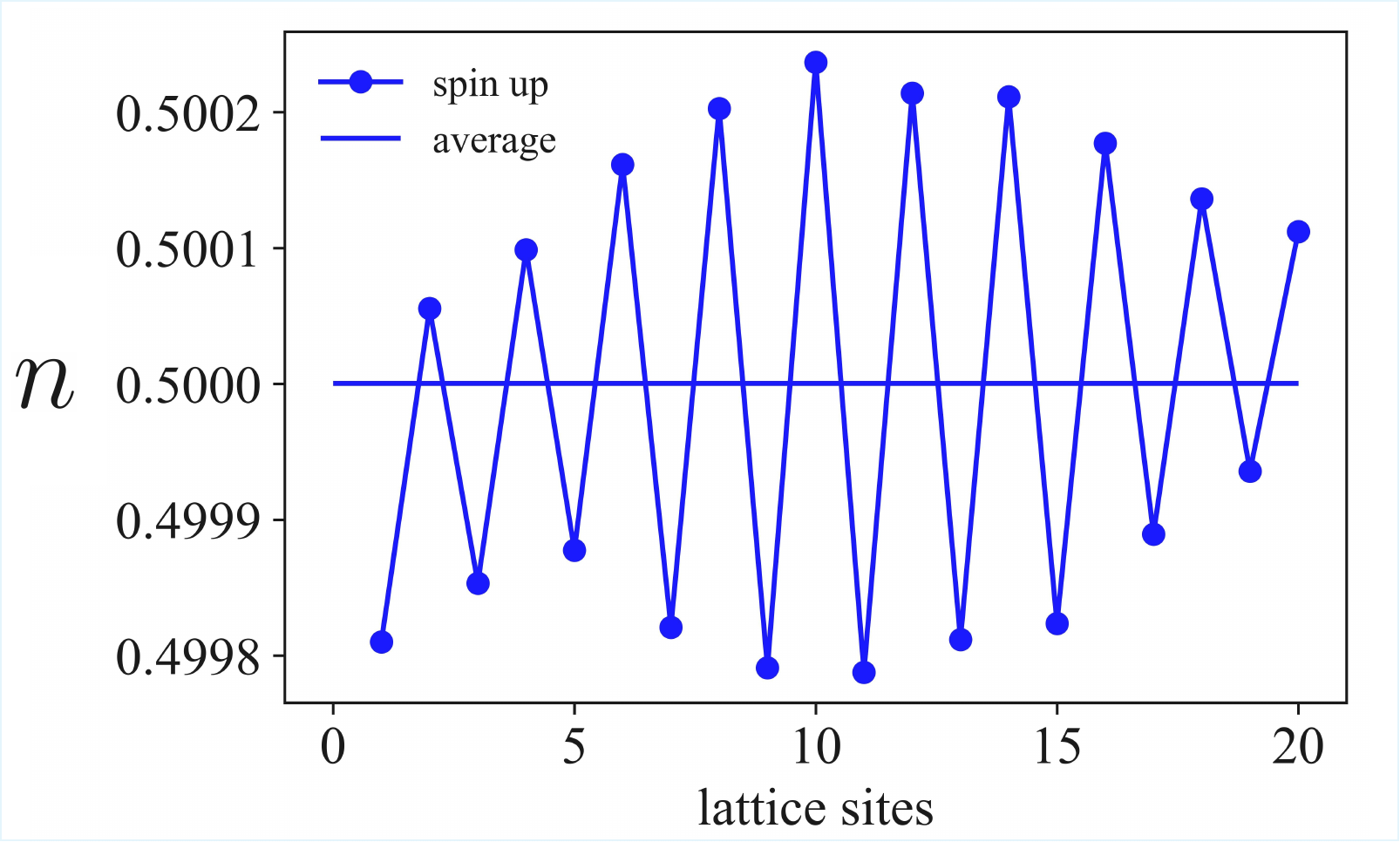}
			\put(3,66){(c)}
		\end{overpic}
		\label{dmrg_Hubbard_1a}
	\end{minipage}
	\begin{minipage}{0.48\textwidth}
		\centering 
		\begin{overpic}[height=4.4cm,width=6.7cm]{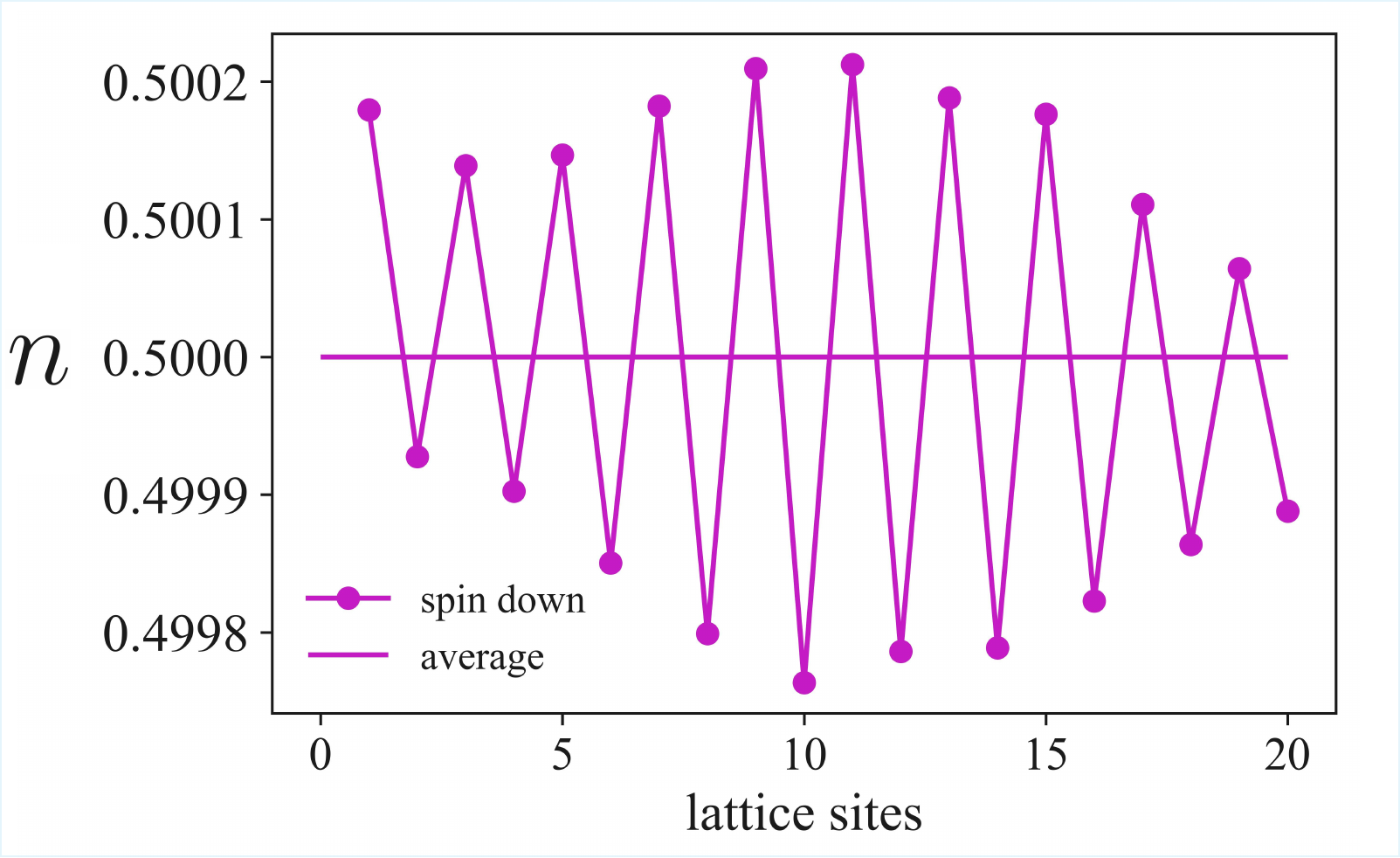}
			\put(-6,66){(d)}
		\end{overpic}
		\label{dmrg_Hubbard_1b}
	\end{minipage}
	\caption{Ground-state properties of the half-filled Hubbard model given by Eq.~(\ref{half_fill_H}) are computed via Grassmann DMRG, with Grassmann MPS bond dimension $\chi=16$. The Grassmann DMRG results are compared with the exact Bethe Ansatz solutions given by Eq.~(\ref{bethe}). (a) Finite-size convergence of the ground-state energy per site at on-site interaction strength $U=9$. (b) Ground-state energy per site as a function of the on-site interaction strength $U$. (c, d) Site-resolved particle number distributions for spin-up (c) and spin-down (d) electrons on a chain of length $m=20$.}
	\label{gdmrg_example1}
\end{figure}

The Grassmann MPS is initialized randomly with a moderate bond dimension $\chi=16$. The Grassmann MPO is constructed using Eqs.~(\ref{W1}-\ref{W23}), (\ref{GMPO_def}), and (\ref{GMPO1}) with minor modifications to incorporate the chemical potential term. The MPS tensors are then optimized sequentially by solving the generalized eigenvalue problem given in Eqs.~(\ref{gdmrg8a}-\ref{gdmrg8b}), with convergence monitored via the relative change in the corresponding eigenvalues. Once the ground-state wave function has reached convergence, we compute the site-resolved energy expectation values $\langle \hat{H}_{i} \rangle$ and the particle number densities $\langle \hat{n}_{i} \rangle$ through:
\begin{eqnarray} \label{gdmrg_example}
\langle \hat{H}_{i} \rangle =
\raisebox{-0.39\height}{\includegraphics[width=0.36\textwidth, page=1]{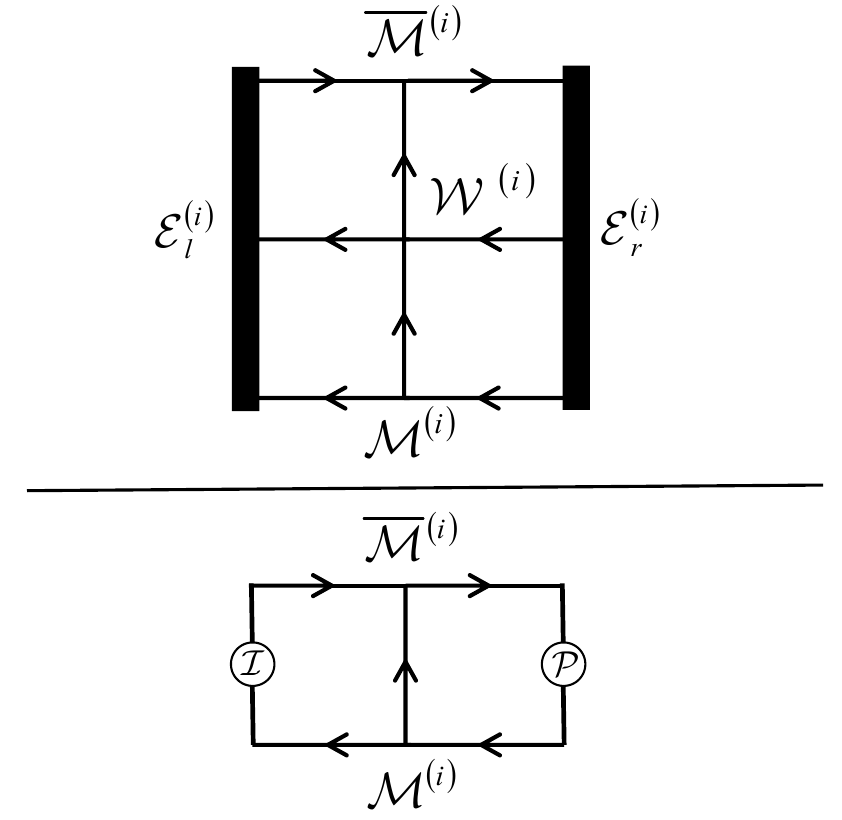}},
\quad
\langle \hat{n}_{i} \rangle =
\raisebox{-0.38\height}{\includegraphics[width=0.34\textwidth, page=1]{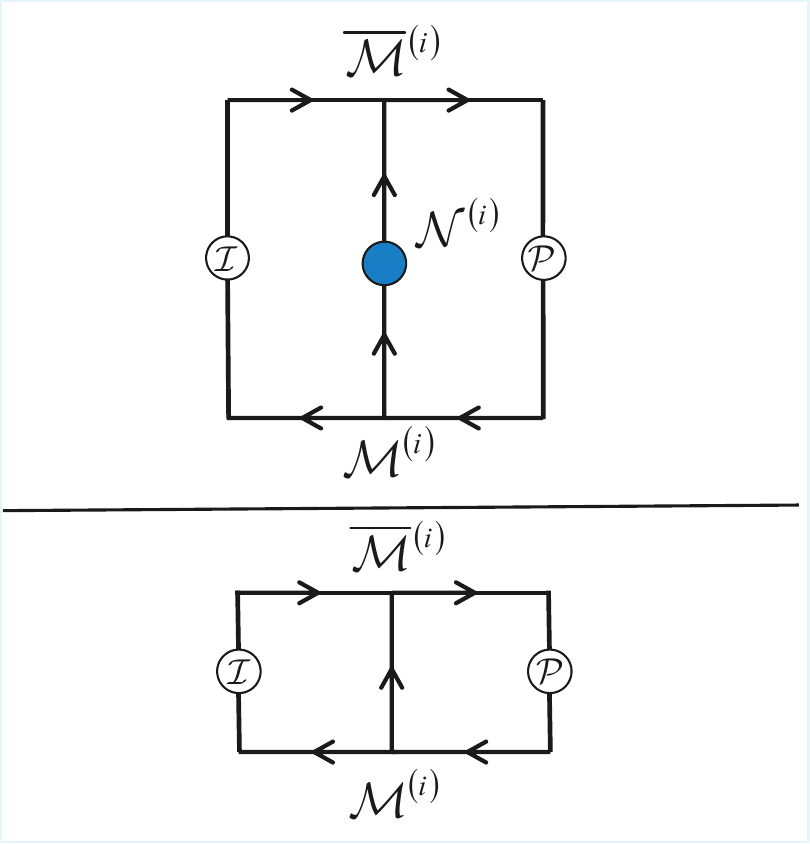}},
\end{eqnarray}
where $\mt{W}^{(i)}$ and $\mt{N}^{(i)}$ are the $i$-th Grassmann MPO tensor and Grassmann tensor representation of the occupation number operator $\hat{n}_{i}$, respectively. The Grassmann MPS is in its site-canonical form centered at site $i$, and $\mt{E}_{l}^{i}$ and $\mt{E}_{r}^{i}$ denote the effective left and right environments at that site. Imposing the right-canonical conditions given in Eqs.~(\ref{gdmrg6c}-\ref{gdmrg6d}) thereby introduces the Grassmann parity matrix $\mt{P}$ in Eq.~(\ref{gdmrg_example}). 

For the half-filled Hubbard Hamiltonian in Eq.~(\ref{half_fill_H}), the exact ground-state energy density in the thermodynamic limit is given by~\cite{Essler20051DHubbard}:
\begin{eqnarray} \label{bethe}
E_{s}^{\textrm{exact}}(U) = -\frac{U}{4} - 4\int_{0}^{\infty}\frac{d\omega}{\omega}\frac{J_{0}(\omega)J_{1}(\omega)}{1+\exp(\omega/2U)},
\end{eqnarray}
where $J_{0}$ and $J_{1}$ denote the zeroth-order and first-order Bessel functions of the first kind, respectively. As shown in Fig.~\ref{gdmrg_example1}(a), the ground-state energy per site $E_{s}$ for $U=9.0$ converges toward the exact Bethe Ansatz solution as the chain length $m$ increases. Furthermore,  Fig.~\ref{gdmrg_example1}(b) demonstrates that the computed energies remain in excellent agreement with the exact values across a broad range of on-site interaction strengths. The particle-number distributions for spin-up and spin-down electrons in a chain of length $m=20$ are displayed in Fig.~\ref{gdmrg_example1}(c) and \ref{gdmrg_example1}(d), respectively. The average occupancy for each spin species is approximately 0.5 electrons per site, in accordance with the expectation for a particle-hole-symmetric Hamiltonian.

\subsection{Grassmann TEBD} \label{GTEBD}

\subsubsection{Method} \label{GTEBD_method}

We present a Grassmann formulation of the classic TEBD algorithm \cite{VidalTEBD2007,RomanTEBD2008}. This formulation enables an accurate approximation of a two-dimensional Grassmann tensor network by solving the dominant eigenvalue problem of a one-dimensional transfer matrix. Consider such a transfer matrix $\mt{M}$ with an $m$-site unit cell (or $m$ sublattices), where each local site corresponds to a rank-4 Grassmann tensor of bond dimension $D$:
\begin{eqnarray}
\mt{M} = 
\raisebox{-0.46\height}{\includegraphics[width=0.56\textwidth, page=1]{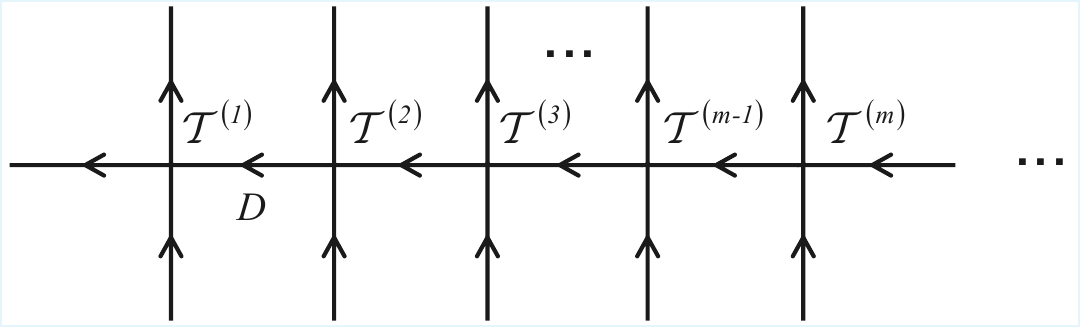}}.
\end{eqnarray}
To approximate the dominant eigenvector of the transfer matrix $\mt{M}$, we employ the power method. The initial state $\mt{C}$ is prepared as an infinite translation-invariant Grassmann MPS with an $m$-site unit cell,
\begin{eqnarray}
\mt{C}
=
\raisebox{-0.18\height}{\includegraphics[width=0.5\textwidth, page=1]{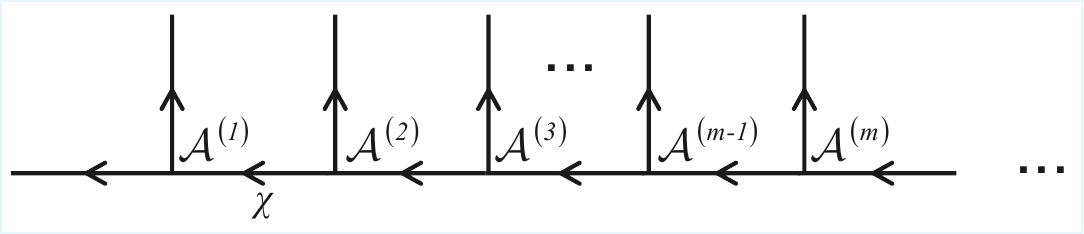}},
\end{eqnarray} 
where $\chi$ is its virtual bond dimension. We iteratively apply the transfer matrix $\mt{M}$ to $\mt{C}$ by contracting their shared physical indices, yielding an updated Grassmann MPS with a virtual dimension $D\chi$. Specifically, each rank-3 local Grassmann tensor, denoted as $\tilde{\mt{A}}^{(n)}$, is constructed in two steps: first, contracting $\mt{T}^{(n)}$ with $\mt{A}^{(n)}$; second, fusing the virtual indices of the resulting rank-5 tensor:
\begin{eqnarray} 
\raisebox{-0.34\height}{\includegraphics[width=0.64\textwidth, page=1]{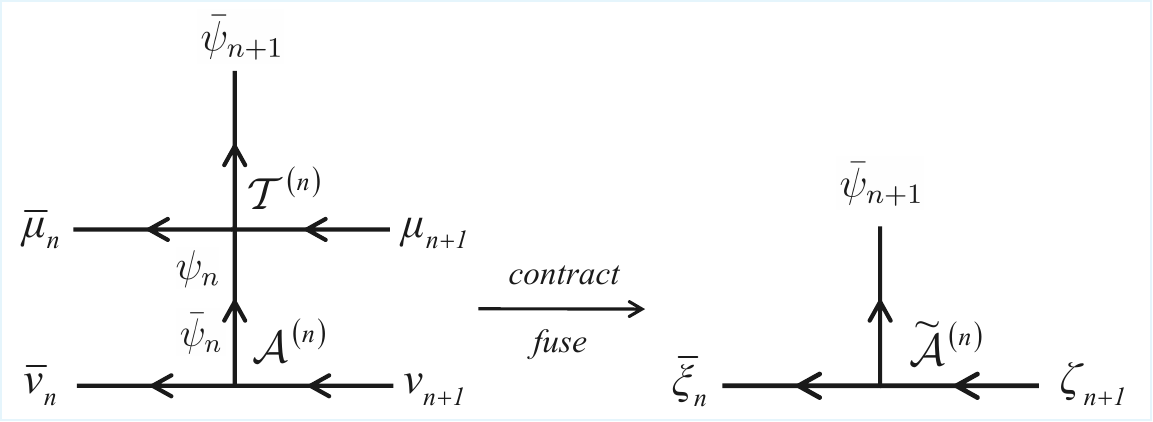}}.
\end{eqnarray} 
To be more precise, we introduce $\mt{T}^{(n)}$ and $\mt{A}^{(n)}$ as
\begin{eqnarray} 
\mt{T}^{(n)}_{k_{n}k_{n+1}i_{n+1}i_{n}}
&=&
T^{(n)}_{k_{n}k_{n+1}i_{n+1}i_{n}}
\bar{\mu}_{n}^{p(k_{n})}
\mu_{n+1}^{p(k_{n+1})}
\bar{\psi}_{n+1}^{p(i_{n+1})}
\psi_{n}^{p(i_{n})},
\label{Eq:tebd_T}
\\ 
\mt{A}^{(n)}_{l_{n}j_{n}l_{n+1}}
&=&
A^{(n)}_{l_{n}j_{n}l_{n+1}}
\bar{\nu}_{n}^{p(l_{n})}
\bar{\psi}_{n}^{p(j_{n})}
\nu_{n+1}^{p(l_{n+1})}.
\label{Eq:tebd_A}
\end{eqnarray}
The $\tilde{\mt{A}}^{(n)}$ tensor is obtained from: 
\begin{eqnarray} \label{gtebd1}
\tilde{\mt{A}}^{(n)}_{
	r_{n}i_{n+1}r_{n+1}
}
=
\tilde{\mt{A}}^{(n)}_{
	(k_{n}l_{n})i_{n+1}(k_{n+1}l_{n+1})
}
= 
\sum_{i_{n}}
\mt{T}^{(n)}_{k_{n}k_{n+1}i_{n+1}i_{n}}
\mt{A}^{(n)}_{l_{n}i_{n}l_{n+1}}.
\end{eqnarray}
In the above equation, the second equal sign  corresponds to a Grassmann contraction, while the first one corresponds to a Grassmann fusion. As discussed in Sec.~\ref{Sec:fusion}, the Grassmann fusion requires a proper definition of composite Grassmann variables
\begin{eqnarray} 
\label{gtebd1a}
\bar{\xi}_{n}^{p(r_{n})} 
\equiv 
\bar{\nu}_{n}^{p(l_{n})}
\bar{\mu}_{n}^{p(k_{n})},
\quad
\zeta_{n+1}^{p(r_{n+1})} 
\equiv  
\mu_{n+1}^{p(k_{n+1})}
\nu_{n+1}^{p(l_{n+1})}.
\end{eqnarray}
The coefficient tensor corresponding to  $\tilde{\mt{A}}^{(n)}$ can be computed:
\begin{eqnarray} 
\tilde{A}^{(n)}_{r_{n}i_{n+1}r_{n+1}}
&=& \label{gtebd3}
\tilde{A}^{(n)}_{
	(k_{n}l_{n})i_{n+1}(k_{n+1}k_{n+1})
}
=
\sum_{i_{n}}
T^{(n)}_{k_{n}k_{n+1}i_{n+1}i_{n}}
A^{(n)}_{l_{n}i_{n}l_{n+1}}
\times
(-1)^{s},
\\ 
s 
&=& \nonumber
p(i_{n})
\times 
p(l_{n})
+
p(l_{n})
\times
\left[
p(k_{n}) + p(k_{n+1}) + p(i_{n+1})
\right]
+
p(i_{n+1}) \times p(k_{n+1})
\\ \label{gtebd3a}
&=&
p(i_{n+1}) \times p(k_{n+1}).
\end{eqnarray}
In the first line of Eq.~(\ref{gtebd3a}), the sign factor $(-1)^{p(i_{n})\times p(l_{n})}$ originates from switching the position of $\psi_{n}$ and $\bar{\nu}_{n}$ to perform the integration over $\psi_{n}$ and $\bar{\psi}_{n}$. The additional sign factors arise from the subsequent permutation of the remaining Grassmann variables
\begin{eqnarray} \nonumber
\left(
\bar{\mu}_{n}^{p(k_{n})}
\mu_{n+1}^{p(k_{n+1})}
\bar{\psi}_{n+1}^{p(i_{n+1})}
\right)
\left(
\bar{\nu}_{n}^{p(l_{n})}
\nu_{n+1}^{p(l_{n+1})}
\right)
\longrightarrow
\left(
\bar{\nu}_{n}^{p(l_{n})}
\bar{\mu}_{n}^{p(k_{n})}
\right)
\bar{\psi}_{n+1}^{p(i_{n+1})}
\left(
\mu_{n+1}^{p(k_{n+1})}
\nu_{n+1}^{p(l_{n+1})}
\right).
\end{eqnarray} 
The Grassmann-even condition of $\mt{T}^{(n)}$ is once again exploited to simplify the sign factor from the first line to the second line in Eq.~(\ref{gtebd3a}).

We seek to optimally truncate the updated Grassmann MPS, composed of local Grassmann tensors $\tilde{\mt{A}}^{(n)}$ for $n = 1, \cdots, m$. This is achieved by inserting a pair of Grassmann isometric matrices $\mt{P}$ and $\mt{Q}$ at each link, as shown below
\begin{eqnarray} \label{gtebd7}
\raisebox{-0.44\height}{\includegraphics[width=0.9\textwidth, page=1]{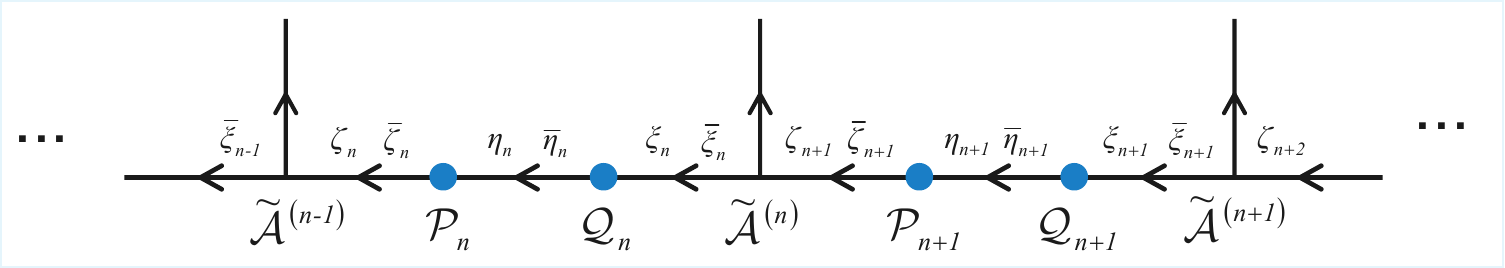}},
\end{eqnarray}
where the subscripts $n$ in $\mt{P}_{n}$ and $\mt{Q}_{n}$ indicate the $n$-th link. A key observation is that, in Eq.~(\ref{gtebd7}), all Grassmann variables to be integrated out are in properly adjacent positions, thereby satisfying the ordering for Grassmann integrals. Consequently, each Grassmann contraction reduces directly to a contraction of corresponding coefficient tensors, without introducing any additional sign factors. In other words, the Grassmann structure simplifies entirely, leaving only an ordinary tensor network contraction at the level of coefficients:
\begin{eqnarray}
\cdots
\tilde{A}^{(n-1)}
\left(
P_{n}
Q_{n}
\right)
\tilde{A}^{(n)}
\left(
P_{n+1}
Q_{n+1}
\right)
\tilde{A}^{(n+1)}
\cdots
\end{eqnarray}
Therefore, the problem reduces to constructing standard isometric matrices $P_{n}$ and $Q_{n}$ that optimally truncate the infinite MPS formed by coefficient tensors $\tilde{A}^{(n)}$ defined in Eq.~(\ref{gtebd3}). This truncation is accomplished via the well-established canonicalization procedure for infinite MPS, an essential technique outlined in Ref.~\cite{RomanTEBD2008}, which we will briefly review below. 

The goal of canonicalization is to apply a set of unitary transformations $\{P_{n}, Q_{n}\}$ that map the local tensors $\{\tilde{A}^{(n)}\}$ to a new set $\{B^{(n)}\}$, such that $\{B^{(n)}\}$ satisfy the left- and right- canonical conditions
\begin{eqnarray}
\raisebox{-0.36\height}{\includegraphics[width=0.8\textwidth, page=1]{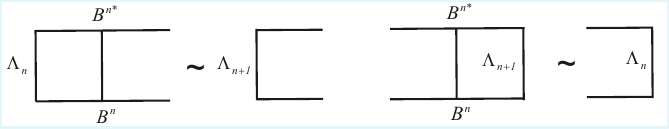}},
\label{infcano}
\end{eqnarray}
where $\{\Lambda_{n}\}$ are semi-positive diagonal matrices placed on each link. Essentially, the canonical form satisfying Eq.~(\ref{infcano}) realizes a Schmidt decomposition on each link, thus $\{\Lambda_{n}\}$ captures the bipartite entanglement entropy across the lattice. The required transformations $\{P_{n}, Q_{n}\}$ and diagonal matrices $\{\Lambda_{n}\}$ can be systematically obtained from block reduced matrices $N^{(n)}$, which are constructed from the infinite MPS through
\begin{eqnarray}
N^{(n)} \equiv 
\raisebox{-0.42\height}{\includegraphics[width=0.54\textwidth, page=1]{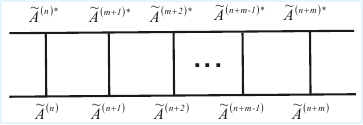}}.
\end{eqnarray}
The left and right dominant eigenvectors of  $N^{(n)}$, denoted as $X_{n}$ and $Y_{n}$ respectively, are then computed and factorized:
\begin{eqnarray}
X_{n}N^{(n)} 
&=& \nonumber
\lambda  X_{n}, \quad
X_{n} = (L_{n})^\dagger L_{n},
\\
N^{(n)}Y_{n} 
&=& \lambda Y_{n}, \quad
Y_{n} = R_{n}(R_{n})^\dagger, \quad \textrm{for}\quad 1 \leq n \leq m.
\end{eqnarray}
The canonical transformation matrices $\{P_{n}, Q_{n}\}$ for the $n$-th link are given by: 
\begin{eqnarray} \label{PQ}
P_{n} = R_{n} V_{n}\left(\Lambda_{n}\right)^{-1/2},
\quad
Q_{n} = 
\left(\Lambda_{n}\right)^{-1/2}
(U_n)^{\dagger}
L_{n},
\end{eqnarray}
where $U_{n}$, $V_{n}$ and $\Lambda_{n}$ are obtained via the decomposition $L_{n}R_{n} = U_{n}\Lambda_{n}(V_{n})^{\dagger}$. The column(row) index of $P$($Q$) can then be truncated by retaining the largest $\chi$ singular values in $\Lambda_{n}$. With the help of $P_{n}$ and $Q_{n}$ in Eq.~(\ref{PQ}), one can obtain the renormalized tensors
\begin{eqnarray}
\mt{B}^{(n)}_{l_{n}i_{n}l_{n+1}}
=
\sum_{r_{n}r_{n+1}}
(\mt{Q}_{n})_{l_{n}r_{n}}
\mt{\tilde{A}}^{(n)}_{r_{n}i_{n}r_{n+1}}
(\mt{P}_{n+1})_{r_{n+1}l_{n+1}},
\end{eqnarray}
where the coefficient tensor of $\mt{B}^{(n)}$ is given by $B^{(n)} = Q_{n}\tilde{A}^{(n)}P_{n+1}$. The leading eigenvector of $\mt{M}$ can be obtained by repeating the canonicalization procedure above until convergence is attained. The Grassmann tensor network, denoted as $Z$, is then evaluated \cite{RomanTEBD2008} via:
\begin{eqnarray} \label{gtebd_Z}
Z = 
\raisebox{-0.43\height}{\includegraphics[width=0.6\textwidth, page=1]{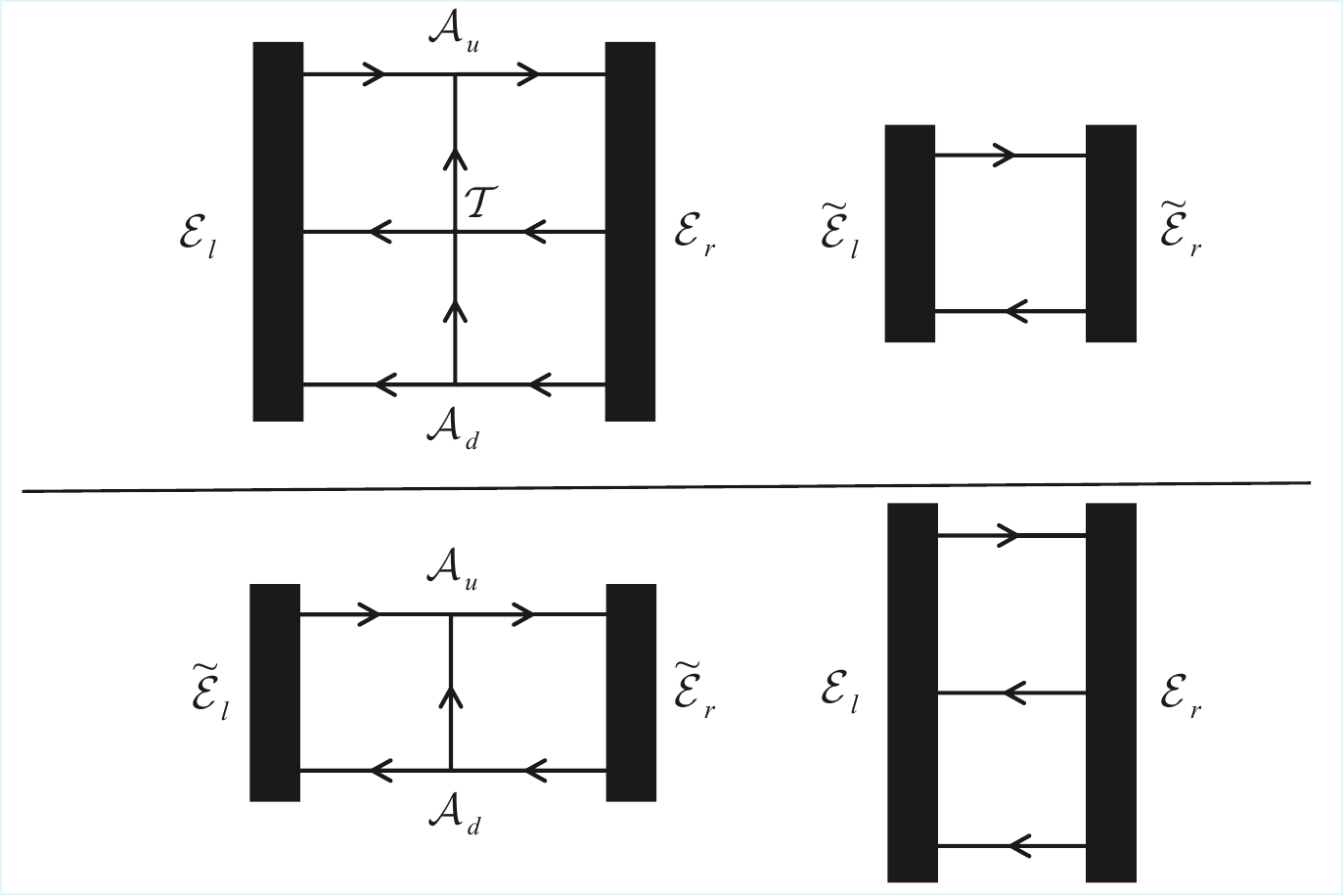}}.
\end{eqnarray}
In the formula above, we have assumed a single-sublattice structure ($m=1$) for simplicity, thus omitting the sublattice labels. The $\mt{A}_{d}$ and $\mt{A}_{u}$ are the converged tensors that generate the right and left dominant eigenvectors of this transfer matrix, respectively. $\mt{E}_{r}$ and $\tilde{\mt{E}}_{r}$ denote the right dominant eigenvectors of the three-row and two-row transfer matrices, respectively. Their explicit forms are given by:
\begin{eqnarray}
\raisebox{-0.44\height}{\includegraphics[width=0.9\textwidth, page=1]{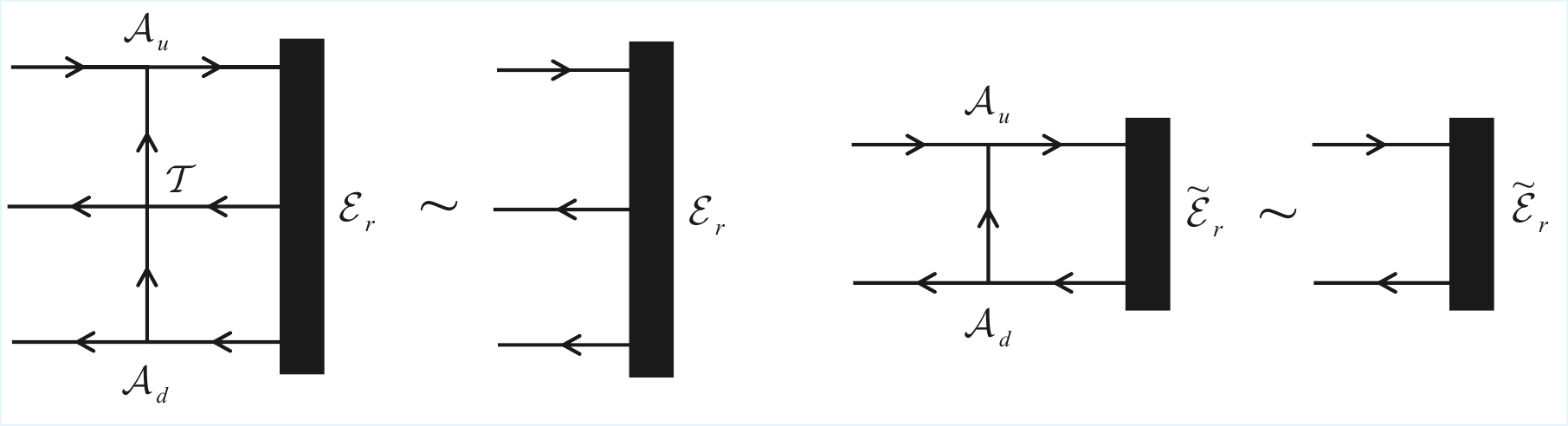}}.
\end{eqnarray}
The left dominant eigenvectors, $\mt{E}_{l}$ and $\tilde{\mt{E}}_{l}$, are available in an analogous manner.

\subsubsection{Example}

We evaluate the finite-temperature free energy of the one-dimensional free-fermion model within the Grassmann path-integral formalism, employing the Grassmann TEBD algorithm introduced in Sec.~\ref{GTEBD_method}.

\begin{eqnarray}\label{gtebd_example}
\raisebox{-0.46\height}{\includegraphics[width=0.48\textwidth, page=1]{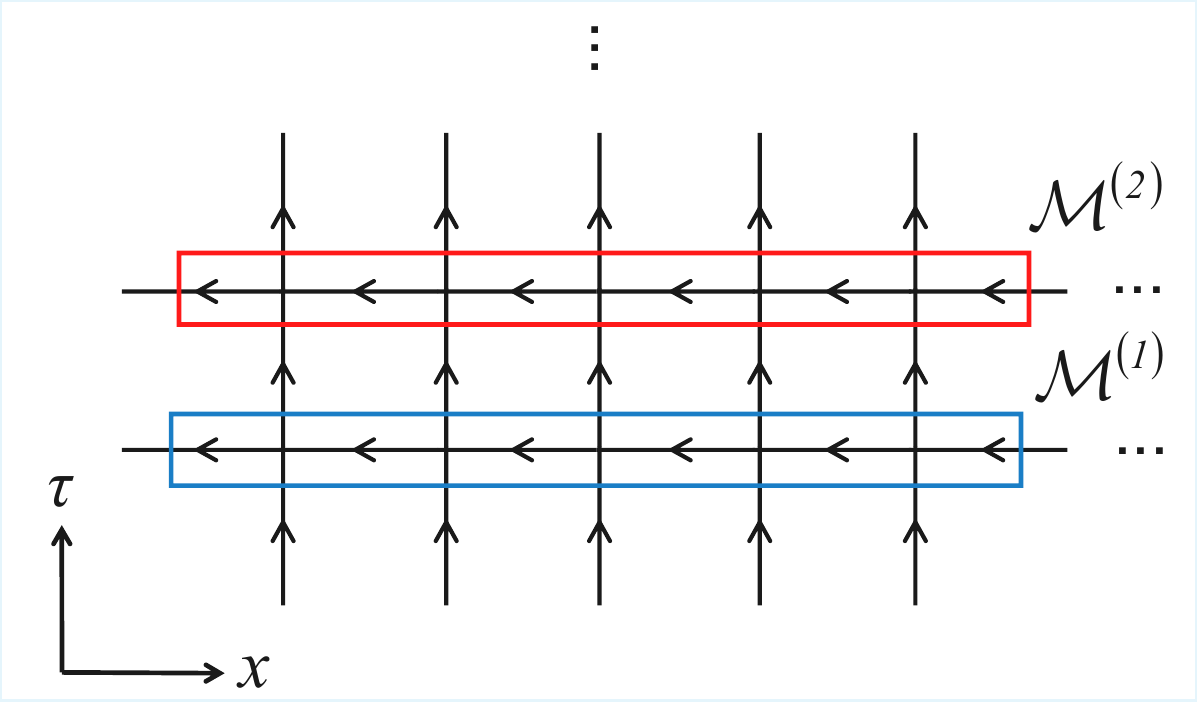}}
\end{eqnarray}

This is accomplished by contracting the corresponding Grassmann tensor network sequentially along the temporal (imaginary-time) direction, as illustrated in Eq.~(\ref{gtebd_example}). The Grassmann tensor network represents the partition function of the one-dimensional free-fermion model, with its local Grassmann tensors defined in Eqs.~(\ref{Tcoef}) and (\ref{Tlocal}). For a network of $M$ rows and $N$ columns, the product $\mt{M}^{[M]} = \mt{M}^{(1)}\mt{M}^{(2)}\cdots\mt{M}^{(M)}$ corresponds to the effective reduced density matrix at temperature $T=1/\beta$, where $\beta = M\epsilon$ and $\epsilon$ is the temporal discretization step. Here, each $\mt{M}^{(i)}$ represents the $i$-th row of the tensor network, containing $N$ Grassmann tensors. 

Consequently, the sequential row-by-row contraction can be interpreted as a cooling process from the infinite-temperature limit down to the target temperature. When applying $\mt{M}^{(n+1)}$ to an intermediate reduced density matrix $\mt{M}^{[n]}$, the local tensors in the resulting reduced density matrix $\mt{M}^{[n+1]}$ can be truncated with the help of the canonicalization method described in Sec.~\ref{GTEBD_method}:
\begin{eqnarray} \label{gtebd_example1}
\raisebox{-0.44\height}{\includegraphics[width=0.82\textwidth, page=1]{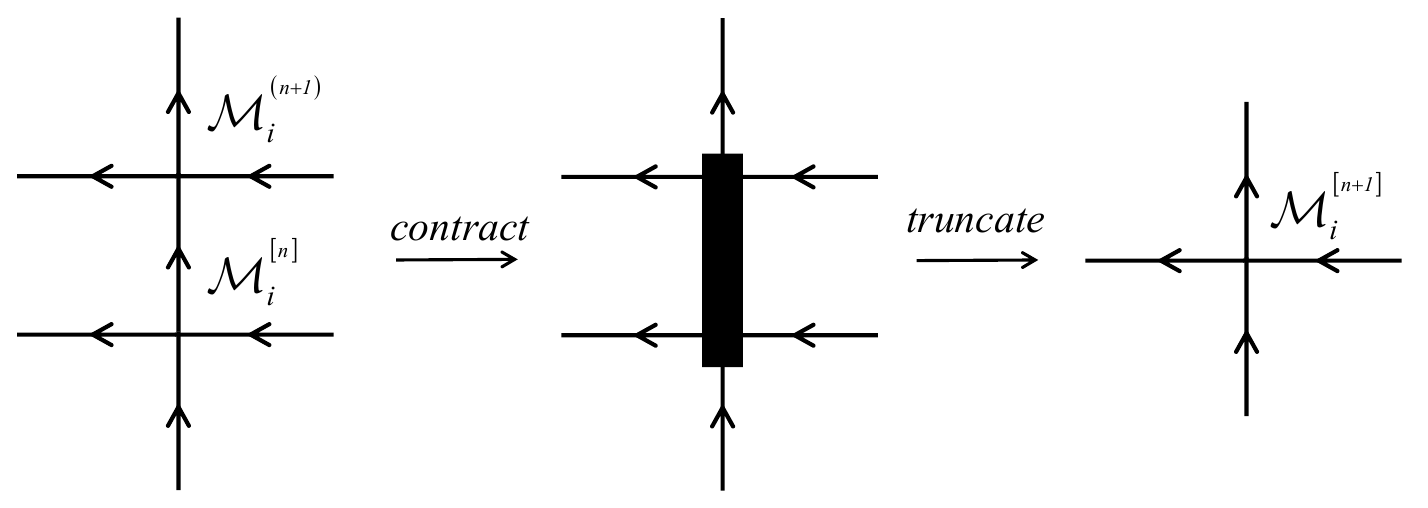}},
\end{eqnarray}
where the subscript $i$ denotes the $i$-th lattice site along the spatial direction. Once $\mt{M}^{[M]}$ has been obtained, we perform the Grassmann tensor trace along the temporal direction. Crucially, this requires imposing the anti-periodic boundary condition given in Eq.~(\ref{APBC_again}),
\begin{eqnarray}
\raisebox{-0.38\height}{\includegraphics[width=0.82\textwidth, page=1]{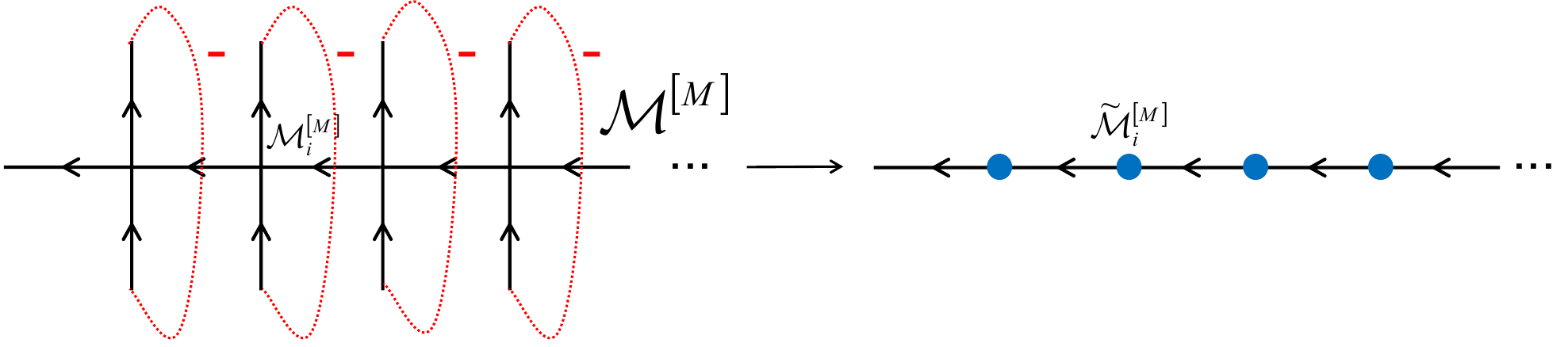}},
\end{eqnarray}
where $\mt{M}_{i}^{[M]}$ denotes the $i$-th local Grassmann tensor within $\mt{M}^{[M]}$, and $\tilde{\mt{M}}_{i}^{[M]}$ is the corresponding Grassmann matrix obtained after performing the Grassmann trace. Furthermore, one easily finds that $\textrm{Tr}\left( \cdots \mt{\tilde{M}}_{i}^{[M]} \cdots \right) = \lambda_{\textrm{max}}^{N}$, where $\lambda_{\textrm{max}}$ is the dominant eigenvalue of the coefficient matrix  $\tilde{M}_{i}$. The free energy density is evaluated via:
\begin{eqnarray}
f = -\dfrac{1}{\beta}\dfrac{1}{N}\log(Z) = -\dfrac{1}{\beta}
\left(
\sum_{i=1}^{M-1} \log(c_{i}) + \log(\lambda_{\textrm{max}})
\right).
\end{eqnarray}
Here, $c_{i}$ denotes the normalization factor arising from each of the $N$ renormalized local tensors in Eq.~(\ref{gtebd_example1}) after applying $\mt{M}^{(i+1)}$ to $\mt{M}^{[i]}$. 

\begin{figure}[htbp]
	\centering
	\includegraphics[height=6.0cm,width=8.4cm]{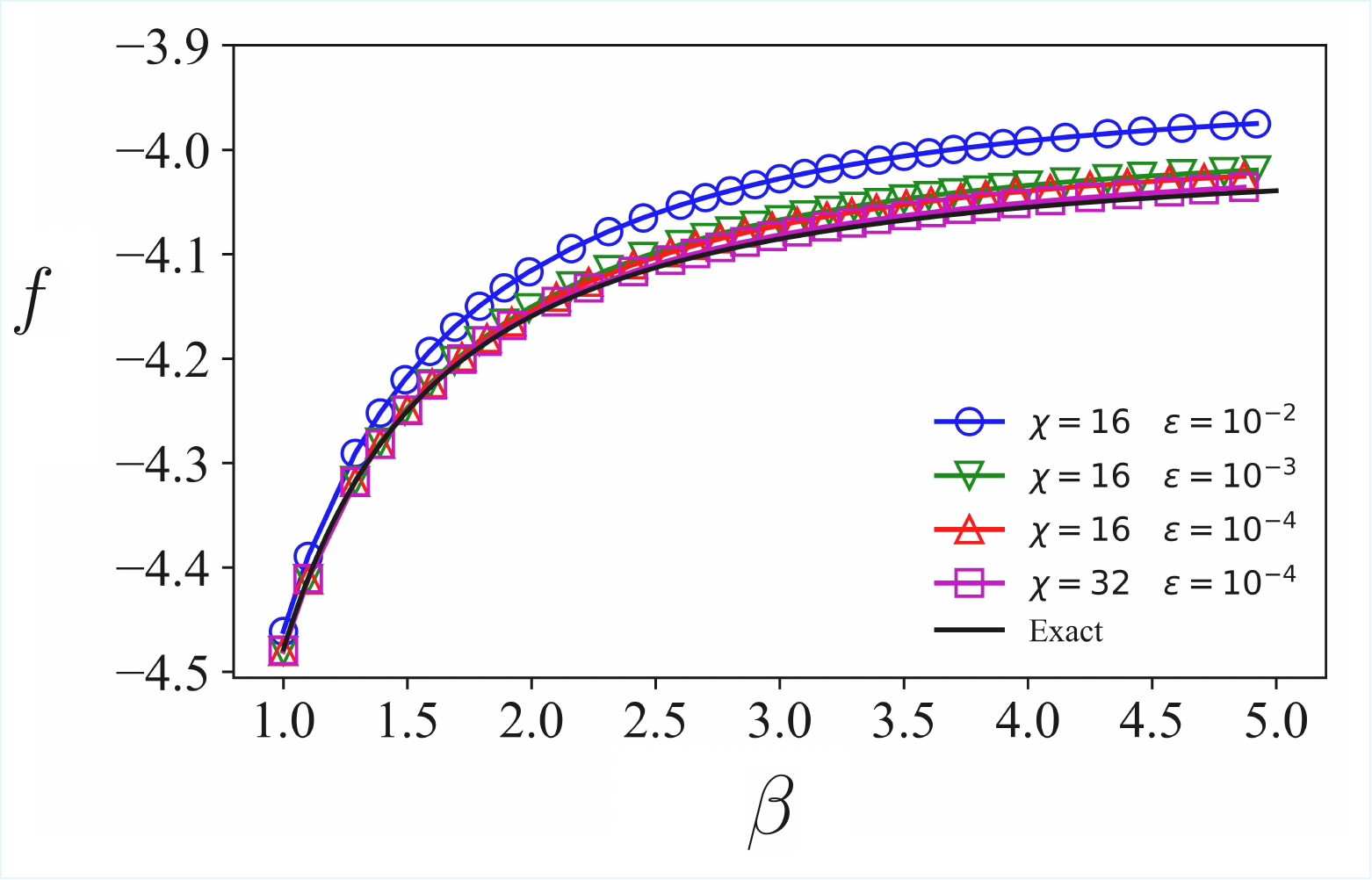}
	\caption{Free energy density $f$ versus inverse temperature $\beta$ for the one-dimensional free-fermion model at $\mu=2.0$, obtained via the Grassmann TEBD algorithm in the coherent-state path-integral formalism. Curves correspond to different temporal steps $\epsilon$ and bond dimensions $\chi$. The black line indicates the exact analytic result.}
	\label{gtebd_example3}
\end{figure}

Fig.~\ref{gtebd_example3} presents the free energy density of the one-dimensional free-fermion model given by Eq.~(\ref{free_fermion_H}) at chemical potential $\mu=2.0$, computed across a range of temporal steps $\epsilon$ and maximum bond dimensions $\chi$. As anticipated, higher accuracy is achieved with finer discretization (smaller $\epsilon$) and a larger bond dimension $\chi$. 

Although the one-dimensional free-fermion example is conceptually simple, to our knowledge, it represents the first application of tensor network methods within the Grassmann coherent-state path-integral framework to investigate the thermodynamics of a one-dimensional fermionic Hamiltonian. Moreover, the Grassmann TEBD algorithm is general: once a Grassmann tensor representation is constructed for a given model, such as the Hubbard model \cite{Akiyama2021Tensor}, it can be applied to compute corresponding finite-temperature properties.

Looking ahead, given that Grassmann tensor networks can also be formulated for two-dimensional quantum systems \cite{Akiyama2022Metal}, a natural and interesting extension would be to investigate the finite-temperature physics of two- and higher-dimensional fermionic models within this path-integral-based tensor network approach.

\subsection{Grassmann CTMRG}
\label{GCTMRG}

\subsubsection{Method} \label{GCTMRG_method}

The CTMRG algorithm was originally developed by Nishino and Okunishi~\cite{Nishino1996Corner} for two-dimensional classical spin models, drawing inspiration from the DMRG~\cite{WhiteDMRG1992} as well as the corner transfer matrix formalism~\cite{baxter2016exactly}. Later adapted and refined within the condensed-matter community~\cite{CorbozTJ2014,Orus2009Simulation,Fishman2018Faster}, it has become a standard technique for accurately computing expectation values in the context of infinite PEPS.

In this work, we present a Grassmann-compatible formulation of a widely used CTMRG variant proposed by Philippe \textit{et al.}~\cite{CorbozTJ2014}. Consider a Grassmann tensor network constructed from a unit cell of $m\times n$ different Grassmann tensors. To approximate its infinite environment, each rank-4 Grassmann tensor $\mt{T}^{(x, y)}$, with bond dimension 
$D$, is associated with four rank-3 edge tensors $\{\mt{E}_{l}^{(x, y)}, \mt{E}_{r}^{(x, y)}, \mt{E}_{u}^{(x, y)}, \mt{E}_{d}^{(x, y)}\}$ and four corner matrices $\{\mt{C}_{lu}^{(x, y)}, \mt{C}_{ru}^{(x, y)}, \mt{C}_{ld}^{(x, y)}, \mt{C}_{rd}^{(x, y)}\}$, as illustrated schematically:
\begin{eqnarray} \label{gctmrg1}
\raisebox{-0.48\height}{\includegraphics[width=0.68\textwidth, page=1]{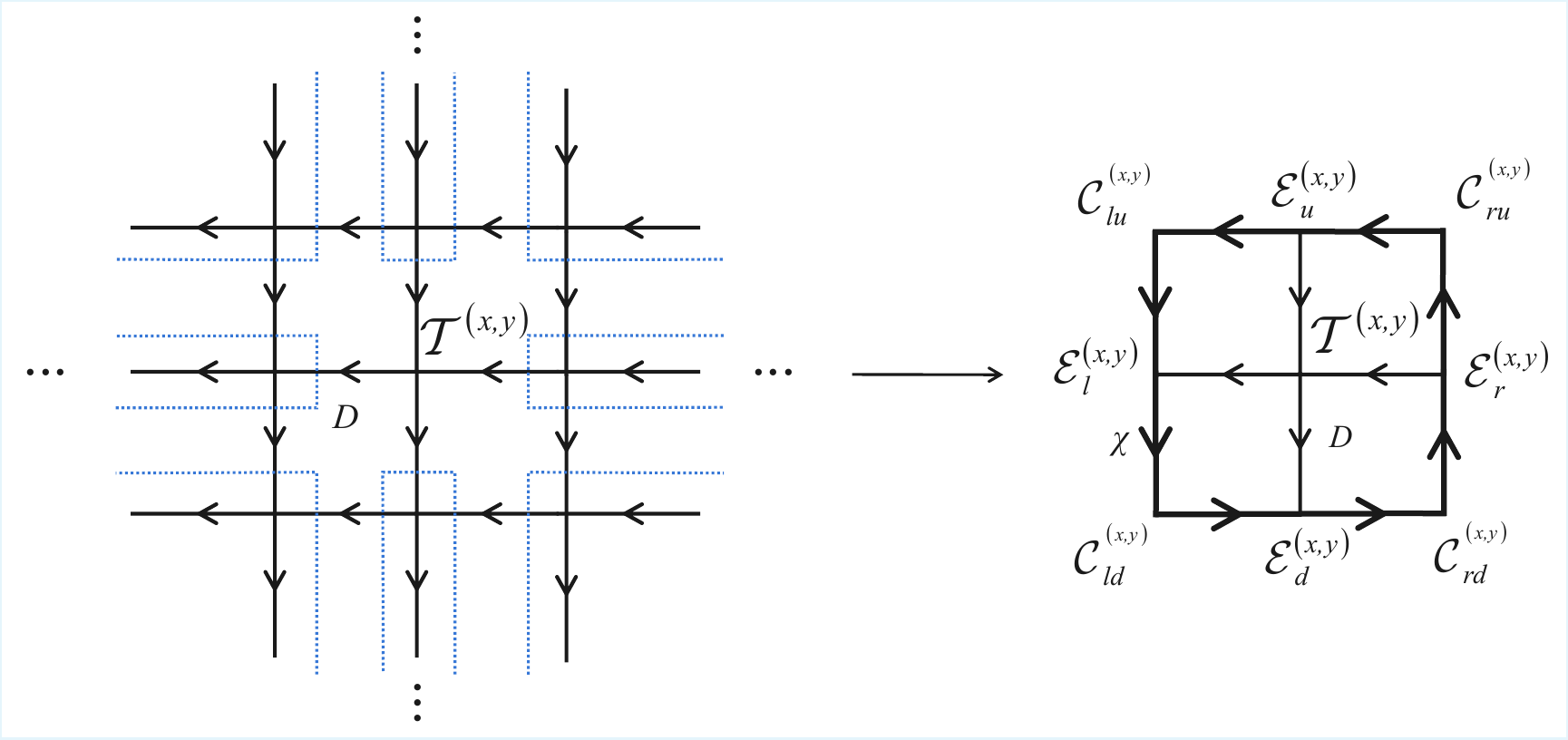}}
\end{eqnarray}
The label $(x,y)$ denotes the coordinate within the $m\times n$ unit cell. The accuracy of the environment approximation is controlled by a hyperparameter $\chi$, which sets the virtual bond dimension of environment tensors (namely, corner matrices and edge tensors). In the limit $\chi\rightarrow \infty$, the left- and right-hand sides of Eq.~(\ref{gctmrg1}) become identical, and larger $\chi$ generally yields better accuracy. Note the arrow conventions assigned to each Grassmann tensor in Eq.~(\ref{gctmrg1}). While initially arbitrary, these conventions must be fixed consistently throughout the calculation, as they determine the sign factors in the subsequent construction of the projectors.

The goal of the CTMRG algorithm is to iteratively optimize the environment tensors through four directional \textit{moves}—left, right, up, and down move—until convergence is reached. For a $m\times n$ unit cell, this entails optimizing $8mn$ environment tensors. 

\begin{eqnarray} \label{gctmrg2}
\raisebox{-0.46\height}{\includegraphics[width=0.32\textwidth, page=1]{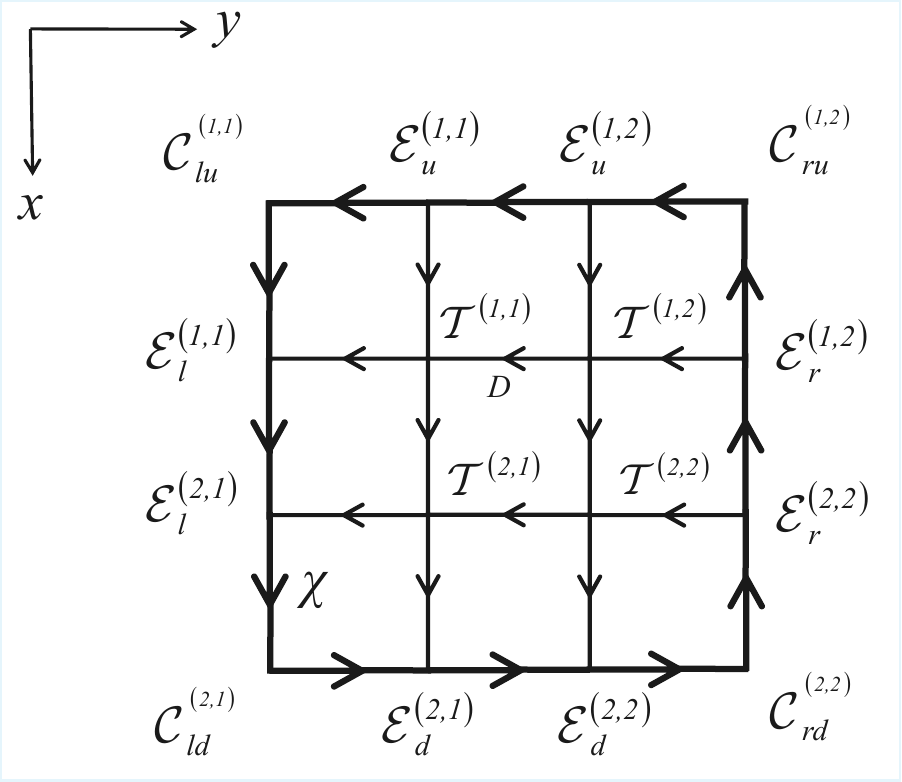}}
\end{eqnarray}

As an illustrative example, we present the \textit{left move} for a $(m, n) = (2, 2)$ unit cell, with the corresponding environment tensors shown in Eq.~\eqref{gctmrg2}. The left move optimizes a set of environment tensors $\{\mt{C}_{lu}^{(x, y)}, \mt{E}_{l}^{(x, y)}, \mt{C}_{ld}^{(x, y)}\}$ located at the left-hand side of each $\mt{T}^{(x, y)}$. This update starts by inserting an entire unit cell from the left, as indicated by the green arrow in Eq.~(\ref{gctmrg3}), and consists of the following sequential steps:

(i) Absorption: The first column of the inserted unit cell is absorbed into the left environments.
\begin{eqnarray} \label{gctmrg3}
\raisebox{-0.5\height}{\includegraphics[width=0.44\textwidth, page=1]{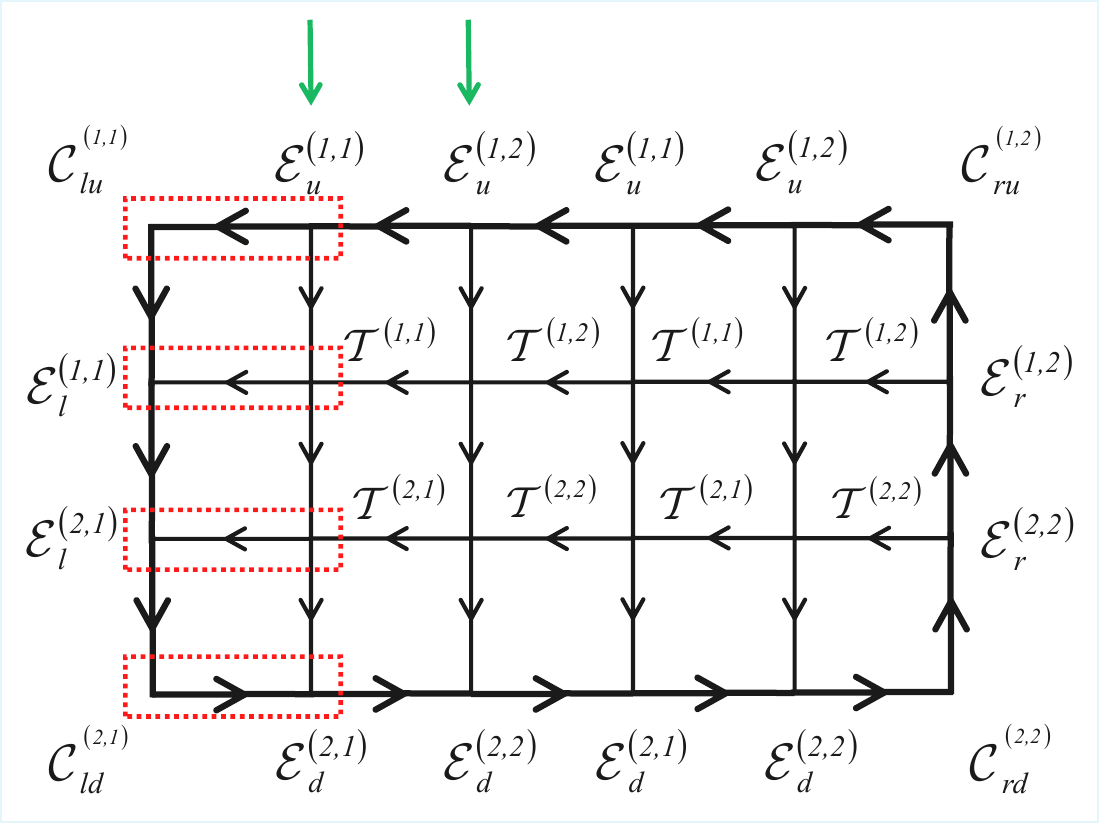}}
\raisebox{-0.56\height}{\includegraphics[width=0.52\textwidth, page=1]{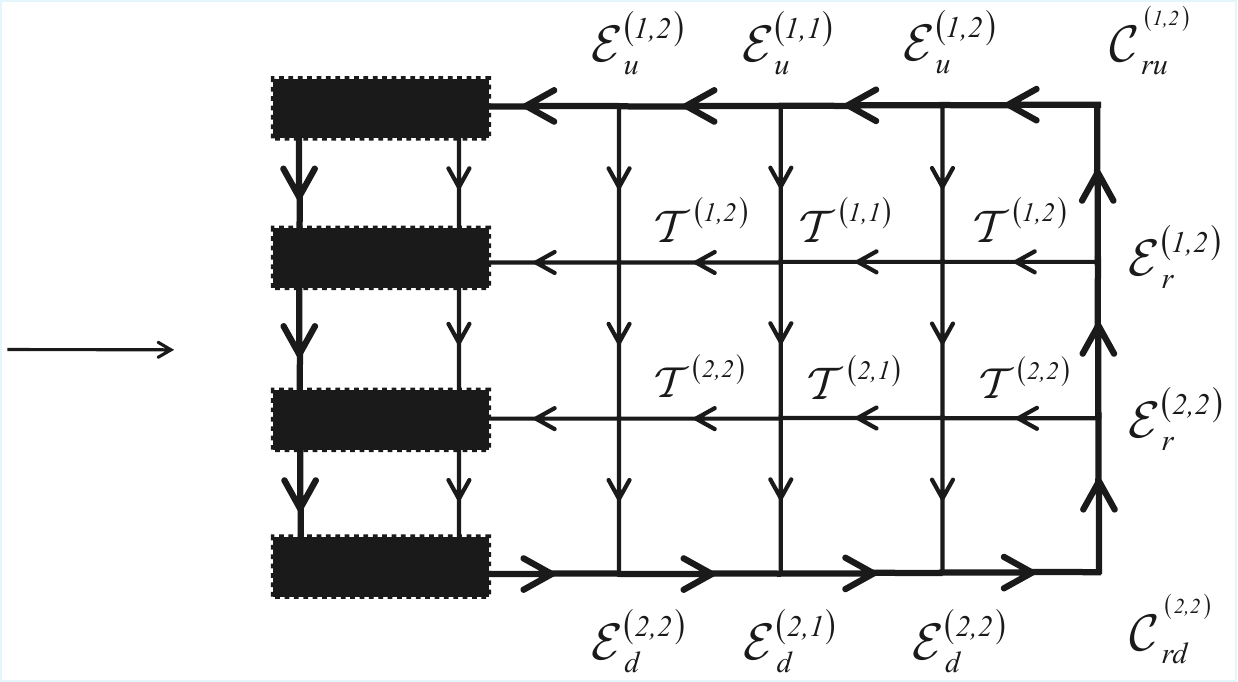}}
\end{eqnarray}

(ii) Truncation and Renormalization: Following the absorption step, the bond dimension of the  environment tensors on the left increases from $\chi$ to $D\chi$. We then construct and insert a pair of Grassmann isometries, $\mt{P}$ and $\mt{Q}$, at corresponding virtual links. This operation simultaneously truncates the bond dimension back to $\chi$ and renormalizes the left environment tensors.
\begin{eqnarray} \label{gctmrg5}
\raisebox{-0.5\height}{\includegraphics[width=0.8\textwidth, page=1]{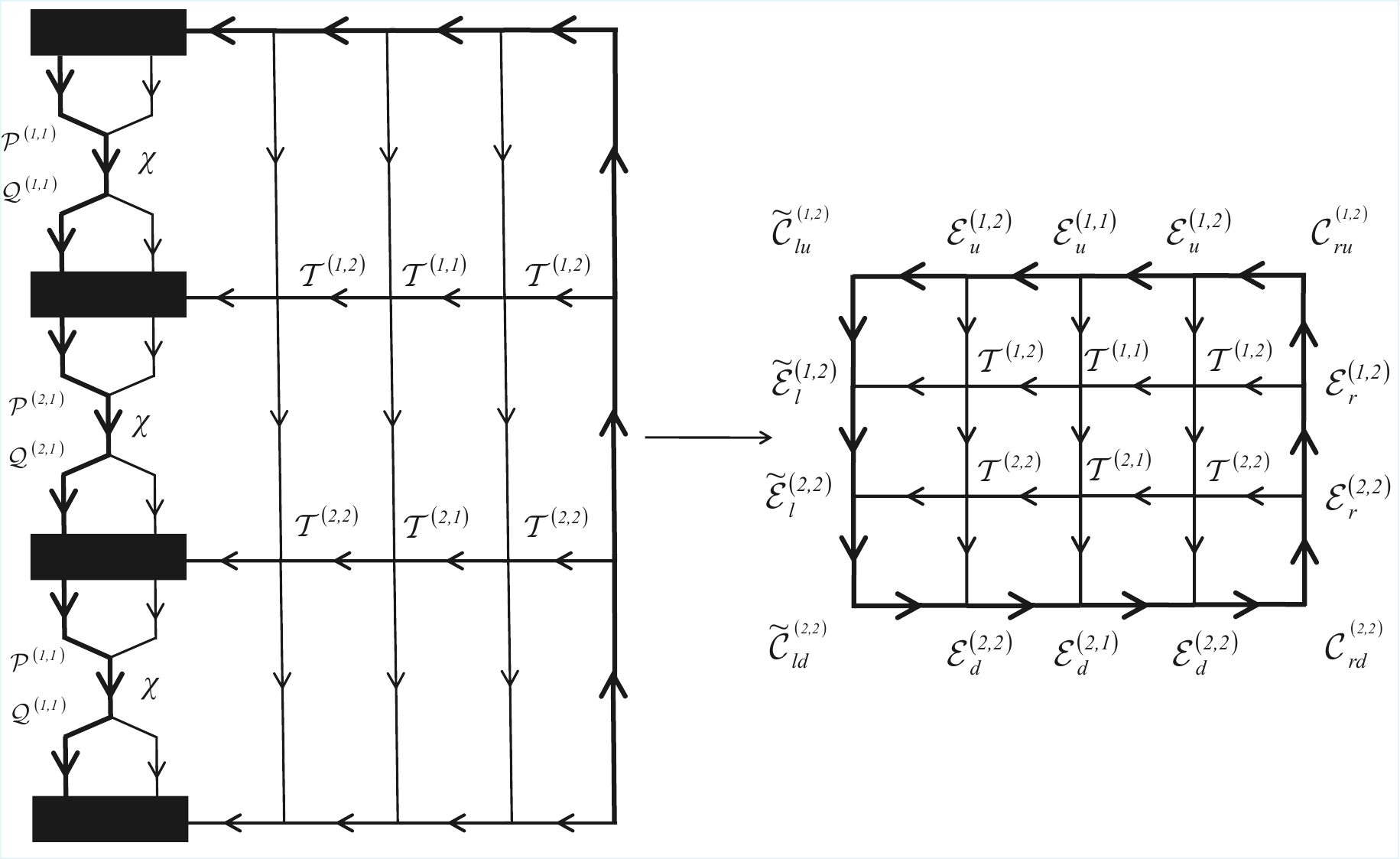}}
\end{eqnarray}
Thus, the left environment tensors for $\mt{T}^{(1, 2)}$ and $\mt{T}^{(2, 2)}$ have been updated. The same procedure then updates corresponding environments for $\mt{T}^{(1, 1)}$ and $\mt{T}^{(2, 1)}$ by absorbing the second column of the inserted unit cell. The left move is finished once the entire unit cell has been absorbed, and all left environments have been updated. It is important to note that, up to this point, the overall procedure mirrors the bosonic CTMRG algorithm, with the essential upgrade being the use of Grassmann tensor contractions in place of ordinary ones to correctly account for fermionic signs.

The most critical step in the CTMRG algorithm is the construction of $\mt{P}$ and $\mt{Q}$ that optimally truncate the enlarged bond. In the following, we detail the construction of the specific projectors $\mt{P}^{(2, 1)}$ and $\mt{Q}^{(2, 1)}$ shown in Eq.~(\ref{gctmrg5}). To this end, we consider a relevant $2\times2$ cell of $\mt{T}$ tensors together with its surrounding environments, as illustrated schematically:
\begin{eqnarray} \label{gctmrg6}
\raisebox{-0.46\height}{\includegraphics[width=0.38\textwidth, page=1]{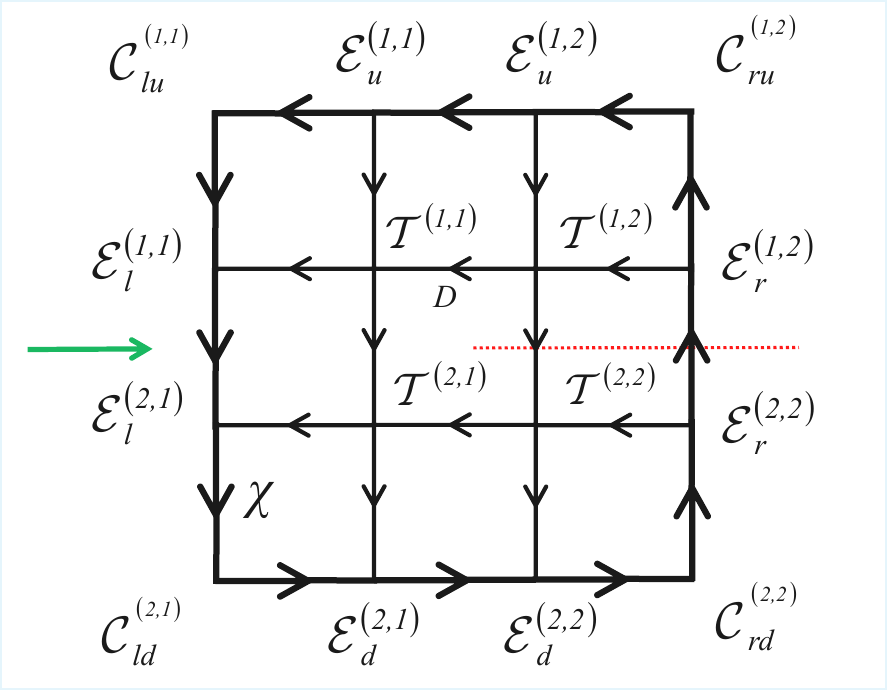}}
\end{eqnarray}
In Eq.~(\ref{gctmrg6}), the position for inserting $\mt{P}^{(2, 1)}$ and $\mt{Q}^{(2, 1)}$ is indicated by the green arrow. By cutting the $4\times4$ cluster on its right side, we obtain a rank-4 Grassmann tensor $\mt{W}$, which can be interpreted as the outcome of contracting two rank-4 Grassmann tensors
\begin{eqnarray} \label{gctmrg8}
\mt{W}_{i_{1}i_{2}i_{3}i_{4}}
=
\sum_{i_{5}i_{6}}
\mt{X}_{i_{1}i_{2}i_{5}i_{6}}
\mt{Y}_{i_{5}i_{6}i_{3}i_{4}},
\end{eqnarray}
where the Grassmann tensors $\mt{X}$ and $\mt{Y}$ are defined via (see also Eq.~(\ref{gctmrg7a})) :
\begin{eqnarray} \label{gctmrg8a}
& &
\mt{X}_{i_{1}i_{2}i_{5}i_{6}}
=
X_{i_{1}i_{2}i_{5}i_{6}}
\xi_{1}^{p(i_{1})}
\bar{\xi}_{2}^{p(i_{2})}
\bar{\xi}_{6}^{p(i_{6})}
\bar{\xi}_{5}^{p(i_{5})},
\\ 
& &
\mt{Y}_{i_{5}i_{6}i_{3}i_{4}}
=
Y_{i_{5}i_{6}i_{3}i_{4}}
\xi_{5}^{p(i_{5})}
\xi_{6}^{p(i_{6})}
\xi_{3}^{p(i_{3})}
\bar{\xi}_{4}^{p(i_{4})}.
\end{eqnarray}
Then we can reformulate our goal as:
\begin{eqnarray} 
\label{gctmrg7a}
\raisebox{-0.5\height}{\includegraphics[width=0.88\textwidth, page=1]{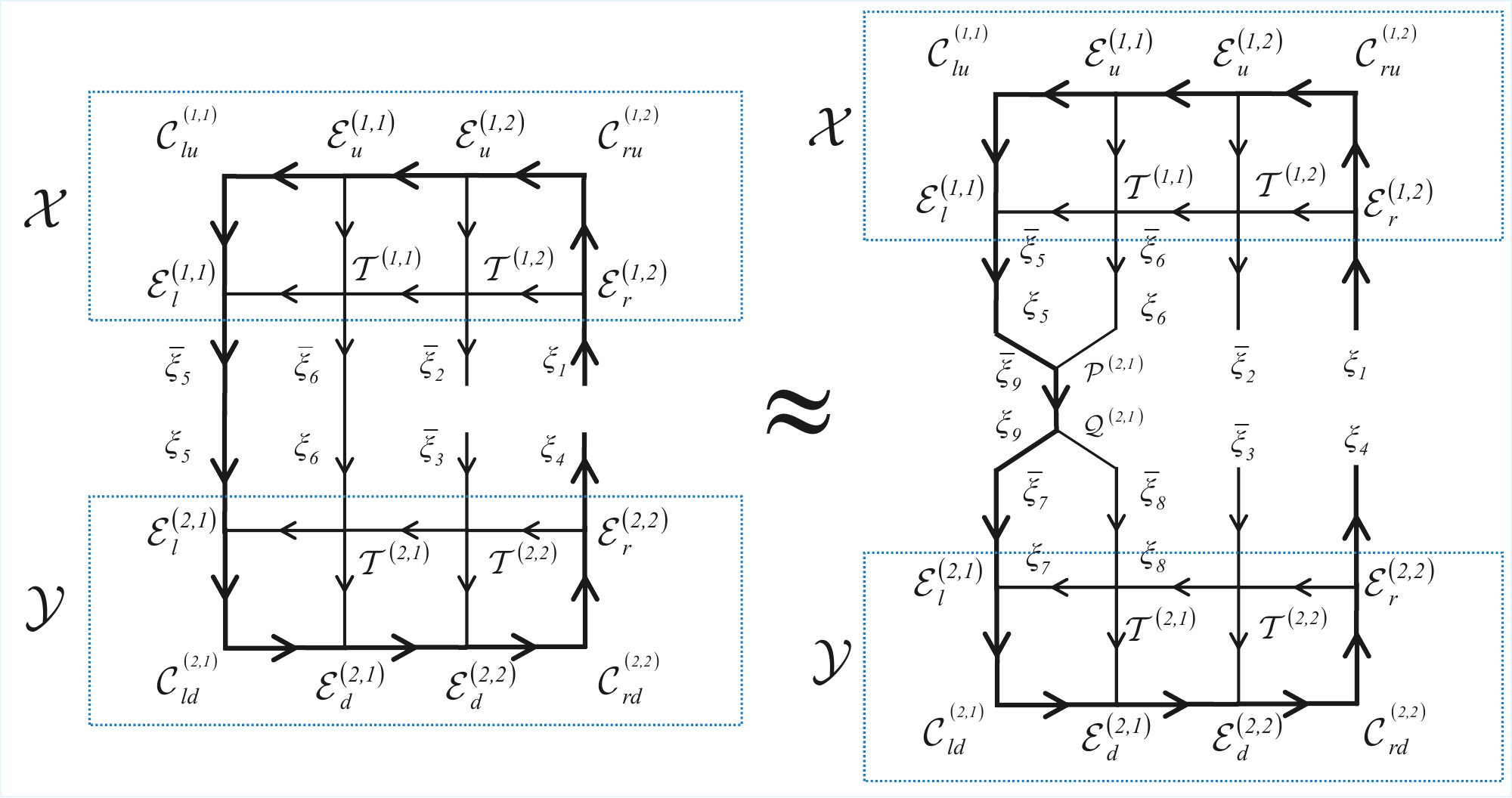}}.
\end{eqnarray}
Here, the $\approx$ indicates the index truncation has been performed within $\mt{P}$ and $\mt{Q}$, leading to certain approximation errors. Expressing Eq.~(\ref{gctmrg7a}) in terms of a Grassmann tensor equation, we have
\begin{eqnarray} \label{gctmrg7b}
\mt{W}_{i_{1}i_{2}i_{3}i_{4}}
\approx
\sum_{i_{5}i_{6}}
\sum_{i_{7}i_{8}}
\sum_{i_{9}}
\mt{X}_{i_{1}i_{2}i_{5}i_{6}}
\left(
\mt{P}_{i_{5}i_{6}i_{9}}
\mt{Q}_{i_{9}i_{7}i_{8}}
\right)
\mt{Y}_{i_{7}i_{8}i_{3}i_{4}},
\end{eqnarray}
with 
\begin{eqnarray} 
\label{gctmrg7c}
\mt{P}_{i_{5}i_{6}i_{9}}
&=&
P_{i_{5}i_{6}i_{9}}
\xi_{5}^{p(i_{5})}
\xi_{6}^{p(i_{6})}
\bar{\xi}_{9}^{p(i_{9})},
\\ 
\label{gctmrg7d}
\mt{Q}_{i_{9}i_{7}i_{8}}
&=&
Q_{i_{9}i_{7}i_{8}}
\xi_{9}^{p(i_{9})}
\bar{\xi}_{8}^{p(i_{8})}
\bar{\xi}_{7}^{p(i_{7})}.
\end{eqnarray}

We circumvent Eq.~\eqref{gctmrg7b} by reformulating the problem in terms of their coefficient tensors. Firstly, Eq.~\eqref{gctmrg8} corresponds to the following coefficient tensor equation:
\begin{eqnarray} \label{gctmrg10}
W_{i_{1}i_{2}i_{3}i_{4}}
=
\sum_{i_{5}i_{6}}
X_{i_{1}i_{2}i_{5}i_{6}}
Y_{i_{5}i_{6}i_{3}i_{4}}
\times
(-1)^{p(i_{5}) + p(i_{6})}
=
\sum_{i_{5}i_{6}}
X_{i_{1}i_{2}i_{5}i_{6}}
\tilde{Y}_{i_{5}i_{6}i_{3}i_{4}}.
\end{eqnarray}
Note the sign factor has been absorbed into $\tilde{Y}$. We then introduce a pair of ordinary  isometries $\tilde{P}$ and $\tilde{Q}$ that provide the optimal approximation of $W$ via $W \approx X(\tilde{P}\tilde{Q})\tilde{Y}$. This optimality condition is equivalent to requiring that the insertion of $\tilde{P}$ and $\tilde{Q}$ implements an SVD of $W$. The solution follows the standard prescription \cite{CorbozTJ2014}:
\begin{eqnarray} \label{gctmrg12b}
\tilde{P} = \tilde{Y} V \Lambda^{-1/2}
= 
\raisebox{-0.4\height}{\includegraphics[width=0.24\textwidth, page=1]{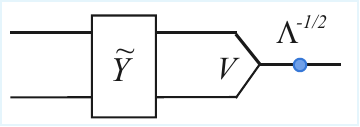}}, 
~
\tilde{Q} = \Lambda^{-1/2}U^{\dagger}X
=
\raisebox{-0.4\height}{\includegraphics[width=0.24\textwidth, page=1]{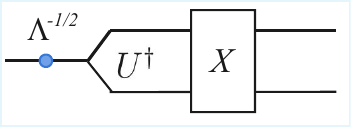}},
\end{eqnarray}
where $U$, $V$ and $\Lambda$ are obtained from the decomposition $W = U\Lambda V^{\dagger}$. In Eq.~(\ref{gctmrg12b}), the absence of arrows on the links indicates that the corresponding tensors are ordinary coefficient tensors. 

Finally, we compute the coefficient form of the Grassmann Eq.~(\ref{gctmrg7b}):
\begin{eqnarray} \label{gctmrg13}
W_{i_{1}i_{2}i_{3}i_{4}}
\approx
\sum_{i_{5}i_{6}}
\sum_{i_{7}i_{8}}
\sum_{i_{9}}
X_{i_{1}i_{2}i_{5}i_{6}}
\left(
P_{i_{5}i_{6};i_{9}}
Q_{i_{9}i_{7}i_{8}}
\right)
\tilde{Y}_{i_{7}i_{8}i_{3}i_{4}}
\times
(-1)^{s_{1}+s_{2}}.
\end{eqnarray}
In Eq.~(\ref{gctmrg13}), a sign factor $(-1)^{p(i_{7})+p(i_{8})}$ arising from the contraction of $\mt{Q}$ and $\mt{Y}$ has been absorbed into $\tilde{Y}$. The contraction of $\mt{X}$ and $\mt{P}$ yields a sign $(-1)^{s_{1}} \equiv (-1)^{p(i_{5})+p(i_{6})}$ , and the contraction of $\mt{P}$ and $\mt{Q}$ contributes $(-1)^{s_{2}} \equiv (-1)^{p(i_{9})}$. 

Crucially, we observe that if these additional sign factors in Eq.~(\ref{gctmrg13}) are incorporated into $\tilde{P}$ and $\tilde{Q}$ introduced in Eq.~(\ref{gctmrg12b}), the optimization of Eq.~(\ref{gctmrg13})—and consequently of the Grassmann Eq.~(\ref{gctmrg7b})—is achieved. The desired $P$ and $Q$ are given by:
\begin{eqnarray}
\label{gctmrg14a}
P_{i_{5}i_{6}i_{9}}
&=&
\sum_{i_{5}i_{6}i_{9}}
\tilde{P}_{i_{5}i_{6}i_{9}}
\times
(-1)^{p(i_{5}) + p(i_{6})},
\\ 
\label{gctmrg14b}
Q_{i_{9}i_{7}i_{8}}
&=&
\sum_{i_{7}i_{8}i_{9}}
\tilde{Q}_{i_{9}i_{7}i_{8}}
\times
(-1)^{p(i_{9})}.
\end{eqnarray}

The Grassmann isometries $\mt{P}$ and $\mt{Q}$, with their coefficient tensors given by Eqs.~(\ref{gctmrg14a}-\ref{gctmrg14b}), can be shown to induce a Grassmann SVD of $\mt{W}$. This can be directly verified by evaluating the corresponding Grassmann integrals, which yield
\begin{eqnarray}
& & \nonumber
\sum_{i_{5}i_{6}}
\sum_{i_{7}i_{8}}
\sum_{i_{9}}
\mt{X}_{i_{1}i_{2}i_{5}i_{6}}
\left(
\mt{P}_{i_{5}i_{6}i_{9}}
\mt{Q}_{i_{9}i_{7}i_{8}}
\right)
\mt{Y}_{i_{7}i_{8}i_{3}i_{4}}
\\ \nonumber
&=&
\sum_{i_{5}i_{6}}
\sum_{i_{7}i_{8}}
\sum_{i_{9}}
X_{i_{1}i_{2}i_{5}i_{6}}
\left(
P_{i_{5}i_{6}i_{9}}
Q_{i_{9}i_{7}i_{8}}
\right)
Y_{i_{7}i_{8}i_{3}i_{4}}
\times
g,
\\ 
&=& \label{svdproof}
\left(
U\Lambda V^{\dagger}
\right)_{i_{1}i_{2}i_{3}i_{4}}
\xi_{1}^{p(i_{1})}
\bar{\xi}_{2}^{p(i_{2})}
\xi_{3}^{p(i_{3})}
\bar{\xi}_{4}^{p(i_{4})},
\end{eqnarray}
where the symbol $g$ denotes
\begin{eqnarray}
& &\nonumber
g \equiv 
\int_{\bar{\xi}_{5}\xi_{5}}
\int_{\bar{\xi}_{6}\xi_{6}}
\int_{\bar{\xi}_{7}\xi_{7}}
\int_{\bar{\xi}_{8}\xi_{8}}
\int_{\bar{\xi}_{9}\xi_{9}}
\left(
\xi_{1}^{p(i_{1})}
\bar{\xi}_{2}^{p(i_{2})}
\bar{\xi}_{6}^{p(i_{6})}
\bar{\xi}_{5}^{p(i_{5})}
\right)
\left(
\xi_{5}^{p(i_{5})}
\xi_{6}^{p(i_{6})}
\bar{\xi}_{9}^{p(i_{9})}
\right)
\\ 
&\times& 
\left(
\xi_{9}^{p(i_{9})}
\bar{\xi}_{8}^{p(i_{8})}
\bar{\xi}_{7}^{p(i_{7})}
\right)
\left(
\xi_{7}^{p(i_{7})}
\xi_{8}^{p(i_{8})}
\xi_{3}^{p(i_{3})}
\bar{\xi}_{4}^{p(i_{4})}
\right).
\end{eqnarray}
Therefore, the constructed $\mt{P}$ and $\mt{Q}$ are optimal in the sense that they implement the Grassmann SVD of $\mt{W}$. The truncation can be performed by discarding the columns of $P$ and rows of $Q$ that correspond to the smallest singular values in $\Lambda$. Grassmann isometries for the right, up, and down move are constructed in an analogous manner. Once the environment tensors are converged via these iterative updates, the infinite Grassmann tensor network can be efficiently evaluated through \cite{Jahromi2018Infinite,Liu2022Variational,Okunishi2022Developments},
\begin{eqnarray} \label{gctmrg_Z}
Z = \raisebox{-0.48\height}{\includegraphics[width=0.48\textwidth, page=1]{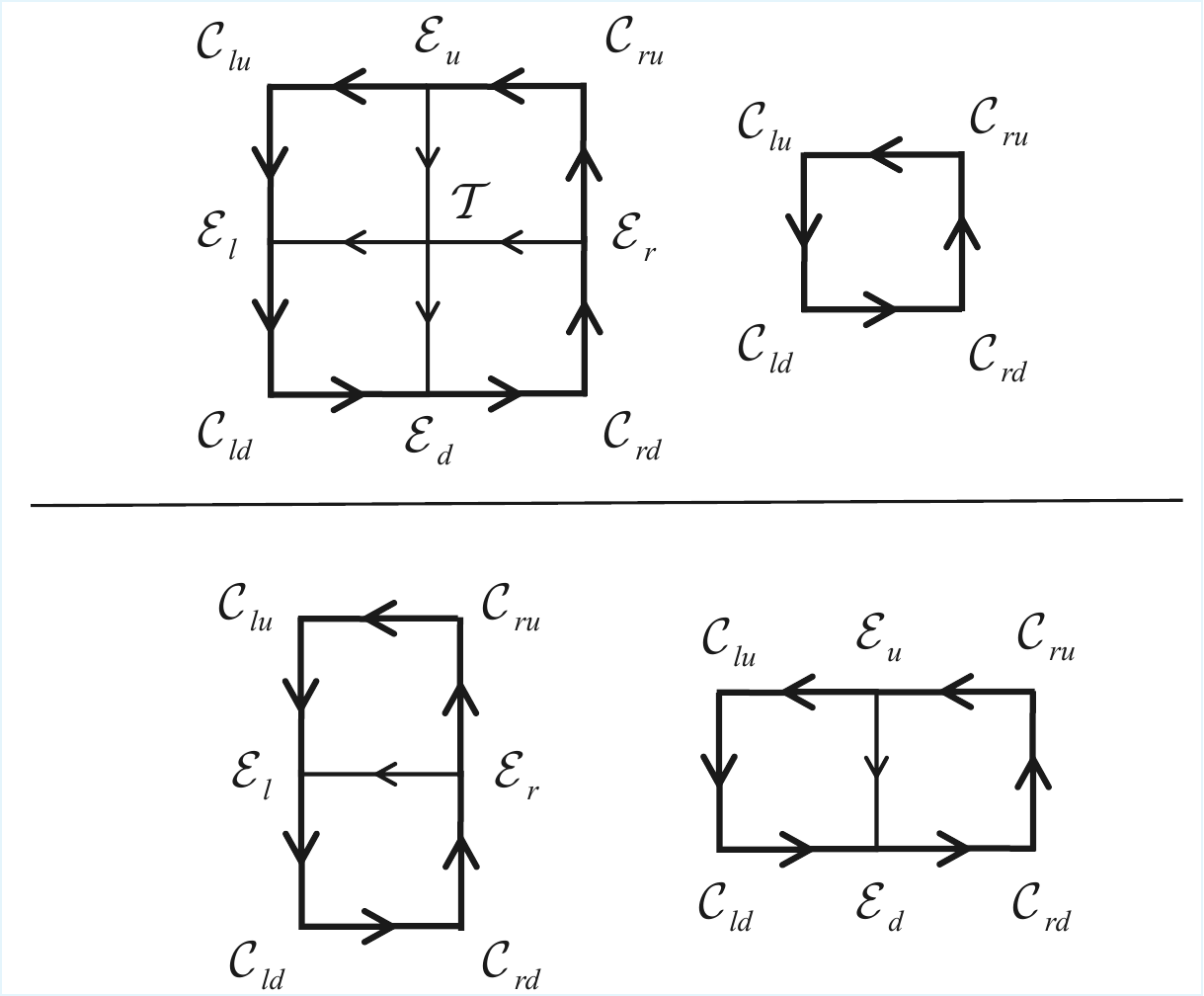}}
\end{eqnarray} 

Note the formula above holds for any sublattice $(x, y)$ with $x = 1, \dots, m$ and $y = 1, \cdots, n$. 

\subsubsection{Example}

The Gross-Neveu model~\cite{Gross1974Dynamical} is a (1+1)-dimensional relativistic quantum field theory of interacting Dirac fermions. It holds significant importance in high-energy physics owing to its features, such as asymptotic freedom, which it shares with (3+1)-dimensional quantum chromodynamics. It also serves as a low-energy effective field theory for condensed matter systems hosting Dirac fermions on honeycomb lattices~\cite{Herbut2006Interactions,Herbut2009Theory}.

For non-perturbative calculations, the quantum field theory must be discretized on a lattice. A commonly used regularization scheme is the Gross-Neveu–Wilson (GNW) model~\cite{Wilson1975Quarks}, which employs Wilson fermions to circumvent the fermion doubling problem. In this subsection, we employ Grassmann CTMRG and TEBD algorithms to compute the free energy density of the (1+1)-dimensional, single-flavor GNW model. Its full lattice action reads
\begin{eqnarray} \label{GNaction}
S 
&=&
-\dfrac{1}{2}\sum_{n\in\Lambda_{2}}\sum_{\nu=1,2}
\left[
\bar{\psi}(n)
\left(
r\mathds{1} - \gamma_{\nu}
\right)
\psi(n+\hat{\nu})
+
\bar{\psi}(n+\hat{\nu})
\left(
r\mathds{1} + \gamma_{\nu}
\right)
\psi(n)
\right]
\\ 
&+& \nonumber
(m+2r)
\sum_{n}
\bar{\psi}(n)
\psi(n)
-
\dfrac{g^{2}_{\sigma}}{2}
\sum_{n}
\left(
\bar{\psi}(n)
\psi(n)
\right)^{2}
-
\dfrac{g^{2}_{\pi}}{2}
\sum_{n}
\left(
\bar{\psi}(n)
i\gamma_{5}
\psi(n)
\right)^{2},
\end{eqnarray}
where $\psi(n) = \left(\psi_{1}(n),\psi_{2}(n)\right)^{T}$ and $\bar{\psi}(n) = \left(\bar{\psi}_{1}(n),\bar{\psi}_{2}(n)\right)$ are  two-component Grassmann fields living on the two-dimensional spacetime lattice site $n$. The model parameters are the bare mass $m$, the Wilson parameter $r$ (set to $r=1$ in our computations), and two kinds of four-fermion coupling constants $g^{2}_{\sigma}$ and $g^{2}_{\pi}$. The Dirac $\gamma$-matrices are represented by Pauli matrices: $\gamma_{1} = \hat{\sigma}_{x}$, $\gamma_{2} = \hat{\sigma}_{y}$, and $\gamma_{5} = \hat{\sigma}_{z}$. 

The single‑flavor GNW model~\cite{Aoki1984New,Kenna2001Weakly,Bermudez2018Gross} exhibits a rich phase structure as a function of bare mass and coupling, including a parity‑broken Aoki phase \cite{Aoki1984New,Kenna2001Weakly}, a trivially gapped phase, and a symmetry‑protected topological phase \cite{Bermudez2018Gross}. 
The exploration of these phases within a Lagrangian formulation using sign-problem-free tensor network methods remains largely unexplored. Recently, we reported the first numerical determination of the complete phase diagram of the model described by Eq.~\eqref{GNaction}, and characterized the associated phase boundaries via Grassmann CTMRG~\cite{Kong2026Phase}. In particular, contrary to large-$N_{f}$ prediction~\cite{Aoki1984New}, we found that the parity‑broken phase is entirely bounded by phase boundaries with central charge $c = 1/2$, and does not extend into the strong-coupling region.
In this work, we focus on the non-interacting case ($g_{\sigma} = g_{\pi} = 0$) as a benchmark for numerical methods. 

The partition function corresponding to the action in Eq.~(\ref{GNaction}) is represented by a Grassmann tensor network. Its elementary building block is a rank-8 Grassmann tensor:
\begin{eqnarray} \label{GN1}
\mt{T}_{
	i_{1}j_{1}
	i_{2}j_{2}
	k_{1}l_{1}
	k_{2}l_{2}} 
= 
T_{
	i_{1}j_{1}
	i_{2}j_{2}
	k_{1}l_{1}
	k_{2}l_{2}} 
\xi_{1}^{i_{1}}
\zeta_{1}^{j_{1}}
\xi_{2}^{i_{2}}
\zeta_{2}^{j_{2}}
\bar{\zeta}_{1}^{l_{1}}
\bar{\xi}_{1}^{k_{1}}
\bar{\zeta}_{2}^{l_{2}}
\bar{\xi}_{2}^{k_{2}},
\end{eqnarray}
where each index of the coefficient tensor $T$ takes integer values 0 or 1. Its explicit form for Eq.~(\ref{GN1}) can be derived following a procedure analogous to that outlined in Sec.~\ref{fpf}, the full expression with $g_{\sigma}^{2} = g_{\pi}^{2} = g^{2}$ is provided in Ref.~\cite{Akiyama2023Implementation}. For numerical implementations, the rank-8 Grassmann tensor is fused into a rank-4 Grassmann tensor as shown below,
\begin{eqnarray} 
\raisebox{-0.38\height}{\includegraphics[width=0.56\textwidth, page=1]{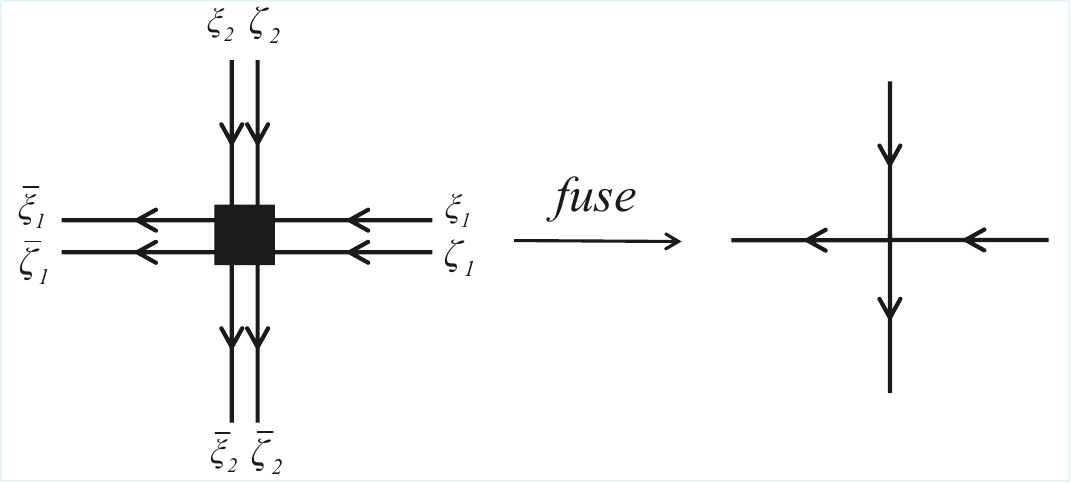}}.
\end{eqnarray}

We compute the free energy density $f = -\ln(Z)/V$ of the GNW model at $g_{\sigma} = g_{\pi} = 0$ and mass $m=-1.2$ using Grassmann tensor network methods. The partition function $Z$ can be obtained using Eq.~(\ref{gtebd_Z}) and Eq.~(\ref{gctmrg_Z}) for Grassmann TEBD and CTMRG, respectively. The accuracy is quantified by the relative error $\delta f = \large| f - f_{\textrm{exact}} \large| / \large| f_{\textrm{exact}} \large|$, where $f_{\textrm{exact}}$ is the exact value for this non-interacting model. For comparison, we also implement the Grassmann version of the original TRG method~\cite{Levin2007}.

\begin{figure}[htbp]
	\centering
	\includegraphics[height=6.0cm,width=9.0cm]{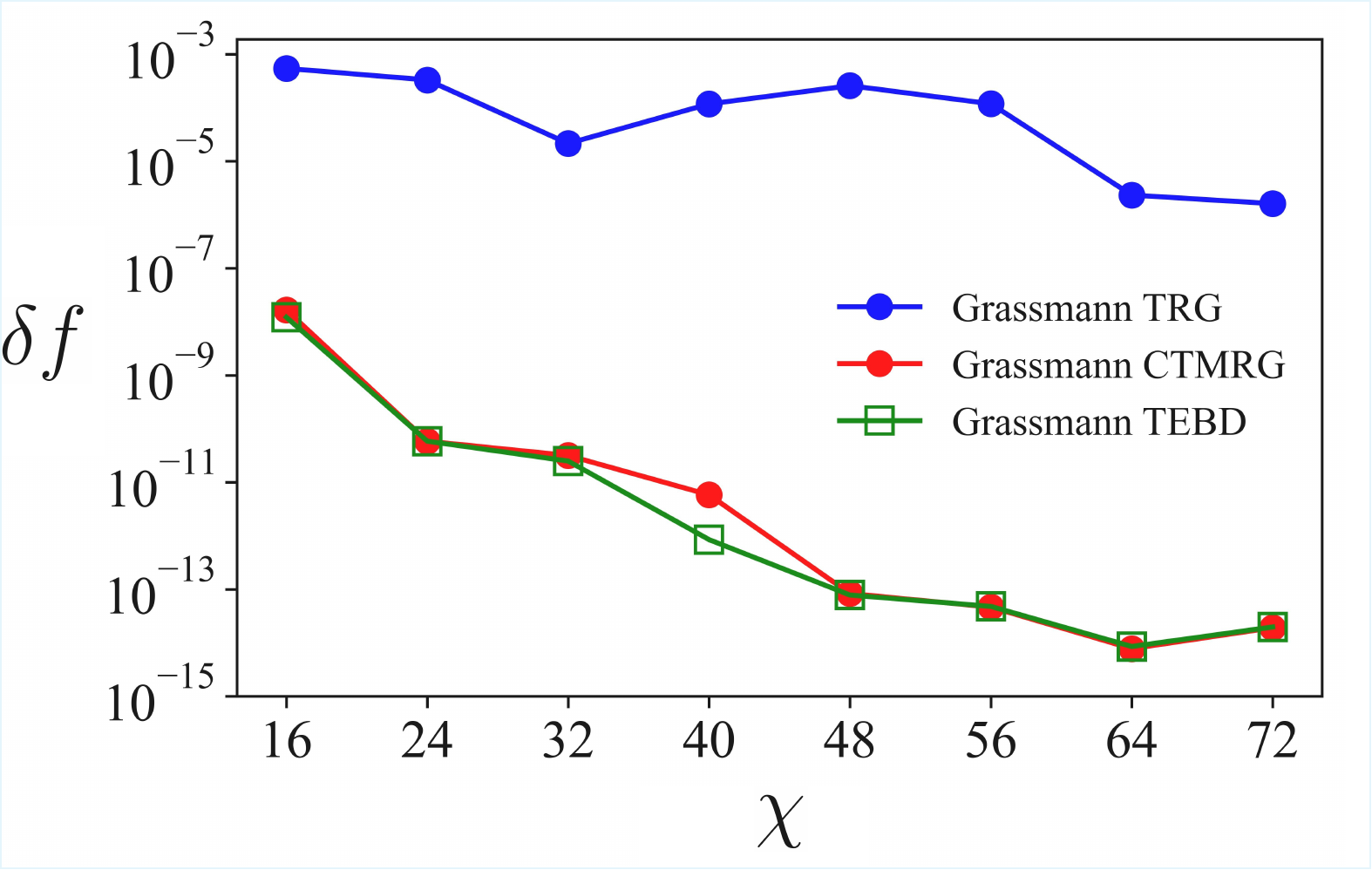}
	\caption{Relative error $\delta f$ of the free energy density for the single-flavor, non-interacting GNW model at $m = -1.2$, comparing Grassmann CTMRG, Grassmann TEBD, and Grassmann TRG methods at various bond dimensions $\chi$.}
	\label{gctmrg_example2}
\end{figure} 

Fig.~\ref{gctmrg_example2} clearly establishes that both Grassmann CTMRG and Grassmann TEBD achieve accurate computations for two-dimensional Grassmann tensor networks. The two methods yield consistent results, in agreement with earlier observations for quantum spin models~\cite{Orus2009Simulation}. As anticipated, Grassmann CTMRG and Grassmann TEBD surpass Grassmann TRG in accuracy by orders of magnitude at comparable bond dimensions, reaching nearly exact value within double-precision accuracy already at $\chi = 64$. This success establishes a pathway for deploying state-of-the-art tensor network techniques from condensed matter physics (such as the variational uniform matrix product state~\cite{Fishman2018Faster}) to address non-perturbative problems in two-dimensional lattice field theories. As a final remark, we note that Grassmann CTMRG has also proven effective at quantitatively characterizing the ground-state phase diagram of one-dimensional fermionic Hubbard model within the path-integral formalism~\cite{Kong2026Hubbard}.

\subsection{Grassmann imaginary-time evolution} \label{gsu}

\subsubsection{Method} \label{gsu_method}

The ground state of a fermionic Hamiltonian $\hat{H}$ can be prepared via imaginary-time evolution. Let $\vert \Psi_{0} \rangle$ be an arbitrary (non-orthogonal) initial state, then
\begin{eqnarray}
\vert \Psi \rangle = 
\lim_{\beta\rightarrow\infty}
{\rm e}^{
	-\hat{H}\beta
}
\vert
\Psi_{0}
\rangle
=
\lim_{m\rightarrow\infty}
\left(
{\rm e}^{
	-\hat{H}\Delta\tau
}
\right)^{m}
\vert
\Psi_{0}
\rangle, \quad
\beta = m\Delta\tau.
\end{eqnarray}
For sufficiently large imaginary time $\beta$, the evolution operator effectively projects $\vert \Psi_{0} \rangle$ onto the ground state $\vert \Psi \rangle$. If the $\hat{H}$ consists solely of local interactions, the many-body evolution operator can be factorized via a Trotter decomposition. To illustrate, take a nearest-neighbor fermionic Hamiltonian on a honeycomb lattice. It can be divided into three non-commuting groups $\hat{H} = \hat{H}_{\alpha} + \hat{H}_{\beta} + \hat{H}_{\gamma}$, each associated with a distinct lattice direction. Then we have
\begin{eqnarray} 
{\rm e}^{
	-\hat{H}\Delta\tau
}
~=~ 
{\rm e}^{
	-\hat{H}_{\alpha}\Delta\tau
}
{\rm e}^{
	-\hat{H}_{\beta}\Delta\tau
}
{\rm e}^{
	-\hat{H}_{\gamma}\Delta\tau
}
+
O(\Delta\tau^{2}).
\end{eqnarray}
Evolving the state by an imaginary time step is achieved by sequentially applying a set of commuting local bond operators to the state:
\begin{eqnarray} \label{bond_op}
\left(
\prod_{j}
{\rm e}^{
	-\hat{H}_{\alpha}^{j}\Delta\tau
}
\prod_{k}
{\rm e}^{
	-\hat{H}_{\beta}^{k}\Delta\tau
}
\prod_{l}
{\rm e}^{
	-\hat{H}_{\gamma}^{l}\Delta\tau
}
\right)
\vert
\Psi_{0}
\rangle.
\end{eqnarray}
Within the fermionic coherent-state representation, each local action corresponds to a Grassmann contraction involving three tensors: a rank-4 Grassmann tensor $\mt{G}$, representing a single bond operator in Eq.~(\ref{bond_op}), and two adjacent rank-4 Grassmann PEPS tensors, $\mt{A}$ and $\mt{B}$. For instance, if we denote the physical and virtual bond dimension of the Grassmann PEPS by $d$ and $D$, respectively, the contraction on an $\alpha$ bond is:
\begin{eqnarray} \label{gsu1}
\raisebox{-0.42\height}{\includegraphics[width=0.76\textwidth, page=1]{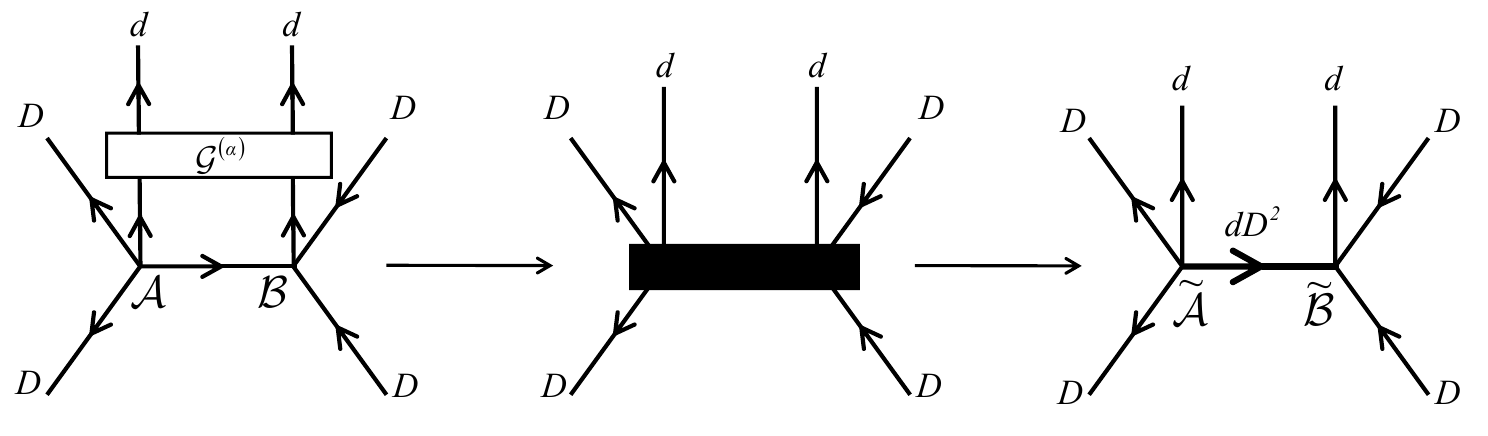}}.
\end{eqnarray}
The bond dimension is increased to $dD^{2}$ and must be truncated after each gate application to maintain computational tractability. Several truncation schemes have been developed \cite{VidalTEBD2003, Jiang2008Accurate,Wang2011Cluster,Jordan2008Classical,Dziarmaga2022Time,Phien2015Infinite}. Among these, the simple update (SU) \cite{Jiang2008Accurate, Xie2014Tensor} offers a particularly simple and efficient approach that incorporates environment effects via bond weights, essentially serving as an entanglement mean-field approximation. In the context of a Grassmann PEPS on a honeycomb lattice, this is implemented by introducing positive diagonal matrices $\{ \Lambda_{\alpha},\Lambda_{\beta},\Lambda_{\gamma}\}$ on every virtual link. These bond matrices are efficiently updated via SVDs, thus minimizing local truncation errors. 

\begin{figure}[htbp]
	\centering
	\includegraphics[height=8.0cm,width=14.0cm]{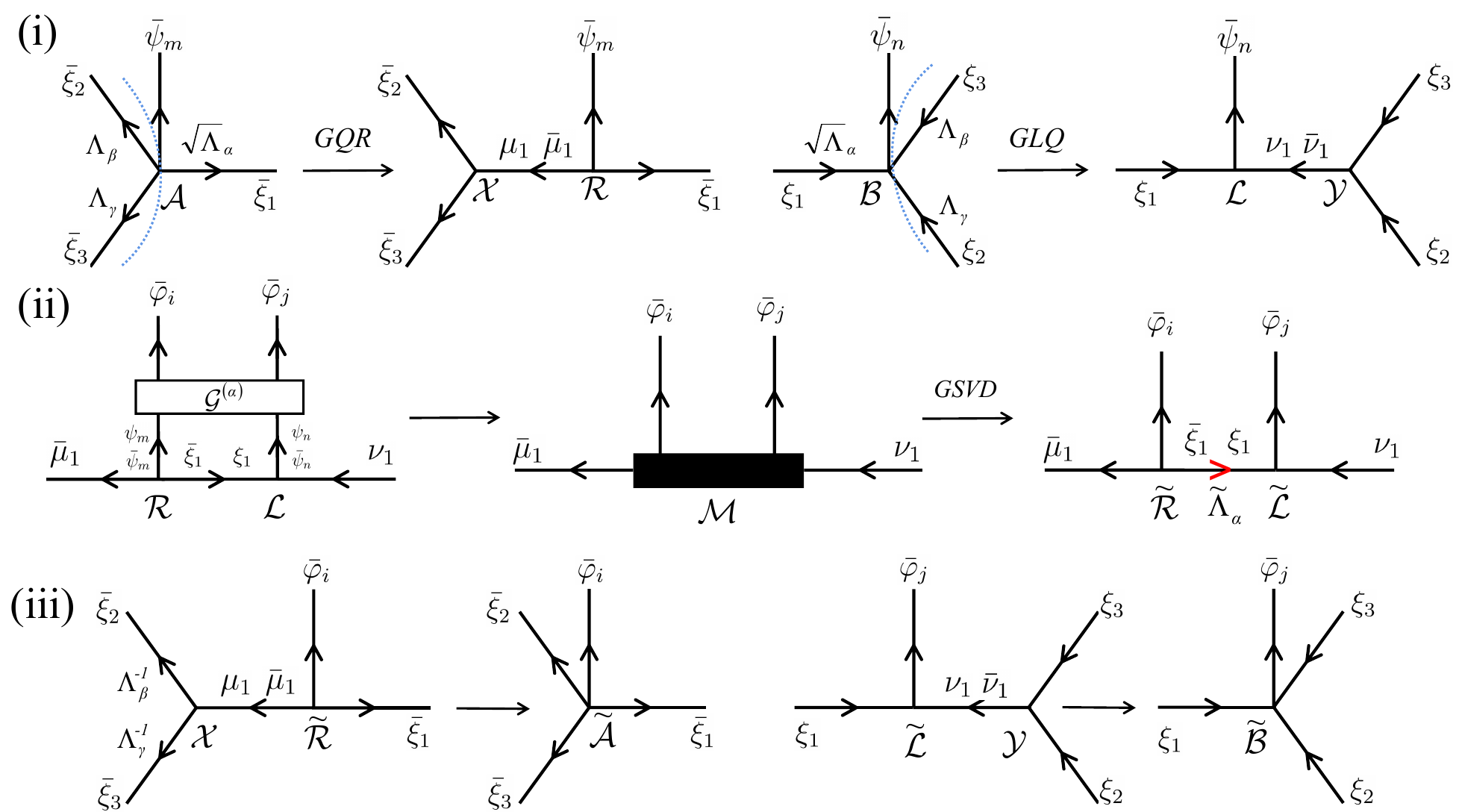}
	\caption{Applying the Grassmann gate $\mt{G}^{(\alpha)}$ to bond $\alpha$ in the Grassmann SU algorithm involves the following steps: (i) preprocessing, (ii) contraction and truncation and (iii) renormalization, leading to updated local tensors $\mt{A}$, $\mt{B}$ and $\Lambda_{\alpha}$.}
	\label{gsu0}
\end{figure}

As an illustrative example, consider applying an imaginary-time evolution gate $\mt{G}^{(\alpha)}$ to update the local Grassmann PEPS tensors $\mt{A}$, $\mt{B}$, and the associated bond matrix $\Lambda_{\alpha}$. The Grassmann SU algorithm, schematically depicted in Fig.~\ref{gsu0}, proceeds through the following steps:

(i) Preprocessing: Absorb the surrounding bond weights into the local tensors, and perform Grassmann QR and LQ decompositions on the rank-4 Grassmann PEPS tensors $\mt{A}$ and $\mt{B}$. This simple bond projection trick~\cite{Benedikt2021Beginner} reduces the computational cost of subsequent SVDs. To elaborate, we define Grassmann tensors $\mt{A}$, $\mt{B}$ and $\mt{G}^{(\alpha)}$ as}
\begin{eqnarray}
\mt{A}_{mk_{1}k_{2}k_{3}} 
&=& 
A_{mk_{1}k_{2}k_{3}}
\bar{\psi}_{m}^{p(m)}
\bar{\xi}_{1}^{p(k_{1})}
\bar{\xi}_{2}^{p(k_{2})}
\bar{\xi}_{3}^{p(k_{3})},
\\
\mt{B}_{nk_{1}l_{2}l_{3}} 
&=& 
B_{nk_{1}l_{2}l_{3}}
\bar{\psi}_{n}^{p(n)}
\xi_{1}^{p(k_{1})}
\xi_{2}^{p(l_{2})}
\xi_{3}^{p(l_{3})},
\\
 \label{E_def}
\mt{G}_{ijmn}^{(\alpha)}
&=&
G_{ijmn}^{(\alpha)}
\bar{\varphi}_{i}^{p(i)}
\bar{\varphi}_{j}^{p(j)}
\psi_{n}^{p(n)}
\psi_{m}^{p(m)}.
\end{eqnarray}
Notice the ordering of Grassmann numbers for $\mt{G}^{(\alpha)}$ in Eq.~(\ref{E_def}), as explained in Sec.~\ref{Sec:ferop}. To prepare for the Grassmann decomposition, the bond weights are first absorbed into local tensors, and the Grassmann variables are then reordered,
\begin{eqnarray}
\mt{A}^{'}_{k_{2}k_{3}mk_{1}} 
&=& 
A^{'}_{k_{2}k_{3}mk_{1}}
\bar{\xi}_{2}^{p(k_{2})}
\bar{\xi}_{3}^{p(k_{3})}
\bar{\psi}_{m}^{p(m)}
\bar{\xi}_{1}^{p(k_{1})},
\\
\mt{B}^{'}_{nk_{1}l_{2}l_{3}} 
&=& 
B^{'}_{nk_{1}l_{2}l_{3}}
\bar{\psi}_{n}^{p(n)}
\xi_{1}^{p(k_{1})}
\xi_{2}^{p(l_{2})}
\xi_{3}^{p(l_{3})},
\end{eqnarray}
where the $A^{'}$ and $B^{'}$ are
\begin{eqnarray} \label{coef_At}
A^{'}_{k_{2}k_{3}mk_{1}}
&=&
A^{'}_{mk_{1}k_{2}k_{3}}
(\sqrt{\Lambda_{\alpha}})_{k_{1}}
(\Lambda_{\beta})_{k_{2}}
(\Lambda_{\gamma})_{k_{3}}
\times
(-1)^{
	p(k_{2}) + p(k_{3})
},
\\
B^{'}_{nk_{1}l_{2}l_{3}} 
&=& 
B^{'}_{nk_{1}l_{2}l_{3}}
(\sqrt{\Lambda_{\alpha}})_{k_{1}}
(\Lambda_{\beta})_{l_{2}}
(\Lambda_{\gamma})_{l_{3}}.
\end{eqnarray}
The sign factor in Eq.~(\ref{coef_At}) has been simplified by exploiting the Grassmann-even property of $\mt{A}$, namely $\mmod{p(m) + p(k_{1}), 2} = \mmod{p(k_{2}) + p(k_{3}), 2}$, leading to:
\begin{eqnarray}
& & \nonumber
\mmod{\left[
	p(m) + p(k_{1})
	\right]\times
	\left[
	p(k_{2}) + p(k_{3})
	\right]
	, 2}
\\ 
&=&  \nonumber
\mmod{\left[
	p(k_{2}) + p(k_{3})
	\right]
	\times
	\left[
	p(k_{2}) + p(k_{3})
	\right]
	, 2},
\\
&=&
\mmod{
	p(k_{2}) + p(k_{3}), 2}.
\end{eqnarray}
Then we perform the Grassmann QR and LQ decompositions on $\mt{A}^{'}$ and $\mt{B}^{'}$ respectively,
\begin{eqnarray}
\mt{A}^{'}_{k_{2}k_{3}mk_{1}}
&=&
\sum_{a_{1}}
\mt{X}_{k_{2}k_{3}a_{1}}
\mt{R}_{a_{1}mk_{1}},
\\
\mt{B}^{'}_{nk_{1}l_{2}l_{3}}
&=&
\sum_{b_{1}}
\mt{L}_{nk_{1}b_{1}}
\mt{Y}_{b_{1}l_{2}l_{3}}.
\end{eqnarray}
Here, the corresponding coefficient tensors $X$, $R$ and $L$, $Y$, are obtained via the ordinary QR and LQ decompositions via $A^{'} = XR$ and $B^{'} = LY$.

(ii) Contraction and Truncation: Contract the gate $\mt{G}^{(\alpha)}$, $\mt{R}$, and $\mt{L}$, 
\begin{eqnarray}
\mt{M}_{ia_{1}jb_{1}}
&=&
\sum_{mn}
\sum_{k_{1}}
\mt{G}_{ijmn}^{(\alpha)}
\mt{R}_{a_{1}mk_{1}}
\mt{L}_{nk_{1}b_{1}}
=
M_{ia_{1}jb_{1}}
\bar{\varphi}_{i}^{p(i)}
\bar{\mu}_{1}^{p(a_{1})}
\bar{\varphi}_{j}^{p(j)}
\nu_{1}^{p(b_{1})},
\\ 
\label{coef_M}
M_{ia_{1}jb_{1}}
&=&
\sum_{mn}
\sum_{k_{1}}
G_{ijmn}^{(\alpha)}
R_{a_{1}mk_{1}}
L_{nk_{1}b_{1}}
\times
(-1)^{
	s_{1} + s_{2}
},
\\ 
s_{1}
&=& \nonumber
p(m)
\times 
\left[
p(a_{1}) + p(n)
\right]
+
p(m) + p(n) + p(k_{1})
,
\quad
s_{2}
=
p(a_{1})
\times 
p(j).
\end{eqnarray}
The sign factors $(-1)^{s_{1}}$ and $(-1)^{s_{2}}$ arise from the Grassmann contraction and reordering, respectively. We then perform the Grassmann SVD on  $\mt{M}$, 
\begin{eqnarray} \label{MSVD}
\mt{M}_{ia_{1}jb_{1}} = 
\sum_{k_{1}}
\mt{U}_{ia_{1}k_{1}}
(\tilde{\Lambda}_{\alpha})_{k_{1}}
\mt{V}_{k_{1}jb_{1}},
\end{eqnarray}
where $\tilde{\Lambda}_{\alpha}$ is regarded as a new weight matrix on the bond $\alpha$. The $\mt{\tilde{R}}$ and $\mt{\tilde{L}}$ is then introduced as
\begin{eqnarray}
\mt{\tilde{R}}_{ia_{1}k_{1}} 
=
\tilde{R}_{ia_{1}k_{1}}
\bar{\varphi}_{i}^{p(i)}
\bar{\mu}_{1}^{p(a_{1})}
\bar{\xi}_{1}^{p(k_{1})},
\quad
\mt{\tilde{L}}_{k_{1}jb_{1}}
=
\tilde{L}_{k_{1}jb_{1}}
\xi_{1}^{p(k_{1})}
\bar{\varphi}_{j}^{p(j)}
\nu_{1}^{p(b_{1})},
\end{eqnarray}
where the coefficient tensors $\tilde{R}$ and $\tilde{L}$ are defined as
\begin{eqnarray} \label{RLdef}
\tilde{R}_{ia_{1}k_{1}}
\equiv
U_{ia_{1}k_{1}}
\times
(-1)^{p(k_{1})}
,
\quad
\tilde{L}_{k_{1}jb_{1}}
\equiv
V_{k_{1}jb_{1}}.
\end{eqnarray}
The matrices $U$, $V$ and $\tilde{\Lambda}_{\alpha}$ are obtained from an ordinary SVD of $M$ defined in Eq.~(\ref{coef_M}). The link with dimension $dD^{2}$ is then truncated by retaining only the largest $D$ singular values of $\tilde{\Lambda}_{\alpha}$. Crucially, the sign factor $(-1)^{p(k_{1})}$ in $\tilde{R}$ results from reordering Grassmann variables when inserting a resolution of identity,
\begin{eqnarray}
\int_{\bar{\xi}_{1}\xi_{1}}
\xi_{1}^{p(k_{1})}
\bar{\xi}_{1}^{p(k_{1})}
=
\int_{\bar{\xi}_{1}\xi_{1}}
\bar{\xi}_{1}^{p(k_{1})}
\xi_{1}^{p(k_{1})}
\times
(-1)^{p(k_{1})}.
\end{eqnarray}
This adjustment, also mentioned in Sec.~\ref{Sec:GTD}, ensures a consistent arrow convention throughout the computation: all virtual links of $\mt{A}$ point outward, and those of $\mt{B}$ point inward. 

(iii) Renormalization. Finally, we combine the $\mt{\tilde{R}}$ with $\mt{X}$, and $\mt{\tilde{L}}$ with $\mt{Y}$. The previously absorbed bond matrices,  $\Lambda_{\beta}$ and $\Lambda_{\gamma}$, are then split off from the resulting tensors. The updated local tensors $\tilde{\mt{A}}$ and $\tilde{\mt{B}}$ are
\begin{eqnarray}
\mt{\tilde{A}}_{mk_{1}k_{2}k_{3}} 
&=& 
\sum_{a_{1}}
\mt{X}_{k_{2}k_{3}a_{1}}
\mt{\tilde{R}}_{ma_{1}k_{1}},
\\
\mt{\tilde{B}}_{nk_{1}l_{2}l_{3}} 
&=& 
\sum_{b_{1}}
\mt{\tilde{L}}_{k_{1}nb_{1}}
\mt{Y}_{b_{1}l_{2}l_{3}}.
\end{eqnarray}
We have renamed the index $i \rightarrow m$ and $j \rightarrow n$ in $\mt{\tilde{R}}$ and $\mt{\tilde{L}}$, respectively. The coefficient tensors $A$ and $B$ are given by
\begin{eqnarray}
\label{A_coef_new}
\tilde{A}_{mk_{1}k_{2}k_{3}}
&=&
\sum_{a_{1}}
X_{k_{2}k_{3}a_{1}}
\tilde{R}_{ma_{1}k_{1}}
(\Lambda_{\beta}^{-1})_{k_{2}}
(\Lambda_{\gamma}^{-1})_{k_{3}}
\times
(-1)^{
	p(a_{1})
	\times 
	p(m) 
	+ 
	p(k_{2}) + p(k_{3})
},
\\ 
 \label{B_coef_new}
\tilde{B}_{nk_{1}l_{2}l_{3}} 
&=& 
\sum_{b_{1}}
\tilde{L}_{k_{1}nb_{1}}
Y_{b_{1}l_{2}l_{3}}
(\Lambda_{\beta}^{-1})_{l_{2}}
(\Lambda_{\gamma}^{-1})_{l_{3}}
\times
(-1)^{p(n)\times p(k_{1})}.
\end{eqnarray}
The sign factors in Eqs.~(\ref{A_coef_new}-\ref{B_coef_new}) again originate from performing Grassmann integrals and reordering Grassmann variables. This completes one full step of imaginary-time evolution. The Grassmann SU algorithm iterates these steps on different bonds of the honeycomb lattice until the bond matrices $\{\Lambda_{\alpha}, \Lambda_{\beta}, \Lambda_{\gamma}\}$ converge.

We now describe an adapted Grassmann imaginary-time evolution algorithm based on the higher-order SVD(HOSVD)~\cite{Xie2012HOTRG,Xie2014Tensor}. Rather than dividing the Hamiltonian terms by lattice direction, we now consider operators defined on upward- and downward-pointing triangles centered on the A and B sublattices of the honeycomb lattice:
\begin{eqnarray}
\hat{H} = \hat{H}_{\vartriangle} + \hat{H}_{\triangledown}. 
\end{eqnarray}
This decomposition scheme is particularly advantageous for Hamiltonians that include next-nearest-neighbor interactions on the honeycomb lattice. In such cases, the imaginary-time evolution gates within the fermionic coherent-state representation are rank-8 Grassmann tensors $\mt{G}^{(\vartriangle)}$ and $\mt{G}^{(\triangledown)}$:
\begin{eqnarray} \label{gsu6}
{\rm e}^{-\hat{H}\Delta\tau} 
\approx
{\rm e}^{-\hat{H}_{\vartriangle}\Delta\tau}
{\rm e}^{-\hat{H}_{\triangledown}\Delta\tau}  
\longrightarrow
\raisebox{-0.42\height}{\includegraphics[width=0.4\textwidth, page=1]{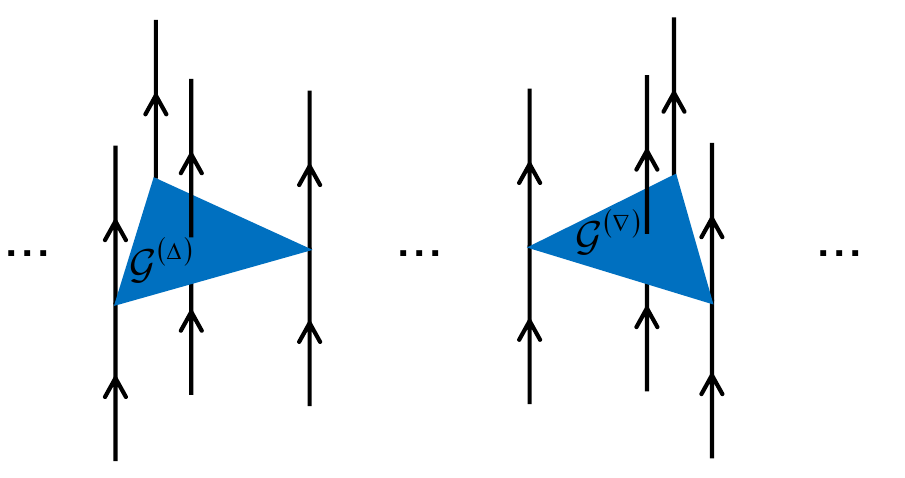}}.
\end{eqnarray}

We perform the imaginary-time evolution of an initial state using the gates defined in Eq.~(\ref{gsu6}), combined with the Grassmann SU truncation scheme. The overall procedure closely resembles the one described previously, we therefore outline only the key steps below. Consider applying the gate $\mt{G}^{(\vartriangle)}$ to update the four local Grassmann PEPS tensors $\mt{A}$, $\mt{B}_{1}$, $\mt{B}_{2}$, $\mt{B}_{3}$ and the surrounding bond weights of $\mt{A}$, it starts by:

(i) Preprocessing: Perform Grassmann QR and LQ decompositions on the three nearest-neighbor $\mt{B}$ tensors surrounding the central $\mt{A}$ tensor, after absorbing all relevant bond weights
\begin{eqnarray}
\raisebox{-0.42\height}{\includegraphics[width=0.83\textwidth, page=1]{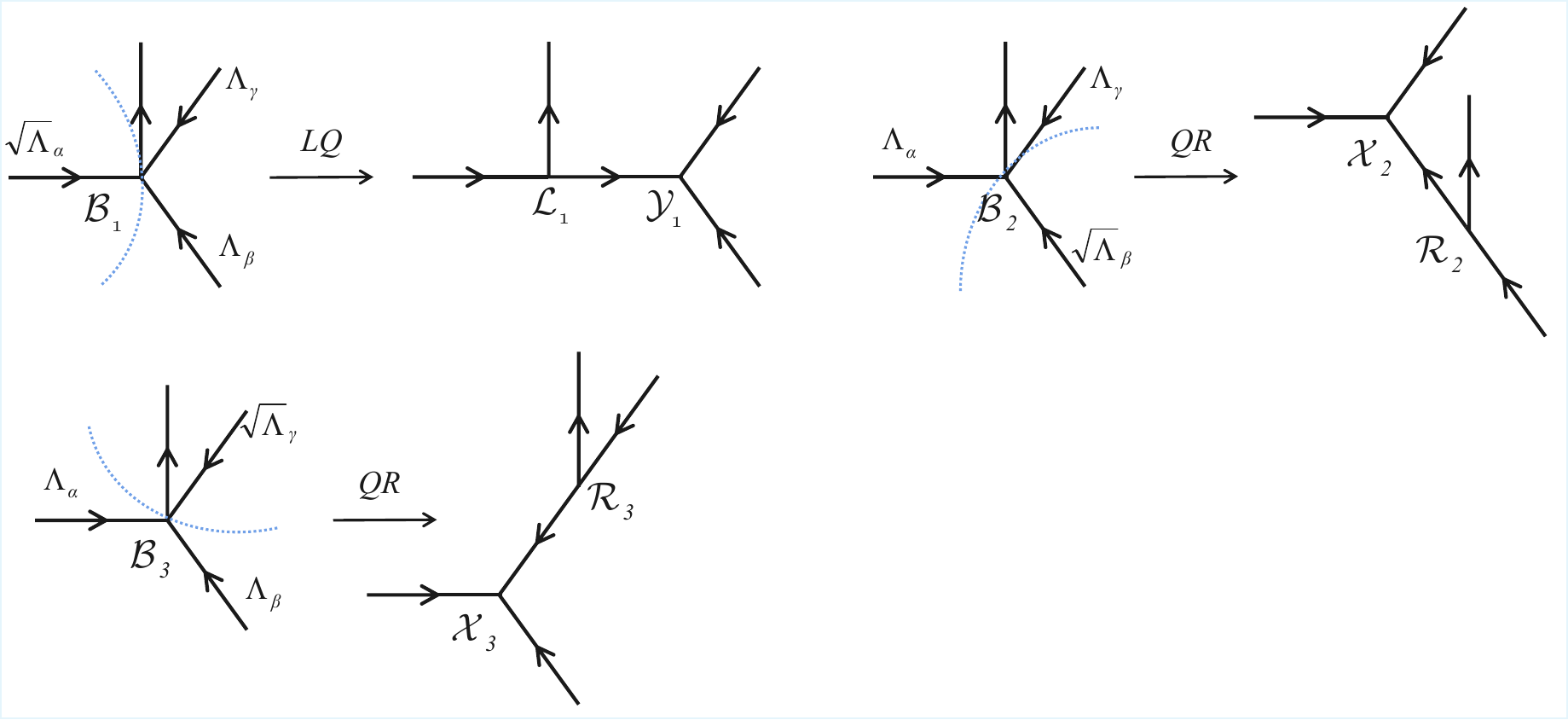}}
\end{eqnarray}

(ii) Contraction and Truncation: Apply the Grassmann gate tensor $\mt{G}^{(\vartriangle)}$ to the cluster formed by the central tensor $\mt{A}$ and its three neighbors $\mt{L}_{1}$, $\mt{R}_{2}$ and $\mt{R}_{3}$, yielding a rank-7 Grassmann tensor $\mt{M}$:
\begin{eqnarray} \label{def_M}
\raisebox{-0.42\height}{\includegraphics[width=0.68\textwidth, page=1]{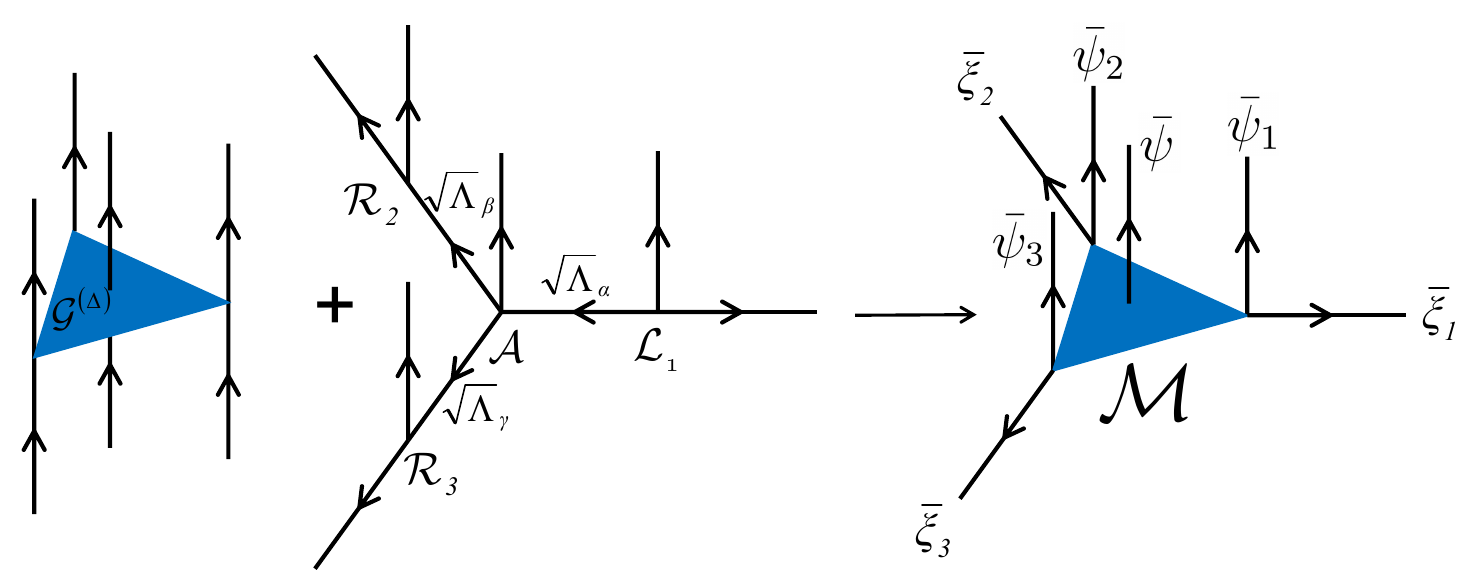}}
\end{eqnarray}
At this stage, the links with increased dimension can be truncated simultaneously using Grassmann HOSVD. This is achieved by constructing and inserting a pair of Grassmann isometric tensors at each link—an operation mathematically equivalent to performing HOSVD on the original tensor. To construct these Grassmann isometric tensors, we examine the explicit form of the Grassmann tensor $\mt{M}$ (see also Eq.~(\ref{def_M})):
\begin{eqnarray} \nonumber
\mt{M}_{ii_{1}k_{1}i_{2}k_{2}i_{3}k_{3}}
=
M_{ii_{1}k_{1}i_{2}k_{2}i_{3}k_{3}}
\bar{\psi}^{p(i)}
\bar{\psi}_{1}^{p(i_{1})}
\bar{\xi}_{1}^{p(k_{1})}
\bar{\psi}_{2}^{p(i_{2})}
\bar{\xi}_{2}^{p(k_{2})}
\bar{\psi}_{3}^{p(i_{3})}
\bar{\xi}_{3}^{p(k_{3})}.
\end{eqnarray}
The required Grassmann isometric tensors are obtained by performing following Grassmann SVD:
\begin{eqnarray}
\mathcal{M}_{i_{1}k_{1};ii_{2}k_{2}i_{3}k_{3}}^{(1)}
&=& \label{U1}
\sum_{a_{1}}
\mathcal{U}^{(1)}_{i_{1}k_{1}a_{1}}
(\tilde{\Lambda}_{\alpha})_{a_{1}}
\mathcal{V}^{(1)}_{a_{1}ii_{2}k_{2}i_{3}k_{3}},
\\
\mathcal{M}_{i_{2}k_{2};ii_{1}k_{1}i_{3}k_{3}}^{(2)}
&=& \label{U2}
\sum_{a_{2}}
\mathcal{U}^{(2)}_{i_{2}k_{2}a_{2}}
(\tilde{\Lambda}_{\beta})_{a_{2}}
\mathcal{V}^{(2)}_{a_{2}ii_{1}k_{1}i_{3}k_{3}},
\\
\mathcal{M}_{i_{3}k_{3};ii_{1}k_{1}i_{2}k_{2}}^{(3)}
&=& \label{U3}
\sum_{a_{3}}
\mathcal{U}^{(3)}_{i_{3}k_{3}a_{3}}
(\tilde{\Lambda}_{\gamma})_{a_{3}}
\mathcal{V}^{(3)}_{a_{3}ii_{1}k_{1}i_{2}k_{2}}.
\end{eqnarray}
The tensors $\mt{M}^{(1,2,3)}$ result from permuting the indices of the original tensor $\mt{M}$. Consequently, their coefficient tensors acquire a sign factor determined by the reordering of Grassmann variables:
\begin{eqnarray}
M_{i_{1}k_{1};ii_{2}k_{2}i_{3}k_{3}}^{(1)}
&=&
M_{ii_{1}k_{1}i_{2}k_{2}i_{3}k_{3}}
\times
(-1)^{\left(
	p(i_{1}) + p(k_{1})	
	\right)
	\times
	p(i)
},
\\ 
M^{(2)}_{i_{2}k_{2};ii_{1}k_{1}i_{3}k_{3}}
&=&
M_{ii_{1}k_{1}i_{2}k_{2}i_{3}k_{3}}
\times
(-1)^{\left(
	p(i_{2}) + p(k_{2})	
	\right)
	\times
	\left(
	p(i_{1}) + p(k_{1}) + p(i)
	\right)
},
\\ 
M_{i_{3}k_{3};ii_{1}k_{1}i_{2}k_{2}}^{(3)}
&=&
M_{ii_{1}k_{1}i_{2}k_{2}i_{3}k_{3}}
\times
(-1)^{
	p(i_{3}) + p(k_{3})	
}.
\end{eqnarray}
To implement the truncation, we therefore insert the following pairs of Grassmann tensors generated from Eqs.~(\ref{U1}-\ref{U3}):
\begin{eqnarray} \label{gsu8}
& &
\sum_{a_{1}}
\mathcal{U}^{(1)}_{i_{1}k_{1}a_{1}}
\bar{\mathcal{U}}^{(1)}_{a_{1}j_{1}l_{1}},
\quad
\sum_{a_{2}}
\mathcal{U}^{(2)}_{i_{2}k_{2}a_{2}}
\bar{\mathcal{U}}^{(2)}_{a_{2}j_{2}l_{2}},
\quad
\sum_{a_{3}}
\mathcal{U}^{(3)}_{i_{3}k_{3}a_{3}}
\bar{\mathcal{U}}^{(3)}_{a_{3}j_{3}l_{3}},
\end{eqnarray}
where $\mt{U}^{(s)}$ and $\bar{\mt{U}}^{(s)}$ are defined as:
\begin{eqnarray}
\label{gsu7a}
\mathcal{U}^{(s)}_{i_{s}k_{s}a_{s}}
&=&
U^{(s)}_{i_{s}k_{s}a_{s}}
\bar{\psi}_{s}^{p(i_{s})}
\bar{\xi}_{s}^{p(k_{a})}
\zeta_{s}^{p(a_{s})},
\\ 
\label{gsu7b}
\bar{\mathcal{U}}^{(s)}_{a_{s}j_{s}l_{s}}
&=&
U^{(s)*}_{a_{s}j_{s}l_{s}}
\bar{\zeta}_{s}^{p(a_{s})}
\xi_{s}^{p(l_{s})}
\psi_{s}^{p(j_{s})}, 
\quad 
s = 1, 2, 3.
\end{eqnarray}

The index $a_{s}$ ($s = 1, 2, 3$) can be truncated based on the largest singular values of the new bond matrices $\tilde{\Lambda}_{\alpha}$, $\tilde{\Lambda}_{\beta}$, and $\tilde{\Lambda}_{\gamma}$, respectively. The truncation of each enlarged bond of $\mt{M}$ is then performed via Grassmann contractions with  corresponding $\bar{\mt{U}}^{(s)}$, yielding the updated tensor $\tilde{\mt{A}}$
\begin{eqnarray}
\raisebox{-0.42\height}{\includegraphics[width=0.5\textwidth, page=1]{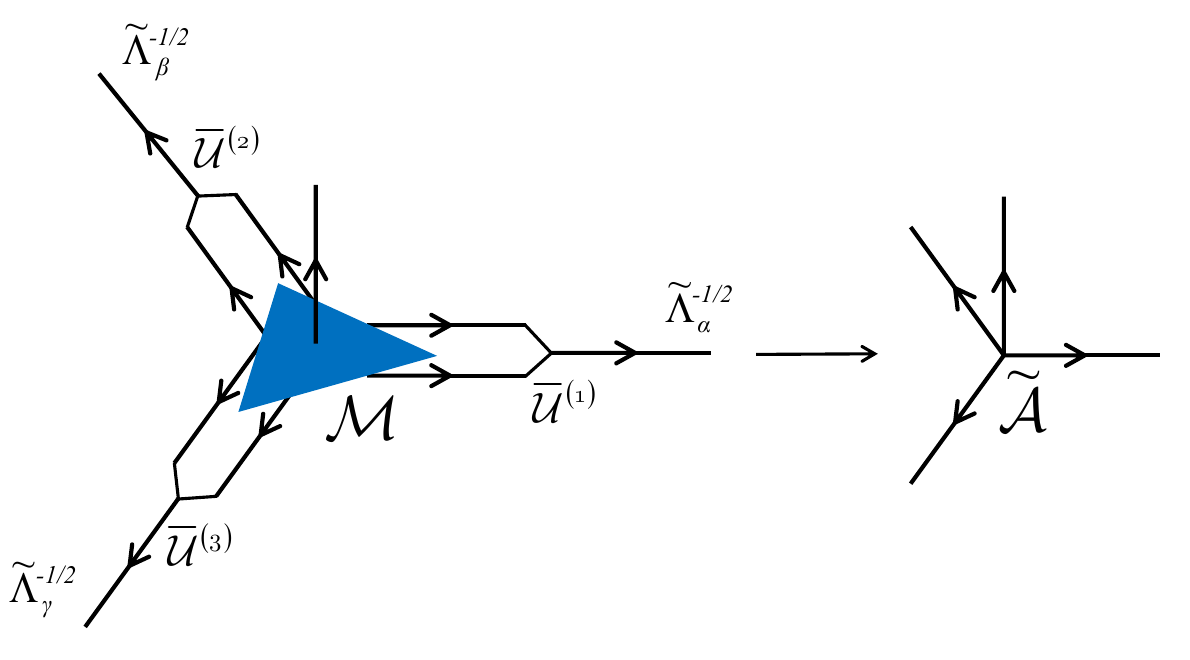}}
\label{A_HOSVD}
\end{eqnarray}

The three $\mt{B}$ tensors are finally renormalized  via:
\begin{eqnarray}
\raisebox{-0.42\height}{\includegraphics[width=0.88\textwidth, page=1]{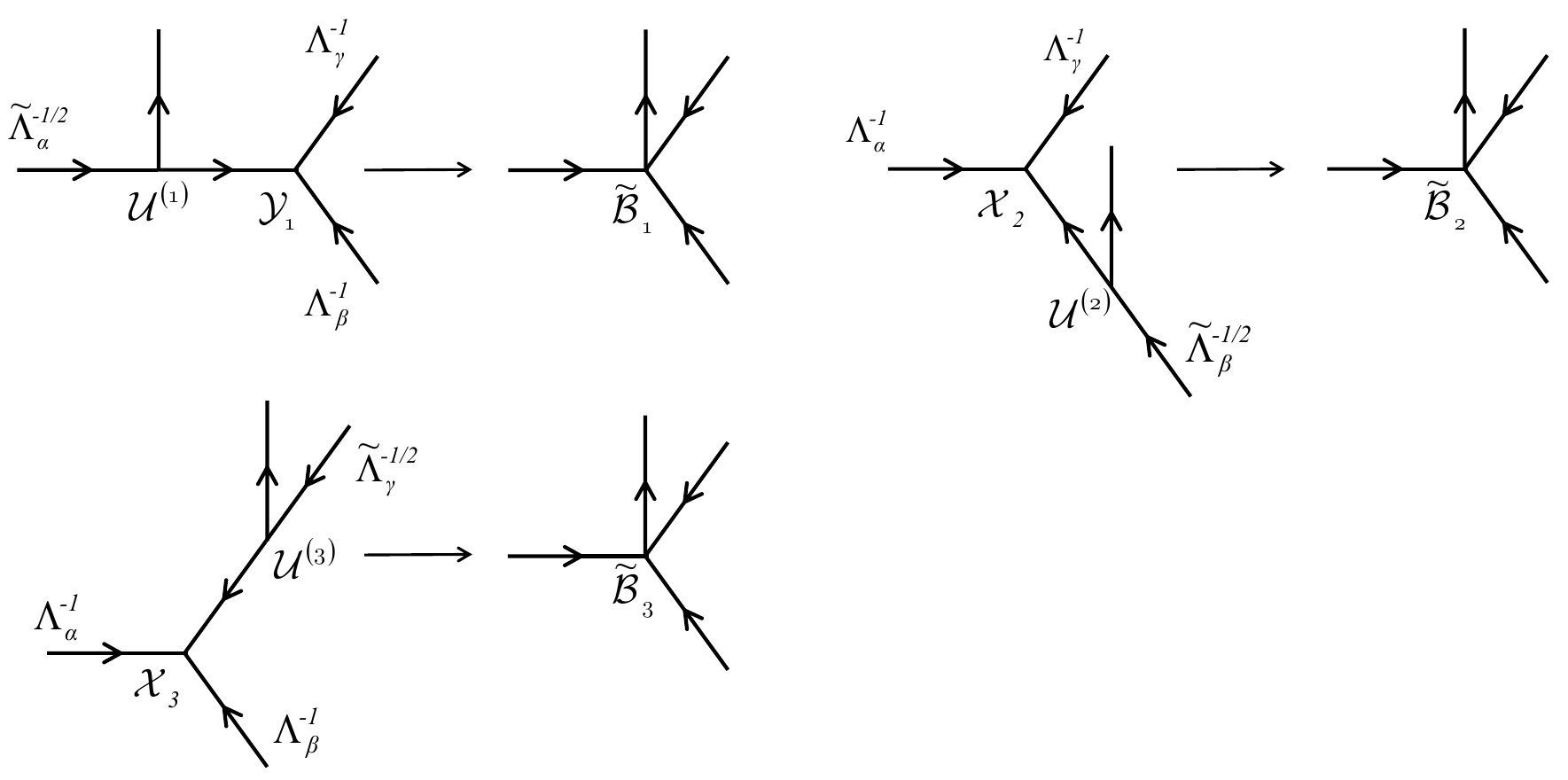}}
\label{B_HOSVD}
\end{eqnarray}
Notice that in Eqs.~(\ref{A_HOSVD}-\ref{B_HOSVD}), we have also incorporated $\tilde{\Lambda}_{\alpha}^{-1/2}$, $\tilde{\Lambda}_{\beta}^{-1/2}$, and $\tilde{\Lambda}_{\gamma}^{-1/2}$ in the renormalization step to properly introduce the new bond matrices $\tilde{\Lambda}_{s}$ through

\begin{eqnarray}
\mathbf{1} 
=
\tilde{\Lambda}_{s}
\tilde{\Lambda}_{s}^{-1} 
=
\tilde{\Lambda}_{s}^{-1/2}
\tilde{\Lambda}_{s}
\tilde{\Lambda}_{s}^{-1/2},
\quad
s = \alpha, \beta, \gamma,
\end{eqnarray}
where $\mathbf{1}$ is the identity matrix of the same dimension as the $\tilde{\Lambda}$ matrices.

\subsubsection{Example}

\begin{figure}[htbp]
	\centering
	\includegraphics[height=6.0cm,width=9.2cm]{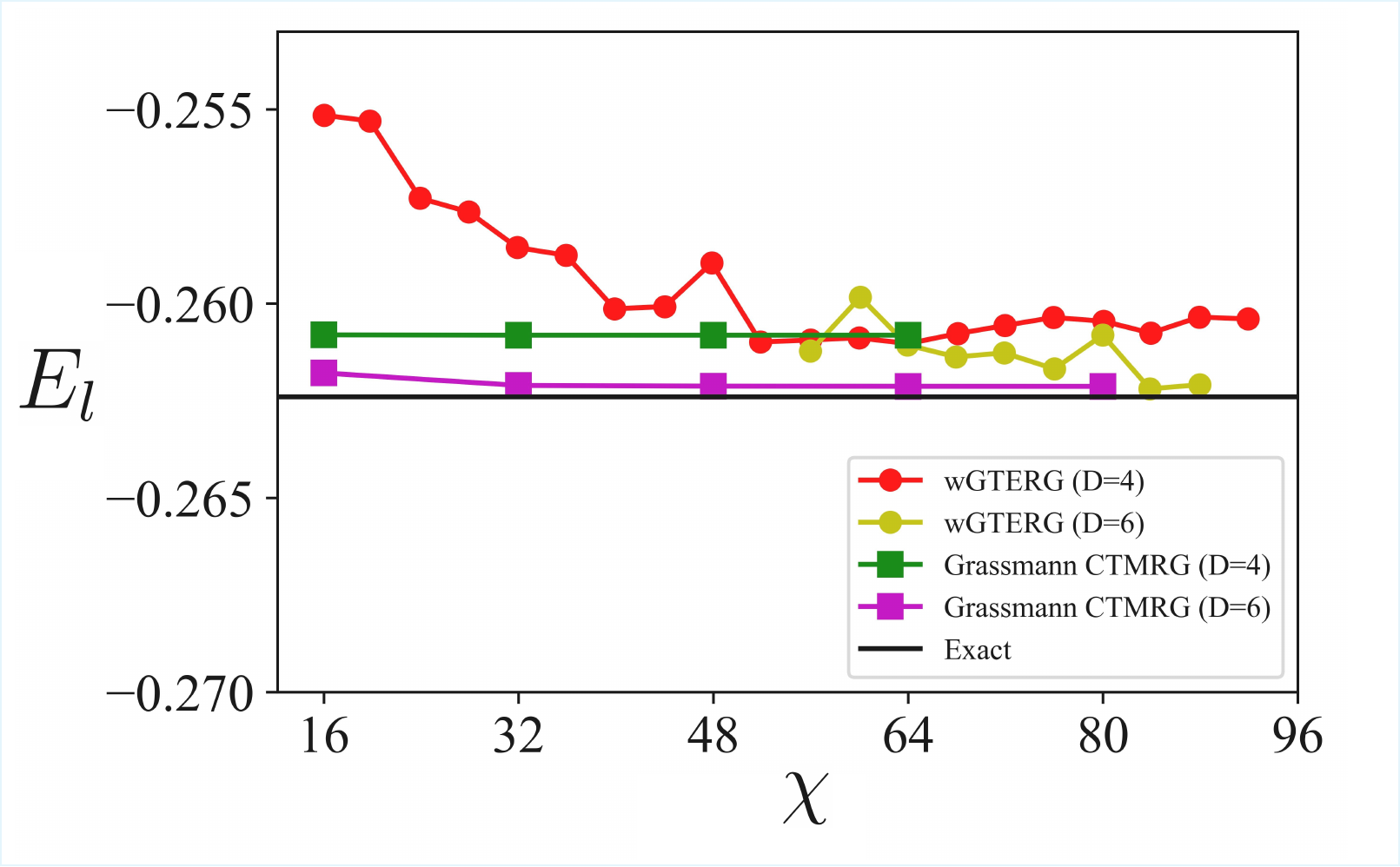}
	\caption{Energy per link for the free-fermion model in Eq.~(\ref{free_pairing}) at chemical potential $\mu=0.0$, computed via the Grassmann SU and Grassmann CTMRG methods for different bond dimensions $D$ and $\chi$. The wGTERG data are reproduced from Ref.~\cite{Gu2013Efficient}.}
	\label{gsu_example4}
\end{figure}

In this section, we apply Grassmann tensor network methods to two-dimensional fermionic Hamiltonians. Specifically, we first consider the free spinless fermion model with nearest-neighbour pairing on a honeycomb lattice~\cite{Gu2013Efficient}:
\begin{eqnarray} \label{free_pairing}
\hat{H} = -2\Delta \sum_{\langle ij \rangle}
\left(
\hat{c}_{i}^{\dagger}\hat{c}_{j}^{\dagger}
+
h.c.
\right)
+
\mu
\sum_{i}
\hat{n}_{i},
\end{eqnarray}
where $\Delta$ denotes the pairing strength and $\mu$ is the chemical potential. The ground-state wave function is obtained via the Grassmann SU algorithm detailed in Sec.~\ref{gsu_method}. Norms and expectation values, such as local bond energies, are then evaluated using the Grassmann CTMRG algorithm after deforming the honeycomb lattice to a square lattice. For instance, the energy associated with an $\alpha$ bond, $\langle \hat{H}_{b}^{(\alpha)} \rangle$, is given by:
\begin{eqnarray} \label{ctmrg_exp1}
\langle \hat{H}_{b}^{(\alpha)} \rangle = 
\raisebox{-0.48\height}{\includegraphics[width=0.44\textwidth, page=1]{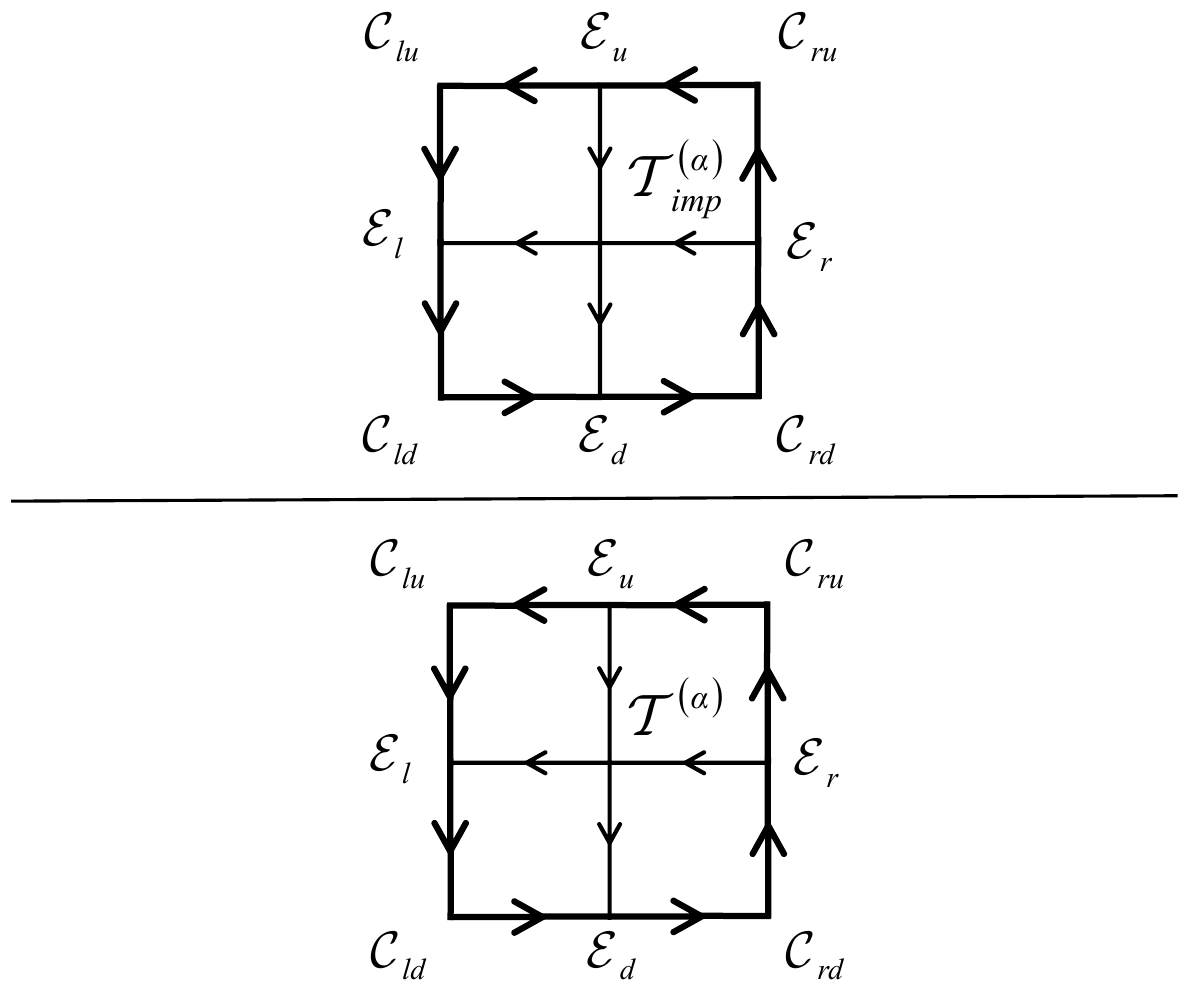}},
\end{eqnarray}
where $\hat{H}_{b}^{(\alpha)} = -2\Delta\left( \hat{c}^{\dagger}_{i}\hat{c}^{\dagger}_{j} + \hat{c}_{i}\hat{c}_{j} \right) + \dfrac{\mu}{3}\left( \hat{n}_{i} + \hat{n}_{j} \right)$. The Grassmann tensors $\mt{T}^{(\alpha)}$ and $\mt{T}_{\textrm{imp}}^{(\alpha)}$ appearing in Eq.~(\ref{ctmrg_exp1}) are constructed as follows:
\begin{eqnarray} \label{ctmrg_exp2a}
\raisebox{-0.46\height}{\includegraphics[width=0.64\textwidth, page=1]{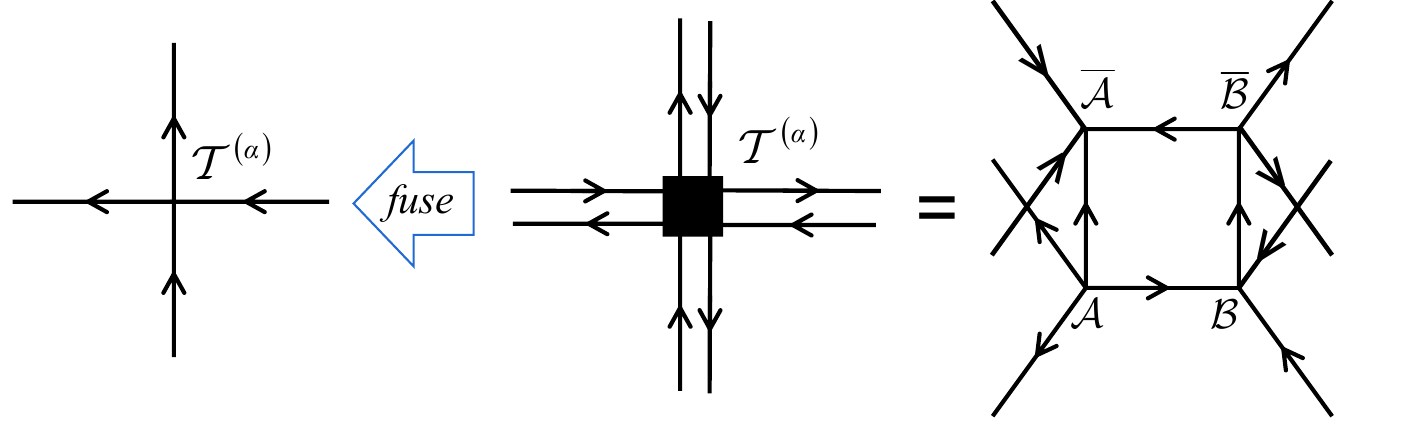}}
\end{eqnarray}
\begin{eqnarray} \label{ctmrg_exp2b}
\raisebox{-0.46\height}{\includegraphics[width=0.64\textwidth, page=1]{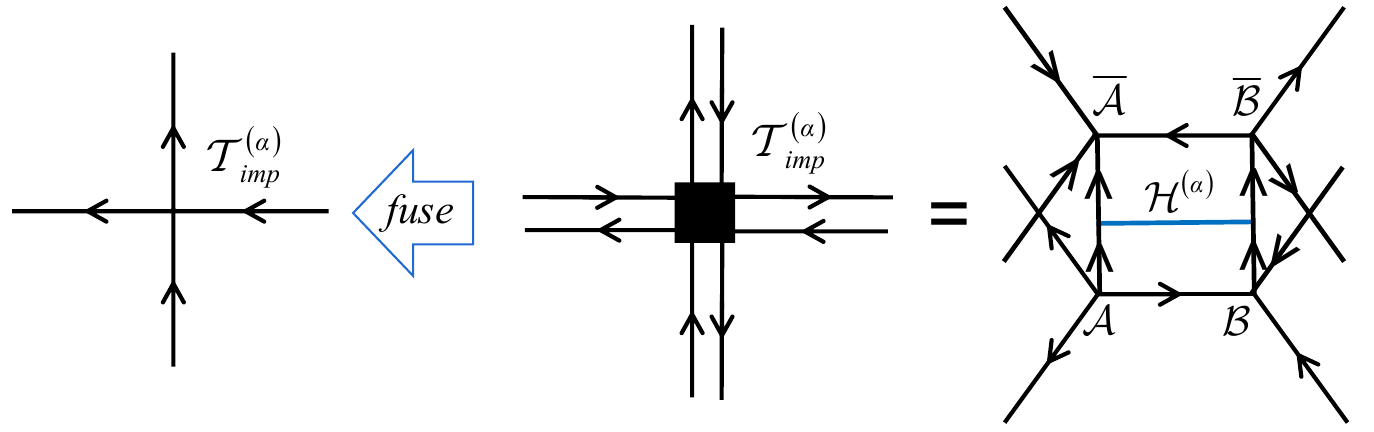}}
\end{eqnarray}
where $\mt{H}^{(\alpha)}$ is the Grassmann tensor representation of the bond operator $\hat{H}^{(\alpha)}_{b}$. The expectation values $\langle\hat{H}_{b}^{(\beta)}\rangle$ and $\langle\hat{H}_{b}^{(\gamma)}\rangle$ are evaluated similarly. The ground-state energy per link is defined as 
\begin{eqnarray}
E_{l} = 
\dfrac{1}{3}
\left(
\langle\hat{H}_{b}^{(\alpha)}\rangle + \langle\hat{H}_{b}^{(\beta)}\rangle + \langle\hat{H}_{b}^{(\gamma)}\rangle
\right).
\end{eqnarray}
We calculate $E_{l}$ for the free-fermion model in Eq.~(\ref{free_pairing}) at $\mu=0.0$, with various Grassmann PEPS bond dimensions $D$ and environment bond dimensions $\chi$. The results are presented in Fig.~\ref{gsu_example4}.

We observe in Fig.~\ref{gsu_example4} that Grassmann CTMRG indeed converges more rapidly towards the exact energy for $D=4$ and $D=6$ compared with the weighted Grassmann tensor entanglement‑filtering renormalization group (wGTERG) data from Ref.~\cite{Gu2013Efficient}.

Next, we consider a more challenging system: the $t$-$V_{1}$-$V_{2}$ model on a honeycomb lattice:
\begin{eqnarray} \label{t-V1-V2}
\hat{H} = -t \sum_{\langle ij \rangle}
\left(
\hat{c}_{i}^{\dagger}\hat{c}_{j}
+
h.c.
\right)
+
V_{1}
\sum_{\langle ij \rangle}
\hat{n}_{i}\hat{n}_{j}
+
V_{2}
\sum_{\langle\langle ij \rangle\rangle}
\hat{n}_{i}\hat{n}_{j}
-
\mu
\sum_{i}
\hat{n}_{i}.
\end{eqnarray}
The first term in Eq.~(\ref{t-V1-V2}) describes fermion hopping, while the second and third terms represent nearest-neighbor (NN) and next-nearest-neighbor (NNN) density-density interactions, respectively. At half-filling with repulsive interactions $V_{1}, V_{2}>0$, the model has been extensively studied to determine whether it hosts the intriguing quantum anomalous Hall phase, see Ref.~\cite{Capponi2016Phase} and references therein.

A recent DMRG study has revealed a pair density wave (PDW) state in this model~\cite{Jiang2024Pair}, renewing theoretical interest. A PDW state is characterized by a spatially modulated superconducting order parameter~\cite{Agterberg2020Physics}. In the parameter regime $V_{1}<0, V_{2}>0$ and away from half-filling, the model in Eq.~(\ref{t-V1-V2}) is believed to host such a phase. 

To investigate the model in the parameter regime relevant for the PDW state, that is $V_{1}=-1.0$ and $V_{2}=0.5$~\cite{Jiang2024Pair}, we construct a Grassmann PEPS ansatz with a minimal $2\times2$ unit cell to accommodate the NNN interactions,
\begin{eqnarray}
\raisebox{-0.56\height}{\includegraphics[width=0.5\textwidth, page=1]{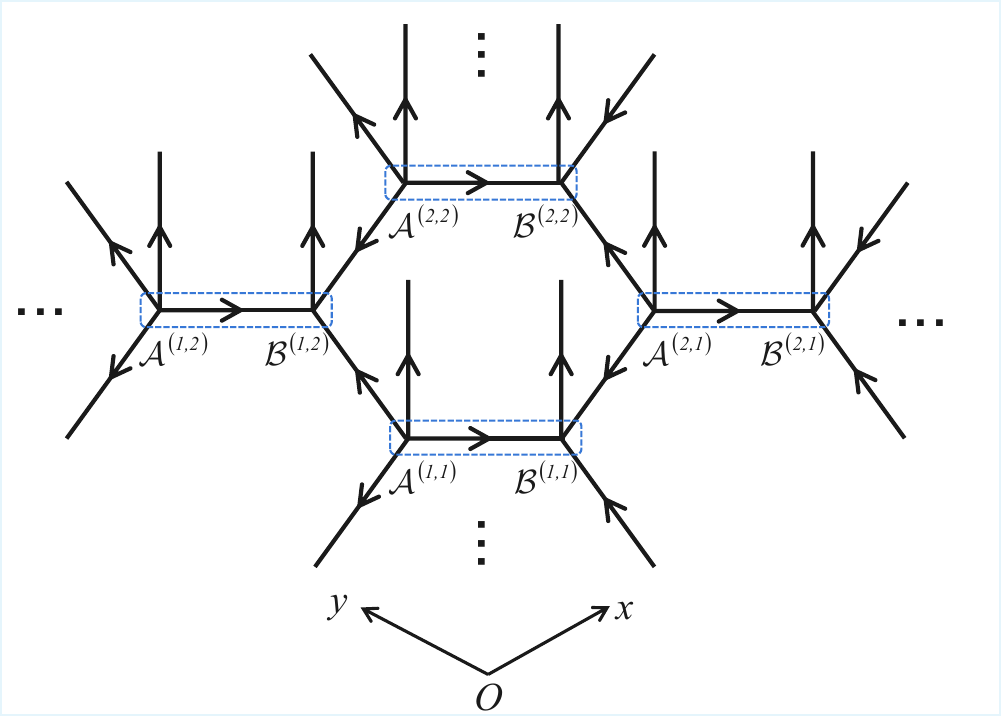}}
\end{eqnarray}
The Grassmann PEPS wavefunction, composed of 8 different rank-4 Grassmann tensors and 12 weight matrices, is optimized using the Grassmann SU method with HOSVD truncation, as detailed in Sec.~\ref{gsu_method}. To compute the energy expectation value, we note that each lattice site has three nearest neighbors and three next-nearest neighbors. Therefore, the Hamiltonian within a $2\times2$ unit cell can be divided into 12 distinct NN terms $\hat{H}^{NN}_{b}$ and 24 distinct NNN terms $\hat{H}^{NNN}_{b}$, where
\begin{eqnarray}
& &
\hat{H}_{b}^{NN} =
-t
\left(
\hat{c}_{i}^{\dagger}\hat{c}_{j}
+
h.c.
\right)
+
V_{1}
\hat{n}_{i}\hat{n}_{j}
-
\dfrac{\mu}{3}
\left(
\hat{n}_{i}
+
\hat{n}_{j}
\right),
\quad
\hat{H}_{b}^{NNN} =
V_{2}
\hat{n}_{i}\hat{n}_{j}.
\end{eqnarray}
The expectation value of NN terms is evaluated with the help of converged CTMRG environments via Eqs.~(\ref{ctmrg_exp1}-\ref{ctmrg_exp2b}). A single NNN term is computed as follows:
\begin{eqnarray}
\langle \hat{H}_{b}^{NNN} \rangle
=
\raisebox{-0.48\height}{\includegraphics[width=0.4\textwidth, page=1]{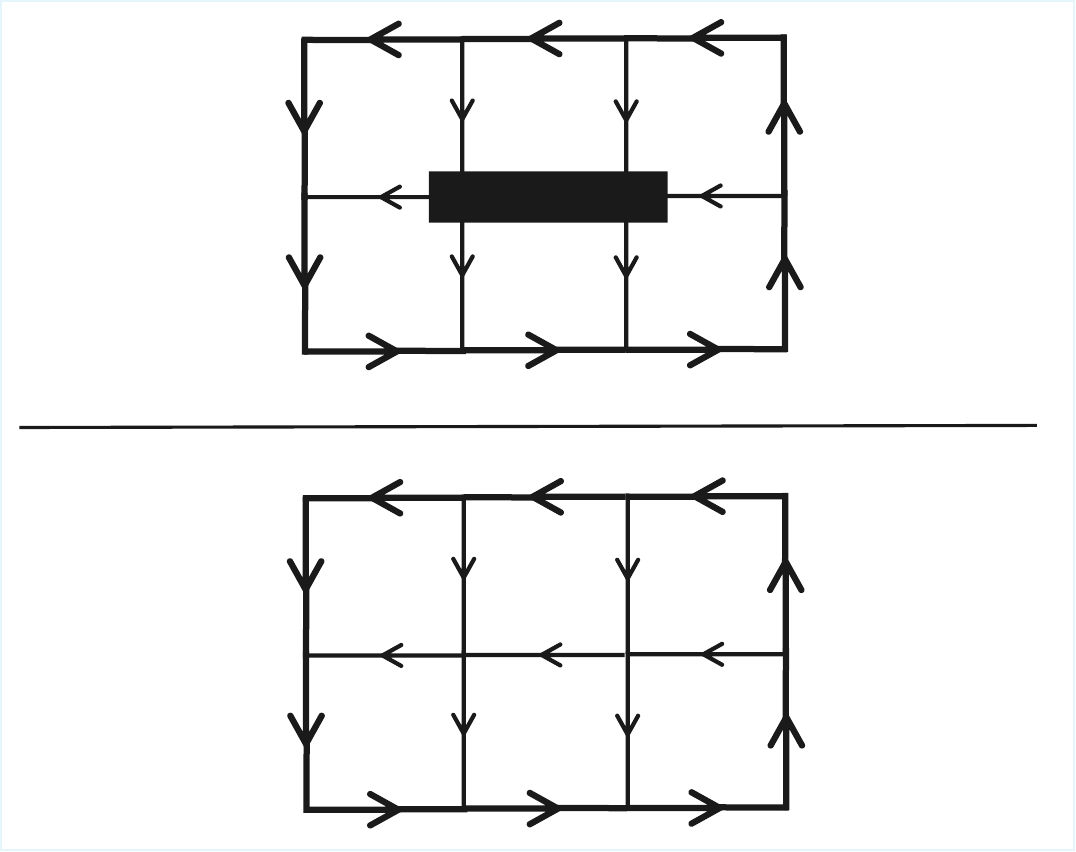}}
\end{eqnarray}
For simplicity, all Grassmann tensor labels are omitted in the diagram above. The corresponding impurity tensor is constructed through:
\begin{eqnarray}
\raisebox{-0.56\height}{\includegraphics[width=0.8\textwidth, page=1]{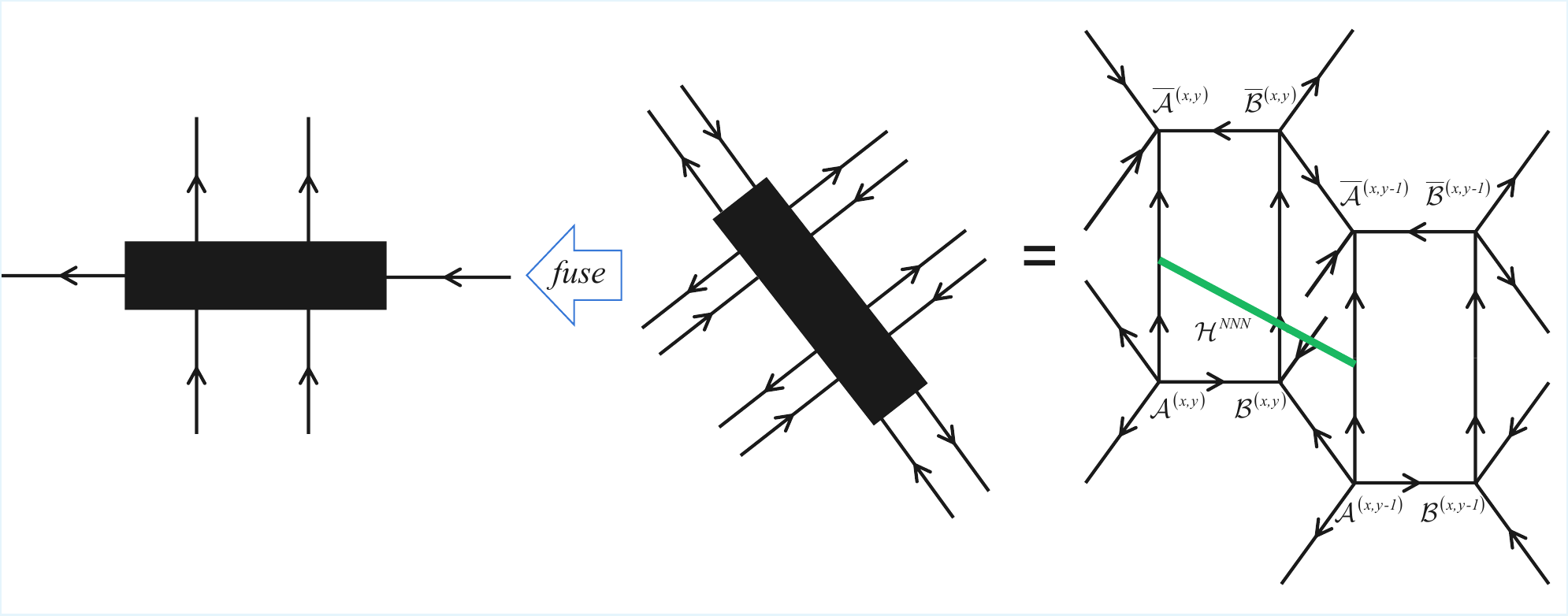}}
\end{eqnarray}

The ground-state energy per site $E_{s}$ is obtained by:
\begin{eqnarray}
E_{s} = \dfrac{
	\langle \hat{H}_{b}^{NN1} \rangle + \cdots + 
	\langle \hat{H}_{b}^{NN12} \rangle + \langle \hat{H}_{b}^{NNN1} \rangle + \cdots + \langle \hat{H}_{b}^{NNN24} \rangle
}{8}.
\end{eqnarray}

Furthermore, we can control the average particle number $n$ within the grand canonical ensemble by tuning the chemical potential $\mu$ in Eq.~(\ref{t-V1-V2}). The average particle number is defined as:  
\begin{eqnarray}
n(\mu) = \dfrac{
	\langle \hat{n}_{(1,1)}^{A} \rangle + \langle \hat{n}_{(1,1)}^{B} \rangle + \cdots +
	\langle \hat{n}_{(2,2)}^{A} \rangle + \langle \hat{n}_{(2,2)}^{B} \rangle
}{8},
\end{eqnarray}
where the subscripts $(i, j)$ denote the coordinates within the unit cell with $i, j=1, 2$, and the superscripts $A$(or $B$) indicate that the particle number operator $\hat{n}$ is evaluated on the $A$(or $B$) sublattice. 

\begin{figure}[htbp]
	\centering
	\begin{minipage}{0.48\textwidth}
		\centering
		\begin{overpic}[height=4.8cm,width=7.2cm]{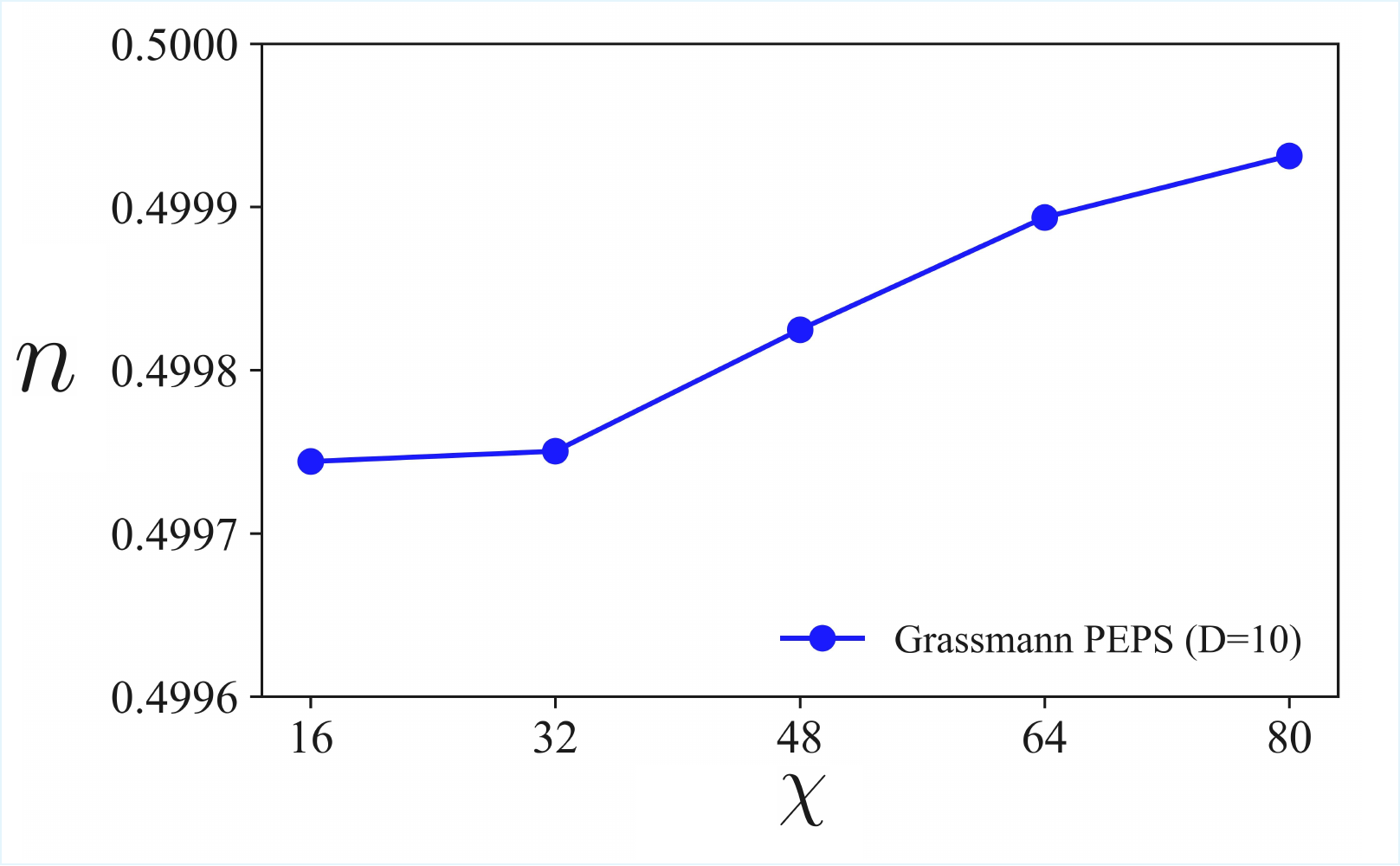}
			\put(-2,66){(a)}
		\end{overpic}
	\end{minipage}
	\begin{minipage}{0.48\textwidth}
		\centering
		\begin{overpic}[height=4.8cm,width=7.2cm]{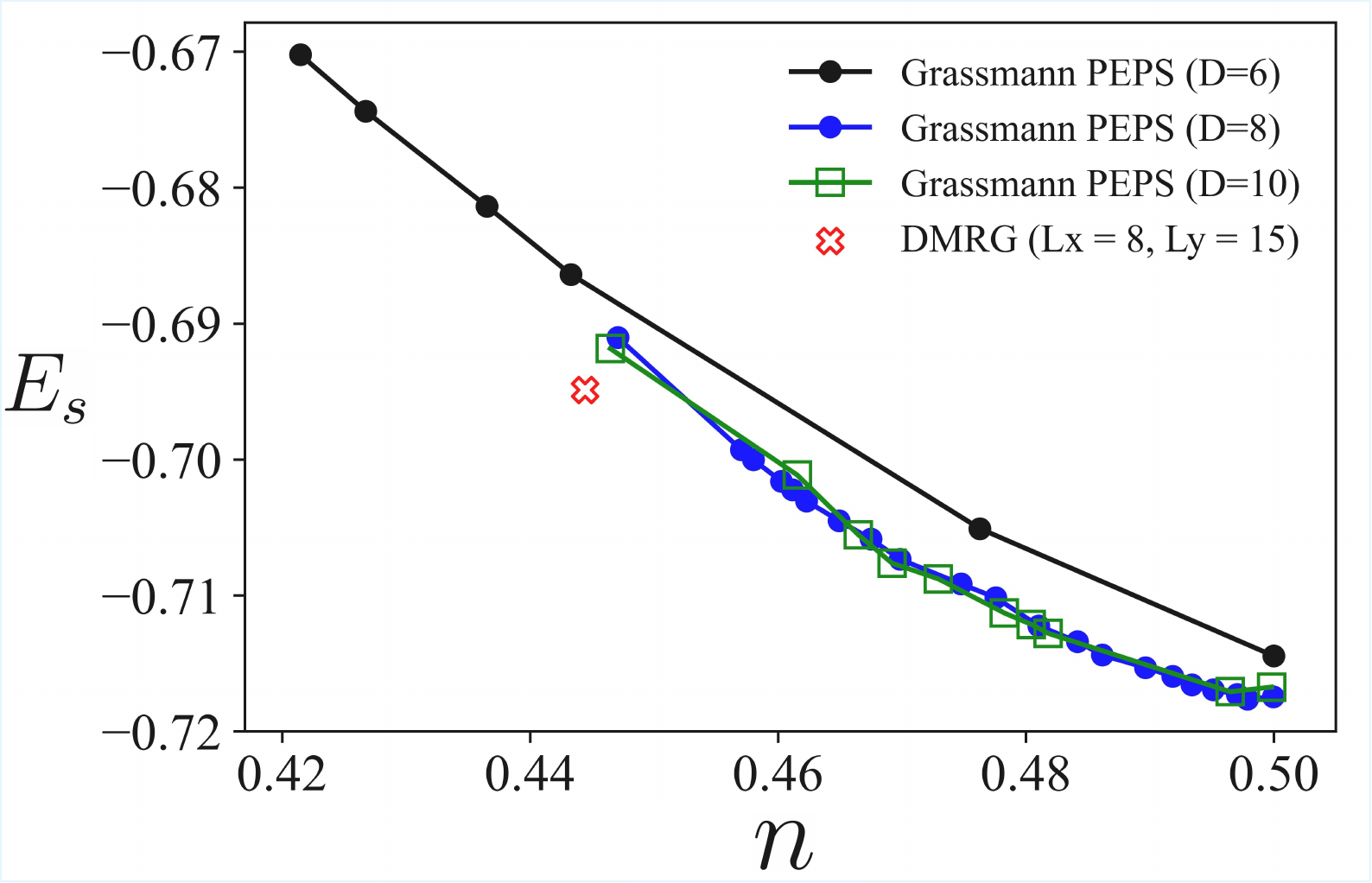}
			\put(-2,66){(b)}
		\end{overpic}
	\end{minipage}
	\caption{ 
		For the $t$-$V_{1}$-$V_{2}$ model at $V_{1}=-1.0$ and $V_{2}=0.5$: (a) Convergence of the particle density $n$ with Grassmann CTMRG environment dimension $\chi$ at $\mu=0$. (b) Energy per site $E_{s}$ versus particle density $n$. Red cross denotes data from the large-scale (non-Grassmann) DMRG calculation.}
	\label{gsu_example10}
\end{figure}

At $\mu=0.0$, the particle-hole symmetry of the Hamiltonian dictates an exact average density of $n=0.5$. This is confirmed numerically, as shown in Fig.~\ref{gsu_example10}(a), where the computed particle number density converges towards $0.5$ with increasing environment dimension $\chi$. Figure \ref{gsu_example10}(b) presents the energy per site $E_{s}$ of the $t$-$V_{1}$-$V_{2}$ model as a function of particle density $n$, obtained from Grassmann tensor network computations. The Grassmann PEPS energies show good convergence for bond dimensions $D=8$ and $D=10$. They are slightly higher than, yet remain in reasonable agreement with results from large-scale (non-Grassmann) DMRG calculations at $n \approx 0.444$. The DMRG computations were performed on an $L_{x}=15$, $L_{y}=8$ cylinder with open boundary conditions, retaining up to 33000 states. 

While the accuracy of the present Grassmann PEPS calculations may be improved by employing larger unit cells and bond dimensions, these preliminary results successfully establish a foundation for investigating the superconducting properties of the $t$-$V_{1}$-$V_{2}$ model within the Grassmann tensor network framework.

\section{Conclusions and Outlook} \label{conclusion}

Developing efficient and accurate numerical methods to study two-dimensional strongly correlated systems poses a great challenge in physics. Tensor network states have drawn increasing attention over the last two decades, due to their advantages for characterizing lowly entangled states and for addressing sign-problematic systems \cite{PEPS2004,RomanEPJB2014,CiracRMP2021, RomanAOP2014, RomanNRP2019}. However, most applications of tensor networks, especially in the condensed-matter physics community, focus on spin or bosonic systems, while the vast field of fermionic systems is relatively under-addressed. Grassmann tensor networks based on the fermionic coherent-state representation can provide an elegant and systematic route to studying fermionic models and have begun to play an important role in understanding the correlation physics of fermionic systems, ranging from condensed matter to particle physics.  

This work aims to present the fermionic tensor network methods based on the Grassmann algebra in great detail \cite{Gu2010Grassmann, Gu2013Efficient, ShimizuGrassmann2014, Akiyama2021More}. Starting from the most essential building block, i.e., the Grassmann tensor, on which we introduce basic tensor operations, such as Grassmann tensor contraction, fusion, and decomposition. The Grassmann tensor network representation of fermionic states, operators, and partition functions is subsequently derived, enabling us to solve physical problems using Grassmann tensor network algorithms. The equivalence between the Grassmann tensor networks and other proposals, such as the $\mathbb{Z}_{2}$-graded space approach and the swap-gate fermionic PEPS approach, is rigorously established in the most commonly used settings. The Grassmannization of some tensor network algorithms is discussed in the context of Grassmann DMRG, Grassmann TEBD, Grassmann CTMRG, and an adapted imaginary-time evolution of a two-dimensional Grassmann state. As applications, typical examples based on these Grassmann tensor network algorithms are presented, including the ground state of the 1D fermionic Hubbard model, the thermodynamics of 1D free fermions, the (1+1)-dimensional Gross-Neveu-Wilson model, and the ground state of a two-dimensional spinless fermion model. Promising results are also discussed.

In this work, we have tried our best to provide sufficient but necessary detail to explain most of the topics, and some of them have not been discussed in this way elsewhere. Although only a few typical algorithms are Grassmannized in this work, extending the procedure to other tensor network methods should be straightforward. As a prospect in the future, Grassmannization of other existing advanced methods, such as excited state calculations based on based on the single-mode approximation \cite{Haegeman2012Variational,Ponsioen2020Excitations,Boris2022Automatic} and the thermal tensor network methods \cite{Li2011Linearized,Li2023Tangent,Sinha2022Finite}, are of great importance to simulate fermionic matter from dynamical and thermodynamical views, instead of the ground state feature only, and provide numerical data comparable with real experiments. Especially, it would also be interesting to explore the Grassmann tensor network methods in the path integral formulation, both in the imaginary time setting \cite{Akiyama2022Metal,Akiyama2021Tensor,Kong2026Hubbard} and real time setting \cite{Chen2024Real,Xu2024Grassmann,Chen2025Path}, which may help to clarify the rich physics in Kondo impurity systems and related issues in fermionic open systems.

\section*{Acknowledgements}

We are grateful to Dr. Shinichiro Akiyama for helpful discussions. We appreciate valuable discussions with Professor Tao Xiang. We thanks Yi-Fan Jiang for providing the DMRG data regarding the $t$-$V_{1}$-$V_{2}$ model. This work was supported by the National R\&D Program of China (Grants No. 2023YFA1406500, No. 2024YFA1408604), National Natural Science Foundation of China (Grants No. 12274458), and the the Natural Science Foundation of Chongqing, China (Grants No. CSTB2025NSCQ-GPX1272).

\bibliographystyle{unsrt}
\bibliography{bib}

@article{Hubbard1963Electron,
    author = {Hubbard, J.},
    title = {Electron correlations in narrow energy bands},
    journal = {Proceedings of the Royal Society of London. A. Mathematical and Physical Sciences},
    volume = {276},
    number = {1365},
    pages = {238-257},
    year = {1963},
    month = {11},
    issn = {0080-4630},
    doi = {10.1098/rspa.1963.0204},
    url = {https://doi.org/10.1098/rspa.1963.0204},
    eprint = {https://royalsocietypublishing.org/rspa/article-pdf/276/1365/238/54456/rspa.1963.0204.pdf},
}

@article{HubbardPRL1959,
  title = {Calculation of Partition Functions},
  author = {Hubbard, J.},
  journal = {Phys. Rev. Lett.},
  volume = {3},
  issue = {2},
  pages = {77--78},
  numpages = {0},
  year = {1959},
  month = {Jul},
  publisher = {American Physical Society},
  doi = {10.1103/PhysRevLett.3.77},
  url = {https://link.aps.org/doi/10.1103/PhysRevLett.3.77}
}

@misc{BLAS,
howpublished = {\url{https://netlib.org/blas/}}
}

@misc{IntelMKL,
howpublished = {\url{https://www.intel.com/content/www/us/en/developer/tools/oneapi/onemkl.html}}
}

@article{BerezinInt1966,
    author = "Berezin, F. A.",
    title = "{The method of second quantization}",
    journal = "Pure Appl. Phys.",
    volume = "24",
    pages = "1--228",
    year = "1966"
}

@article{Zhang1988Effective,
  title = {Effective Hamiltonian for the superconducting Cu oxides},
  author = {Zhang, F. C. and Rice, T. M.},
  journal = {Phys. Rev. B},
  volume = {37},
  issue = {7},
  pages = {3759--3761},
  numpages = {0},
  year = {1988},
  month = {Mar},
  publisher = {American Physical Society},
  doi = {10.1103/PhysRevB.37.3759},
  url = {https://link.aps.org/doi/10.1103/PhysRevB.37.3759}
}

@article{ClockLZQ2020,
  title = {Critical properties of the two-dimensional $q$-state clock model},
  author = {Li, Zi-Qian and Yang, Li-Ping and Xie, Z. Y. and Tu, Hong-Hao and Liao, Hai-Jun and Xiang, T.},
  journal = {Phys. Rev. E},
  volume = {101},
  issue = {6},
  pages = {060105},
  numpages = {5},
  year = {2020},
  month = {Jun},
  publisher = {American Physical Society},
  doi = {10.1103/PhysRevE.101.060105},
  url = {https://link.aps.org/doi/10.1103/PhysRevE.101.060105}
}

@article{Anderson1961Localized,
  title = {Localized Magnetic States in Metals},
  author = {Anderson, P. W.},
  journal = {Phys. Rev.},
  volume = {124},
  issue = {1},
  pages = {41--53},
  numpages = {0},
  year = {1961},
  month = {Oct},
  publisher = {American Physical Society},
  doi = {10.1103/PhysRev.124.41},
  url = {https://link.aps.org/doi/10.1103/PhysRev.124.41}
}

@article{Kondo1964,
  title = {Resistance Minimum in Dilute Magnetic Alloys},
  author = {Kondo, Jun},
  journal = {Progress of Theoretical Physics},
  volume = {32},
  issue = {1},
  pages = {41--53},
  numpages = {0},
  year = {1964},
  month = {July},
  doi = {10.1143/PTP.32.37},
  url = {https://doi.org/10.1143/PTP.32.37}
}

@book{AssaBook1994,
  title={Interacting Electrons and Quantum Magnetism},
  author={Assa Auerbach},
  year={1994},
  publisher={Springer New York, NY}
}

@book{ColemanBook2015,
  title={Introduction to many-body physics},
  author={Piers Coleman},
  year={2015},
  publisher={Cambridge University Press}
}

@book{Negele2018Quantum,
  title={Quantum Many Particle Systems},
  author={John W Negele},
  year={2018},
  publisher={CRC Press}
}

@book{Pierre2001,
  title={Field Theory : A Modern Primer},
  author={Pierre Ramond},
  year={2001},
  publisher={Westview Press}
}

@article{Wilson1974,
  title = {Confinement of quarks},
  author = {Wilson, Kenneth G.},
  journal = {Phys. Rev. D},
  volume = {10},
  issue = {8},
  pages = {2445--2459},
  numpages = {0},
  year = {1974},
  month = {Oct},
  publisher = {American Physical Society},
  doi = {10.1103/PhysRevD.10.2445},
  url = {https://link.aps.org/doi/10.1103/PhysRevD.10.2445}
}

@article{Nambu1961,
  title = {Dynamical Model of Elementary Particles Based on an Analogy with Superconductivity. I},
  author = {Nambu, Y. and Jona-Lasinio, G.},
  journal = {Phys. Rev.},
  volume = {122},
  issue = {1},
  pages = {345--358},
  numpages = {0},
  year = {1961},
  month = {Apr},
  publisher = {American Physical Society},
  doi = {10.1103/PhysRev.122.345},
  url = {https://link.aps.org/doi/10.1103/PhysRev.122.345}
}

@article{SRGreview2010,
  title = {Renormalization of tensor-network states},
  author = {Zhao, H. H. and Xie, Z. Y. and Chen, Q. N. and Wei, Z. C. and Cai, J. W. and Xiang, T.},
  journal = {Phys. Rev. B},
  volume = {81},
  issue = {17},
  pages = {174411},
  numpages = {17},
  year = {2010},
  month = {May},
  publisher = {American Physical Society},
  doi = {10.1103/PhysRevB.81.174411},
  url = {https://link.aps.org/doi/10.1103/PhysRevB.81.174411}
}

@book{AltlandQFT2010, 
place={Cambridge}, 
edition={2}, 
title={Condensed Matter Field Theory}, 
publisher={Cambridge University Press}, 
author={Altland, Alexander and Simons, Ben D.}, 
year={2010},
}

@article{PEPS2004,
      title={Renormalization algorithms for Quantum-Many Body Systems in two and higher dimensions}, 
      author={F. Verstraete and J. I. Cirac},
      year={2004},
      eprint={cond-mat/0407066},
      archivePrefix={arXiv},
      primaryClass={cond-mat.str-el},
      journal = {Arxiv},
      volume = {abs/0407066},
      url={https://arxiv.org/abs/cond-mat/0407066}
}

@book{RanBook2020,
  title={Tensor Network Contractions},
  author={Shi-Ju Ran and Emanuele Tirrito and Cheng Peng and Xi Chen and Luca Tagliacozzo and Gang Su and Maciej Lewenstein},
  year={2020},
  publisher={Springer}
}

@book{XiangBook2023,
  title={Density Matrix and Tensor Network Renormalization},
  author={Tao Xiang},
  year={2023},
  publisher={Cambridge University Press}
}

@article{Kagome2017,
  title = {Gapless Spin-Liquid Ground State in the $S=1/2$ Kagome Antiferromagnet},
  author = {Liao, H. J. and Xie, Z. Y. and Chen, J. and Liu, Z. Y. and Xie, H. D. and Huang, R. Z. and Normand, B. and Xiang, T.},
  journal = {Phys. Rev. Lett.},
  volume = {118},
  issue = {13},
  pages = {137202},
  numpages = {6},
  year = {2017},
  month = {Mar},
  publisher = {American Physical Society},
  doi = {10.1103/PhysRevLett.118.137202},
  url = {https://link.aps.org/doi/10.1103/PhysRevLett.118.137202}
}

@article{LWY2024,
  title = {Tensor network study of the spin-$\frac{1}{2}$ square-lattice ${J}_{1}\text{\ensuremath{-}}{J}_{2}\text{\ensuremath{-}}{J}_{3}$ model: Incommensurate spiral order, mixed valence-bond solids, and multicritical points},
  author = {Liu, Wen-Yuan and Poilblanc, Didier and Gong, Shou-Shu and Chen, Wei-Qiang and Gu, Zheng-Cheng},
  journal = {Phys. Rev. B},
  volume = {109},
  issue = {23},
  pages = {235116},
  numpages = {15},
  year = {2024},
  month = {Jun},
  publisher = {American Physical Society},
  doi = {10.1103/PhysRevB.109.235116},
  url = {https://link.aps.org/doi/10.1103/PhysRevB.109.235116}
}

@article{LY2025,
  title = {Quantum dynamics in a spin-$\frac{1}{2}$ square lattice ${J}_{1}\text{\ensuremath{-}}{J}_{2}\text{\ensuremath{-}}\ensuremath{\delta}$ altermagnet},
  author = {Liu, Yang and Shao, Shiqi and He, Saisai and Xie, Z. Y. and Mei, Jia-Wei and Luo, Hong-Gang and Zhao, Jize},
  journal = {Phys. Rev. B},
  volume = {111},
  issue = {24},
  pages = {245117},
  numpages = {7},
  year = {2025},
  month = {Jun},
  publisher = {American Physical Society},
  doi = {10.1103/PhysRevB.111.245117},
  url = {https://link.aps.org/doi/10.1103/PhysRevB.111.245117}
}

@article{Nobert2013,
  title = {Topological Order in the Projected Entangled-Pair States Formalism: Transfer Operator and Boundary Hamiltonians},
  author = {Schuch, Norbert and Poilblanc, Didier and Cirac, J. Ignacio and P\'erez-Garc\'{\i}a, David},
  journal = {Phys. Rev. Lett.},
  volume = {111},
  issue = {9},
  pages = {090501},
  numpages = {5},
  year = {2013},
  month = {Aug},
  publisher = {American Physical Society},
  doi = {10.1103/PhysRevLett.111.090501},
  url = {https://link.aps.org/doi/10.1103/PhysRevLett.111.090501}
}

@article{WR2024,
  title = {Susceptibility indicator for chiral topological orders emergent from correlated fermions},
  author = {Wang, Rui and Yang, Tao and Xie, Z. Y. and Wang, Baigeng and Xie, X. C.},
  journal = {Phys. Rev. B},
  volume = {109},
  issue = {24},
  pages = {L241113},
  numpages = {6},
  year = {2024},
  month = {Jun},
  publisher = {American Physical Society},
  doi = {10.1103/PhysRevB.109.L241113},
  url = {https://link.aps.org/doi/10.1103/PhysRevB.109.L241113}
}

@article{CMPS2010,
  title = {Continuous Matrix Product States for Quantum Fields: An Energy Minimization Algorithm},
  author = {Ganahl, Martin and Rinc\'on, Juli\'an and Vidal, Guifre},
  journal = {Phys. Rev. Lett.},
  volume = {118},
  issue = {22},
  pages = {220402},
  numpages = {6},
  year = {2017},
  month = {Jun},
  publisher = {American Physical Society},
  doi = {10.1103/PhysRevLett.118.220402},
  url = {https://link.aps.org/doi/10.1103/PhysRevLett.118.220402}
}

@article{CPEPS2019,
  title = {Continuous Tensor Network States for Quantum Fields},
  author = {Tilloy, Antoine and Cirac, J. Ignacio},
  journal = {Phys. Rev. X},
  volume = {9},
  issue = {2},
  pages = {021040},
  numpages = {17},
  year = {2019},
  month = {May},
  publisher = {American Physical Society},
  doi = {10.1103/PhysRevX.9.021040},
  url = {https://link.aps.org/doi/10.1103/PhysRevX.9.021040}
}

@article{Levin2007,
  title = {Tensor Renormalization Group Approach to Two-Dimensional Classical Lattice Models},
  author = {Levin, Michael and Nave, Cody P.},
  journal = {Phys. Rev. Lett.},
  volume = {99},
  issue = {12},
  pages = {120601},
  numpages = {4},
  year = {2007},
  month = {Sep},
  publisher = {American Physical Society},
  doi = {10.1103/PhysRevLett.99.120601},
  url = {https://link.aps.org/doi/10.1103/PhysRevLett.99.120601}
}

@article{YS2023,
  title = {Competing orders in the honeycomb lattice $t\text{\ensuremath{-}}J$ model},
  author = {Xu, Zheng-Tao and Gu, Zheng-Cheng and Yang, Shuo},
  journal = {Phys. Rev. B},
  volume = {108},
  issue = {3},
  pages = {035144},
  numpages = {13},
  year = {2023},
  month = {Jul},
  publisher = {American Physical Society},
  doi = {10.1103/PhysRevB.108.035144},
  url = {https://link.aps.org/doi/10.1103/PhysRevB.108.035144}
}

@article{CorbozHubbard2016,
  title = {Improved energy extrapolation with infinite projected entangled-pair states applied to the two-dimensional Hubbard model},
  author = {Corboz, Philippe},
  journal = {Phys. Rev. B},
  volume = {93},
  issue = {4},
  pages = {045116},
  numpages = {7},
  year = {2016},
  month = {Jan},
  publisher = {American Physical Society},
  doi = {10.1103/PhysRevB.93.045116},
  url = {https://link.aps.org/doi/10.1103/PhysRevB.93.045116}
}

@article{CorbozTJ2014,
  title = {Competing States in the $t$-$J$ Model: Uniform $d$-Wave State versus Stripe State},
  author = {Corboz, Philippe and Rice, T. M. and Troyer, Matthias},
  journal = {Phys. Rev. Lett.},
  volume = {113},
  issue = {4},
  pages = {046402},
  numpages = {5},
  year = {2014},
  month = {Jul},
  publisher = {American Physical Society},
  doi = {10.1103/PhysRevLett.113.046402},
  url = {https://link.aps.org/doi/10.1103/PhysRevLett.113.046402}
}

@book{NightBook1999,
  title={Quantum Monte Carlo Methods in Physics and Chemistry},
  author={M. P. Nightingale, J. C. Umrigar},
  year={1999},
  publisher={Springer}
}

@article{SchRev2011,
title = {The density-matrix renormalization group in the age of matrix product states},
journal = {Annals of Physics},
volume = {326},
number = {1},
pages = {96-192},
year = {2011},
note = {January 2011 Special Issue},
issn = {0003-4916},
doi = {https://doi.org/10.1016/j.aop.2010.09.012},
url = {https://www.sciencedirect.com/science/article/pii/S0003491610001752},
author = {Ulrich Schollwöck}
}

@article{WhiteDMRG1992,
  title={Density matrix formulation for quantum renormalization groups},
  author={White, Steven R},
  journal={Physical review letters},
  volume={69},
  number={19},
  pages={2863},
  year={1992},
  publisher={APS}
}

@article{Jordan1928JW,
  title={{\"U}ber das paulische {\"a}quivalenzverbot},
  author={Jordan, Pascual and Wigner, Eugene},
  journal={Zeitschrift f{\"u}r Physik},
  volume={47},
  number={9},
  pages={631--651},
  year={1928},
  url = {https://doi.org/10.1007/BF01331938},
  doi = {10.1007/BF01331938},
  publisher={Springer}
}

@article{Pineda2010Unitary,
  title = {Unitary circuits for strongly correlated fermions},
  author = {Pineda, Carlos and Barthel, Thomas and Eisert, Jens},
  journal = {Phys. Rev. A},
  volume = {81},
  issue = {5},
  pages = {050303},
  numpages = {4},
  year = {2010},
  month = {May},
  publisher = {American Physical Society},
  doi = {10.1103/PhysRevA.81.050303},
  url = {https://link.aps.org/doi/10.1103/PhysRevA.81.050303}
}

@article{Corboz2010Simulation,
  title = {Simulation of interacting fermions with entanglement renormalization},
  author = {Corboz, Philippe and Evenbly, Glen and Verstraete, Frank and Vidal, Guifr\'e},
  journal = {Phys. Rev. A},
  volume = {81},
  issue = {1},
  pages = {010303},
  numpages = {4},
  year = {2010},
  month = {Jan},
  publisher = {American Physical Society},
  doi = {10.1103/PhysRevA.81.010303},
  url = {https://link.aps.org/doi/10.1103/PhysRevA.81.010303}
}

@article{Kraus2010Fermionic,
  title = {Fermionic projected entangled pair states},
  author = {Kraus, Christina V. and Schuch, Norbert and Verstraete, Frank and Cirac, J. Ignacio},
  journal = {Phys. Rev. A},
  volume = {81},
  issue = {5},
  pages = {052338},
  numpages = {6},
  year = {2010},
  month = {May},
  publisher = {American Physical Society},
  doi = {10.1103/PhysRevA.81.052338},
  url = {https://link.aps.org/doi/10.1103/PhysRevA.81.052338}
}

@article{Barthel2009Contraction,
  title = {Contraction of fermionic operator circuits and the simulation of strongly correlated fermions},
  author = {Barthel, Thomas and Pineda, Carlos and Eisert, Jens},
  journal = {Phys. Rev. A},
  volume = {80},
  issue = {4},
  pages = {042333},
  numpages = {12},
  year = {2009},
  month = {Oct},
  publisher = {American Physical Society},
  doi = {10.1103/PhysRevA.80.042333},
  url = {https://link.aps.org/doi/10.1103/PhysRevA.80.042333}
}

@article{Pivzorn2010Fermionic,
  title = {Fermionic implementation of projected entangled pair states algorithm},
  author = {Pi\ifmmode \check{z}\else \v{z}\fi{}orn, Iztok and Verstraete, Frank},
  journal = {Phys. Rev. B},
  volume = {81},
  issue = {24},
  pages = {245110},
  numpages = {8},
  year = {2010},
  month = {Jun},
  publisher = {American Physical Society},
  doi = {10.1103/PhysRevB.81.245110},
  url = {https://link.aps.org/doi/10.1103/PhysRevB.81.245110}
}

@article{Corboz2009Fermionic,
  title = {Fermionic multiscale entanglement renormalization ansatz},
  author = {Corboz, Philippe and Vidal, Guifr\'e},
  journal = {Phys. Rev. B},
  volume = {80},
  issue = {16},
  pages = {165129},
  numpages = {12},
  year = {2009},
  month = {Oct},
  publisher = {American Physical Society},
  doi = {10.1103/PhysRevB.80.165129},
  url = {https://link.aps.org/doi/10.1103/PhysRevB.80.165129}
}

@article{Corboz2010SimulationPEPS,
  title = {Simulation of strongly correlated fermions in two spatial dimensions with fermionic projected entangled-pair states},
  author = {Corboz, Philippe and Or\'us, Rom\'an and Bauer, Bela and Vidal, Guifr\'e},
  journal = {Phys. Rev. B},
  volume = {81},
  issue = {16},
  pages = {165104},
  numpages = {22},
  year = {2010},
  month = {Apr},
  publisher = {American Physical Society},
  doi = {10.1103/PhysRevB.81.165104},
  url = {https://link.aps.org/doi/10.1103/PhysRevB.81.165104}
}

@article{Corboz2011Stripes,
  title = {Stripes in the two-dimensional $t$-$J$ model with infinite projected entangled-pair states},
  author = {Corboz, Philippe and White, Steven R. and Vidal, Guifr\'e and Troyer, Matthias},
  journal = {Phys. Rev. B},
  volume = {84},
  issue = {4},
  pages = {041108},
  numpages = {5},
  year = {2011},
  month = {Jul},
  publisher = {American Physical Society},
  doi = {10.1103/PhysRevB.84.041108},
  url = {https://link.aps.org/doi/10.1103/PhysRevB.84.041108}
}

@article{Corboz2012Comment,
doi = {10.1209/0295-5075/98/27005},
url = {https://doi.org/10.1209/0295-5075/98/27005},
year = {2012},
month = {apr},
publisher = {},
volume = {98},
number = {2},
pages = {27005},
author = {Corboz, P. and Capponi, S. and Läuchli, A. M. and Bauer, B. and Orús, R.},
title = {Comment on “Topological quantum phase transitions of attractive spinless fermions in a honeycomb lattice” by Poletti D. et al.},
journal = {Europhysics Letters}
}

@article{Wang2014Fermionic,
doi = {10.1088/1367-2630/16/10/103008},
url = {https://doi.org/10.1088/1367-2630/16/10/103008},
year = {2014},
month = {oct},
publisher = {IOP Publishing},
volume = {16},
number = {10},
pages = {103008},
author = {Wang, Lei and Corboz, Philippe and Troyer, Matthias},
title = {Fermionic quantum critical point of spinless fermions on a honeycomb lattice},
journal = {New Journal of Physics}
}

@article{Zheng2017Stripe,
author = {Bo-Xiao Zheng  and Chia-Min Chung  and Philippe Corboz  and Georg Ehlers  and Ming-Pu Qin  and Reinhard M. Noack  and Hao Shi  and Steven R. White  and Shiwei Zhang  and Garnet Kin-Lic Chan },
title = {Stripe order in the underdoped region of the two-dimensional Hubbard model},
journal = {Science},
volume = {358},
number = {6367},
pages = {1155-1160},
year = {2017},
doi = {10.1126/science.aam7127},
URL = {https://www.science.org/doi/abs/10.1126/science.aam7127},
eprint = {https://www.science.org/doi/pdf/10.1126/science.aam7127}
}

@article{Ponsioen2019Period,
  title = {Period 4 stripe in the extended two-dimensional Hubbard model},
  author = {Ponsioen, Boris and Chung, Sangwoo S. and Corboz, Philippe},
  journal = {Phys. Rev. B},
  volume = {100},
  issue = {19},
  pages = {195141},
  numpages = {7},
  year = {2019},
  month = {Nov},
  publisher = {American Physical Society},
  doi = {10.1103/PhysRevB.100.195141},
  url = {https://link.aps.org/doi/10.1103/PhysRevB.100.195141}
}

@article{Benedikt2021Beginner,
	title={{A beginner's guide to non-abelian iPEPS for correlated fermions}},
	author={Benedikt Bruognolo and Jheng-Wei Li and Jan von Delft and Andreas Weichselbaum},
	journal={SciPost Phys. Lect. Notes},
	pages={25},
	year={2021},
	publisher={SciPost},
	doi={10.21468/SciPostPhysLectNotes.25},
	url={https://scipost.org/10.21468/SciPostPhysLectNotes.25},
}

@article{Li2021Study,
  title = {Study of spin symmetry in the doped $t\ensuremath{-}J$ model using infinite projected entangled pair states},
  author = {Li, Jheng-Wei and Bruognolo, Benedikt and Weichselbaum, Andreas and von Delft, Jan},
  journal = {Phys. Rev. B},
  volume = {103},
  issue = {7},
  pages = {075127},
  numpages = {6},
  year = {2021},
  month = {Feb},
  publisher = {American Physical Society},
  doi = {10.1103/PhysRevB.103.075127},
  url = {https://link.aps.org/doi/10.1103/PhysRevB.103.075127}
}

@article{Ponsioen2023Superconducting,
  title = {Superconducting stripes in the hole-doped three-band Hubbard model},
  author = {Ponsioen, Boris and Chung, Sangwoo S. and Corboz, Philippe},
  journal = {Phys. Rev. B},
  volume = {108},
  issue = {20},
  pages = {205154},
  numpages = {7},
  year = {2023},
  month = {Nov},
  publisher = {American Physical Society},
  doi = {10.1103/PhysRevB.108.205154},
  url = {https://link.aps.org/doi/10.1103/PhysRevB.108.205154}
}

@article{Chung2019SU3,
  title = {SU(3) fermions on the honeycomb lattice at $\frac{1}{3}$ filling},
  author = {Chung, Sangwoo S. and Corboz, Philippe},
  journal = {Phys. Rev. B},
  volume = {100},
  issue = {3},
  pages = {035134},
  numpages = {6},
  year = {2019},
  month = {Jul},
  publisher = {American Physical Society},
  doi = {10.1103/PhysRevB.100.035134},
  url = {https://link.aps.org/doi/10.1103/PhysRevB.100.035134}
}

@article{Chen2024Orbital,
  title = {Orbital-selective superconductivity in the pressurized bilayer nickelate ${\mathrm{La}}_{3}{\mathrm{Ni}}_{2}{\mathrm{O}}_{7}$: An infinite projected entangled-pair state study},
  author = {Chen, Jialin and Yang, Fan and Li, Wei},
  journal = {Phys. Rev. B},
  volume = {110},
  issue = {4},
  pages = {L041111},
  numpages = {7},
  year = {2024},
  month = {Jul},
  publisher = {American Physical Society},
  doi = {10.1103/PhysRevB.110.L041111},
  url = {https://link.aps.org/doi/10.1103/PhysRevB.110.L041111}
}

@article{Zhang2025Frustration,
  title = {Frustration-Induced Superconductivity in the $t\text{\ensuremath{-}}{t}^{\ensuremath{'}}$ Hubbard Model},
  author = {Zhang, Changkai and Li, Jheng-Wei and Nikolaidou, Dimitra and von Delft, Jan},
  journal = {Phys. Rev. Lett.},
  volume = {134},
  issue = {11},
  pages = {116502},
  numpages = {7},
  year = {2025},
  month = {Mar},
  publisher = {American Physical Society},
  doi = {10.1103/PhysRevLett.134.116502},
  url = {https://link.aps.org/doi/10.1103/PhysRevLett.134.116502}
}

@article{Singh2011Tensor,
  title = {Tensor network states and algorithms in the presence of a global U(1) symmetry},
  author = {Singh, Sukhwinder and Pfeifer, Robert N. C. and Vidal, Guifre},
  journal = {Phys. Rev. B},
  volume = {83},
  issue = {11},
  pages = {115125},
  numpages = {22},
  year = {2011},
  month = {Mar},
  publisher = {American Physical Society},
  doi = {10.1103/PhysRevB.83.115125},
  url = {https://link.aps.org/doi/10.1103/PhysRevB.83.115125}
}

@article{Singh2012Tensor,
  title = {Tensor network states and algorithms in the presence of a global SU(2) symmetry},
  author = {Singh, Sukhwinder and Vidal, Guifre},
  journal = {Phys. Rev. B},
  volume = {86},
  issue = {19},
  pages = {195114},
  numpages = {31},
  year = {2012},
  month = {Nov},
  publisher = {American Physical Society},
  doi = {10.1103/PhysRevB.86.195114},
  url = {https://link.aps.org/doi/10.1103/PhysRevB.86.195114}
}

@article{Marek2025YASTN,
	title={YASTN: Yet another symmetric tensor networks; A Python library for Abelian symmetric tensor network calculations},
	author={Marek M. Rams and Gabriela Wójtowicz and Aritra Sinha and Juraj Hasik},
	journal={SciPost Phys. Codebases},
	pages={52},
	year={2025},
	publisher={SciPost},
	doi={10.21468/SciPostPhysCodeb.52},
	url={https://scipost.org/10.21468/SciPostPhysCodeb.52},
}

@article{Lukas2025TensorKit,
      title={TensorKit.jl: A Julia package for large-scale tensor computations, with a hint of category theory}, 
      author={Lukas Devos and Jutho Haegeman},
      year={2025},
      eprint={2508.10076},
      archivePrefix={arXiv},
      primaryClass={cs.MS},
      url={https://arxiv.org/abs/2508.10076}, 
      journal = {Arxiv},
      volume = {abs/2508.10076}
}

@article{Gu2010Grassmann,
      title={Grassmann tensor network states and its renormalization for strongly correlated fermionic and bosonic states}, 
      author={Zheng-Cheng Gu and Frank Verstraete and Xiao-Gang Wen},
      year={2010},
      eprint={1004.2563},
      archivePrefix={arXiv},
      primaryClass={cond-mat.str-el},
      url={https://arxiv.org/abs/1004.2563}, 
      journal = {Arxiv},
      volume = {abs/1004.2563}
}

@article{Gu2008Tensor,
  title = {Tensor-entanglement renormalization group approach as a unified method for symmetry breaking and topological phase transitions},
  author = {Gu, Zheng-Cheng and Levin, Michael and Wen, Xiao-Gang},
  journal = {Phys. Rev. B},
  volume = {78},
  issue = {20},
  pages = {205116},
  numpages = {11},
  year = {2008},
  month = {Nov},
  publisher = {American Physical Society},
  doi = {10.1103/PhysRevB.78.205116},
  url = {https://link.aps.org/doi/10.1103/PhysRevB.78.205116}
}

@article{Gu2013Efficient,
  title = {Efficient simulation of Grassmann tensor product states},
  author = {Gu, Zheng-Cheng},
  journal = {Phys. Rev. B},
  volume = {88},
  issue = {11},
  pages = {115139},
  numpages = {12},
  year = {2013},
  month = {Sep},
  publisher = {American Physical Society},
  doi = {10.1103/PhysRevB.88.115139},
  url = {https://link.aps.org/doi/10.1103/PhysRevB.88.115139}
}

@article{Gu2013Time,
  title = {Time-reversal symmetry breaking superconducting ground state in the doped Mott insulator on the honeycomb lattice},
  author = {Gu, Zheng-Cheng and Jiang, Hong-Chen and Sheng, D. N. and Yao, Hong and Balents, Leon and Wen, Xiao-Gang},
  journal = {Phys. Rev. B},
  volume = {88},
  issue = {15},
  pages = {155112},
  numpages = {7},
  year = {2013},
  month = {Oct},
  publisher = {American Physical Society},
  doi = {10.1103/PhysRevB.88.155112},
  url = {https://link.aps.org/doi/10.1103/PhysRevB.88.155112}
}

@article{Gu2014Symmetry,
  title = {Symmetry-protected topological orders for interacting fermions: Fermionic topological nonlinear $\ensuremath{\sigma}$ models and a special group supercohomology theory},
  author = {Gu, Zheng-Cheng and Wen, Xiao-Gang},
  journal = {Phys. Rev. B},
  volume = {90},
  issue = {11},
  pages = {115141},
  numpages = {59},
  year = {2014},
  month = {Sep},
  publisher = {American Physical Society},
  doi = {10.1103/PhysRevB.90.115141},
  url = {https://link.aps.org/doi/10.1103/PhysRevB.90.115141}
}

@article{Gu2020Emergence,
  title = {Emergence of $p+ip$ superconductivity in two-dimensional doped Dirac systems},
  author = {Gu, Zheng-Cheng and Jiang, Hong-Chen and Baskaran, G.},
  journal = {Phys. Rev. B},
  volume = {101},
  issue = {20},
  pages = {205147},
  numpages = {8},
  year = {2020},
  month = {May},
  publisher = {American Physical Society},
  doi = {10.1103/PhysRevB.101.205147},
  url = {https://link.aps.org/doi/10.1103/PhysRevB.101.205147}
}

@article{Anderson1987Resonating,
author = {P. W. Anderson},
title = {The Resonating Valence Bond State in La2CuO4 and Superconductivity},
journal = {Science},
volume = {235},
number = {4793},
pages = {1196-1198},
year = {1987},
doi = {10.1126/science.235.4793.1196},
URL = {https://www.science.org/doi/abs/10.1126/science.235.4793.1196},
eprint = {https://www.science.org/doi/pdf/10.1126/science.235.4793.1196}
}

@article{Miao2025Spin,
  title = {Spin-charge separation and unconventional superconductivity in the $t\text{\ensuremath{-}}J$ model on the honeycomb lattice},
  author = {Miao, Jian-Jian and Yue, Zheng-Yuan and Zhang, Hao and Chen, Wei-Qiang and Gu, Zheng-Cheng},
  journal = {Phys. Rev. B},
  volume = {111},
  issue = {17},
  pages = {174518},
  numpages = {20},
  year = {2025},
  month = {May},
  publisher = {American Physical Society},
  doi = {10.1103/PhysRevB.111.174518},
  url = {https://link.aps.org/doi/10.1103/PhysRevB.111.174518}
}

@article{Yue2024Pseudogap,
      title={Pseudogap phase as fluctuating pair density wave}, 
      author={Zheng-Yuan Yue and Zheng-Tao Xu and Shuo Yang and Zheng-Cheng Gu},
      year={2024},
      eprint={2404.16770},
      archivePrefix={arXiv},
      primaryClass={cond-mat.str-el},
      url={https://arxiv.org/abs/2404.16770}, 
      journal = {Arxiv},
      volume = {abs/2404.16770}
}

@article{Jie2015Combining,
      title={Combining Grassmann algebra with entanglement renormalization method}, 
      author={Jie Lou and Yan Chen},
      year={2015},
      eprint={1506.03716},
      archivePrefix={arXiv},
      primaryClass={cond-mat.str-el},
      url={https://arxiv.org/abs/1506.03716}, 
      journal = {Arxiv},
      volume = {abs/1506.03716}
}

@article{ShimizuGrassmann2014,
  title = {Grassmann tensor renormalization group approach to one-flavor lattice Schwinger model},
  author = {Shimizu, Yuya and Kuramashi, Yoshinobu},
  journal = {Phys. Rev. D},
  volume = {90},
  issue = {1},
  pages = {014508},
  numpages = {10},
  year = {2014},
  month = {Jul},
  publisher = {American Physical Society},
  doi = {10.1103/PhysRevD.90.014508},
  url = {https://link.aps.org/doi/10.1103/PhysRevD.90.014508}
}

@article{Xie2012HOTRG,
  title = {Coarse-graining renormalization by higher-order singular value decomposition},
  author = {Xie, Z. Y. and Chen, J. and Qin, M. P. and Zhu, J. W. and Yang, L. P. and Xiang, T.},
  journal = {Phys. Rev. B},
  volume = {86},
  issue = {4},
  pages = {045139},
  numpages = {9},
  year = {2012},
  month = {Jul},
  publisher = {American Physical Society},
  doi = {10.1103/PhysRevB.86.045139},
  url = {https://link.aps.org/doi/10.1103/PhysRevB.86.045139}
}

@article{Sakai2017Higher,
    author = {Sakai, Ryo and Takeda, Shinji and Yoshimura, Yusuke},
    title = {Higher-order tensor renormalization group for relativistic fermion systems},
    journal = {Progress of Theoretical and Experimental Physics},
    volume = {2017},
    number = {6},
    pages = {063B07},
    year = {2017},
    month = {06},
    issn = {2050-3911},
    doi = {10.1093/ptep/ptx080},
    url = {https://doi.org/10.1093/ptep/ptx080},
    eprint = {https://academic.oup.com/ptep/article-pdf/2017/6/063B07/17990708/ptx080.pdf}
}

@article{Akiyama2021Restoration,
  title={Restoration of chiral symmetry in cold and dense Nambu-Jona-Lasinio model with tensor renormalization group},
  author={Akiyama, Shinichiro and Kuramashi, Yoshinobu and Yamashita, Takumi and Yoshimura, Yusuke},
  journal={Journal of High Energy Physics},
  volume={2021},
  number={1},
  pages={1--17},
  year={2021},
  url = {https://doi.org/10.1007/JHEP01(2021)121},
  doi = {10.1007/JHEP01(2021)121},
  publisher={Springer}
}

@article{Akiyama2021More,
  title={More about the Grassmann tensor renormalization group},
  author={Akiyama, Shinichiro and Kadoh, Daisuke},
  journal={Journal of High Energy Physics},
  volume={2021},
  number={10},
  pages={1--16},
  year={2021},
  url = {https://doi.org/10.1007/JHEP10(2021)188},
  doi = {10.1007/JHEP10(2021)188},
  publisher={Springer}
}

@article{Akiyama2024Tensor,
doi = {10.1088/1361-648X/ad4760},
url = {https://doi.org/10.1088/1361-648X/ad4760},
year = {2024},
month = {may},
publisher = {IOP Publishing},
volume = {36},
number = {34},
pages = {343002},
author = {Akiyama, Shinichiro and Meurice, Yannick and Sakai, Ryo},
title = {Tensor renormalization group for fermions},
journal = {Journal of Physics: Condensed Matter}
}

@article{Akiyama2021Tensor,
  title = {Tensor renormalization group approach to ($1+1$)-dimensional Hubbard model},
  author = {Akiyama, Shinichiro and Kuramashi, Yoshinobu},
  journal = {Phys. Rev. D},
  volume = {104},
  issue = {1},
  pages = {014504},
  numpages = {8},
  year = {2021},
  month = {Jul},
  publisher = {American Physical Society},
  doi = {10.1103/PhysRevD.104.014504},
  url = {https://link.aps.org/doi/10.1103/PhysRevD.104.014504}
}

@article{Akiyama2022Metal,
    author = {Akiyama, Shinichiro and Kuramashi, Yoshinobu and Yamashita, Takumi},
    title = {Metal–insulator transition in the (2+1)-dimensional Hubbard model with the tensor renormalization group},
    journal = {Progress of Theoretical and Experimental Physics},
    volume = {2022},
    number = {2},
    pages = {023I01},
    year = {2022},
    month = {01},
    issn = {2050-3911},
    doi = {10.1093/ptep/ptac014},
    url = {https://doi.org/10.1093/ptep/ptac014},
    eprint = {https://academic.oup.com/ptep/article-pdf/2022/2/023I01/42931059/ptac014.pdf},
}

@article{Adachi2020Anisotropic,
  title = {Anisotropic tensor renormalization group},
  author = {Adachi, Daiki and Okubo, Tsuyoshi and Todo, Synge},
  journal = {Phys. Rev. B},
  volume = {102},
  issue = {5},
  pages = {054432},
  numpages = {7},
  year = {2020},
  month = {Aug},
  publisher = {American Physical Society},
  doi = {10.1103/PhysRevB.102.054432},
  url = {https://link.aps.org/doi/10.1103/PhysRevB.102.054432}
}

@article{Gu2014,
  title = {Lattice model for fermionic toric code},
  author = {Gu, Zheng-Cheng and Wang, Zhenghan and Wen, Xiao-Gang},
  journal = {Phys. Rev. B},
  volume = {90},
  issue = {8},
  pages = {085140},
  numpages = {9},
  year = {2014},
  month = {Aug},
  publisher = {American Physical Society},
  doi = {10.1103/PhysRevB.90.085140},
  url = {https://link.aps.org/doi/10.1103/PhysRevB.90.085140}
}

@article{Bultinck2017fMPS,
  title = {Fermionic matrix product states and one-dimensional topological phases},
  author = {Bultinck, Nick and Williamson, Dominic J. and Haegeman, Jutho and Verstraete, Frank},
  journal = {Phys. Rev. B},
  volume = {95},
  issue = {7},
  pages = {075108},
  numpages = {20},
  year = {2017},
  month = {Feb},
  publisher = {American Physical Society},
  doi = {10.1103/PhysRevB.95.075108},
  url = {https://link.aps.org/doi/10.1103/PhysRevB.95.075108}
}

@article{Bultinck2017fPEPS,
doi = {10.1088/1751-8121/aa99cc},
url = {https://doi.org/10.1088/1751-8121/aa99cc},
year = {2017},
month = {dec},
publisher = {IOP Publishing},
volume = {51},
number = {2},
pages = {025202},
author = {Bultinck, Nick and Williamson, Dominic J and Haegeman, Jutho and Verstraete, Frank},
title = {Fermionic projected entangled-pair states and topological phases},
journal = {Journal of Physics A: Mathematical and Theoretical}
}

@article{Zheng2025Competing,
  title={Competing pair density wave orders in the square lattice t-J model},
  author={Zheng, Wayne and Yue, Zheng-Yuan and Zhang, Jian-Hao and Gu, Zheng-Cheng},
  journal={Communications Physics},
  url = {https://doi.org/10.1038/s42005-025-02346-0},
  doi = {10.1038/s42005-025-02346-0},
  volume = {8},
  number = {456},
  publisher = {Nature},
  month = {nov},
  year={2025}
}

@article{Zheng2025Revealing,
  title = {Revealing quantum phase string effect in a doped Mott insulator: A tensor network state study},
  author = {Zheng, Wayne and Zhang, Jia-Xin and Yue, Zheng-Yuan and Gu, Zheng-Cheng and Weng, Zheng-Yu},
  journal = {Phys. Rev. B},
  volume = {112},
  issue = {21},
  pages = {214514},
  numpages = {12},
  year = {2025},
  month = {Dec},
  publisher = {American Physical Society},
  doi = {10.1103/2f9b-h135},
  url = {https://link.aps.org/doi/10.1103/2f9b-h135}
}

@article{Liu2025Accurate,
  title = {Accurate Simulation of the Hubbard Model with Finite Fermionic Projected Entangled Pair States},
  author = {Liu, Wen-Yuan and Zhai, Huanchen and Peng, Ruojing and Gu, Zheng-Cheng and Chan, Garnet Kin-Lic},
  journal = {Phys. Rev. Lett.},
  volume = {134},
  issue = {25},
  pages = {256502},
  numpages = {8},
  year = {2025},
  month = {Jun},
  publisher = {American Physical Society},
  doi = {10.1103/r4q9-4yvj},
  url = {https://link.aps.org/doi/10.1103/r4q9-4yvj}
}

@article{Quinten2025Fermionic,
	title={{Fermionic tensor network methods}},
	author={Quinten Mortier and Lukas Devos and Lander Burgelman and Bram Vanhecke and Nick Bultinck and Frank Verstraete and Jutho Haegeman and Laurens Vanderstraeten},
	journal={SciPost Phys.},
	volume={18},
	pages={012},
	year={2025},
	publisher={SciPost},
	doi={10.21468/SciPostPhys.18.1.012},
	url={https://scipost.org/10.21468/SciPostPhys.18.1.012},
}

@article{Xie2014Tensor,
  title = {Tensor Renormalization of Quantum Many-Body Systems Using Projected Entangled Simplex States},
  author = {Xie, Z. Y. and Chen, J. and Yu, J. F. and Kong, X. and Normand, B. and Xiang, T.},
  journal = {Phys. Rev. X},
  volume = {4},
  issue = {1},
  pages = {011025},
  numpages = {12},
  year = {2014},
  month = {Feb},
  publisher = {American Physical Society},
  doi = {10.1103/PhysRevX.4.011025},
  url = {https://link.aps.org/doi/10.1103/PhysRevX.4.011025}
}

@article{Xu2024Global,
      title={Global phase diagram of doped quantum spin liquid on the Kagome lattice}, 
      author={Zheng-Tao Xu and Zheng-Cheng Gu and Shuo Yang},
      year={2024},
      eprint={2404.05685},
      archivePrefix={arXiv},
      primaryClass={cond-mat.str-el},
      url={https://arxiv.org/abs/2404.05685}, 
      journal = {Arxiv},
      volume = {abs/2404.05685}
}

@inbook{Shimizu2014Application,
title = {Application of Matrix Product States to the Hubbard Model in One Spatial Dimension},
author={Shimizu, Yukihiro and Matsuura, Koji and Yahagi, Hikaru},
booktitle = {Proceedings of the 12th Asia Pacific Physics Conference (APPC12)},
chapter = {},
pages={016011},
doi = {10.7566/JPSCP.1.016011},
year={2014},
URL = {https://journals.jps.jp/doi/abs/10.7566/JPSCP.1.016011},
eprint = {https://journals.jps.jp/doi/pdf/10.7566/JPSCP.1.016011}
}

@article{Ostlund1995,
  title = {Thermodynamic Limit of Density Matrix Renormalization},
  author = {\"Ostlund, Stellan and Rommer, Stefan},
  journal = {Phys. Rev. Lett.},
  volume = {75},
  issue = {19},
  pages = {3537--3540},
  numpages = {0},
  year = {1995},
  month = {Nov},
  publisher = {American Physical Society},
  doi = {10.1103/PhysRevLett.75.3537},
  url = {https://link.aps.org/doi/10.1103/PhysRevLett.75.3537}
}

@article{Dukelsky1998Equivalence,
doi = {10.1209/epl/i1998-00381-x},
url = {https://doi.org/10.1209/epl/i1998-00381-x},
year = {1998},
month = {aug},
publisher = {},
volume = {43},
number = {4},
pages = {457},
author = {J. Dukelsky and M. A. Martín-Delgado and T. Nishino and G. Sierra},
title = {Equivalence of the variational matrix product method and the 
density matrix renormalization group applied to spin chains},
journal = {Europhysics Letters}
}

@article{Lanczos1950Iteration,
  title={An iteration method for the solution of the eigenvalue problem of linear differential and integral operators},
  author={Lanczos, Cornelius},
  journal={Journal of research of the National Bureau of Standards},
  volume={45},
  number={4},
  pages={255--282},
  year={1950}, 
  doi = {10.6028/jres.045.026}, 
  url = {https://doi.org/10.6028/jres.045.026}
}

@article{Haegeman2011TDVP,
  title = {Time-Dependent Variational Principle for Quantum Lattices},
  author = {Haegeman, Jutho and Cirac, J. Ignacio and Osborne, Tobias J. and Pi\ifmmode \check{z}\else \v{z}\fi{}orn, Iztok and Verschelde, Henri and Verstraete, Frank},
  journal = {Phys. Rev. Lett.},
  volume = {107},
  issue = {7},
  pages = {070601},
  numpages = {5},
  year = {2011},
  month = {Aug},
  publisher = {American Physical Society},
  doi = {10.1103/PhysRevLett.107.070601},
  url = {https://link.aps.org/doi/10.1103/PhysRevLett.107.070601}
}

@article{Laurens2019Tangent,
	title={Tangent-space methods for uniform matrix product states},
	author={Laurens Vanderstraeten and Jutho Haegeman and Frank Verstraete},
	journal={SciPost Phys. Lect. Notes},
	pages={7},
	year={2019},
	publisher={SciPost},
	doi={10.21468/SciPostPhysLectNotes.7},
	url={https://scipost.org/10.21468/SciPostPhysLectNotes.7},
}

@article{VidalTEBD2003,
  title = {Efficient Classical Simulation of Slightly Entangled Quantum Computations},
  author = {Vidal, Guifr\'e},
  journal = {Phys. Rev. Lett.},
  volume = {91},
  issue = {14},
  pages = {147902},
  numpages = {4},
  year = {2003},
  month = {Oct},
  publisher = {American Physical Society},
  doi = {10.1103/PhysRevLett.91.147902},
  url = {https://link.aps.org/doi/10.1103/PhysRevLett.91.147902}
}

@Article{Atis2023GrassmannTN,
	title={{GrassmannTN: A Python package for Grassmann tensor network computations}},
	author={Atis Yosprakob},
	journal={SciPost Phys. Codebases},
	pages={20},
	year={2023},
	publisher={SciPost},
	doi={10.21468/SciPostPhysCodeb.20},
	url={https://scipost.org/10.21468/SciPostPhysCodeb.20},
}

@article{VidalTEBD2007,
  title = {Classical Simulation of Infinite-Size Quantum Lattice Systems in One Spatial Dimension},
  author = {Vidal, G.},
  journal = {Phys. Rev. Lett.},
  volume = {98},
  issue = {7},
  pages = {070201},
  numpages = {4},
  year = {2007},
  month = {Feb},
  publisher = {American Physical Society},
  doi = {10.1103/PhysRevLett.98.070201},
  url = {https://link.aps.org/doi/10.1103/PhysRevLett.98.070201}
}

@article{RomanTEBD2008,
  title={Infinite time-evolving block decimation algorithm beyond unitary evolution},
  author={Orus, Roman and Vidal, Guifre},
  journal={Physical Review B—Condensed Matter and Materials Physics},
  volume={78},
  number={15},
  pages={155117},
  year={2008},
  publisher={APS}
}

@article{Nishino1996Corner,
author = {Nishino ,Tomotoshi and Okunishi ,Kouichi},
title = {Corner Transfer Matrix Renormalization Group Method},
journal = {Journal of the Physical Society of Japan},
volume = {65},
number = {4},
pages = {891-894},
year = {1996},
doi = {10.1143/JPSJ.65.891},
url = {https://doi.org/10.1143/JPSJ.65.891},
eprint = {https://doi.org/10.1143/JPSJ.65.891}
}

@book{baxter2016exactly,
  title={Exactly solved models in statistical mechanics},
  author={Baxter, Rodney J},
  year={2016},
  publisher={Elsevier}
}

@article{Orus2009Simulation,
  title = {Simulation of two-dimensional quantum systems on an infinite lattice revisited: Corner transfer matrix for tensor contraction},
  author = {Or\'us, Rom\'an and Vidal, Guifr\'e},
  journal = {Phys. Rev. B},
  volume = {80},
  issue = {9},
  pages = {094403},
  numpages = {4},
  year = {2009},
  month = {Sep},
  publisher = {American Physical Society},
  doi = {10.1103/PhysRevB.80.094403},
  url = {https://link.aps.org/doi/10.1103/PhysRevB.80.094403}
}

@article{Fishman2018Faster,
  title = {Faster methods for contracting infinite two-dimensional tensor networks},
  author = {Fishman, M. T. and Vanderstraeten, L. and Zauner-Stauber, V. and Haegeman, J. and Verstraete, F.},
  journal = {Phys. Rev. B},
  volume = {98},
  issue = {23},
  pages = {235148},
  numpages = {17},
  year = {2018},
  month = {Dec},
  publisher = {American Physical Society},
  doi = {10.1103/PhysRevB.98.235148},
  url = {https://link.aps.org/doi/10.1103/PhysRevB.98.235148}
}

@article{Jahromi2018Infinite,
  title = {Infinite projected entangled-pair state algorithm for ruby and triangle-honeycomb lattices},
  author = {Jahromi, Saeed S. and Or\'us, Rom\'an and Kargarian, Mehdi and Langari, Abdollah},
  journal = {Phys. Rev. B},
  volume = {97},
  issue = {11},
  pages = {115161},
  numpages = {12},
  year = {2018},
  month = {Mar},
  publisher = {American Physical Society},
  doi = {10.1103/PhysRevB.97.115161},
  url = {https://link.aps.org/doi/10.1103/PhysRevB.97.115161}
}

@article{Liu2022Variational,
doi = {10.1088/0256-307X/39/6/067502},
url = {https://doi.org/10.1088/0256-307X/39/6/067502},
year = {2022},
month = {jun},
publisher = {Chinese Physical Society and IOP Publishing Ltd},
volume = {39},
number = {6},
pages = {067502},
author = {Liu, X. F. and Fu, Y. F. and Yu, W. Q. and Yu, J. F. and Xie, Z. Y.},
title = {Variational Corner Transfer Matrix Renormalization Group Method for Classical Statistical Models},
journal = {Chinese Physics Letters}
}

@article{Okunishi2022Developments,
author = {Okunishi ,Kouichi and Nishino ,Tomotoshi and Ueda ,Hiroshi},
title = {Developments in the Tensor Network — from Statistical Mechanics to Quantum Entanglement},
journal = {Journal of the Physical Society of Japan},
volume = {91},
number = {6},
pages = {062001},
year = {2022},
doi = {10.7566/JPSJ.91.062001},
URL = { https://doi.org/10.7566/JPSJ.91.062001},
eprint = {https://doi.org/10.7566/JPSJ.91.062001}
}

@article{Jiang2008Accurate,
  title = {Accurate Determination of Tensor Network State of Quantum Lattice Models in Two Dimensions},
  author = {Jiang, H. C. and Weng, Z. Y. and Xiang, T.},
  journal = {Phys. Rev. Lett.},
  volume = {101},
  issue = {9},
  pages = {090603},
  numpages = {4},
  year = {2008},
  month = {Aug},
  publisher = {American Physical Society},
  doi = {10.1103/PhysRevLett.101.090603},
  url = {https://link.aps.org/doi/10.1103/PhysRevLett.101.090603}
}

@article{Wang2011Cluster,
      title={Cluster update for tensor network states}, 
      author={Ling Wang and Frank Verstraete},
      year={2011},
      eprint={1110.4362},
      archivePrefix={arXiv},
      primaryClass={cond-mat.str-el},
      url={https://arxiv.org/abs/1110.4362}, 
      journal = {Arxiv},
      volume = {abs/1110.4362}
}

@article{Dziarmaga2022Time,
  title = {Time evolution of an infinite projected entangled pair state: A gradient tensor update in the tangent space},
  author = {Dziarmaga, Jacek},
  journal = {Phys. Rev. B},
  volume = {106},
  issue = {1},
  pages = {014304},
  numpages = {8},
  year = {2022},
  month = {Jul},
  publisher = {American Physical Society},
  doi = {10.1103/PhysRevB.106.014304},
  url = {https://link.aps.org/doi/10.1103/PhysRevB.106.014304}
}

@article{BeriPRL2011,
  title = {Local Tensor Network for Strongly Correlated Projective States},
  author = {B\'eri, B. and Cooper, N. R.},
  journal = {Phys. Rev. Lett.},
  volume = {106},
  issue = {15},
  pages = {156401},
  numpages = {4},
  year = {2011},
  month = {Apr},
  publisher = {American Physical Society},
  doi = {10.1103/PhysRevLett.106.156401},
  url = {https://link.aps.org/doi/10.1103/PhysRevLett.106.156401}
}

@article{Phien2015Infinite,
  title = {Infinite projected entangled pair states algorithm improved: Fast full update and gauge fixing},
  author = {Phien, Ho N. and Bengua, Johann A. and Tuan, Hoang D. and Corboz, Philippe and Or\'us, Rom\'an},
  journal = {Phys. Rev. B},
  volume = {92},
  issue = {3},
  pages = {035142},
  numpages = {12},
  year = {2015},
  month = {Jul},
  publisher = {American Physical Society},
  doi = {10.1103/PhysRevB.92.035142},
  url = {https://link.aps.org/doi/10.1103/PhysRevB.92.035142}
}

@article{RomanAOP2014,
title = {A practical introduction to tensor networks: Matrix product states and projected entangled pair states},
journal = {Annals of Physics},
volume = {349},
pages = {117-158},
year = {2014},
issn = {0003-4916},
doi = {https://doi.org/10.1016/j.aop.2014.06.013},
url = {https://www.sciencedirect.com/science/article/pii/S0003491614001596},
author = {Román Orús},
}

@article{RomanNRP2019,
title = {Tensor networks for complex quantum systems},
journal = {Nature Reviews Physics},
volume = {1},
pages = {538-550},
year = {2019},
doi = {https://doi.org/10.1038/s42254-019-0086-7},
url = {https://www.nature.com/articles/s42254-019-0086-7#citeas},
author = {Román Orús},
}

@article{RomanEPJB2014,
  title={Advances on tensor network theory: symmetries, fermions, entanglement, and holography.},
  author={Or{\'u}s, Rom{\'a}n},
  journal={Eur. Phys. J. B},
  volume={87},
  number={0},
  pages={280},
  year={2014}
 }

@article{CiracRMP2021,
  title = {Matrix product states and projected entangled pair states: Concepts, symmetries, theorems},
  author = {Cirac, J. Ignacio and P\'erez-Garc\'{\i}a, David and Schuch, Norbert and Verstraete, Frank},
  journal = {Rev. Mod. Phys.},
  volume = {93},
  issue = {4},
  pages = {045003},
  numpages = {65},
  year = {2021},
  month = {Dec},
  publisher = {American Physical Society},
  doi = {10.1103/RevModPhys.93.045003},
  url = {https://link.aps.org/doi/10.1103/RevModPhys.93.045003}
}

@article{Jordan2008Classical,
  title = {Classical Simulation of Infinite-Size Quantum Lattice Systems in Two Spatial Dimensions},
  author = {Jordan, J. and Or\'us, R. and Vidal, G. and Verstraete, F. and Cirac, J. I.},
  journal = {Phys. Rev. Lett.},
  volume = {101},
  issue = {25},
  pages = {250602},
  numpages = {4},
  year = {2008},
  month = {Dec},
  publisher = {American Physical Society},
  doi = {10.1103/PhysRevLett.101.250602},
  url = {https://link.aps.org/doi/10.1103/PhysRevLett.101.250602}
}

@article{Arovas2022Hubbard,
   author = "Arovas, Daniel P. and Berg, Erez and Kivelson, Steven A. and Raghu, Srinivas",
   title = "The Hubbard Model", 
   journal= "Annual Review of Condensed Matter Physics",
   year = "2022",
   volume = "13",
   number = "Volume 13, 2022",
   pages = "239-274",
   doi = "https://doi.org/10.1146/annurev-conmatphys-031620-102024",
   url = "https://www.annualreviews.org/content/journals/10.1146/annurev-conmatphys-031620-102024",
   publisher = "Annual Reviews",
   issn = "1947-5462",
   type = "Journal Article",
  }

@article{Qin2022Hubbard,
   author = "Qin, Mingpu and Schäfer, Thomas and Andergassen, Sabine and Corboz, Philippe and Gull, Emanuel",
   title = "The Hubbard Model: A Computational Perspective", 
   journal= "Annual Review of Condensed Matter Physics",
   year = "2022",
   volume = "13",
   number = "Volume 13, 2022",
   pages = "275-302",
   doi = "https://doi.org/10.1146/annurev-conmatphys-090921-033948",
   url = "https://www.annualreviews.org/content/journals/10.1146/annurev-conmatphys-090921-033948",
   publisher = "Annual Reviews",
   issn = "1947-5462",
   type = "Journal Article",
  }

@article{Lieb1968Absence,
  title = {Absence of Mott Transition in an Exact Solution of the Short-Range, One-Band Model in One Dimension},
  author = {Lieb, Elliott H. and Wu, F. Y.},
  journal = {Phys. Rev. Lett.},
  volume = {20},
  issue = {25},
  pages = {1445--1448},
  numpages = {0},
  year = {1968},
  month = {Jun},
  publisher = {American Physical Society},
  doi = {10.1103/PhysRevLett.20.1445},
  url = {https://link.aps.org/doi/10.1103/PhysRevLett.20.1445}
}

@book{Essler20051DHubbard,
  title={The one-dimensional Hubbard model},
  author={Essler, Fabian HL and Frahm, Holger and G{\"o}hmann, Frank and Kl{\"u}mper, Andreas and Korepin, Vladimir E},
  year={2005},
  publisher={Cambridge University Press}
}

@article{Gross1974Dynamical,
  title = {Dynamical symmetry breaking in asymptotically free field theories},
  author = {Gross, David J. and Neveu, Andr\'e},
  journal = {Phys. Rev. D},
  volume = {10},
  issue = {10},
  pages = {3235--3253},
  numpages = {0},
  year = {1974},
  month = {Nov},
  publisher = {American Physical Society},
  doi = {10.1103/PhysRevD.10.3235},
  url = {https://link.aps.org/doi/10.1103/PhysRevD.10.3235}
}

@article{Herbut2006Interactions,
  title = {Interactions and Phase Transitions on Graphene's Honeycomb Lattice},
  author = {Herbut, Igor F.},
  journal = {Phys. Rev. Lett.},
  volume = {97},
  issue = {14},
  pages = {146401},
  numpages = {4},
  year = {2006},
  month = {Oct},
  publisher = {American Physical Society},
  doi = {10.1103/PhysRevLett.97.146401},
  url = {https://link.aps.org/doi/10.1103/PhysRevLett.97.146401}
}

@article{Herbut2009Theory,
  title = {Theory of interacting electrons on the honeycomb lattice},
  author = {Herbut, Igor F. and Juri\ifmmode \check{c}\else \v{c}\fi{}i\ifmmode \acute{c}\else \'{c}\fi{}, Vladimir and Roy, Bitan},
  journal = {Phys. Rev. B},
  volume = {79},
  issue = {8},
  pages = {085116},
  numpages = {14},
  year = {2009},
  month = {Feb},
  publisher = {American Physical Society},
  doi = {10.1103/PhysRevB.79.085116},
  url = {https://link.aps.org/doi/10.1103/PhysRevB.79.085116}
}

@incollection{Wilson1975Quarks,
  title={Quarks and strings on a lattice},
  author={Wilson, Kenneth G},
  booktitle={New Phenomena in Subnuclear Physics: Part A},
  pages={69--142},
  year={1975},
  publisher={Springer}
}

@article{Aoki1984New,
  title = {New phase structure for lattice QCD with Wilson fermions},
  author = {Aoki, Sinya},
  journal = {Phys. Rev. D},
  volume = {30},
  issue = {12},
  pages = {2653--2663},
  numpages = {0},
  year = {1984},
  month = {Dec},
  publisher = {American Physical Society},
  doi = {10.1103/PhysRevD.30.2653},
  url = {https://link.aps.org/doi/10.1103/PhysRevD.30.2653}
}

@article{Kenna2001Weakly,
  title = {Weakly coupled Gross-Neveu model with Wilson fermions},
  author = {Kenna, R. and Sexton, J. C.},
  journal = {Phys. Rev. D},
  volume = {65},
  issue = {1},
  pages = {014507},
  numpages = {9},
  year = {2001},
  month = {Dec},
  publisher = {American Physical Society},
  doi = {10.1103/PhysRevD.65.014507},
  url = {https://link.aps.org/doi/10.1103/PhysRevD.65.014507}
}

@article{Bermudez2018Gross,
title = {Gross–Neveu–Wilson model and correlated symmetry-protected topological phases},
journal = {Annals of Physics},
volume = {399},
pages = {149-180},
year = {2018},
issn = {0003-4916},
doi = {https://doi.org/10.1016/j.aop.2018.10.007},
url = {https://www.sciencedirect.com/science/article/pii/S0003491618302690},
author = {A. Bermudez and E. Tirrito and M. Rizzi and M. Lewenstein and S. Hands}
}

@article{Kong2026Phase,
      title={Phase diagram of the single-flavor Gross--Neveu--Wilson model from the Grassmann corner transfer matrix renormalization group}, 
      author={Jian-Gang Kong and Shinichiro Akiyama and Tao Shi and Z. Y. Xie},
      year={2026},
      eprint={2602.21705},
      archivePrefix={arXiv},
      primaryClass={hep-lat},
      url={https://arxiv.org/abs/2602.21705},
      journal = {Arxiv},
      volume = {abs/2602.21705}
}

@article{Akiyama2023Implementation,
      title={Implementation of bond weighting method for the Grassmann tensor renormalization group}, 
      author={Shinichiro Akiyama},
      year={2023},
      eprint={2311.17691},
      archivePrefix={arXiv},
      primaryClass={hep-lat},
      url={https://arxiv.org/abs/2311.17691},
      journal = {Arxiv},
      volume = {abs/2311.17691}
}

@article{Kong2026Hubbard,
      title={Grassmann corner transfer-matrix renormalization group approach to one-dimensional fermionic models}, 
      author={Kong, Jian-Gang and Xie, Zhi Yuan},
      year={2026},
      eprint={2604.05582},
      archivePrefix={arXiv},
      primaryClass={hep-lat},
      url={https://arxiv.org/abs/2604.05582},
      journal = {Arxiv},
      volume = {abs/2604.05582}
}

@article{capponi2016phase,
  title={Phase diagram of interacting spinless fermions on the honeycomb lattice},
  author={Capponi, Sylvain},
  journal={Journal of Physics: Condensed Matter},
  volume={29},
  number={4},
  pages={043002},
  year={2016},
  publisher={IOP Publishing}
}

@article{jiang2024pair,
  title={Pair-density-wave superconductivity: A microscopic model on the 2D honeycomb lattice},
  author={Jiang, Yi-Fan and Yao, Hong},
  journal={Physical Review Letters},
  volume={133},
  number={17},
  pages={176501},
  year={2024},
  publisher={APS}
}

@article{agterberg2020physics,
  title={The physics of pair-density waves: cuprate superconductors and beyond},
  author={Agterberg, Daniel F and Davis, JC S{\'e}amus and Edkins, Stephen D and Fradkin, Eduardo and Van Harlingen, Dale J and Kivelson, Steven A and Lee, Patrick A and Radzihovsky, Leo and Tranquada, John M and Wang, Yuxuan},
  journal={Annual Review of Condensed Matter Physics},
  volume={11},
  number={1},
  pages={231--270},
  year={2020},
  publisher={Annual Reviews}
}

@article{Haegeman2012Variational,
  title = {Variational matrix product ansatz for dispersion relations},
  author = {Haegeman, Jutho and Pirvu, Bogdan and Weir, David J. and Cirac, J. Ignacio and Osborne, Tobias J. and Verschelde, Henri and Verstraete, Frank},
  journal = {Phys. Rev. B},
  volume = {85},
  issue = {10},
  pages = {100408},
  numpages = {5},
  year = {2012},
  month = {Mar},
  publisher = {American Physical Society},
  doi = {10.1103/PhysRevB.85.100408},
  url = {https://link.aps.org/doi/10.1103/PhysRevB.85.100408}
}

@article{Ponsioen2020Excitations,
  title = {Excitations with projected entangled pair states using the corner transfer matrix method},
  author = {Ponsioen, Boris and Corboz, Philippe},
  journal = {Phys. Rev. B},
  volume = {101},
  issue = {19},
  pages = {195109},
  numpages = {12},
  year = {2020},
  month = {May},
  publisher = {American Physical Society},
  doi = {10.1103/PhysRevB.101.195109},
  url = {https://link.aps.org/doi/10.1103/PhysRevB.101.195109}
}

@Article{Boris2022Automatic,
	title={{Automatic differentiation applied to excitations with projected entangled pair states}},
	author={Boris Ponsioen and Fakher F. Assaad and Philippe Corboz},
	journal={SciPost Phys.},
	volume={12},
	pages={006},
	year={2022},
	publisher={SciPost},
	doi={10.21468/SciPostPhys.12.1.006},
	url={https://scipost.org/10.21468/SciPostPhys.12.1.006},
}

@article{Li2011Linearized,
  title = {Linearized Tensor Renormalization Group Algorithm for the Calculation of Thermodynamic Properties of Quantum Lattice Models},
  author = {Li, Wei and Ran, Shi-Ju and Gong, Shou-Shu and Zhao, Yang and Xi, Bin and Ye, Fei and Su, Gang},
  journal = {Phys. Rev. Lett.},
  volume = {106},
  issue = {12},
  pages = {127202},
  numpages = {4},
  year = {2011},
  month = {Mar},
  publisher = {American Physical Society},
  doi = {10.1103/PhysRevLett.106.127202},
  url = {https://link.aps.org/doi/10.1103/PhysRevLett.106.127202}
}

@article{Li2023Tangent,
  title = {Tangent Space Approach for Thermal Tensor Network Simulations of the 2D Hubbard Model},
  author = {Li, Qiaoyi and Gao, Yuan and He, Yuan-Yao and Qi, Yang and Chen, Bin-Bin and Li, Wei},
  journal = {Phys. Rev. Lett.},
  volume = {130},
  issue = {22},
  pages = {226502},
  numpages = {8},
  year = {2023},
  month = {Jun},
  publisher = {American Physical Society},
  doi = {10.1103/PhysRevLett.130.226502},
  url = {https://link.aps.org/doi/10.1103/PhysRevLett.130.226502}
}

@article{Sinha2022Finite,
  title = {Finite-temperature tensor network study of the Hubbard model on an infinite square lattice},
  author = {Sinha, Aritra and Rams, Marek M. and Czarnik, Piotr and Dziarmaga, Jacek},
  journal = {Phys. Rev. B},
  volume = {106},
  issue = {19},
  pages = {195105},
  numpages = {16},
  year = {2022},
  month = {Nov},
  publisher = {American Physical Society},
  doi = {10.1103/PhysRevB.106.195105},
  url = {https://link.aps.org/doi/10.1103/PhysRevB.106.195105}
}

@article{Chen2024Real,
  title = {Real-time impurity solver using Grassmann time-evolving matrix product operators},
  author = {Chen, Ruofan and Xu, Xiansong and Guo, Chu},
  journal = {Phys. Rev. B},
  volume = {109},
  issue = {16},
  pages = {165113},
  numpages = {10},
  year = {2024},
  month = {Apr},
  publisher = {American Physical Society},
  doi = {10.1103/PhysRevB.109.165113},
  url = {https://link.aps.org/doi/10.1103/PhysRevB.109.165113}
}

@article{Xu2024Grassmann,
    author = {Xu, Xiansong and Guo, Chu and Chen, Ruofan},
    title = {Grassmann time-evolving matrix product operators: An efficient numerical approach for fermionic path integral simulations},
    journal = {The Journal of Chemical Physics},
    volume = {161},
    number = {15},
    pages = {151001},
    year = {2024},
    month = {10},
    issn = {0021-9606},
    doi = {10.1063/5.0226167},
    url = {https://doi.org/10.1063/5.0226167},
    eprint = {https://pubs.aip.org/aip/jcp/article-pdf/doi/10.1063/5.0226167/20208636/151001_1_5.0226167.pdf},
}

@article{Chen2025Path,
title = {Path integral formalism for quantum open systems},
journal = {Annals of Physics},
volume = {480},
pages = {170083},
year = {2025},
issn = {0003-4916},
doi = {https://doi.org/10.1016/j.aop.2025.170083},
url = {https://www.sciencedirect.com/science/article/pii/S0003491625001654},
author = {Ruofan Chen}}

@article{Andreas2024QSpace,
	title={QSpace - An open-source tensor library for Abelian and non-Abelian symmetries},
	author={Andreas Weichselbaum},
	journal={SciPost Phys. Codebases},
	pages={40},
	year={2024},
	publisher={SciPost},
	doi={10.21468/SciPostPhysCodeb.40},
	url={https://scipost.org/10.21468/SciPostPhysCodeb.40},
}

@article{SCHMOLL2020168232,
title = {A programming guide for tensor networks with global SU(2) symmetry},
journal = {Annals of Physics},
volume = {419},
pages = {168232},
year = {2020},
issn = {0003-4916},
doi = {https://doi.org/10.1016/j.aop.2020.168232},
url = {https://www.sciencedirect.com/science/article/pii/S0003491620301664},
author = {Philipp Schmoll and Sukhbinder Singh and Matteo Rizzi and Román Orús}}

\end{document}